\documentstyle[12pt,epsf,epsfig]{article}
\newcommand{\cminus}{-}
\newcommand{\cplus}{+}
\newcommand{\lepton}{\ell}
\newcommand{\lminus}{\lepton^-}
\newcommand{\lplus}{\lepton^+}

\newcommand{\nue}{\nu_e}

\newcommand{\numu}{\nu_{\mu}}
\newcommand{\muon}{\mu^{\cminus}}
\newcommand{\nutau}{\nu_{\tau}}

\newcommand{\antinu}{\overline{\nu}}
\newcommand{\antinue}{\antinu_e}

\newcommand{\antimuon}{\mu^{\cplus}}

\newcommand{\nuebar}{\antinu_e}

\newcommand{\numubar}{\antinu_{\mu}}

\newcommand{\nubar}{\antinu}

\newcommand{\muminus}{\muon}
\newcommand{\muplus}{\antimuon}
\newcommand{\dzero}{D^0}
\newcommand{\dzerobar}{\overline{D}^0}

\newcommand{\qu}[1]{{\rm #1}}
\newcommand{\aqu}[1]{\overline{{\rm #1}}}

\newcommand{\quark}{\qu{q}}
\newcommand{\antiquark}{\overline{\quark}}
\newcommand{\uquark}{\qu{u}} 
\newcommand{\dquark}{\qu{d}} 
\newcommand{\squark}{\qu{s}} 
\newcommand{\cquark}{\qu{c}} 
\newcommand{\bquark}{\qu{b}} 
 
\newcommand{\antiuquark}{\aqu{u}} 
\newcommand{\antidquark}{\aqu{d}} 
\newcommand{\antisquark}{\aqu{s}} 
\newcommand{\anticquark}{\aqu{c}}

\newcommand{\onehalf}{{1\over 2}}

\newcommand{\thetaW}{\theta_W}
\newcommand{\siniiW}{\sin^2\thetaW}

\newcommand{\fourv}[1]{{\bf #1}}

\newcommand{\fv}[1]{\fourv{p}_{#1}}
\newcommand{\fvk}[1]{\fourv{k}_{#1}}

\newcommand{\mtarget}{M}

\newcommand{\qsq}{Q^2}

\newcommand{\etal}{{\it et al.}}

\newcommand{\GeV}{\hbox{\rm GeV}}
\newcommand{\centi}{\hbox{\rm cm}}
\newcommand{\gt}{\rightarrow}

\def\url#1{\mbox{\href{#1}{\sf #1}}}

\def\pr#1#2#3{\frenchspacing{\it Phys. Rev. D}{\bf #1} (19#3) #2}

\def\prl#1#2#3{\frenchspacing{\it Phys. Rev. Lett.\ }{\bf #1} (19#3) #2}
\def\pl#1#2#3{\frenchspacing{\it Phys. Lett.\ }{\bf #1} (19#3) #2}
\def\prc#1#2#3{\frenchspacing{\it Phys. Rev. C}{\bf #1} (19#3) #2}
\def\np#1#2#3{\frenchspacing{\it Nucl. Phys. }{\bf #1} (19#3) #2}
\def\ib#1#2#3{\frenchspacing{\it ibid. }{\bf #1} (19#3) #2}

\def\rmp#1#2#3{\frenchspacing{\it Rev. Mod. Phys. }{\bf #1}, #2 (19#3)}

\def\aprge{\buildrel > \over {_{\sim}}}
\newcommand{\gtwid}{\mathrel{\raise.3ex\hbox{$>$\kern-.75em\lower1ex
\hbox{$\sim$}}}}
\newcommand{\ltwid}{\mathrel{\raise.3ex\hbox{$<$\kern-.75em\lower1ex
\hbox{$\sim$}}}}
\def\beq{\begin{equation}}
\def\eeq{\end{equation}}

\begin{document}

\begin{flushright}
{\bf FERMILAB-FN-692}\\
\today
\end{flushright}
\vspace{1.0cm}

\begin{center}

\huge
{\bf Physics at a Neutrino Factory}
\vspace{1.0cm}
\large

\vspace{1.0cm}

\font\eightit=cmti8
\def\r#1{\ignorespaces $^{#1}$}
\hfilneg
\begin{sloppypar}
\noindent
C.~Albright,\r{8,22} 
G.~Anderson,\r{14} 
V.~Barger,\r {21} 
R.~Bernstein,\r{8} 
G.~Blazey,\r{8} 
A.~Bodek,\r {20} 
E.~Buckley--Geer,\r{8} 
A.~Bueno,\r{7} 
M.~Campanelli,\r {7} 
D.~Carey,\r{8} 
D.~Casper,\r{18} 
A.~Cervera,\r{6} 
C.~Crisan,\r{8} 
F.~DeJongh,\r{8} 
S.~Eichblatt,\r{14} 
A.~Erner,\r{14} 
R.~Fernow,\r {3} 
D.~Finley,\r{8} 
J.~Formaggio,\r {5} 
J.~Gallardo,\r {3} 
S.~Geer,\r{8} 
M.~Goodman,\r 1 
D.~Harris,\r{8} 
E.~Hawker,\r{11} 
J.~Hill,\r{16} 
R.~Johnson,\r{9} 
D.~Kaplan,\r{9} 
S.~Kahn,\r {3} 
B.~Kayser,\r {13} 
E.~Kearns,\r 2 
B.J.~King,\r {3} 
H.~Kirk,\r {3} 
J.~Krane,\r{10} 
D.~Krop,\r{14} 
Z.~Ligeti,\r{8} 
J.~Lykken,\r{8} 
K.~McDonald,\r{15} 
K.~McFarland,\r {20} 
I.~Mocioiu,\r{16} 
J.~Morfin,\r{8} 
H.~Murayama,\r{17} 
J.~Nelson,\r{19} 
D.~Neuffer,\r{8} 
P.~Nienaber,\r{12} 
R.~Palmer,\r {3} 
S.~Parke,\r{8} 
Z.~Parsa,\r {3} 
R.~Plunkett,\r{8} 
E.~Prebys,\r{15} 
C.~Quigg,\r{8} 
R.~Raja,\r{8} 
S.~Rigolin,\r {4} 
A.~Rubbia,\r {7} 
H.~Schellman,\r{14,8} 
M.~Shaevitz,\r{8} 
P.~Shanahan,\r{8} 
R.~Shrock,\r{3,16} 
P.~Spentzouris,\r{8} 
R.~Stefanski,\r{8} 
J.~Stone,\r 2 
L.~Sulak,\r{2} 
G.~Unel,\r{14} 
M.~Velasco,\r{14} 
K.~Whisnant\r{10}, 
J.~Yu\r{8}, 
E.D.~Zimmerman\r{5}
\end{sloppypar}

\vskip .026in
\begin{center}
\r 1  {\eightit Argonne National Laboratory, Argonne, IL 60439} \\
\r 2  {\eightit Boston University, Boston, MA 02215}\\
\r {3} {\eightit Brookhaven National Laboratory, Upton, NY 11973} \\
\r {4} {\eightit University of Michigan, Ann Arbor, MI 48105} \\
\r {5} {\eightit Columbia University, New York, NY 10027} \\
\r {6} {\eightit Dept. de Fisica Atomica y Nuclear and IFIC, 
Universidad de Valencia, Spain} \\
\r {7}  {\eightit Institut f\"ur Teilchenphysik, ETHZ, CH-8093, Z\"urich, 
Switserland} \\
\r{8} {\eightit Fermi National Accelerator Laboratory, Batavia, IL 
60510} \\
\r{9} {\eightit Illinois Institute of Technology, Chicago, IL 60616} \\
\r{10} {\eightit Iowa State University, Ames, IA  50011} \\
\r{11} {\eightit Los Alamos National Laboratory, Los Alamos, NM 87545} \\
\r{12} {\eightit Marquette University, Milwaukee, WI 53233}\\
\r{13} {\eightit National Science Foundation, Arlington, VA 22230}\\
\r{14} {\eightit Northwestern University, Evanston, IL 60208} \\
\r{15} {\eightit Princeton University, Princeton, NJ 08544}\\
\r{16} {\eightit State University of New York Stony Brook, Stony Brook, NY 
11794}\\
\r{17} {\eightit Univ. of California Berkeley, Berkeley, CA  94720}\\
\r{18} {\eightit University of California Irvine, Irvine, CA 92697} \\
\r{19} {\eightit University of Minnesota, Minneapolis, MN 55455} \\
\r {20} {\eightit University of Rochester, Rochester, NY 14627} \\
\r {21} {\eightit University of Wisconsin, Madison, WI 53706} \\
\r {22} {\eightit Northern Illinois University, DeKalb, IL 60115} \\
\end{center}

\end{center}

\normalsize
\newpage

\begin{center}
\huge
{\bf Preface}
\normalsize
\end{center}

\bigskip
\bigskip
\bigskip

In response to the growing interest in 
building a $Neutrino$ $Factory$ to produce 
high intensity beams of electron- 
and muon-neutrinos and antineutrinos, 
in October 1999 the Fermilab Directorate 
initiated two six-month studies. 
The first study, organized by N.~Holtkamp 
and D.~Finley, was to investigate the 
technical feasibility of an intense 
neutrino source based on a muon storage 
ring. This design study has produced a report 
in which the basic conclusion is that 
a Neutrino Factory is technically feasible, 
although it requires an aggressive R\&D 
program. The second study, which is the 
subject of this report, was to explore 
the physics potential of a Neutrino Factory 
as a function of the muon beam energy and 
intensity, and for oscillation physics, 
the potential as a function of baseline.

The work presented in this report is the 
result of the enthusiastic contributions 
of many people from many institutions. 
This enthusiasm made the organizers job 
fun. We also want to thank our local 
sub--group organizers and sub--editors 
for their many effective contributions, 
ranging from  
running the study groups to editing the 
report: Bob~Bernstein, Debbie~Harris, 
Eric~Hawker, Stephen~Parke, Panagiotis~Spentzouris, 
and Chris~Quigg.

Neutrino Factories seem to have caught the 
imagination of the community. We hope that 
this report goes some way towards documenting why.

\bigskip
\bigskip
\bigskip
\bigskip
\bigskip

\large
\begin{flushright}
Steve Geer and Heidi Schellman
\end{flushright}
\normalsize

\clearpage
\renewcommand{\thefigure}{\Roman{figure}}


\begin{center}
\huge
{\bf Executive Summary}
\normalsize
\end{center}

\bigskip
\bigskip
\bigskip

In the Fall of 1999, the Fermilab Directorate chartered a study group 
to investigate the physics motivation for a \textit{neutrino factory} 
based on a muon storage ring that would operate in the era beyond the 
current set of neutrino-oscillation experiments.  We were charged to 
evaluate the prospective physics program as a function of the stored 
muon energy (up to $50\hbox{ GeV}$), the number of useful muon decays 
per year (in the range from $10^{19}$ to $10^{21}$ decays per year), 
and the distance from neutrino source to detector. 
A companion study 
evaluated the technical feasibility of a neutrino factory and 
identified an R\&D program that would lead to a detailed design.
Our conclusion is that there is a compelling physics case for a 
neutrino factory with a beam energy of about 20~GeV or greater, 
that initially provides at least O($10^{19}$) muon decays per year. 

The principal motivation for a neutrino factory is to provide the 
intense, controlled, high-energy beams that will make possible 
incisive experiments to pursue the mounting evidence for neutrino 
oscillations.  The composition and spectra of intense neutrino beams 
from a muon storage ring will be determined by the charge, momentum, 
and polarization of the stored muons, through the decays $\mu^{-} 
\rightarrow e^{-}\nu_{\mu}\bar{\nu}_{e}$ or $\mu^{+} \rightarrow 
e^{+}\bar{\nu}_{\mu}\nu_{e}$.  There is no other comparable source of 
electron neutrinos and antineutrinos.  
In addition, a neutrino factory would provide well collimated 
muon neutrino and antineutrino beams. The uncertainties 
on the beam composition and flux are expected to be significantly better than 
those for conventional neutrino beams. 
If the neutrino factory energy 
exceeds about 20~GeV the neutrino beam intensity greatly exceeds 
the corresponding intensity provided by conventional wide band beams. 
The neutrino factory therefore offers 
unprecedented opportunities for precise measurements of nucleon 
structure and of electroweak parameters.  The intense muon source 
needed for the neutrino factory would make possible exquisitely 
sensitive searches for muon-electron conversion and other rare 
processes.

Experiments carried out at a neutrino factory within the next decade 
can add crucial new information to our understanding of neutrino 
oscillations. By studying the oscillations of $\nu_{\mu}$, $\nu_{e}$, 
$\bar{\nu}_{\mu}$, and $\bar{\nu}_{e}$, it will be possible to 
measure, or put stringent limits on, all of the appearance modes 
$\nu_e \rightarrow \nu_\tau$, $\nu_e \rightarrow \nu_\mu$, and 
$\nu_\mu \rightarrow \nu_\tau$. This is a necessary step beyond 
the measurements provided by the next generation of neutrino 
experiments, and will provide a basic test of 
our understanding of neutrino oscillations. It will also 
be possible to determine precisely (or place 
stringent limits on) all of the leading oscillation parameters, 
including the mixing angle $\theta_{13}$ which appears to be 
difficult to determine precisely with conventional neutrino beams.
In addition, a neutrino factory would enable us to 
infer the pattern of neutrino masses; and, under the right 
circumstances, to observe \textsf{CP} violation in the lepton sector.  
Baselines greater than about 2000~km will enable a quantitative study 
of matter effects and a determination of the mass hierarchy.  If the 
Mini\textsc{BooNE} experiment confirms the $\nu_{\mu} \leftrightarrow 
\nu_{e}$ effect reported by the LSND experiment, experiments with 
rather short baselines (a few tens of km) could be extremely 
rewarding, and enable, for example, the search for 
$\nu_e \rightarrow \nu_\tau$ oscillations. 

\begin{figure} 
\epsfxsize=3.3in
\centerline{
\epsffile{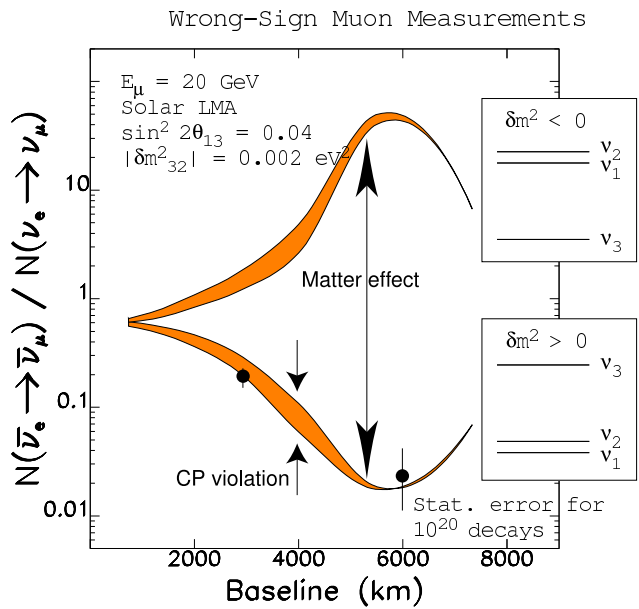}}
\caption{Predicted ratios of 
$\bar\nu_e \to \bar\nu_\mu$ to $\nu_e \to \nu_\mu$ 
rates at a 20~GeV neutrino factory. 
The upper (lower) band 
is for $\delta m^2_{32} < 0$ ($\delta m^2_{32} > 0$).
The range of possible \textsf{CP} violation determines 
the widths of the bands. 
The statistical error shown corresponds to
$10^{20}$ muon decays of each sign and a 50~kt detector. 
Results are from Ref.~\ref{bgrw00}.
}
\label{fig:summary_cp}
\end{figure}

\begin{figure} 
\epsfxsize=3.3in
\centerline{
\epsffile{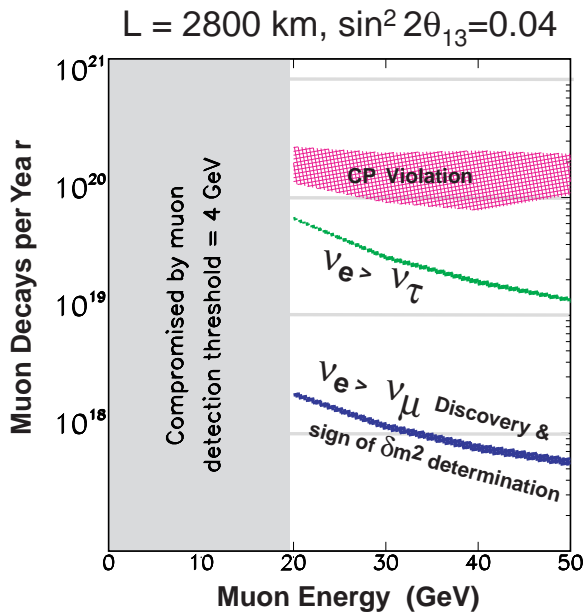}}
\caption{The required number of muon decays needed in a 
neutrino factory to observe $\nu_e \rightarrow \nu_\mu$ 
oscillations in a 50~kt detector and determine the sign of $\delta m^2$, 
and the number of decays needed to observe 
$\nu_e \rightarrow \nu_\tau$ oscillations in a few~kt detector, 
and ultimately 
put stringent limits on (or observe) \textsf{CP} violation in the 
lepton sector with a 50~kt detector. 
Results are from Ref.~\ref{bgrw00}.}
\label{fig:summary_all}
\end{figure}

If the atmospheric neutrino deficit 
is correctly described by three flavor oscillations with
$\delta m^2$ in the range favored by the SuperKamionkande 
data, and if the parameter $\sin^2 2\theta_{13}$ is not 
smaller than $\sim 0.01$, then exciting cutting--edge 
long baseline oscillation physics could begin with an 
$\sim50$~kt detector at a 
neutrino factory with muon energies as low as 20~GeV 
delivering as few as $10^{19}$ muon decays per year. 
This ``entry--level" facility would be able to measure 
$\nu_e \rightarrow \nu_\mu$ and 
$\overline{\nu}_e \rightarrow \overline{\nu}_\mu$ 
oscillations. For baselines of a few thousand km 
the ratio of rates 
$N(\overline{\nu}_e \rightarrow \overline{\nu}_\mu) / 
N(\nu_e \rightarrow \nu_\mu)$ is sensitive to the 
sign of $\delta m^2$, and hence to the pattern of 
neutrino masses (Fig.~\ref{fig:summary_cp}). With $10^{19}$ decays 
and a 50~kt detector a unique and statistically 
significant measurement of the neutrino mass spectrum 
could be made. In addition, the $\nu_e \rightarrow \nu_\mu$ 
event rate is approximately proportional to 
the parameter $\sin^2 2\theta_{13}$, which could therefore 
be measured.

\begin{figure} 
\epsfxsize=3.5in
\centerline{
\epsffile{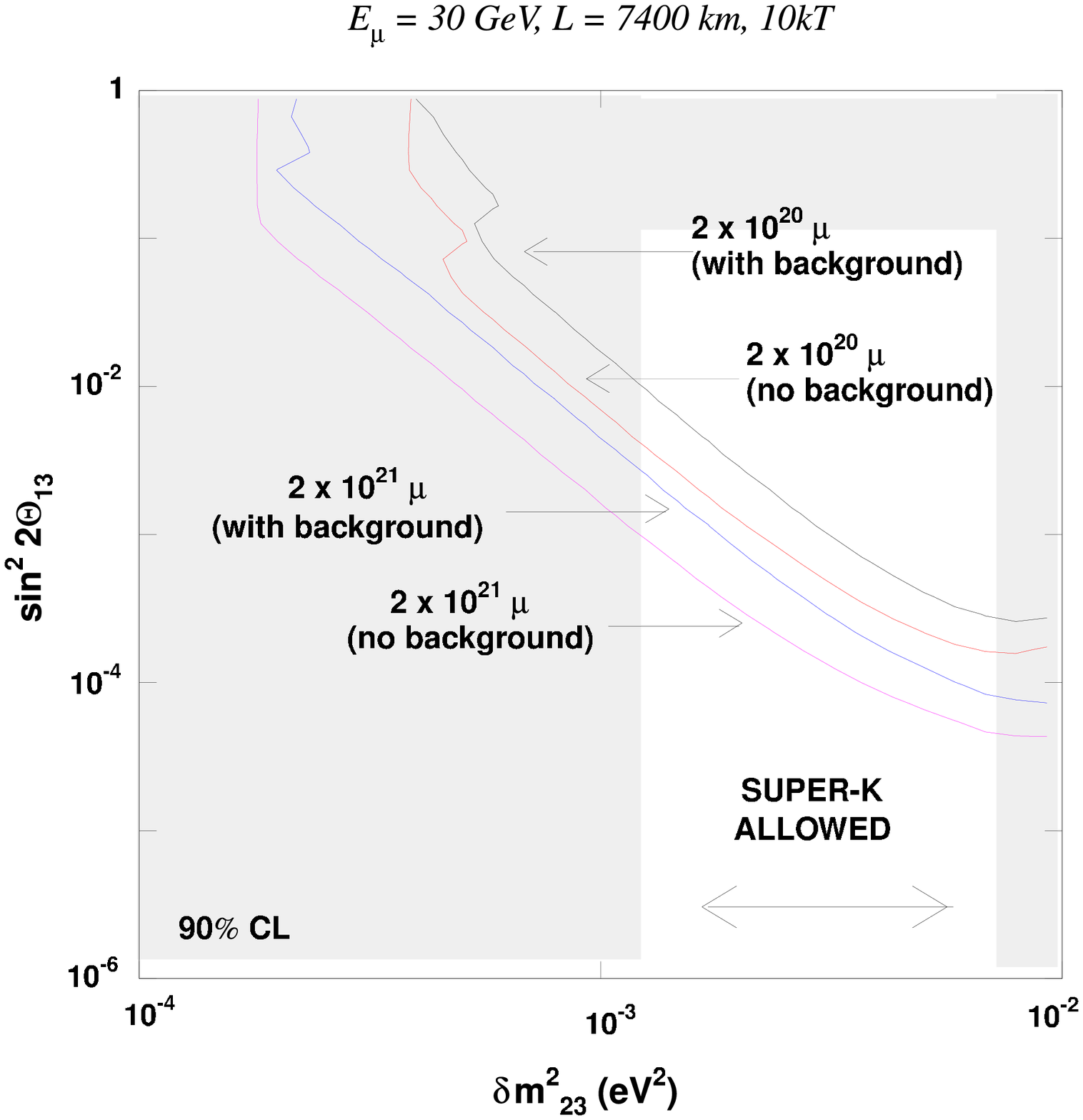}}
\caption{Limits on $\sin^2 2\theta_{13}$ that 
would result from the absence of a $\nu_e \rightarrow \nu_\mu$ 
signal in a 10~kt detector 7400~km downstream of a 
30~GeV neutrino factory in which there are $10^{20}$ and 
$10^{21} \mu^+$ decays, followed by the same number of $\mu^-$ decays.
The limits are shown as a function of $\delta m^2_{32}$. 
The impact of including backgrounds in the analysis is shown. 
Note that 
the unshaded band shows the $\delta m^2$ region favored by the 
SuperK atmospheric neutrino deficit results. 
Results are from Ref.~\ref{camp00}.
}
\label{fig:summary_t13}
\end{figure}

\begin{figure} 
\epsfxsize=5.0in
\centerline{
\epsffile{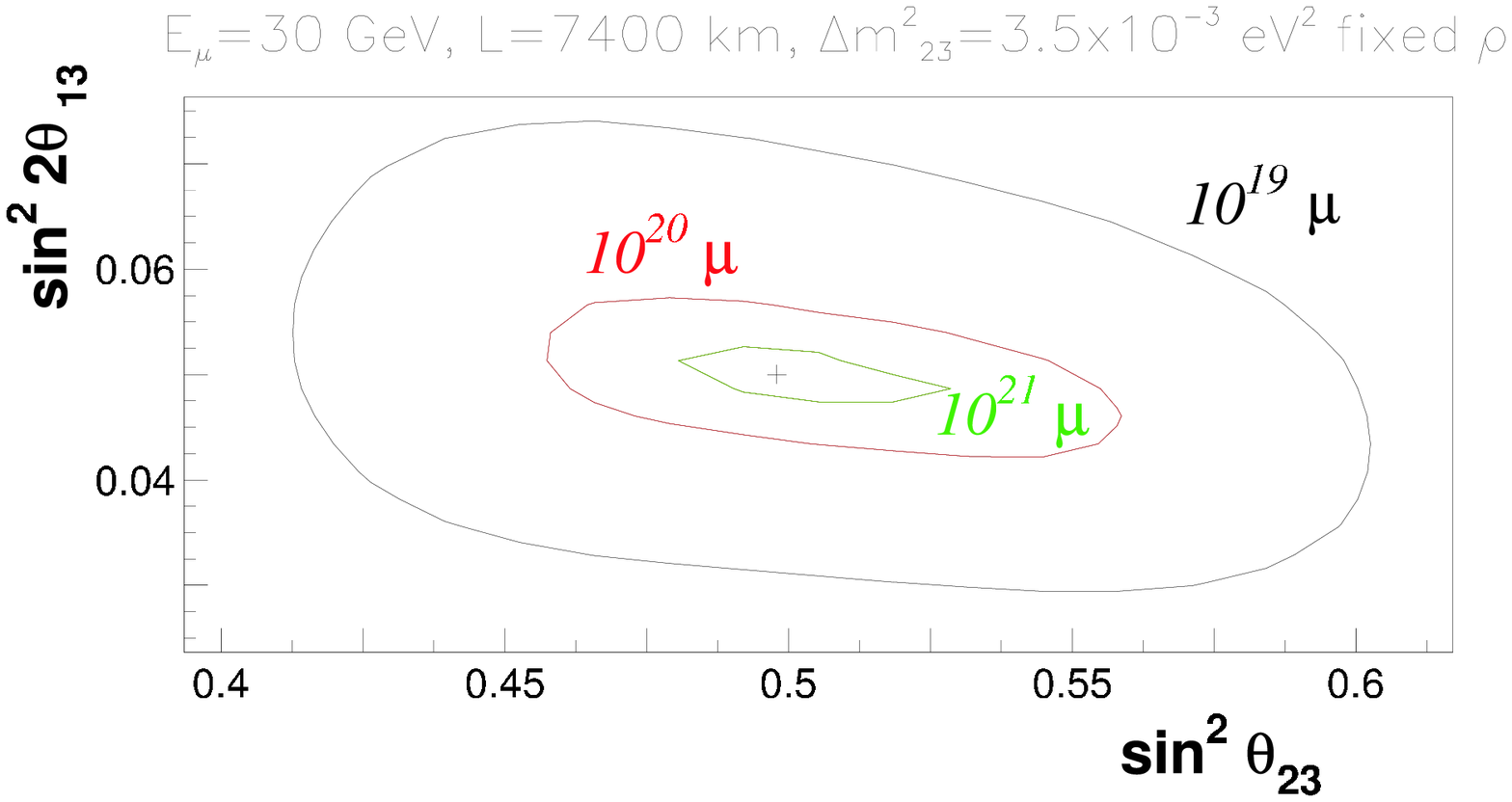}}
\caption{Precision with which the oscillation parameters 
$\sin^2 \theta_{23}$ and $\sin^2 2\theta_{13}$ can be 
measured in a 10~kt detector 7400~km downstream of a 
30~GeV neutrino factory in which there are $10^{19}$, $10^{20}$, and 
$10^{21} \mu^+$ decays.
Results are from Ref.~\ref{camp00}.
}
\label{fig:summary_precision}
\end{figure}

With higher beam intensities and/or higher beam energies the
physics potential of a neutrino factory is enhanced 
(Fig.~\ref{fig:summary_all}). 
In particular, as the intensity is increased to O($10^{20}$) 
decays/year $\nu_e \rightarrow \nu_\tau$ oscillations might 
be measured, and eventually \textsf{CP} violation in the lepton 
sector observed if the large mixing angle MSW solution is the
correct description of the solar neutrino deficit. 
Higher beam intensities would also allow smaller values of 
$\sin^2 2\theta_{13}$ to be probed (Fig. ~\ref{fig:summary_t13}), 
and higher precision 
measurements of the oscillation parameters to be made.
An example of the improvement of measurement precision with 
neutrino factory intensity is shown in
Fig.~\ref{fig:summary_precision} for the 
determinations of $\sin^2 \theta_{23}$ and $\sin^2 2\theta_{13}$.

The physics program at detectors located close to the neutrino factory is
also compelling.  The neutrino fluxes are four orders of magnitude
higher than those from existing beams. Such intense beams make experiments
with high precision detectors and low mass targets feasible for the first
time.Using these detectors and the unique ability of neutrinos to 
probe 
 particular flavors of quarks will allow a precise measurement of the
individual light quark contents of the nucleon in both an isolated and
nuclear environment. 
In addition,
neutrinos provide an elegant tool for probing the spin structure of
the nucleon and may finally enable resolution of the nucleon spin among
its partonic components.
The high event rates at a neutrino factory would 
also enable a new generation of tagged heavy quark production experiments, 
precision measurements of electro-weak 
and strong interaction parameters, and 
searches for exotic phenomena other than oscillations.  

\renewcommand{\thefigure}{\arabic{figure}}
\setcounter{figure}{0}

\subsection*{Recommendations}
The physics program we have explored for a neutrino factory is compelling.
  We recommend a sustained effort to study both the physics 
opportunities and the machine realities.
\begin{description}
\item{(i)} We encourage support for the R\&D needed 
to learn whether a neutrino factory can be a real option in the next
decade. 
\item{(ii)} We propose further studies of 
 detector technologies optimized for a neutrino factory, including
both novel low mass detectors
for near experiments and very high mass detectors for long baselines.
For long baseline experiments  detectors should have
 masses of a few times 10~kt or more that are able to detect and 
measure wrong--sign muons, and detectors of a few kt or more able to 
observe tau--lepton appearance with high efficiency.  It is also 
desirable to identify electrons, and if possible measure the 
sign of their charge. 
Both the detector technologies
themselves and the civil engineering issues 
associated with the construction of such massive
detectors need to be addressed.

\item{(iii)} We recommend continued studies to better compare 
the physics potential of upgraded conventional neutrino beams with 
the corresponding potential at a neutrino factory, and also 
studies to better understand the benefits of muon polarization.

\item{(iv)} The present study concentrated on the muon storage ring as a
 neutrino source and did not cover the additional physics  programs
 which would use the proton driver and the high intensity muon beams. 
We recommend a further study directed at these other facets of physics at a muon 
storage ring facility.
\end{description}

\clearpage

\tableofcontents
\clearpage

\section{Introduction}
    
New accelerator technologies offer the possibility of building, not
too many years in the future, an accelerator complex to accumulate
more than $10^{19}$, and perhaps more than $10^{20}$, 
muons per year~\cite{status_report}.  
It has been proposed~\cite{geer98} to build a \textit{Neutrino Factory} by 
accelerating the muons from this intense source to 
energies of several GeV or more, injecting the muons into a storage ring 
having long straight sections, and exploiting the 
intense neutrino beams that are produced by muons decaying 
in the straight sections. 
If the challenge of producing, 
capturing, accelerating, and storing a millimole of unstable muons can
be met, the decays
\begin{equation}
    \mu^{-}  \rightarrow  e^{-}\nu_{\mu}\bar{\nu}_{e}\; , \qquad 
    \mu^{+}  \rightarrow  e^{+}\bar{\nu}_{\mu}\nu_{e}
    \label{mumpdk}
\end{equation}
offer exciting possibilities for the study of neutrino interactions 
and neutrino properties~\cite{geer98,abp,bgw,suite}.  In a 
Neutrino Factory the composition and spectra of intense 
neutrino beams will be determined by the charge, momentum, and 
polarization of the stored muons.  The prospect of intense, 
controlled, high-energy beams of electron neutrinos and 
antineutrinos---for which we have no other plausible source---is very 
intriguing.

Neutrinos---weakly interacting, nearly massless elementary 
fermions---have long been objects of fascination, as well as reliable 
probes.  One of the most dramatic recent developments in particle 
physics is the growing evidence that neutrinos may oscillate from 
one species to another during propagation, which implies that 
neutrinos have mass.

If neutrinos $\nu_{1}, 
\nu_{2}, \ldots$ have different masses $m_{1}, m_{2}, \ldots$ ,
each neutrino flavor state may be a mixture of different mass states.  Let us 
consider two species for simplicity, and take 
\begin{equation}
\left( 
    \begin{array}{c}
    \nu_{e}  \\
    \nu_{\mu}
    \end{array}
\right) = \left( 
    \begin{array}{cc}
    \cos\theta & \sin\theta  \\
    -\sin\theta & \cos\theta
    \end{array}
\right) \left( 
    \begin{array}{c}
    \nu_{1}\\
    \nu_{2}
    \end{array}
\right)\; .
\end{equation}
The probability for a neutrino born as $\nu_{\mu}$ to oscillate into a 
$\nu_{e}$,
\begin{equation}
P(\nu_{\mu}\rightarrow\nu_{e}) = \sin^{2}2\theta \sin^{2}\left(1.27 \:
\frac{\delta m^{2}}{1\hbox{ eV}^{2}} \cdot \frac{L}{1\hbox{ km}} \cdot 
\frac{1\hbox{ GeV}}{E}\right)\; ,
\end{equation}
depends on two parameters related to experimental conditions: $L$, the 
distance from the neutrino source to the detector, and $E$, the 
neutrino energy.  It also depends on two fundamental neutrino 
parameters: the difference of masses squared, $\delta m^{2} = 
m_{1}^{2} - m_{2}^{2}$, and the neutrino mixing parameter, 
$\sin^{2}2\theta$.  
The probability that a neutrino born as $\nu_{\mu}$ remain a 
$\nu_{\mu}$ at distance $L$ is
\begin{equation}
    P(\nu_{\mu}\rightarrow\nu_{\mu}) = 
1 - \sin^{2}2\theta \sin^{2}\left(1.27\: \frac{\delta m^{2}}{1\hbox{ eV}^{2}} 
\cdot \frac{L}{1\hbox{ km}} \cdot \frac{1\hbox{ GeV}}{E}\right)\; .
\end{equation}

Many experiments have now used natural sources of neutrinos, neutrino 
radiation from fission reactors, and neutrino beams generated in 
particle accelerators to look for evidence of neutrino oscillation.  
The positive indications for neutrino oscillations fall into three 
classes:\cite{janetc}
\begin{enumerate}
    \item  Five solar-neutrino experiments report deficits with respect 
    to the predictions of the standard solar model: Kamiokande and 
    Super-Kamiokande (SuperK) using water-Cerenkov techniques, SAGE and GALLEX 
    using chemical recovery of germanium produced in neutrino 
    interactions with gallium, and Homestake using radiochemical 
    separation of argon produced in neutrino interactions with 
    chlorine.  These results suggest the oscillation $\nu_{e} 
    \rightarrow \nu_{x}$, with $|\delta m^{2}|_{\mathrm{solar}} \approx 
    10^{-5}\hbox{ eV}^{2}$ and $\sin^{2}2\theta_{\mathrm{solar}}\approx 1\hbox{ 
    or a few}\times 10^{-3}$, or $|\delta m^{2}|_{\mathrm{solar}} \approx 
    10^{-10}\hbox{ eV}^{2}$ and $\sin^{2}2\theta_{\mathrm{solar}}\approx 1$.

    \item Five atmospheric-neutrino experiments report anomalies in the
    arrival of muon neutrinos: Kamiokande, IMB, and SuperK using
    water-Cerenkov techniques, and Soudan~2 and MACRO using sampling
    calorimetry.  The most striking result is the zenith-angle dependence
    of the $\nu_{\mu}$ rate reported last year by SuperK
    \cite{SKatm,SKLyon}.  These results suggest the oscillation $\nu_{\mu}
    \rightarrow \nu_{\tau}\hbox{ or }\nu_{s}$, with
    $\sin^{2}2\theta_{\mathrm{atm}} \approx 1$ and $|\delta
    m^{2}|_{\mathrm{atm}} = 10^{-3}\hbox{ to }10^{-2}\hbox{ eV}^{2}$.  The 
    oscillation $\nu_{\mu} \rightarrow \nu_{\tau}$ is increasingly 
    the favored interpretation.

    \item The LSND experiment~\cite{s1LSND} reports the observation of
    $\bar{\nu}_{e}$-like events in what should be an essentially pure
    $\bar{\nu}_{\mu}$ beam produced at the Los Alamos Meson Physics
    Facility, suggesting the oscillation $\bar{\nu}_{\mu} \rightarrow
    \bar{\nu}_{e}$.  This result has not yet been reproduced by any other
    experiment.  The favored region lies along a band from
    $(\sin^{2}2\theta_{\mathrm{LSND}} = 10^{-3},|\delta
    m^{2}|_{\mathrm{LSND}} \approx 1\hbox{ eV}^{2})$ to 
    $(\sin^{2}2\theta_{\mathrm{LSND}} = 1,|\delta
    m^{2}|_{\mathrm{LSND}} \approx 7 \times 10^{-2}\hbox{ eV}^{2})$. 
\end{enumerate}
A host of other experiments have failed to turn up evidence for neutrino 
oscillations in the regimes of their sensitivity.  These results limit 
neutrino mass-squared differences and mixing angles.  In more than a 
few cases, positive and negative claims are in conflict, or at least 
face off against each other.  Over the next five years, many 
experiments will seek to verify, further quantify, and extend these 
claims. If all of the current experimental indications of neutrino 
oscillation survive, there are apparently three different 
mass-squared-difference scales, which cannot be accommodated with 
only three neutrino types. New \textit{sterile} neutrinos may be 
required. This would be a profound discovery.

From the era of the celebrated two-neutrino experiment \cite{twonu} to modern 
times, high-energy neutrino beams have played a decisive role in the 
development of our understanding of the constituents of matter and the 
fundamental interactions among them.  Major landmarks include the 
discovery of weak neutral-current interactions \cite{weaknc}, and 
incisive studies of the structure of the proton and the quantitative 
verification of perturbative quantum chromodynamics as the theory of 
the strong interactions \cite{rmpnurev}.  The determinations of the 
weak mixing parameter $\sin^{2}\theta_{W}$ and the strong coupling 
constant $\alpha_{s}$ in deeply inelastic neutrino interactions are 
comparable in precision to the best current measurements.  Though 
experiments with neutrino beams have a long history, beams of 
greatly enhanced intensity would bring opportunities for dramatic 
improvements.  Because weak-interaction cross sections are small, 
high-statistics studies have required massive targets and 
coarse-grained detectors.  Until now, it has been impractical to 
consider precision neutrino experiments using short liquid hydrogen 
targets, or polarized targets, or active semiconductor 
target-detectors.  All of these options are opened by a muon storage 
ring, which would produce neutrinos at approximately $10^{4}$ times 
the flux of existing neutrino beams.

At the energies best suited for the study of neutrino 
oscillations---tens of GeV, by our current estimates---the muon 
storage ring is compact.  We could build it at one laboratory, pitched 
at a deep angle, to illuminate a laboratory on the other side of the 
globe with a neutrino beam whose properties we can control with great 
precision.  By choosing the right combination of energy and 
destination, we can tune future neutrino-oscillation experiments to 
the physics questions we will need to answer, by specifying the ratio 
of path length to neutrino energy and determining the amount of matter 
the neutrinos traverse.  Although we can use each muon decay only 
once, and we will not be able to select many destinations, we may be 
able to illuminate two or three well-chosen sites from a 
muon-storage-ring neutrino source.  That possibility---added to the 
ability to vary the muon charge, polarization, and energy---may give 
us just the degree of experimental control it will take to resolve the 
outstanding questions about neutrino oscillations.  
Experiments at a Neutrino Factory would seek to verify the number 
of neutrino types participating in the oscillations, precisely 
determine the mixing parameters that relate the flavor states 
to the mass states, determine the pattern of neutrino masses, 
and look for CP violation in the lepton sector.

The prodigious flux of neutrinos close to the muon storage ring raises 
the prospect of neutrino-scattering experiments of unprecedented 
sensitivity and delicacy.  
Experiments that might be pursued at a Neutrino Factory include 
precise measurements of the nucleon structure (including 
changes that occur in a nuclear environment), 
measurements of the spin structure of the nucleon using a 
new and powerful technique, 
charm measurements with several million tagged particles, 
precise measurements of Standard Model parameters, 
and searches for exotic phenomena. 

We believe that the physics program at a Neutrino Factory is 
compelling and encourage support for a vigorous R\&D program 
to make neutrino factories a real option for the future.

\clearpage
\section{Beam properties}

Consider an ensemble of polarized negatively-charged muons. 
When the muons decay they produce muon neutrinos with a distribution 
of energies and angles  in the muon rest--frame described 
by~\cite{gaisser}:
\begin{eqnarray}
\frac{d^2N_{\nu_\mu}}{dxd\Omega_{cm}} &\propto& {2x^2\over4\pi}
 \left[ (3-2x) + (1-2x) P_\mu \cos\theta_{cm} \right] \, ,
\label{eq:n_numu}
\end{eqnarray}
where $x\equiv 2E_\nu/m_\mu$, $\theta_{cm}$ is the angle between the neutrino
momentum vector and the muon spin direction, and $P_\mu$ is the average muon
polarization along the beam direction. 
The electron antineutrino distribution is given by: 
\begin{eqnarray}
\frac{d^2N_{\bar\nu_e}}{dxd\Omega_{cm}} &\propto&  {12x^2\over4\pi}
\left[ (1-x) + (1-x) P_\mu\cos\theta_{cm} \right] \, ,
\label{eq:n_nue}
\end{eqnarray}
and the corresponding distributions for
$\bar\nu_\mu$ and $\nu_e$ from $\mu^+$ decay are obtained by
the replacement $P_{\mu} \to -P_{\mu}$. 
Only neutrinos and antineutrinos emitted in the forward
direction ($\cos\theta_{lab}\simeq1$) are relevant to the neutrino flux for
long-baseline experiments; in this limit
$E_\nu = x E_{max}$ and at high energies the maximum $E_\nu$ in the 
laboratory frame is given by 
$E_{max} = \gamma   (1 + \beta \cos\theta_{cm})m_{\mu}/2 $, 
where $\beta$ and $\gamma$ are the usual relativistic factors. 
The $\nu_\mu$ and $\overline{\nu}_{e}$ distributions as a function
of the laboratory frame variables are then given by:
 \begin{eqnarray}
\frac{d^2N_{\nu_{\mu}}}{dxd\Omega_{lab}} &\propto&  
{1\over \gamma^2 (1- \beta\cos\theta_{lab})^2}\frac{2x^2}{4\pi}
\left[ (3-2x) + (1-2x)P_{\mu}\cos\theta_{cm} \right] , 
\label{eq:numu}
\end{eqnarray}
and
\begin{eqnarray}
 \frac{d^2N_{\overline{\nu}_{e}}}{dxd\Omega_{lab}} &\propto&
{1\over \gamma^2 (1- \beta\cos\theta_{lab})^2}\frac{12x^2}{4\pi}
\left[ (1-x) + (1-x)P_{\mu}\cos\theta_{cm} \right] \; .
\label{eq:nue}
\end{eqnarray}

\begin{figure}
\epsfxsize3.5in
\centerline{\epsffile{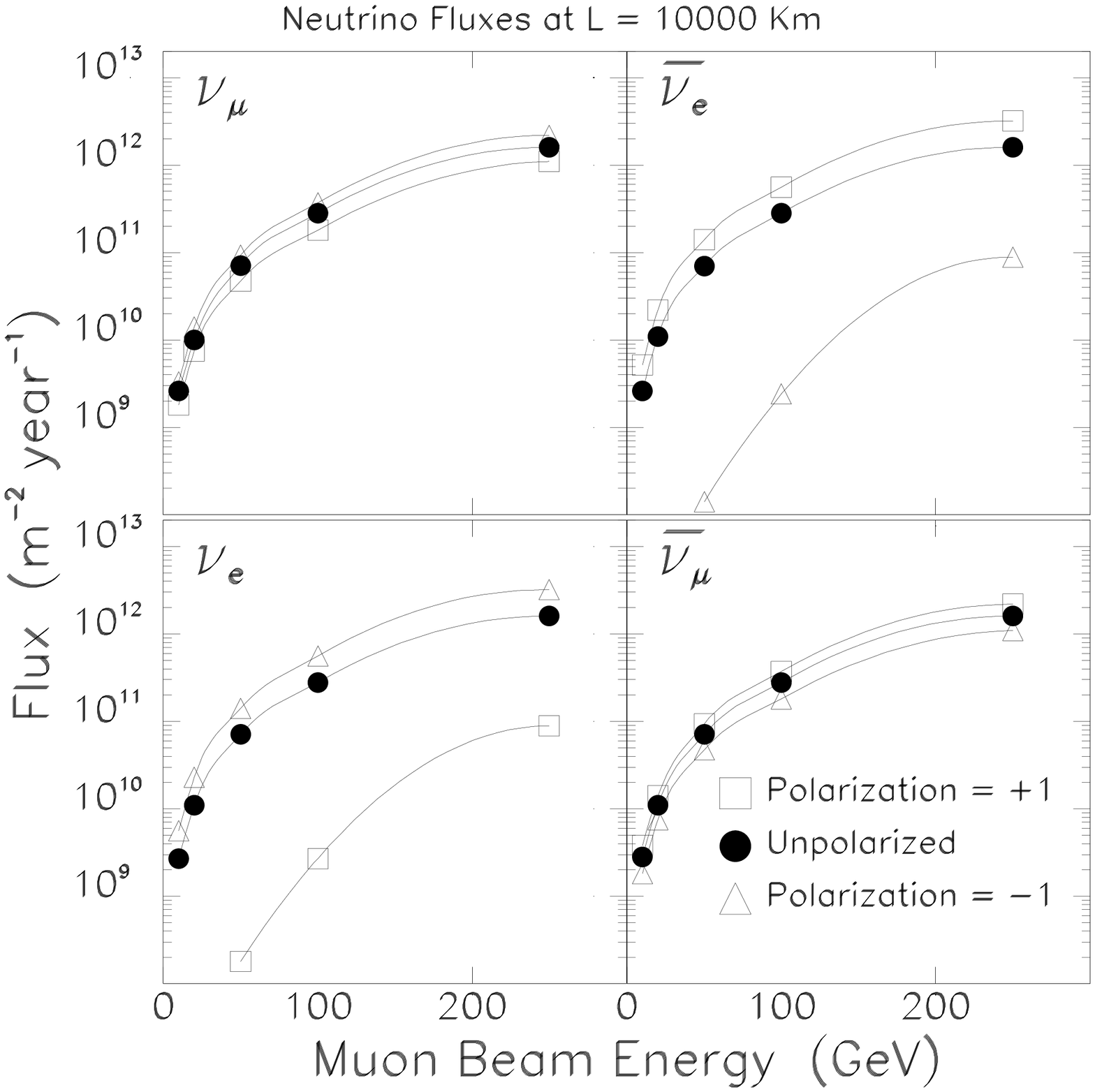}}
\caption{Calculated $\nu$ and $\overline{\nu}$ fluxes in the
absence of oscillations at a far site located 
10000 km from a neutrino factory in which 
$2 \times 10^{20}$ muons have decayed in the beam--forming straight section. 
The fluxes are shown as a function of the energy of the stored 
muons for negative muons (top two plots) 
and positive muons (bottom two plots), and for three muon polarizations
as indicated. The calculated fluxes are averaged over a circular area 
of radius 1~km at the far site. 
Calculation from Ref.~\ref{geer98}.}
\label{fluxes}
\end{figure}

Thus, for a high energy muon beam with no beam divergence, 
the neutrino and antineutrino energy-- and angular--
distributions depend upon the parent muon
energy, the decay angle, and the direction of the
muon spin vector.
With the muon beam intensities that could be provided by a 
muon--collider type muon source~\cite{status_report} 
the resulting neutrino fluxes 
at a distant site would be large. For example, Fig.~\ref{fluxes} shows 
as a function of muon energy and polarization, 
the computed fluxes per $2\times 10^{20}$ muon decays at a site on the 
other side of the Earth ($L = 10000$~km).
Note that the $\nu_e$ ($\overline{\nu}_e$) fluxes are suppressed 
when the muons have $P = +1$ (-1). This can be understood by examining 
Eq.~(\ref{eq:nue}) and noting that for $P = -1$ the two terms cancel 
in the forward direction for all $x$.

\subsection{Interaction rates}

\begin{figure}
\epsfxsize3.in
\centerline{\epsffile{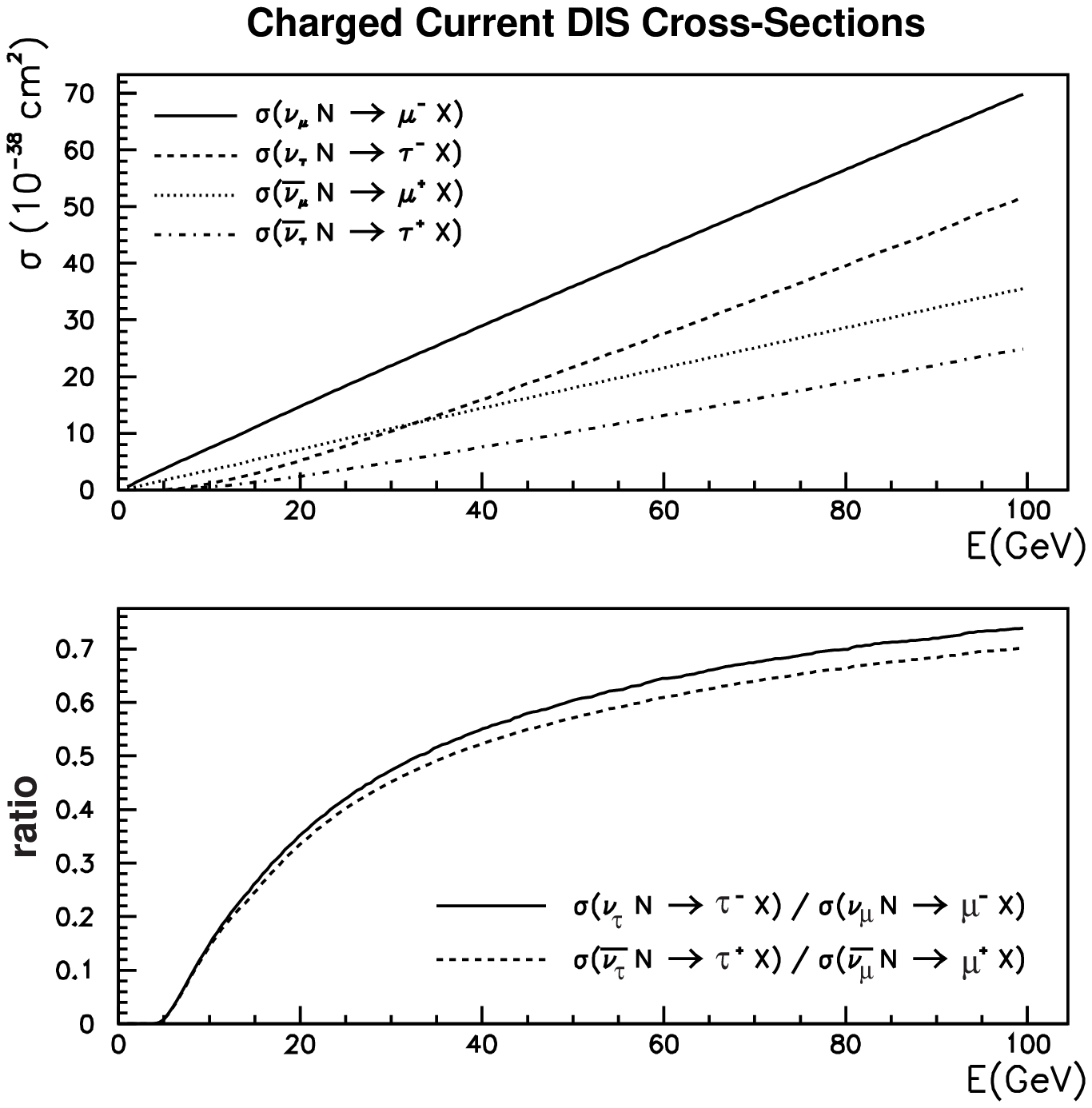}}
\caption{The total cross section for charged current neutrino scattering
by muon and tau neutrinos (top plot), and  
the ratio of tau to muon neutrino cross sections
as a function of neutrino energy (bottom plot).}
\label{tau_fig}
\end{figure}
Neutrino charged current (CC) scattering cross-sections are shown as a function of 
energy in Fig.~\ref{tau_fig}. 
At low energies the neutrino scattering cross section is dominated by
quasi-elastic scattering and resonance production. 
However, if $E_\nu$ is greater than $\sim10$~GeV, 
the total cross section is  dominated by deep inelastic scattering 
and is approximately~\cite{CCFRsigma}:
\begin{eqnarray}
\sigma(\nu +N \gt \lepton^- + X) &\approx& 0.67\times 10^{-38} \;
\centi^2\times (E_{\nu},
\GeV) \, , \\
\sigma(\antinu +N \gt \lepton^+ + X) &\approx& 0.34\times
 10^{-38} \; \centi^2\times (E_{\antinu},
\GeV) \; .
\end{eqnarray}
The number of $\nu$ and $\overline{\nu}$ CC events per incident neutrino
 observed 
in an isoscalar target is given by:
\begin{eqnarray}
N(\nu +N \gt \lepton^- + X) 
&=& 4.0 \times 10^{-15} (E_{\nu}, \GeV) \;
\hbox{events per gr/cm$^2$} \; , \\
N(\antinu +N \gt \lepton^+ + X) 
&=& 2.0 \times 10^{-15}( E_{\antinu}, \GeV) \;
\hbox{events per gr/cm$^2$} \; .
\end{eqnarray}

\begin{figure}
\epsfxsize3.in
\centerline{\epsffile{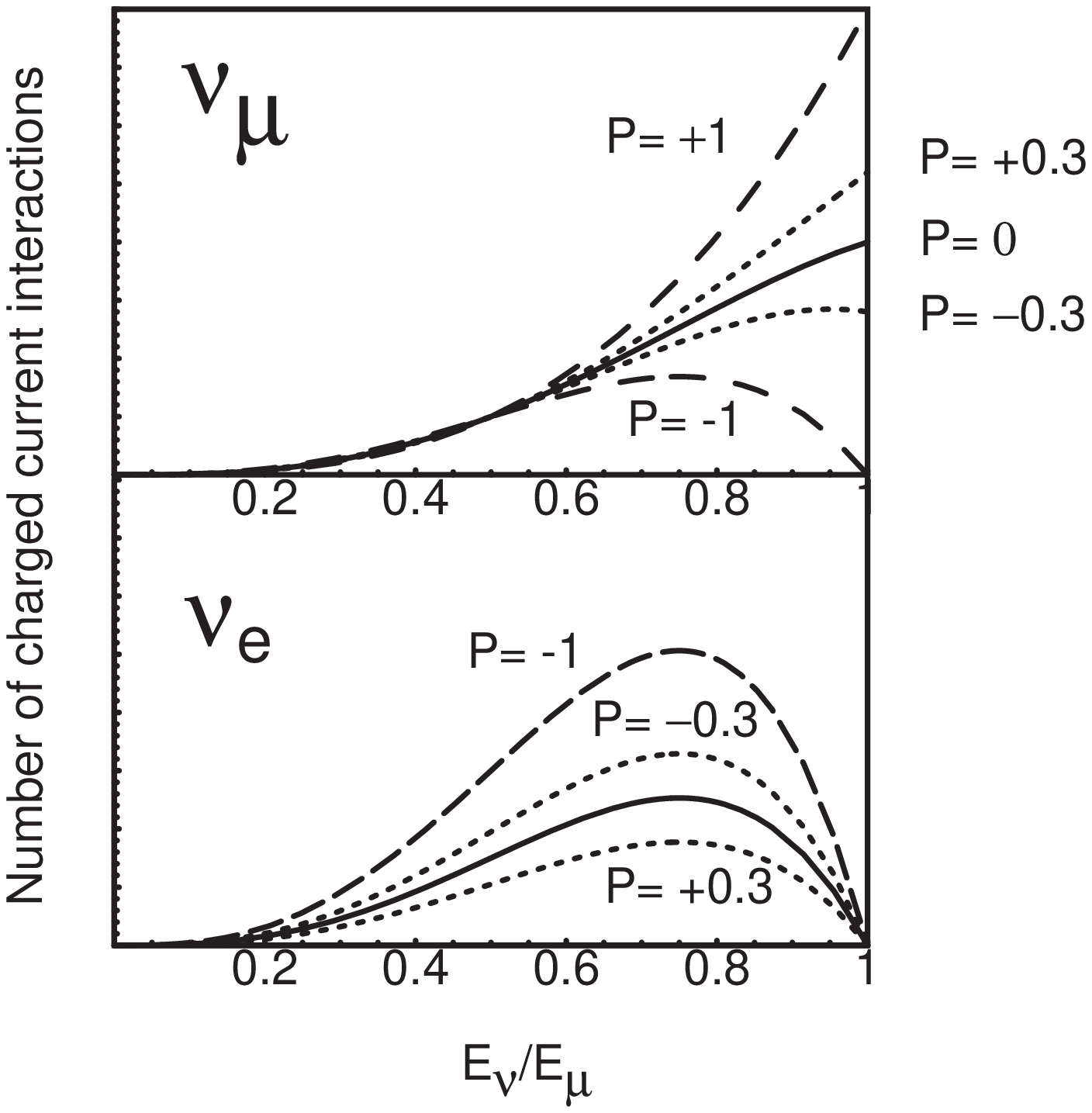}}
\caption{Charged current event spectra at a far detector. 
The solid lines
indicate zero polarization, the dotted lines indicate polarization of $\pm 0.3$ and
the dashed lines indicate full polarization.  The $P=1$ case for electron neutrinos
results in no events and is hidden by the $x$ axis.}
\label{polarization}
\end{figure}

\begin{figure}
\begin{center}
\mbox{
\epsfxsize=6.5truecm
\epsfysize=5.6truecm
\epsffile{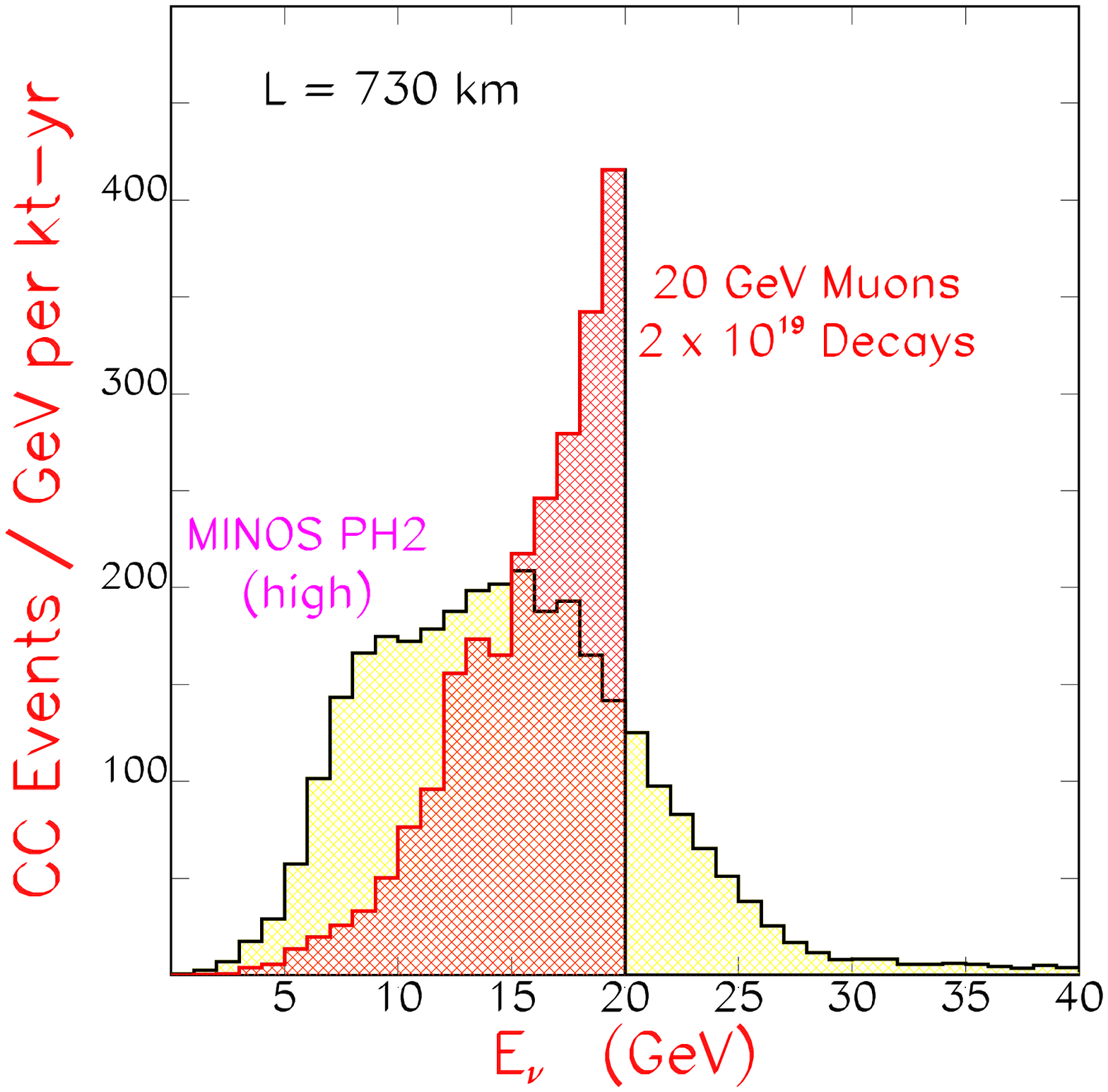}}
\mbox{
\epsfxsize=6.5truecm
\epsfysize=5.6truecm
\epsffile{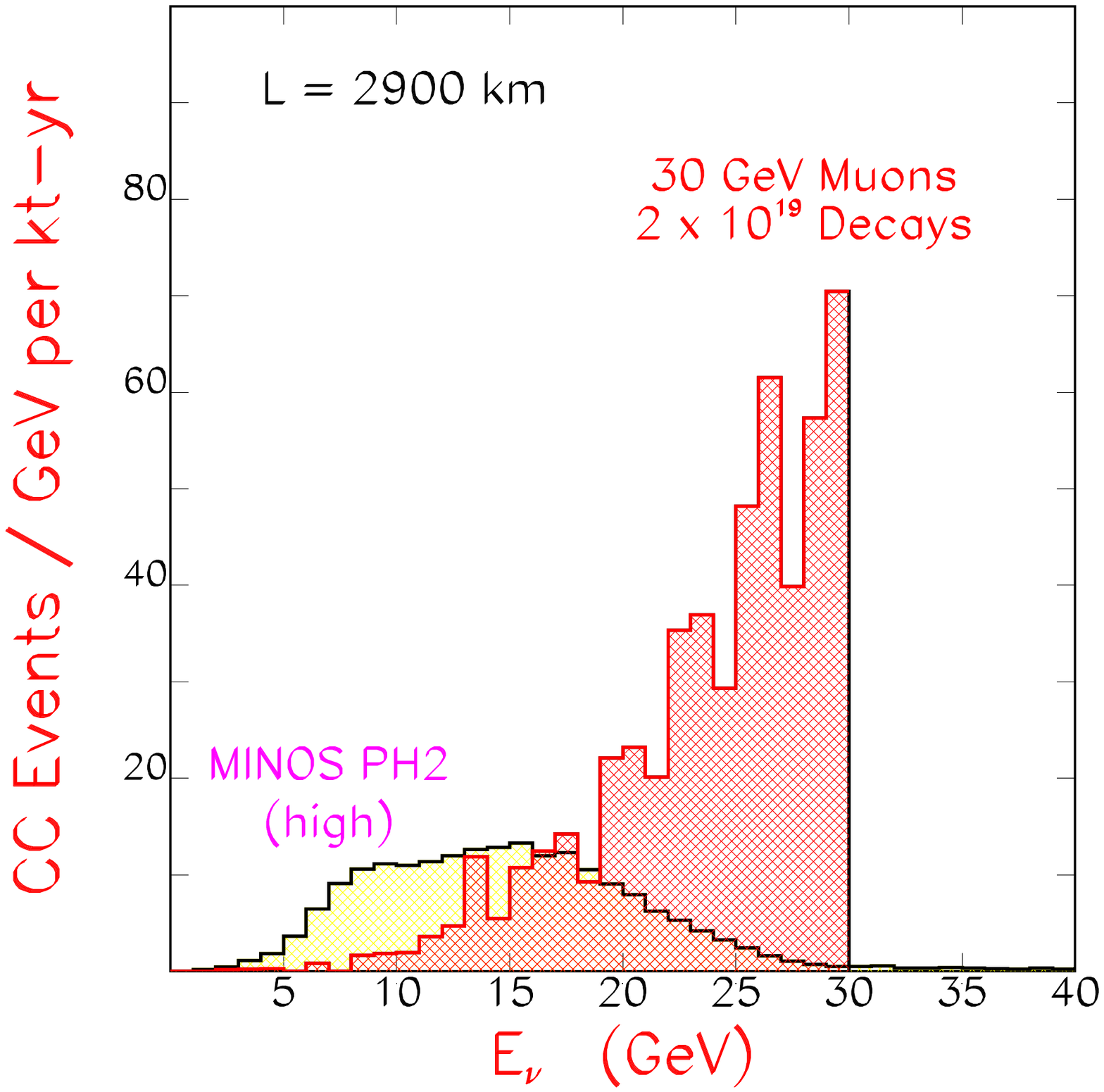}}
\end{center}
\caption{Comparison of interacting $\nu_\mu$ energy distributions for
the NUMI high energy wide band beam (Ref.~\ref{numi}) with
a 20~GeV neutrino factory beam (Ref.~\ref{geer98}) at $L = 730$~km and
a 30~GeV neutrino factory beam at $L = 2900$~km.
The neutrino factory distributions have been calculated based on
Eq.~(\ref{eq:n_numu}) (no approximations), and include realistic
muon beam divergences and energy spreads.
}
\label{minos_wbb}
\end{figure}

Using this simple form for the energy dependence of the cross section, 
the predicted energy distributions for $\nu_e$ and $\nu_\mu$ 
interacting in a far detector ($\cos\theta = 1$) at a neutrino 
factory are shown in Fig.~\ref{polarization}. The interacting $\nu_\mu$ 
energy distribution is compared in Fig.~\ref{minos_wbb} 
with the corresponding distribution arising from the high--energy 
NUMI wide band beam. Note that neutrino beams from a 
neutrino factory can be considered narrow band beams. 
%
\begin{figure}
\epsfxsize3.5in
\centerline{\epsffile{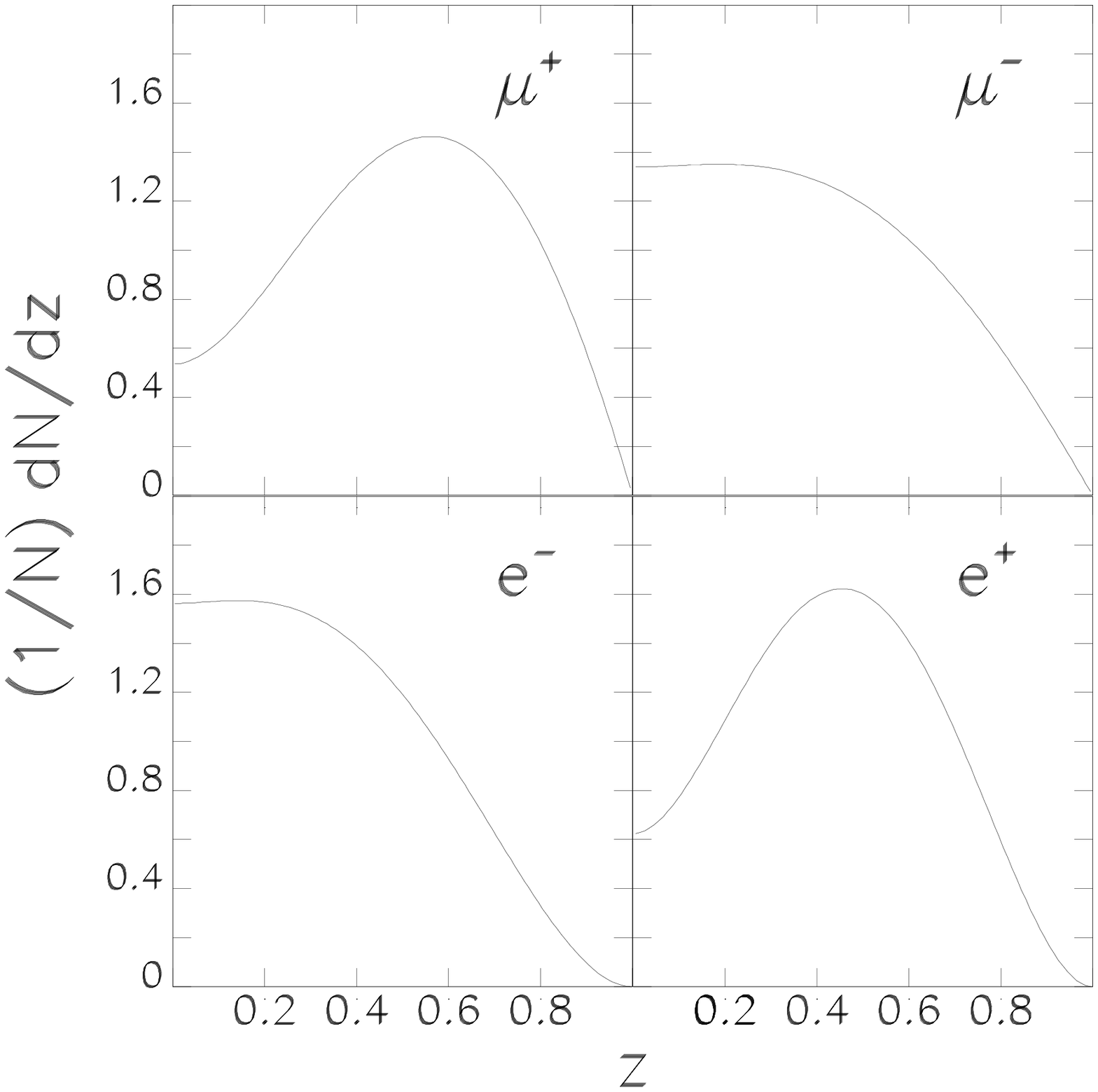}}
\caption{Lepton energy spectra for CC $\overline{\nu}_\mu$ (top left),
$\nu_\mu$ (top right), $\nu_e$ (bottom left), and $\overline{\nu}_e$
(bottom right) interactions. Note that $z$ is the energy normalized
to the primary muon energy  $z = E_{\lepton}/E_\mu$. 
Calculation from Ref.~\ref{bgw99}.}
\label{fig:elept}
\end{figure}
In practice, CC interactions can only be cleanly 
identified when the final 
state lepton exceeds a threshold energy. The calculated final state lepton 
distributions are shown in Fig.~\ref{fig:elept}. 
Integrating over the energy distribution, the total $\nu$ and $\overline{\nu}$ 
interaction rates per muon decay are given by: 
\begin{eqnarray}
N_\nu &=& 1.2 \times 10^{-14} \; \biggr[{(E_{\mu}, \GeV)^3\over (L, km)^2}\biggl] 
\times C(\nu) \;\; \hbox {events per kt}
\end{eqnarray}
and
\begin{eqnarray}
N_{\overline{\nu}}&=&0.6\times10^{-14} \;
\biggr[{(E_{\mu}, \GeV)^3\over (L, km)^2}\biggl] 
\times C(\nu) \;\; \hbox{events per kt} \, ,
\end{eqnarray}
where
\begin{eqnarray}
C(\nu_{\mu})&=& {7\over 10} + P_{\mu} {3\over 10}, \ \ \ \ 
C(\nu_{e}) ={6\over 10} - P_{\mu} {6\over 10}\\
\end{eqnarray}

\begin{figure}
\epsfxsize 3.in
\centerline{\epsffile{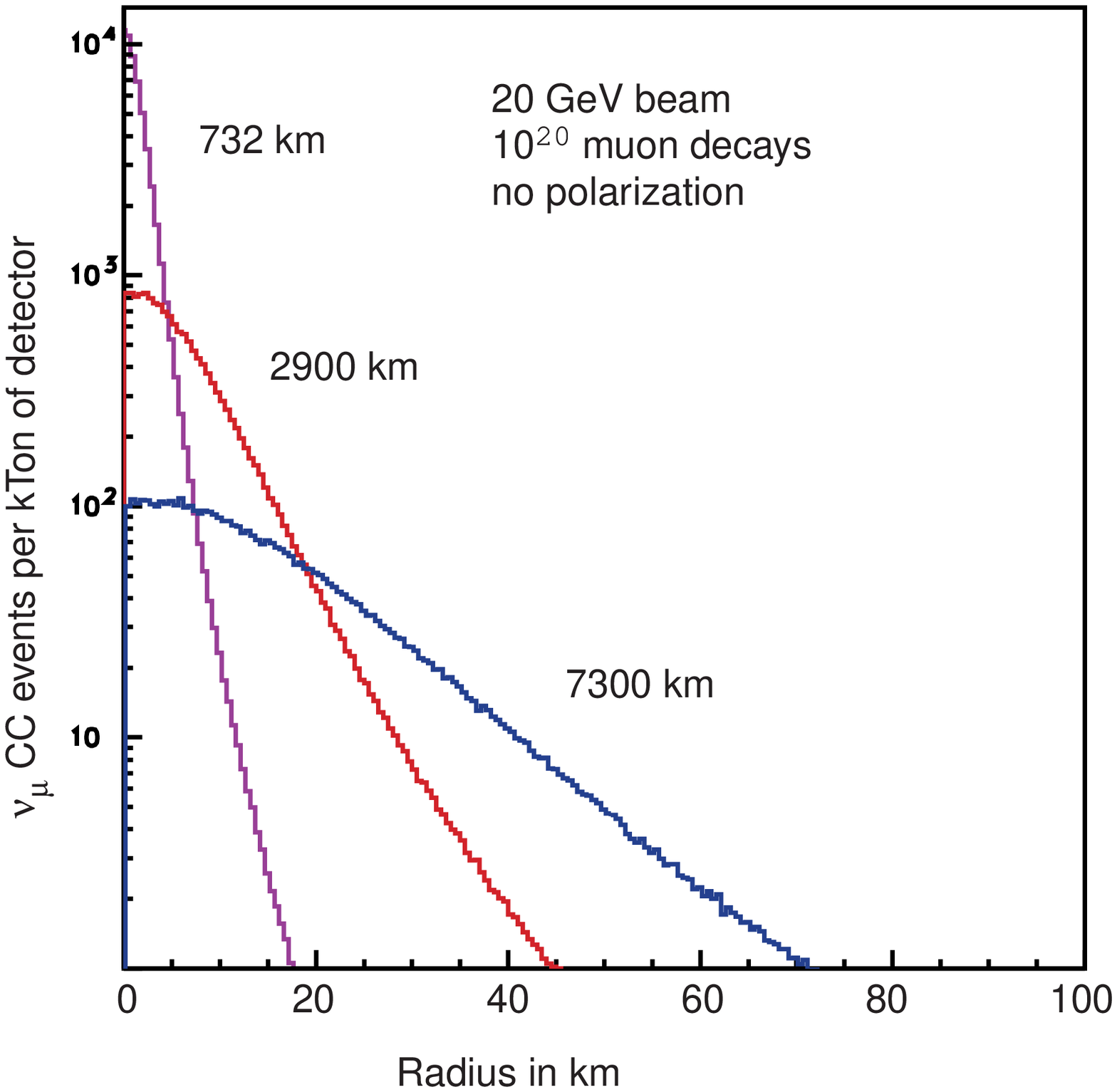}}
\caption{Events/kT of detector as a function of distance from the beam center
for a 20 GeV muon beam.}
\label{fig:radial}
\end{figure}

The calculated $\nu_e$ and $\nu_\mu$ CC interaction rates resulting from 
$10^{20}$ muon 
decays in the beam--forming straight--section of a neutrino factory 
are compared in Table~\ref{table:rates} with 
expectations for the corresponding rates at the next generation 
of accelerator--based neutrino experiments. Note that event rates 
at a neutrino factory increase as $E_\mu^3$, and are significantly 
larger than expected for the next generation of approved experiments 
if $E_\mu > 20$~GeV.
The radial dependence 
of the event rate is shown in Fig.~\ref{fig:radial} for a 20~GeV 
neutrino factory and three baselines. 

\begin{table}
\caption{\label{table:rates}
Muon neutrino and electron antineutrino CC interaction rates in the
absence of oscillations, calculated for baseline length $L = 732$~km
(FNAL $\rightarrow$ Soudan), for MINOS using the wide band beam and a
muon storage ring delivering $10^{20}$ decays 
with $E_\mu=10, 20$, and $50$~GeV at 3 baselines.
The neutrino factory calculation includes a realistic muon beam divergence 
and energy spread.}
\begin{center}
\begin{tabular}{|ccc|cc|cc|cc|cc|cc}
\hline
& &Baseline & $\langle E_{\nu_\mu} \rangle$ & $\langle E_{\bar \nu_e} \rangle$
& N($\nu_\mu$ CC) & N($\bar\nu_e$ CC) \\
Experiment & &(km) & (GeV) & (GeV) & (per kt--yr) & (per kt--yr) \\
\hline
MINOS& Low energy    &732&  3 & -- &  458 & 1.3 \\
     & Medium energy &732&  6 & -- & 1439 & 0.9 \\
     & High energy   &732& 12 & -- & 3207 & 0.9 \\
\hline
Muon ring & $E_\mu$ (GeV) & & & & & \\
\hline
&  10 &732& 7.5 & 6.6 & 1400              & 620 \\
&  20 &732&  15 &  13 & 12000             & 5000\\
&  50 &732&  38 &  33 & 1.8$\times$10$^5$ & 7.7$\times$10$^4$ \\
\hline
Muon ring& $E_\mu$ (GeV)& & & & & \\
\hline
&  10 &2900& 7.6 & 6.5 & 91             &41\\
&  20 &2900&  15 &  13 & 740          & 330\\
&  50 &2900&  38 &  33 & 11000& 4900 \\
\hline 
Muon ring& $E_\mu$ (GeV)& & & & & \\
\hline
&  10 &7300& 7.5 & 6.4 & 14              & 6  \\
&  20 &7300&  15 &  13 & 110            & 51 \\
&  50 &7300&  38 &  33 & 1900 & 770 \\
\hline
\end{tabular}
\end{center}
\end{table}

Finally, for an isoscalar target the neutral current (NC) cross sections 
are approximately 0.4 of the CC cross sections\cite{neutralcurrent}, and are given by:
\begin{eqnarray}
\sigma(\nu +N \gt \nu + X) &\approx& 0.3\times 10^{-38} \;
\centi^2\times (E_{\nu},
\GeV) \, , \\
\sigma(\antinu +N \gt \overline{\nu} + X) &\approx& 0.15\times
 10^{-38} \; \centi^2\times (E_{\antinu},\GeV) \; .
\end{eqnarray}

\subsection{Tau neutrino interactions}
Tau neutrino CC interaction rates are substantially less than the corresponding 
$\nu_e$ and $\nu_\mu$ rates, 
especially near the tau production threshold of $\sim 3.3$~GeV.
The NC rates should be the same as those for electron
and muon neutrinos.
Figure~\ref{tau_fig} shows the calculated~\cite{goodman} 
ratio of $\nu_\tau / \nu_\mu$ CC 
interaction rates as a function of the neutrino energy. 
Near threshold, contributions
from quasi--elastic and resonance production dominate.
If the $\nu_\tau$ cross sections from 
Ref.~\cite{casper} are used, the predicted event rates are 5--7\%
lower.

\begin{table}
\renewcommand{\baselinestretch}{1}
\begin{center}
\label{tab:com}
\vspace{0.6 cm}
\caption{Dependence of predicted charged current  event rates on muon 
beam properties at a neutrino factory. The last column 
lists the required precisions with which each beam property must 
be determined if the uncertainty on the neutrino flux at the 
far site is to be less than $\sim1$\%.  Here $\Delta$ denotes uncertainty
while $\sigma$ denotes the spread in a variable. Table from Ref.~\ref{cg00}.
}
\vspace{0.2cm}
\begin{tabular}{c|c|cc} \hline 
Muon Beam & Beam & Rate & Target\\
property & Type & Dependence & Precision \\
\hline
Energy ($E_\mu$) & $\nu$ (no osc) 
  & $\Delta N / N = 3 \; \Delta E_\mu/E_\mu$ 
  & $\Delta(E_\mu)/E_\mu < 0.003$ \\
  & $\nu_{e} \rightarrow \nu_{\mu}$ 
  &$\Delta N / N = 2 \; \Delta E_\mu/E_\mu$ 
  & $\Delta(E_\mu)/E_\mu < 0.005$ \\
\hline
Direction ($\Delta\theta$) & $\nu$ (no osc) 
  & $\Delta N/N \leq 0.01$ 
  & $\Delta\theta < 0.6 \; \sigma_\theta$ \\
  &  & (for $\Delta\theta < 0.6\; \sigma_\theta$) & \\
\hline
Divergence ($\sigma_\theta$)
  & $\nu$ (no osc) 
  & $\Delta N / N \sim 0.03 \; \Delta\sigma_\theta / \sigma_\theta$
  & $\Delta\sigma_\theta / \sigma_\theta < 0.2$ \\
  & & (for $\sigma_\theta \sim 0.1/\gamma$) 
    & (for $\sigma_\theta \sim 0.1/\gamma$)\\
\hline
Momentum spread ($\sigma_p$) 
  & $\nu$ (no osc) 
  & $\Delta N / N \sim 0.06 \; \Delta\sigma_p / \sigma_p$
  & $\Delta\sigma_p / \sigma_p < 0.17$ \\
\hline
Polarization ($P_\mu$) 
  & $\nu_e$ (no osc) 
  & $\Delta N_{\nu_e} / N_{\nu_e} =  \Delta P_\mu$
  & $\Delta P_\mu < 0.01$ \\
  & $\nu_{\mu}$ (no osc) 
  & $\Delta N_{\nu_\mu} / N_{\nu_\mu} = 0.4 \; \Delta P_\mu$
  & $\Delta P_\mu < 0.025$ \\ 
\hline
\end{tabular}
\label{tab:flux}
\end{center}
\end{table}

\subsection{Systematic uncertainties on the muon beam and neutrino flux}

\begin{figure}
\epsfxsize3.5in
\centerline{\epsffile{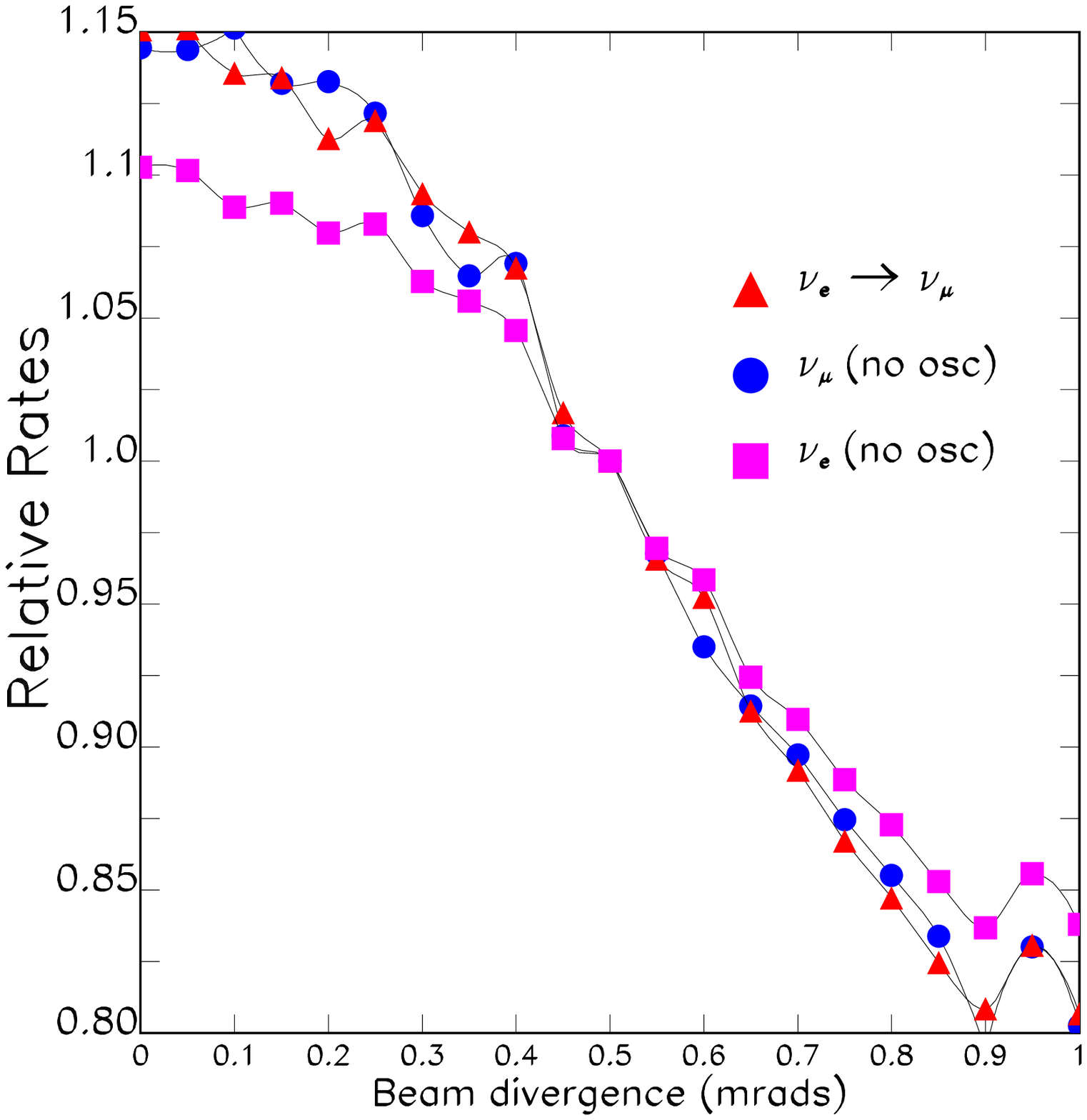}}
\vspace{-2.3cm}
\caption{
Dependence of CC interaction rates on the muon beam divergence
for a detector located at
$L = 2800$~km from a muon storage ring containing 30~GeV unpolarized muons.
Rates are shown
for $\nu_e$ (boxes) and $\nu_\mu$ (circles) beams
in the absence of oscillations,
and for $\nu_e \rightarrow \nu_\mu$ oscillations (triangles) with the
three--flavor oscillation parameters IA1.
The calculation is from Ref.~\ref{cg00}.
}
\label{fig:flux_xy}
\end{figure}

\begin{figure}
\epsfxsize3.5in
\centerline{\epsffile{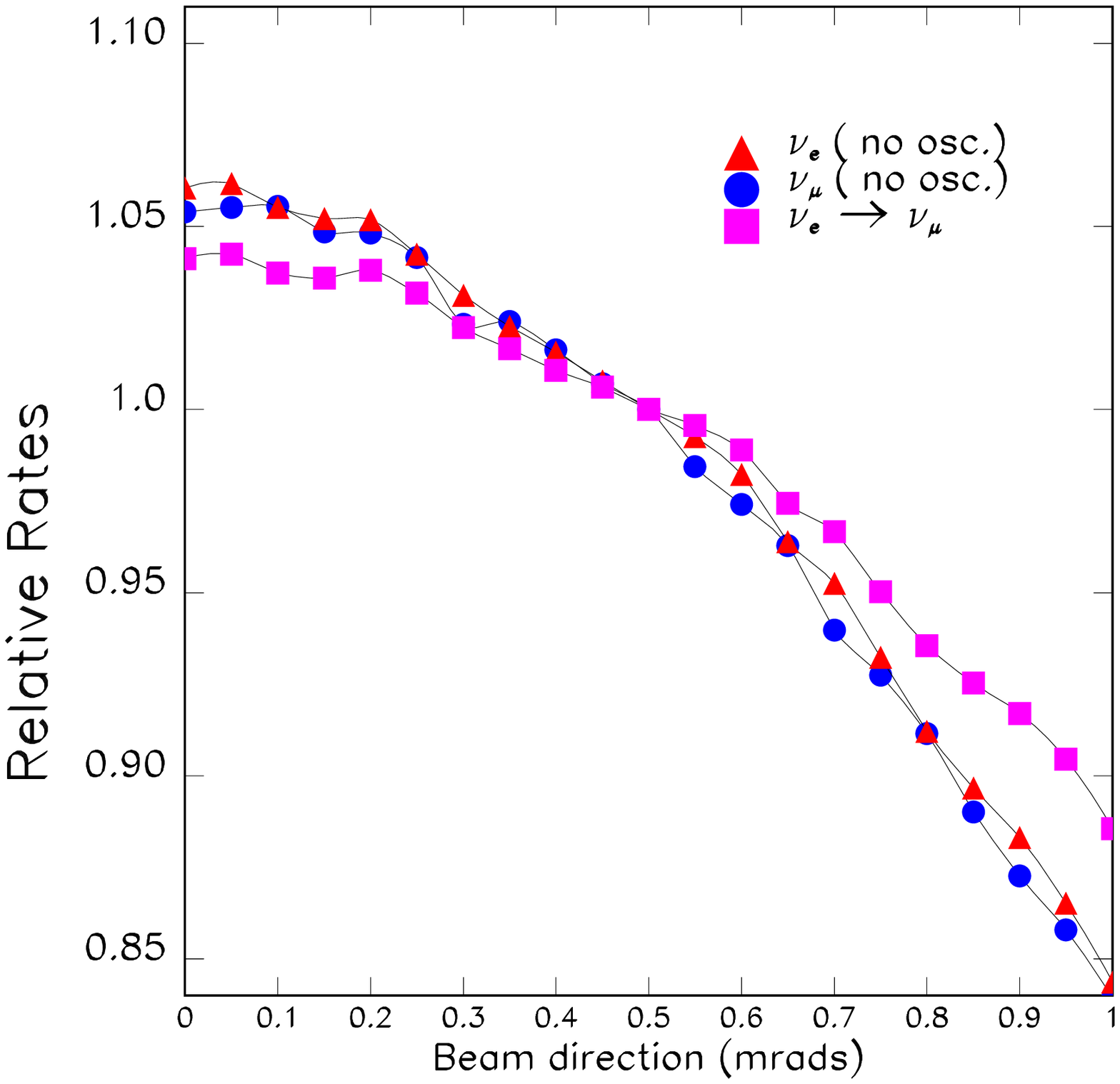}}
\vspace{-2.0cm}
\caption{
Dependence of CC interaction rates on the neutrino beam direction. 
Relative rates are shown for 
a detector at a far site located downstream of 
a storage ring containing 30~GeV unpolarized muons, and 
a muon beam divergence of 0.33~mr. Rates are shown
for $\nu_e$ (triangles) and $\nu_\mu$ (circles) beams
in the absence of oscillations,
and for $\nu_e \rightarrow \nu_\mu$ oscillations (boxes) with the
three--flavor oscillation parameters IA1.
The calculation is from Ref.~\ref{cg00}.
}
\label{fig:flux_d}
\end{figure}

In the neutrino beam--forming straight section the muon beam 
is expected to have an average divergence given by 
$\sigma_\theta =$~O($0.1/\gamma$). The neutrino beam divergence will 
therefore be dominated by muon decay kinematics, and uncertainties 
on the beam direction and divergence will yield only small 
uncertainties in the neutrino flux at a far site. However, if 
precise knowledge of the flux is required, the uncertainties 
on $\theta$ and $\sigma_\theta$ must be taken into account, along 
with uncertainties on the flux arising from uncertainties on 
the muon energy distribution and polarization. 
The relationships between the uncertainties on the muon beam 
properties and the resulting uncertainties on the neutrino flux 
are summarized in Table~\ref{tab:flux}. If, for example, we wanted 
to know the $\nu_e$ and $\nu_{\mu}$ fluxes at a far site with a 
precision of 1\%, we would need to know the beam divergence $\sigma_\theta$ 
to 20\% (Fig.~\ref{fig:flux_xy}), and ensure that the beam direction 
was within $0.6\;\sigma_\theta$ of the nominal direction~\cite{cg00}  
(Fig.~\ref{fig:flux_d}).

\subsection{Event distributions at a near site}

\begin{figure}
\epsfxsize 3.in
\centerline{\epsffile{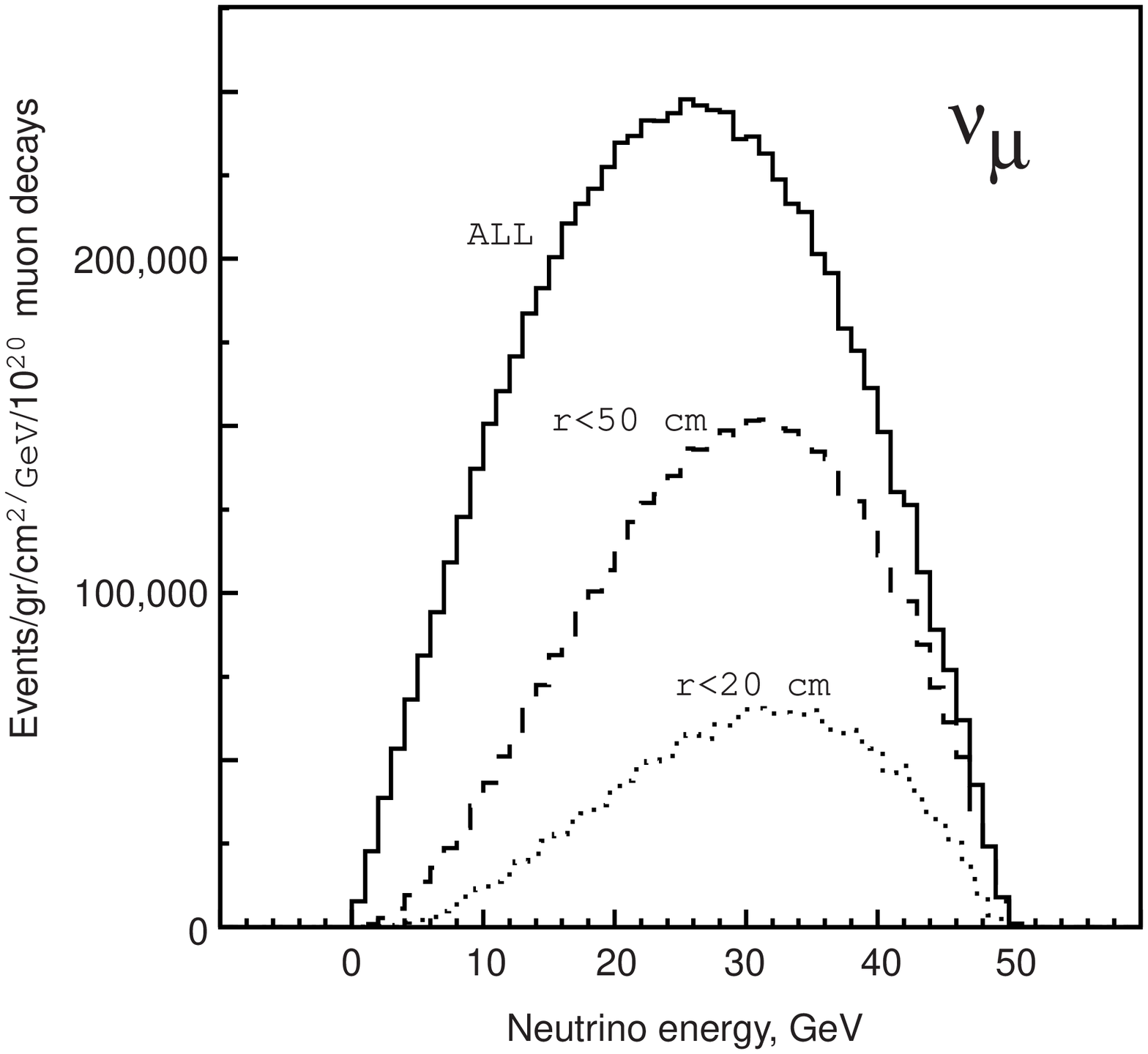}}
\caption{Events per gr/cm$^2$ per GeV for a detector 40~m from a
muon storage ring with a 600 m straight section.  The 3 curves show
all events and those falling within 50 and 20~cm of the beam center. }
\label{nearspectra}
\end{figure}

\begin{figure}
\epsfxsize 3.in
\centerline{\epsffile{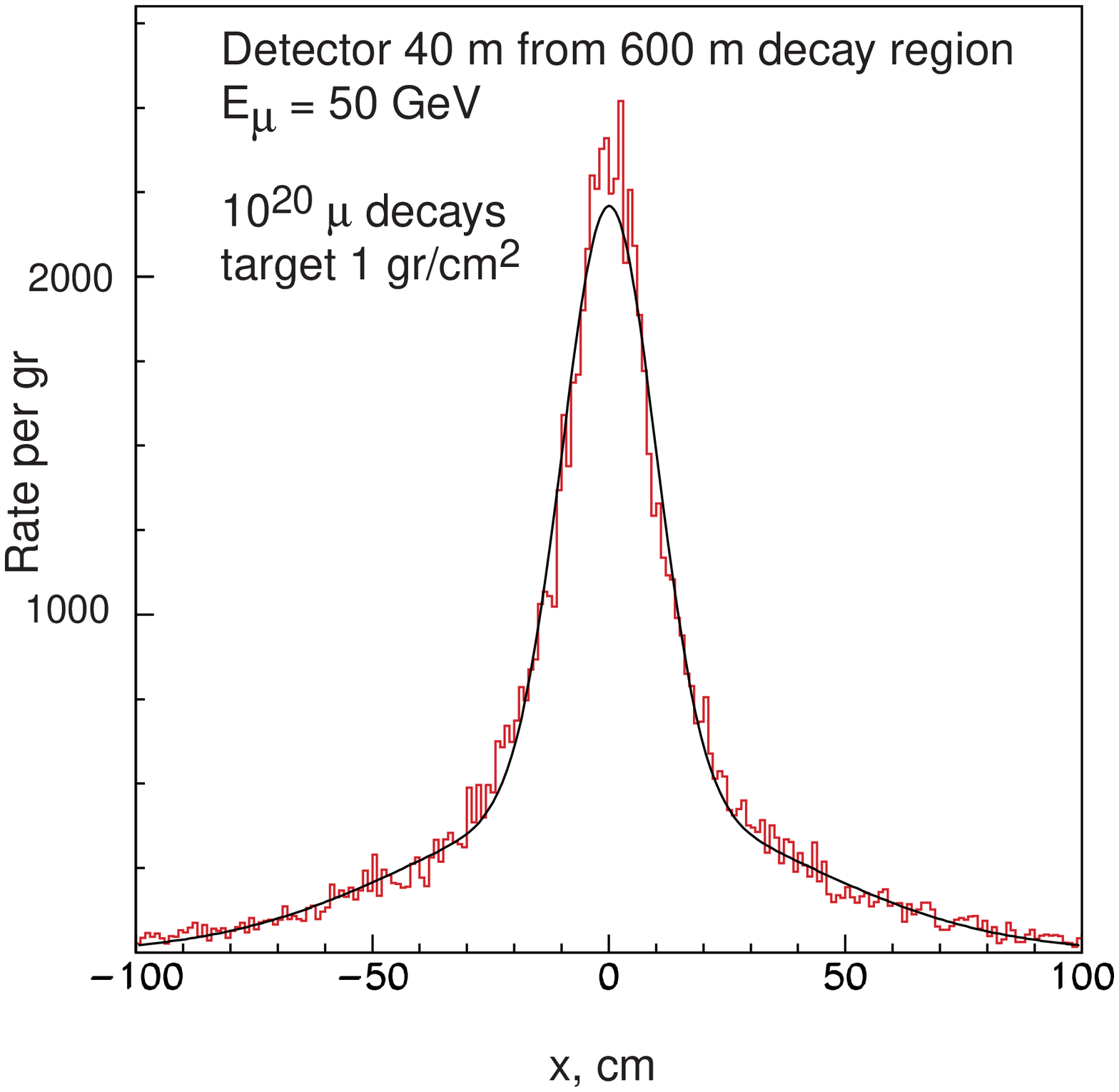}}
\caption{Events per gr/cm$^2$  as a function of the 
transverse coordinate x 50~m downstream of a 50~GeV neutrino 
factory providing $10^{20}$ muon decays. 
The central peak is mainly due to decays 
in the last hundred meters of the decay pipe
while the large tails are due to upstream decays.}
\label{xplot}
\end{figure}

The event distributions measured in a detector close to the neutrino 
factory will be quite different from the corresponding distributions 
at a far site. There are two main reasons for this difference. 
First, the near detector accepts neutrinos over a large range of 
muon decay angles $\theta$, not just those neutrinos traveling in the 
extreme forward direction. This results in a broader neutrino energy 
distribution that is sensitive to the radial size of the 
detector (Fig.~\ref{nearspectra}). 
Second, if the distance of the detector from the end of the beam forming 
straight section is of the order of the straight section length, 
then the $\theta$ acceptance of the detector varies with the position of the
muon decay along the straight section. This results in a more complicated 
radial flux distribution than expected for a far detector. However,
since the dominant effects are decay length and muon decay kinematics,
 it should be modeled quite
accurately.
(Fig.~\ref{xplot}).


\begin{figure}
\epsfxsize 3.in
\centerline{\epsffile{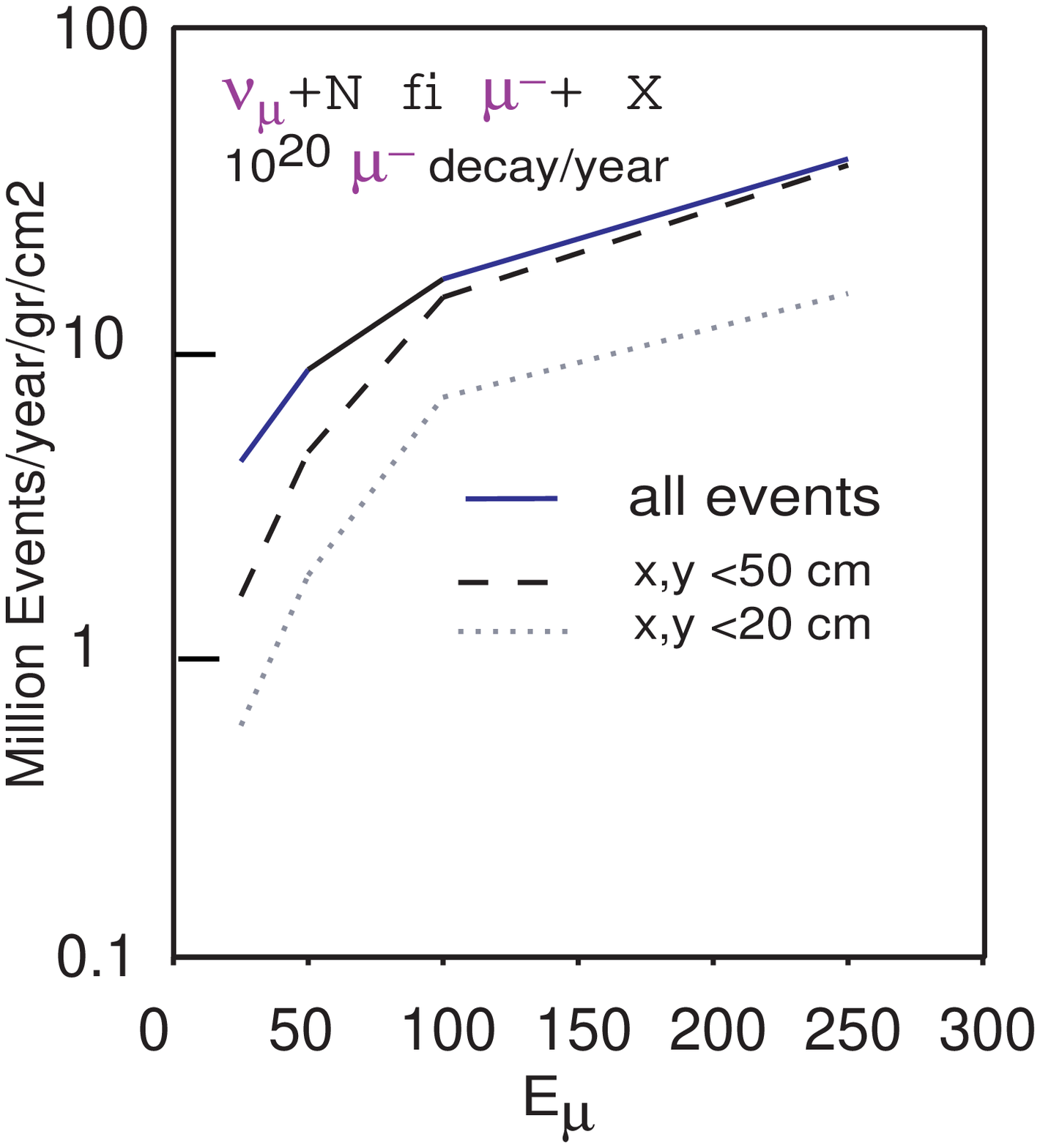}}
\caption{Events per gr/cm$^2$ at a near detector as a function of muon beam energy.  The curves
indicate (solid) all events, the dashed and dotted curves show the effects of
radial position cuts.}
\label{eventrates}
\end{figure}

Note that, even in a limited angular range, the event rates in a 
near detector are very high.  Figure~\ref{eventrates} illustrates the 
event rates per gram/cm$^2$ as a function of energy.  Because
most of the neutrinos produced forward in the center of mass traverse the 
detector fiducial volume, 
the factor of $\gamma^2$
present in the  flux for $\theta \sim  0$ is lost and the event rate 
increases linearly with $E_{\mu}$.  For a 50~GeV
muon storage ring, the interaction rate per 10$^{20}$ muon decays is 
7~million events/gram/cm$^2$.  Rate calculations are discussed further
in the context of specific experiments in the section on 
non--oscillation experiments.
Finally, in the absence of special magnetized shielding, 
the high neutrino event rates in any material
upstream of the detector will cause substantial backgrounds.  
The event rate in the last 3 interaction lengths (300~gr/cm$^2$) 
of the shielding between
the detector and the storage ring would be 30 interactions per beam 
spill at a 15 Hz
machine delivering $2\times 10^{20}$ muon decays per year.  
These high background rates will require clever magnetized shielding 
designs and fast detector
readout to avoid overly high accidental rates in low mass experiments.

\clearpage
\section{Oscillation physics}

The recent impressive atmospheric neutrino results from the SuperK 
experiment have gone a long way towards establishing the existence 
of neutrino oscillations~\cite{superk}. 
Up to the present era, neutrino oscillation 
experiments at accelerators were searches for a phenomenon that might 
or might not be within experimental reach. The situation now is quite 
different. The atmospheric neutrino deficit defines for us the 
$\delta m^2$ and oscillation amplitude that future long-baseline oscillation 
experiments must be sensitive to, namely $\delta m^2 = 
$~\cal{O}($10^{-3}$)~eV$^2$ 
and $\sin^2 2\theta =$~\cal{O}(1). 
Experiments that achieve these 
sensitivities are guaranteed an excellent physics program that addresses 
fundamental physics questions. We can hope that future neutrino 
oscillation experiments will provide the keys we need to understand 
really fundamental questions, for example: the origin of the minute 
neutrino masses and the reason why there are three lepton families.
We cannot guarantee that these insights will be forthcoming from 
neutrino oscillation measurements, but they might be. For this reason 
it is important to understand how our community can get 
detailed experimental information on the neutrino oscillation scheme, 
the mass splittings between the neutrino mass eigenstates, and the 
leptonic mixing matrix that controls the oscillation probabilities. 
A neutrino factory would be a new tool, providing a beam of energetic 
electron neutrinos. In the following we address how this new tool might 
be exploited to go well beyond the capabilities of the next generation 
of neutrino oscillation experiments.


In this section we begin by describing the theoretical basis for 
neutrino oscillations, and then define a selection of oscillation 
parameter sets that can be used in assessing the physics program 
at a neutrino factory. This is followed by a summary of the current 
experimental status and how
it can be expected to change in the next few years. 
We then discuss the parameters and the performance 
of candidate detectors at a neutrino factory. 
The section is completed 
with a survey of the physics measurements that can be 
performed at a neutrino factory as a function of beam energy, 
intensity, and baseline, and finally, a summary of our conclusions.

\subsection{Theoretical framework}
\label{theory}

There exist three known flavors of active neutrinos which
form left-handed doublets with their associated charged leptons. 
The interaction of these active neutrinos with the 
electroweak gauge bosons is described by the Standard Model (SM).  
In principle there can be additional flavors of neutrino 
which are singlets under the electroweak gauge group. 
These electroweak singlet neutrinos do not have electroweak 
couplings, and their interactions are not described by 
the SM. Let us denote the flavor 
vector of the SU(2) $\times$ U(1) active neutrinos as $\nu = 
(\nu_e,\nu_\mu,\nu_\tau)$ and the vector of electroweak-singlet neutrinos 
as $\chi = (\chi_1,..,\chi_{n_s})$.  The Dirac
and Majorana neutrino mass terms can then be written compactly as 
\beq 
-{\cal L}_m = 
{1 \over 2}(\bar\nu_L \ \overline{\chi^c}_L) \left( \begin{array}{cc}
M_L & M_D \\ (M_D)^T & M_R \end{array} \right )\left( \begin{array}{c}
\nu^{c}_R \\ \chi_R \end{array} \right ) + h.c.
\label{numass}
\eeq
where $M_L$ is the $3 \times 3$ left-handed Majorana mass matrix, $M_R$ is a
$n_s \times n_s$ right-handed Majorana mass matrix, and $M_D$ is the 3-row by
$n_s$-column Dirac mass matrix.  In general, all of these are complex, and
$(M_L)^T = M_L \ , \quad (M_R)^T = M_R$.  Without further theoretical input,
the number $n_s$ of ``sterile" electroweak-singlet neutrinos is 
not determined.  
For example, in the SM, minimal supersymmetric standard model (MSSM),
or minimal SU(5) grand unified theory (GUT), $n_s=0$, while in the SO(10) 
GUT, $n_s=3$.  (This is true for both the original non-supersymmetric and the
current supersymmetric versions of these GUTs.) 
Since the terms $\chi_{jR}^TC \chi_{k R}$ are electroweak singlets, the 
elements of the matrix $M_R$, would not be expected to be related
to the electroweak symmetry breaking scale, but instead, would be expected to
be much larger, plausibly of the order of the GUT scale. 

Mechanisms involving $M_L$ only for the generation of neutrino masses without 
the presence of electroweak-singlet neutrinos exist. 
The simplest scenarios, in which one or more Higgs triplets are introduced to 
couple to a pair of left-handed neutrinos, 
are excluded by measurements of the $\rho$ parameter. 
Therefore, other extensions of the SM must be considered, 
for example the addition of one or more Higgs singlets, non-renormalizable
terms involving a large mass scale such as the GUT scale, or R-parity-violating 
terms in the context of supersymmetry.  

The most natural explanation for the three known ultra-light neutrino 
masses is generally regarded to be the seesaw mechanism~\cite{seesaw}, 
which involves $M_R$, and arises from Eq.~(\ref{numass}) 
in the case of $n_s = 3$ electroweak singlet neutrinos.  This leads to 
neutrino masses generically of order 
\beq
m_\nu \sim \frac{m_D^2}{m_R}
\label{seesaw}
\eeq
where $m_D$ and $m_R$ denote typical elements of the corresponding
matrices.  With $m_D \sim m_t$ and $m_R \sim 10^{16}$~GeV,
as suggested in a (supersymmetric) SO(10) grand unified theory framework,
a scale of
$m_\nu \sim 10^{-3}$~eV is readily obtained.
In this case 
the three light neutrino 
masses are obtained by diagonalization of the effective $3 \times 3$ 
light neutrino mass matrix
\beq
M_\nu = - M_D M_R^{-1} M_D^T
\label{meffective}
\eeq
while the super-heavy neutrinos are determined from the right-handed Majorana
matrix $M_R$.

Additional electroweak-singlet neutrinos may arise in string theory 
with the existence of supersymmetric partners of moduli fields, 
resulting in the appearance of $n_\ell$ light sterile neutrinos.
But the presence of these light sterile neutrinos may undermine the
seesaw mechanism and, for this reason, is not very appealing. 
However, if one tries to fit all of the data from the oscillation experiments, 
to obtain a reasonable $\chi^2$ it is necessary to include light sterile
neutrinos. We shall illustrate some of the effects of sterile neutrinos with a
toy model in which one studies the minimal number, $n_\ell=1$.

\subsubsection{Neutrino Oscillations in Vacuum}

The presence of non-zero masses for the light neutrinos introduces a leptonic
mixing matrix, $U$, which is the analogue of the CKM quark mixing matrix, 
and which in general is not expected to be diagonal. 
The matrix $U$ connects the flavor eigenstates 
with the mass eigenstates: 

\begin{equation}
	|\nu_\alpha\rangle = \sum_i U_{\alpha i}|\nu_i\rangle,
\end{equation}

\noindent
where $\alpha$ denotes one of the active neutrino flavors, $e,\ \mu$ or $\tau$
or one of the $n_\ell$ light sterile flavors, while $i$ runs over 
the light mass eigenstate labels.  The number of flavor states considered
here is equal to the number of light mass eigenstates, so $U$ is a square
unitary matrix.  

The neutrino mass differences and the mixing parameters can be probed by 
studying oscillations between different flavors of neutrinos, 
as a function of the neutrino energy $E$ and the distance 
traversed $L$.  
The oscillation probability $P(\nu_\alpha \rightarrow \nu_\beta)$ 
is given by the absolute square of the overlap of 
the observed flavor state, $|\nu_\beta\rangle$, with the time-evolved
initially-produced flavor state, $|\nu_\alpha\rangle$.  In vacuum, the 
evolution operator involves just the Hamiltonian $H_0$ of a free particle, 
yielding the well-known result:
\begin{equation}
\begin{array}{rl}
	P(\nu_\alpha \rightarrow \nu_\beta) =&\left|\langle\nu_\beta | 
		e^{-iH_0L}|\nu_\alpha\rangle\right|^2 
	      =	\sum_{i,j} U_{\alpha i}U^*_{\beta i}U^*_{\alpha j}U_{\beta j}
		e^{-i\delta m^2_{ij}L/2E}\\[0.1in]
	=&P_{\rm CP-even}(\nu_\alpha \rightarrow \nu_\beta) 
		+ P_{\rm CP-odd}(\nu_\alpha \rightarrow \nu_\beta) \; . \\[0.1in]
\end{array}
\end{equation}

\noindent
The CP-even and CP-odd contributions are

\begin{equation}
\begin{array}{rl}
	P_{\rm CP-even}(\nu_\alpha \rightarrow \nu_\beta) =&P_{\rm CP-even}(
		\bar{\nu}_\alpha \rightarrow \bar{\nu}_\beta)\\[0.1in]
  	=&\delta_{\alpha\beta} -4\sum_{i>j}\ Re\ (U_{\alpha i}
		U^*_{\beta i}U^*_{\alpha j}U_{\beta j})\sin^2 
		({{\delta m^2_{ij}L}\over{4E}})\\[0.1in]
	P_{\rm CP-odd}(\nu_\alpha \rightarrow \nu_\beta) =&-P_{\rm CP-odd}(
		\bar{\nu}_\alpha \rightarrow \bar{\nu}_\beta)\\[0.1in]
        =&2\sum_{i>j}\ Im\ (U_{\alpha i}U^*_{\beta i}U^*_{\alpha j}
          U_{\beta j})\sin ({{\delta m^2_{ij}L}\over{2E}})\\[0.1in]
\end{array}
\label{cprels}
\end{equation}
so that
\beq
P(\bar\nu_\alpha \to \bar\nu_\beta)= P(\nu_\beta \to \nu_\alpha) = 
P_{\rm CP-even}(\nu_\alpha \rightarrow \nu_\beta) -
P_{\rm CP-odd}(\nu_\alpha \rightarrow \nu_\beta)
\label{cprels2}
\eeq
where, by CPT invariance, $P(\nu_\alpha \to \nu_\beta) = 
P(\bar\nu_\beta \to \bar\nu_\alpha)$. 
In vacuum the CP-even and CP-odd contributions are even 
and odd, respectively, under time reversal: $\alpha \leftrightarrow \beta$.  
In Eq.~(\ref{cprels}),
 $\delta m^2_{ij} = m(\nu_i)^2 - m(\nu_j)^2$, and the combination 
$\delta m^2_{ij}L/(4E)$ in $\hbar = c = 1$ units can be replaced
by $1.2669 \cdots \delta m^2_{ij}L/E$ with $\delta m^2_{ij}$ 
in ${\rm eV^2}$ and $(L,\ E)$ in $({\rm km,\ GeV})$. 
In  disappearance experiments $\beta = \alpha$ and 
no CP-violation can appear since the product of the mixing matrix
elements is inherently real.  At distances $L$ large compared to all the
individual oscillation lengths, 
$\lambda_{ij}^{\rm osc} \sim E/\delta m^2_{ij}$, the sine
squared terms in $P_{\rm CP-even}$ average to 0.5 whereas the sine terms in
$P_{\rm CP-odd}$ average to zero. 
Therefore CP violating effects are largest and hence easiest to observe
at distances between the smallest and largest oscillation lengths.


\subsubsection{Three Active Neutrinos Only}

With three neutrinos, the mixing matrix $U$ is the $3\times3$ 
unitary Maki-Nagawa-Sakata (MNS) matrix~\cite{mns}. We
parameterize $U$ by
\begin{equation}
U
= \left( \begin{array}{ccc}
  c_{13} c_{12}       & c_{13} s_{12}  & s_{13} e^{-i\delta} \\
- c_{23} s_{12} - s_{13} s_{23} c_{12} e^{i\delta}
& c_{23} c_{12} - s_{13} s_{23} s_{12} e^{i\delta}
& c_{13} s_{23} \\
    s_{23} s_{12} - s_{13} c_{23} c_{12} e^{i\delta}
& - s_{23} c_{12} - s_{13} c_{23} s_{12} e^{i\delta}
& c_{13} c_{23} \\
\end{array} \right) \,,
\end{equation}
where $c_{jk} \equiv \cos\theta_{jk}$ and $s_{jk} \equiv \sin\theta_{jk}$.
For Majorana neutrinos, $U$ contains two further multiplicative phase
factors, but these do not enter in oscillation phenomena.

With the plausible hierarchical neutrino mass spectrum 
$m_1 < m_2 \ll m_3$ 
and the assumption that the LSND effect is not a neutrino oscillation
phenomena, 
we can identify the largest $\delta m^2$ scale 
with the atmospheric neutrino deficit: 
$\delta M^2 = \delta m^2_{atm} 
= \delta m^2_{32} \simeq \delta m^2_{31}$. 
In the approximation that we neglect oscillations 
driven by the small $\delta m^2$ scale, 
the $\nu_e$ oscillation probabilities can be written as
\begin{equation}
\begin{array}{rl}
        P(\nu_e \rightarrow \nu_e) \simeq & 1 
          -4|U_{e3}|^2 (1 - |U_{e3}|^2)
	\sin^2 \Bigl (\frac{\delta m^2_{atm}L}{4E} \Bigr ) \\[0.05in]
	= & 1 - \sin^2(2\theta_{13}) 
	\sin^2 \Bigl (\frac{\delta m^2_{atm}L}{4E} \Bigr ) \; ,
\end{array}
\end{equation}

\beq
\begin{array}{rl}
P(\nu_e \to \nu_\mu) \simeq& 4|U_{e3}|^2 |U_{\mu 3}|^2 
\sin^2 \Bigl ( \frac{\delta m^2_{atm}L}{4E} \Bigr ) \\[0.05in]
=& \sin^2(2\theta_{13})\sin^2(\theta_{23}) 
\sin^2 \Bigl (\frac{\delta m^2_{atm}L}{4E} \Bigr ) \; ,
\end{array}
\label{pnuenumu}
\eeq

\beq
\begin{array}{rl}
P(\nu_e \to \nu_\tau) \simeq& 4|U_{\tau 3}|^2 |U_{e3}|^2
\sin^2 \Bigl ( \frac{\delta m^2_{atm}L}{4E} \Bigr ) \\[0.05in]
=& \sin^2(2\theta_{13})\cos^2(\theta_{23}) 
\sin^2 \Bigl (\frac{\delta m^2_{atm}L}{4E} \Bigr ) \; 
\end{array}
\label{pnuenutau}
\eeq

\noindent
and 
the $\nu_\mu$ oscillation probabilities are

\beq 
\begin{array}{rl}
         P(\nu_\mu \rightarrow \nu_\mu) \simeq &  
       1 -   4|U_{\mu 3}|^2 (1 - |U_{\mu 3}|^2 ) 
    \sin^2 ({{\delta m^2_{atm} L} \over{4E}}) \\[0.05in]
 =& 1 -  4 \sin^2(\theta_{23})\cos^2(\theta_{13}) 
(1 - \sin^2(\theta_{23})\cos^2(\theta_{13})) 
\sin^2 \Bigl (\frac{\delta m^2_{atm}L}{4E} \Bigr ) \; ,
\end{array}
\eeq

\beq
\begin{array}{rl}
P(\nu_\mu \to \nu_e) \simeq& 4|U_{e3}|^2 |U_{\mu 3}|^2 
\sin^2 \Bigl ( \frac{\delta m^2_{atm}L}{4E} \Bigr ) \\[0.05in]
=& \sin^2(2\theta_{13})\sin^2(\theta_{23}) 
\sin^2 \Bigl (\frac{\delta m^2_{atm}L}{4E} \Bigr ) \; ,
\end{array}
\label{pnuenumu}
\eeq

\beq 
\begin{array}{rl}
        P(\nu_\mu \rightarrow \nu_\tau) \simeq & 
          4|U_{\mu 3}|^2 |U_{\tau 3}|^2 
    \sin^2 ({{\delta m^2_{atm} L} \over{4E}}) \\[0.05in]
 =& \sin^2(2\theta_{23})\cos^4(\theta_{13}) 
\sin^2 \Bigl (\frac{\delta m^2_{atm}L}{4E} \Bigr ) \; .
\end{array}
\eeq

The CP-odd contribution to the atmospheric neutrino oscillation probability
vanishes in the one-mass-scale-dominant approximation.
However if we include the effects of the small mass scale, $\delta m^2_{21}$,
then

\begin{equation}
\begin{array}{rl}
        P_{\rm CP-odd}(\nu_\mu \rightarrow & \nu_\tau) 
          = -4c_{12}c^2_{13}c_{23}s_{12}s_{13}s_{23}(\sin \delta)
\\[0.05in]
           & \left [ \sin(\frac{\delta m^2_{21}L}{2E})
           \sin^2(\frac{\delta m^2_{atm}L}{4E}) 
           + \sin(\frac{\delta m^2_{atm}L}{2E})
           \sin^2(\frac{\delta m^2_{21}L}{4E}) \right ] . 
  \end{array}
\end{equation}
At distances significantly larger than the atmospheric neutrino oscillation
length, $ E/\delta m^2_{atm}$, the second term in brackets
averages to zero whereas the $\sin$ squared part of the first term
averages to one half, leaving
\begin{equation}
\begin{array}{rl}
        P_{\rm CP-odd}(\nu_\mu \rightarrow & \nu_\tau) 
          \simeq -2c_{12}c^2_{13}c_{23}s_{12}s_{13}s_{23}(\sin \delta)
  \sin(\frac{\delta m^2_{21}L}{2E}).
   \end{array}
\end{equation}
The Jarlskog factor~\cite{jarlskog}, J, is given by
$J=c_{12}c^2_{13}c_{23}s_{12}s_{13}s_{23}(\sin \delta)$
and is a convenient measure of the size of the CP violation.

\begin{figure}
\epsfxsize3.5in
\centerline{\epsffile{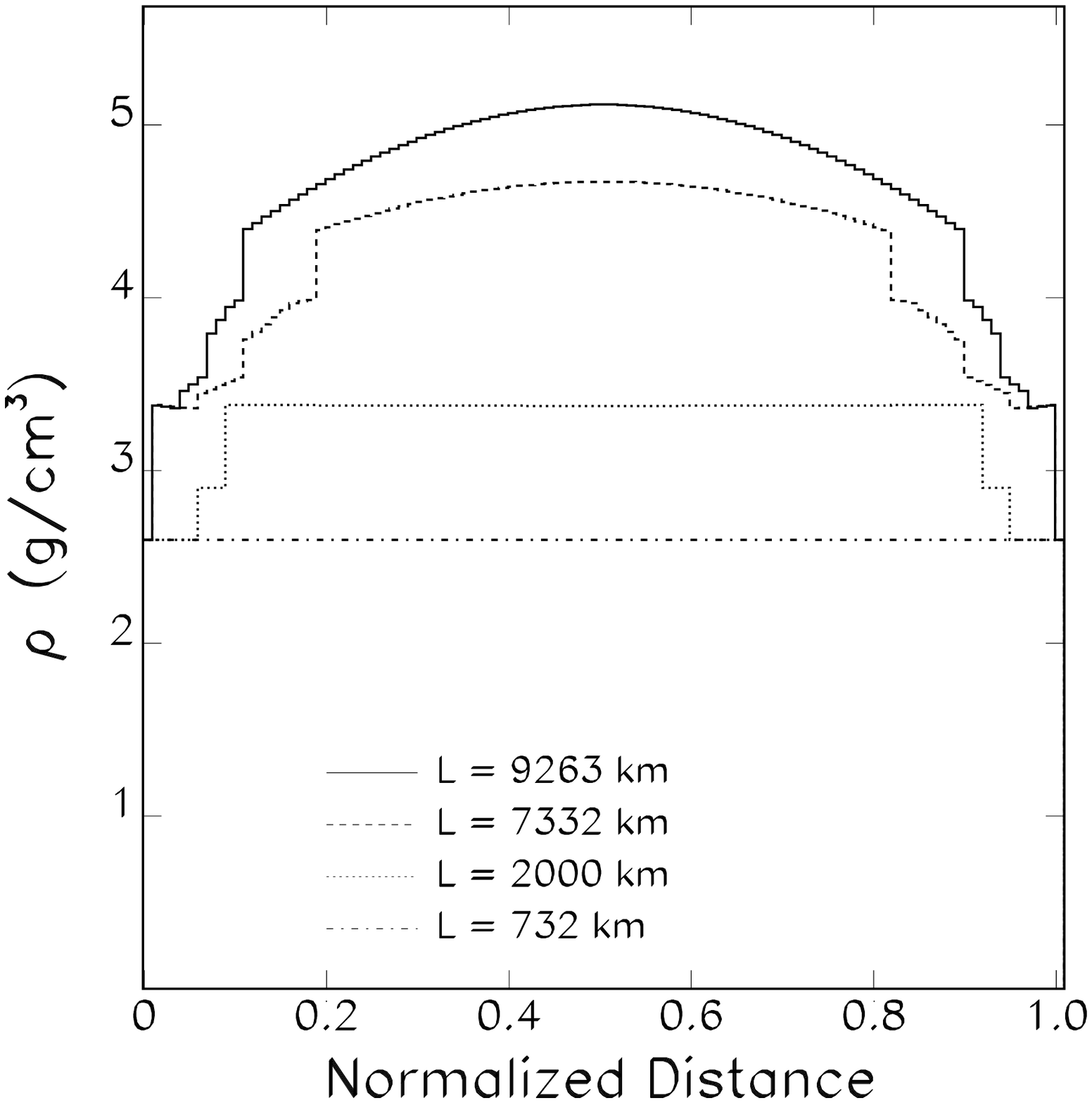}}
\caption{Density profiles for trajectories through the Earth. 
Calculation from Ref.~\ref{bgw99}.
}
\label{profiles}
\end{figure}

If the neutrinos propagate through matter, these expressions must 
be modified.
The propagation of neutrinos through matter is described by the evolution
equation
\begin{equation}
i{d\nu_\alpha\over dt} = \sum_\beta \left[ \left( \sum_j U_{\alpha j} U_{\beta
j}^* {m_j^2\over 2E_\nu} \right) + {A\over 2E_\nu} \delta_{\alpha e}
\delta_{\beta e} \right] \nu_\beta \,,  \label{eq:prop}
\end{equation}
where $A/(2E_\nu)$ is the amplitude for
coherent forward charged-current scattering of $\nu_e$ on electrons,
\begin{equation}
A = 2\sqrt2 G_F N_e E_\nu = 1.52 \times 10^{-4}{\rm\,eV^2} Y_e
\rho({\rm\,g/cm^3}) E({\rm\,GeV}) \,
\label{eq:defnA}
\end{equation}
(for $\bar{\nu_e}$ A is replaced with -A).
Here $Y_e$ is the electron fraction and $\rho(t)$ is the matter density. 
Density profiles through the 
earth can be calculated using the Earth Model~\cite{prem}, 
and are shown 
in Fig.~\ref{profiles}. For
neutrino trajectories through the earth's crust, the density is typically of
order 3~gm/cm$^3$, and $Y_e \simeq 0.5$. 
For very long baselines a constant density approximation is not sufficient
and oscillation calculations must explicitly take account of $\rho(t)$.
However the constant density approximation is very useful to understand 
the physics of neutrinos propagating through the earth since the variation
of the earth's density is not large.

The propagation Eq.~(\ref{eq:prop})
can be re-expressed in terms of mass-squared differences:
\begin{equation}
i{d\nu_\alpha\over dt} = \sum_\beta {1\over2E_\nu} \left[
  \delta m_{31}^2 U_{\alpha 3} U_{\beta 3}^*
+ \delta m_{21}^2 U_{\alpha 2} U_{\beta 2}^*
+ A \delta_{\alpha e} \delta_{\beta e} \right]
\nu_\beta\,.  \label{eq:prop2}
\end{equation}
This evolution equation can be solved numerically for given input values of the
$\delta m^2$ and mixing matrix elements.

In the approximation where we neglect oscillations 
driven by the small $\delta m^2$ scale, the evolution equations are:
\begin{equation}
i {d\over dt}
\left( \begin{array}{c} \nu_e \\ \nu_\mu \\ \nu_\tau  \end{array} \right)
= {\delta m^2\over 2E}
\left( \begin{array}{ccc}
{A\over \delta m^2} + |U_{e3}|^2 & U_{e3}U_{\mu3}^* & U_{e3}U_{\tau3}^* \\
U_{e3}^*U_{\mu3} & |U_{\mu3}|^2 & U_{\mu3}U_{\tau3}^* \\
U_{e3}^*U_{\tau3} & U_{\mu3}^*U_{\tau3} & |U_{\tau3}|^2
\end{array} \right)
\left( \begin{array}{c} \nu_e \\ \nu_\mu \\ \nu_\tau \end{array} \right)
\,.
\end{equation}
For propagation through matter of constant density, the flavor eigenstates are
related to the mass eigenstates $\nu_j^m$ by
\begin{equation}
\nu_\alpha = \sum U_{\alpha j}^m | \nu_j^m \rangle \,,
\end{equation}
where
\begin{equation}
U^m = \left( \begin{array}{ccc}
      0 &  c_{13}^m & s_{13}^m \\
-c_{23} & -s_{13}^m s_{23} & c_{13}^m s_{23} \\
 s_{23} & -s_{13}^m c_{23} & c_{13}^m c_{23}
\end{array} \right)
\label{eq:matter}
\end{equation}
and $\theta_{13}^m$ is related to $\theta_{13}$ by
\begin{equation}
\tan 2\theta_{13}^m = {\sin2\theta_{13}} / \left( {\cos2\theta_{13}
- {A\over \delta m^2}}\right) \,. \label{eq:tan}
\end{equation}
We note that $U^m$ has the form of the vacuum $U$ with the substitutions
\begin{equation}
\theta_{13}\to\theta_{13}^m\,, \quad \theta_{23} \to\theta_{23}\,, \quad
\theta_{12}\to\pi/2\,,\quad \delta = 0 \,.
\end{equation}
Equation~(\ref{eq:tan})
implies that
\begin{equation}
\sin^2 2\theta_{13}^m = \sin^22\theta_{13} /
\left({ \left( {A\over\delta m^2} - \cos 2\theta_{13} \right)^2
+ \sin^2 2\theta_{13}} \right) \,. \label{eq:sin}
\end{equation}
Thus there is a resonant enhancement for
\begin{equation}
A = \delta m^2 \cos2\theta_{13}
\end{equation}
or equivalently
\begin{equation}
E_\nu \approx 15{\rm\ GeV} \left(\delta m^2 \over 3.5\times
10^{-3}{\rm\,eV^2}\right) \left( 1.5{\rm\ g/cm^3}\over \rho Y_e \right)
\cos2\theta_{13} \,. \label{eq:Enu}
\end{equation}
The resonance occurs only for positive $\delta m^2$ for neutrinos 
and only for negative $\delta m^2$ for 
anti-neutrinos.\footnote{If the LSND effect is due to neutrino oscillations
then $\delta m^2 >>$~O($10^{-3}$)~eV$^2$ and 
the  resonance occurs at energies much higher than those of interest 
at the currently invisioned neutrino factory.}
For negative
$\delta m^2$ the oscillation amplitude in Eq.~(\ref{eq:sin}) is smaller than the
vacuum oscillation amplitude. Thus the matter effects give us a way in
principle to determine the sign of $\delta m^2$.

\begin{figure}
\begin{center}
\mbox{
\epsfxsize=6.5truecm
\epsfysize=5.6truecm
\epsffile{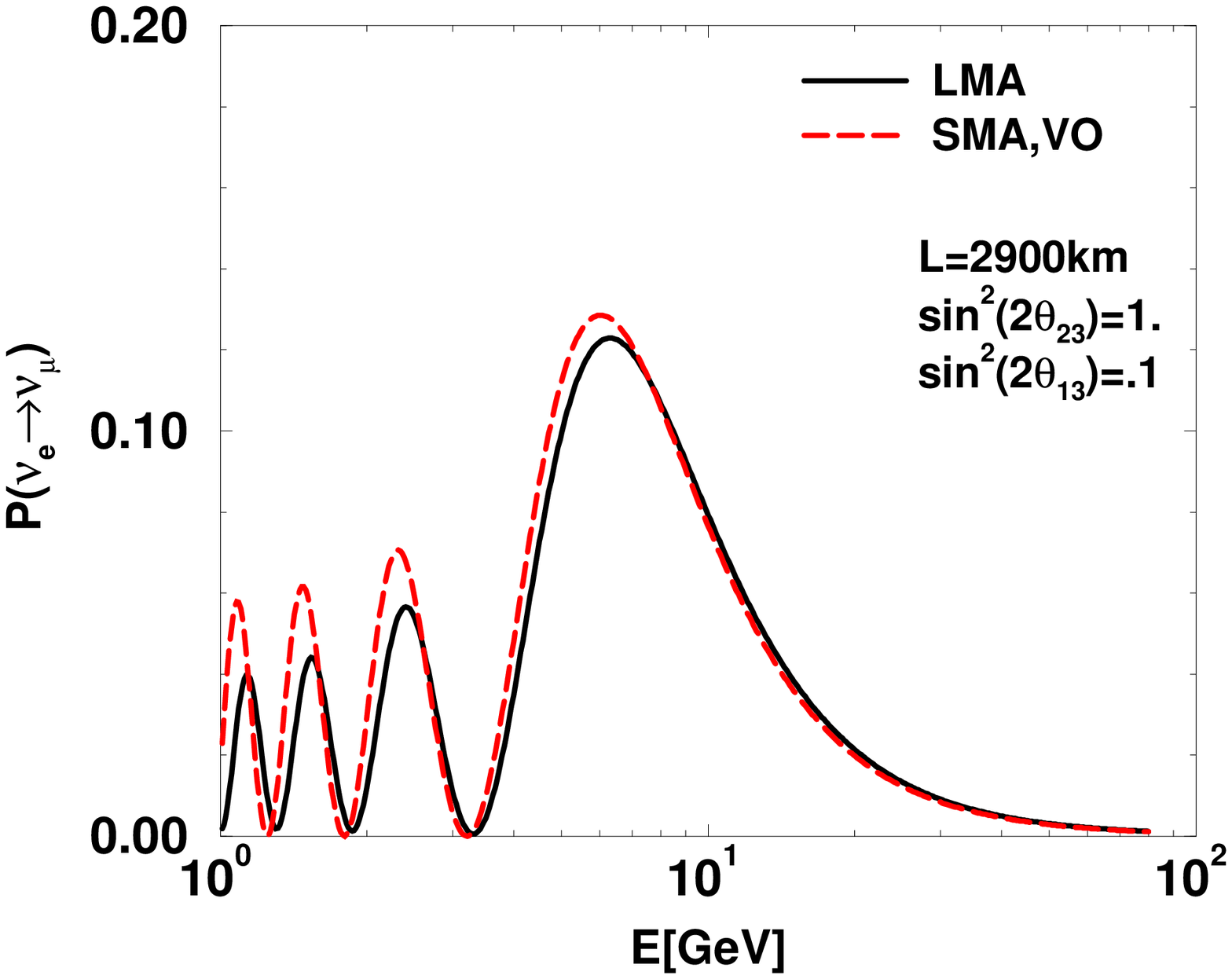}}
\mbox{
\epsfxsize=6.5truecm
\epsfysize=5.6truecm
\epsffile{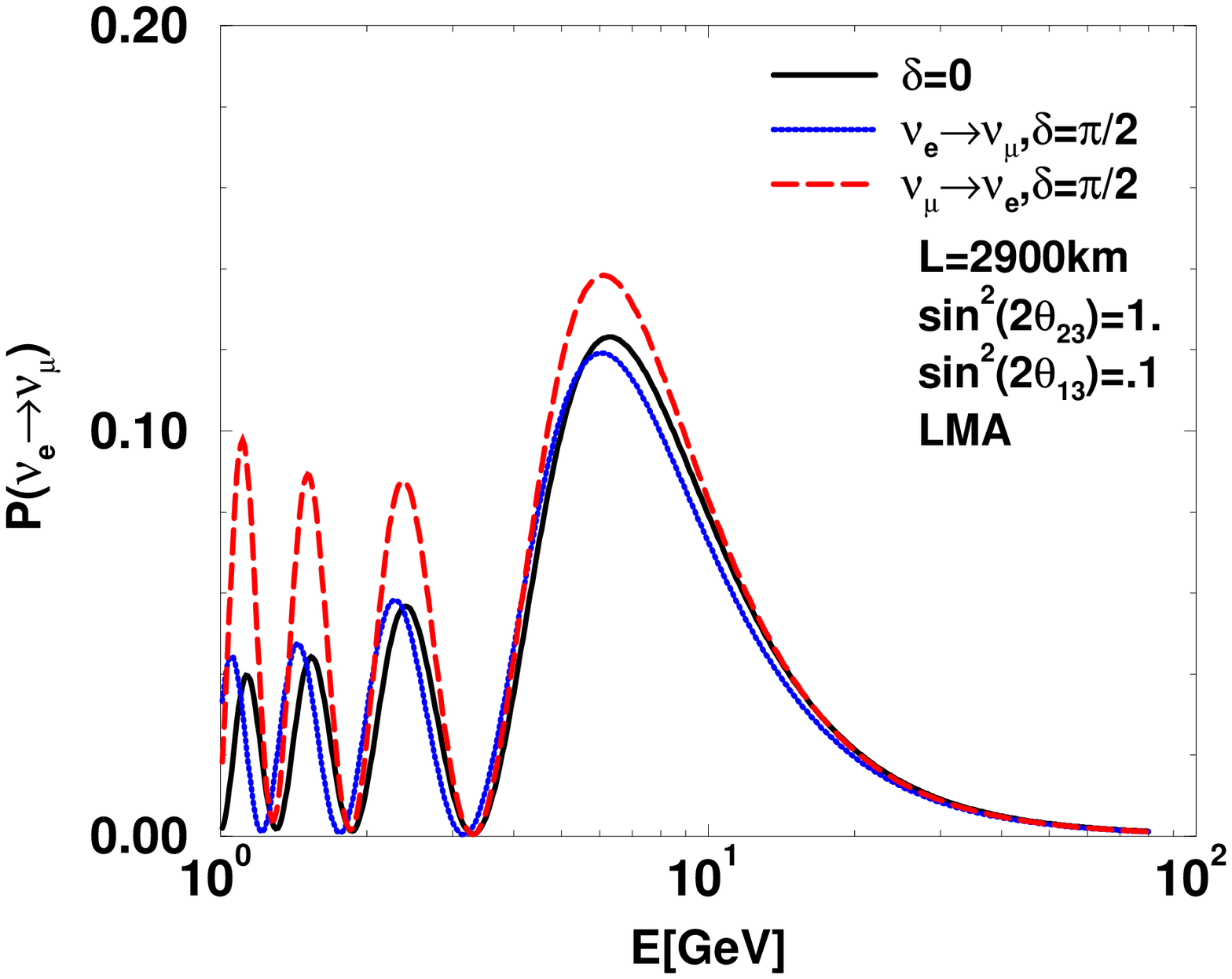}}

\mbox{\epsfxsize=6.5truecm
\epsfysize=6truecm
\epsffile{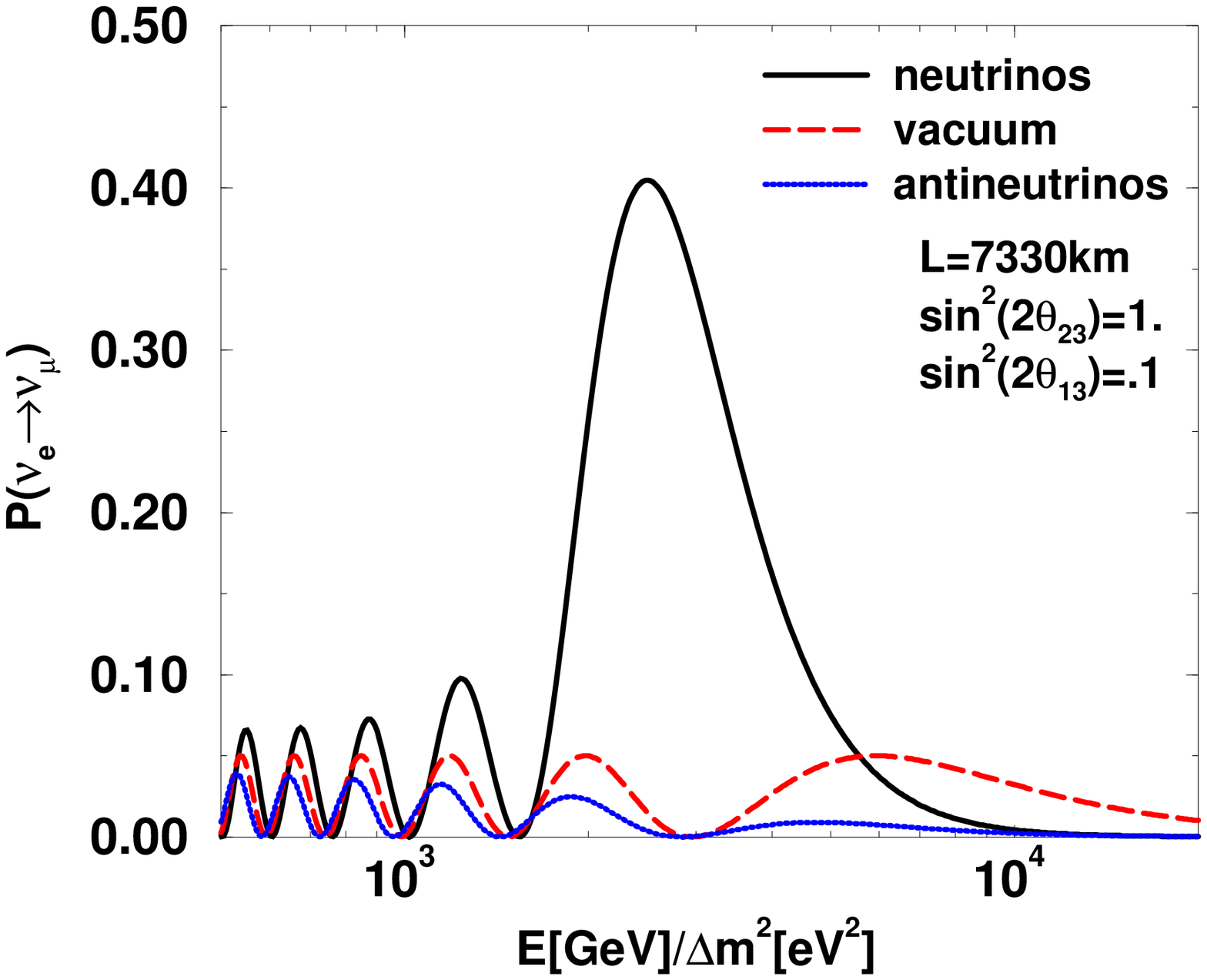}}
\mbox{\epsfxsize=6.5truecm
\epsfysize=6truecm
\epsffile{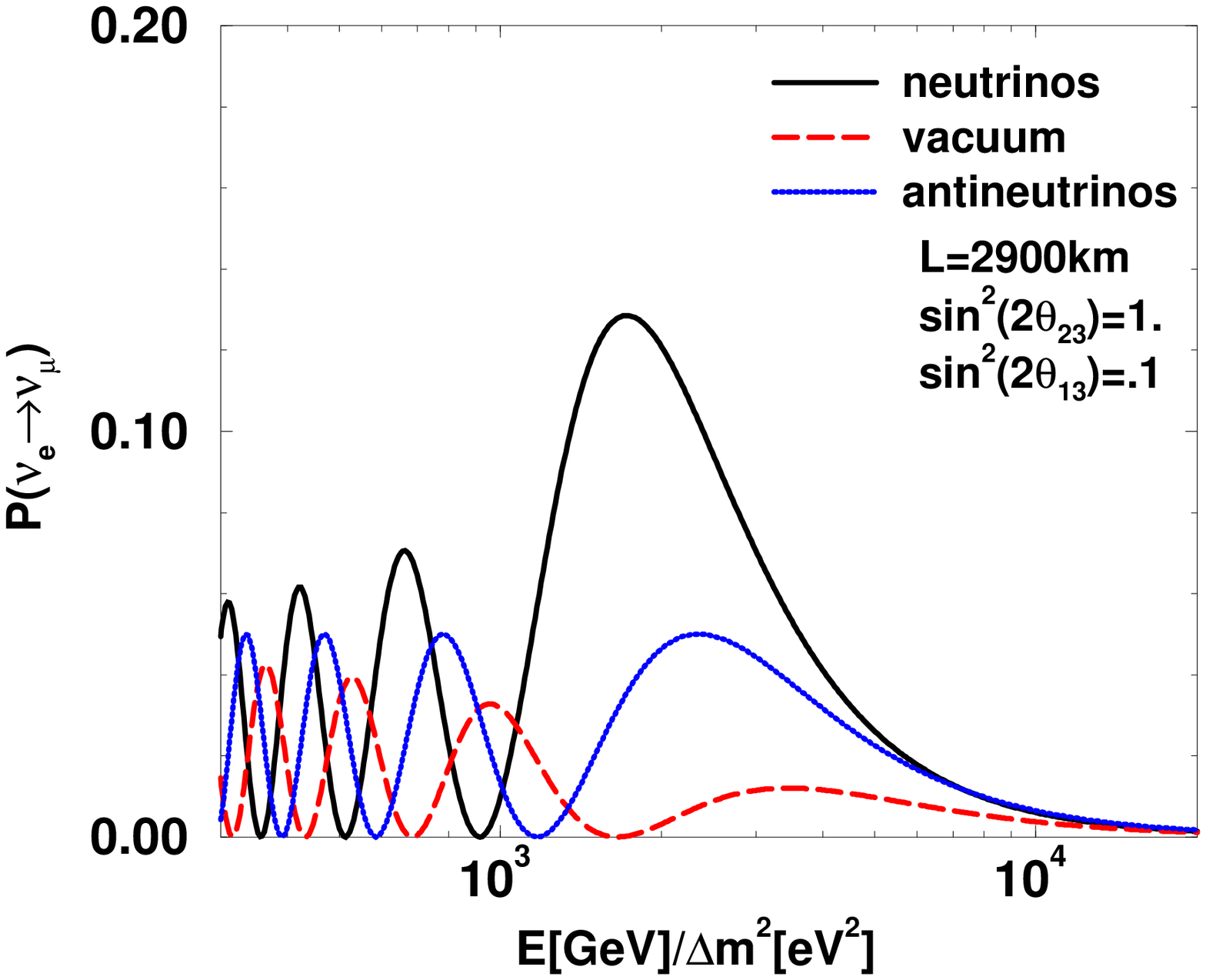}}

\mbox{\epsfxsize=6.5truecm
\epsfysize=6truecm
\epsffile{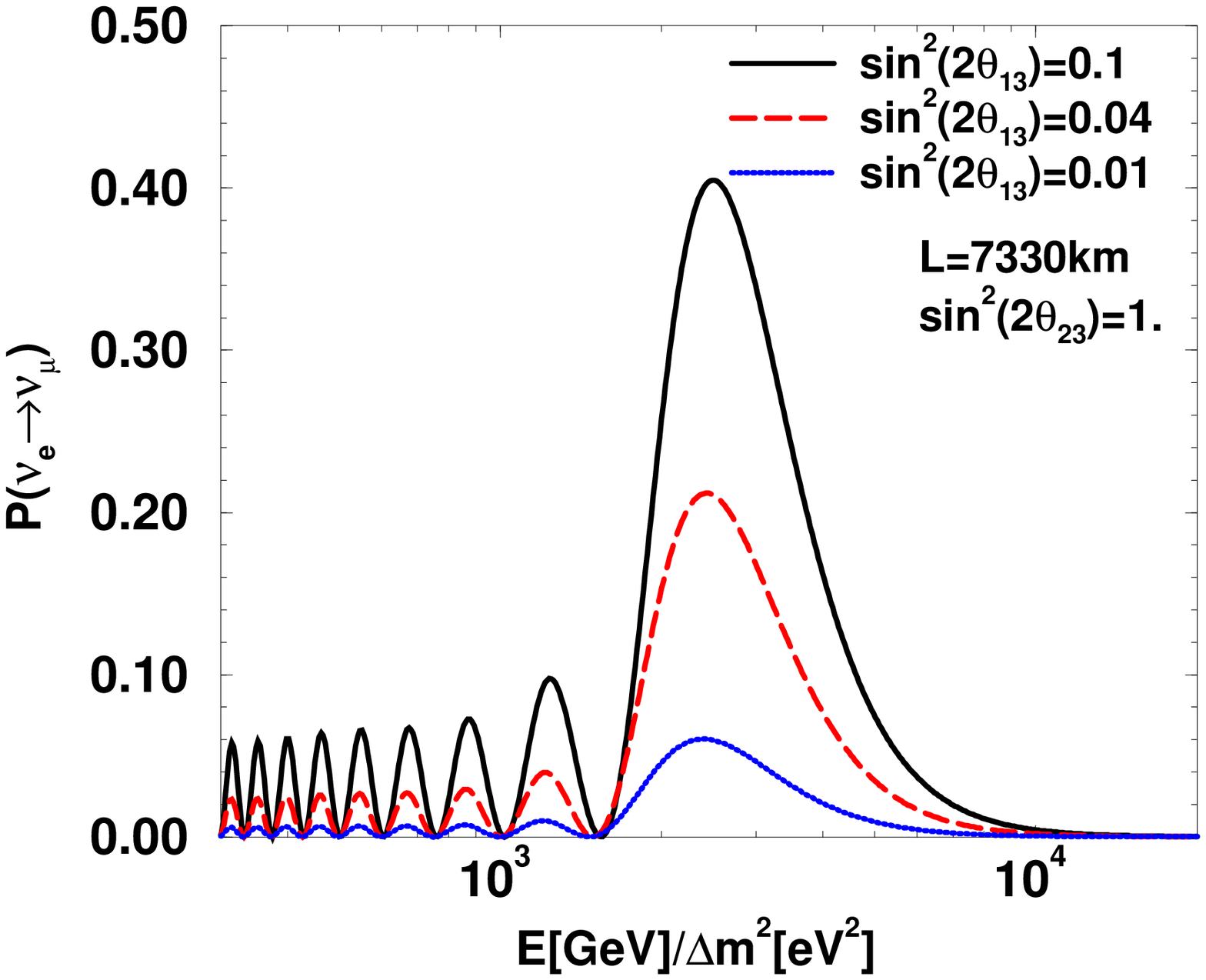}}
\mbox{\epsfxsize=6.5truecm
\epsfysize=6truecm
\epsffile{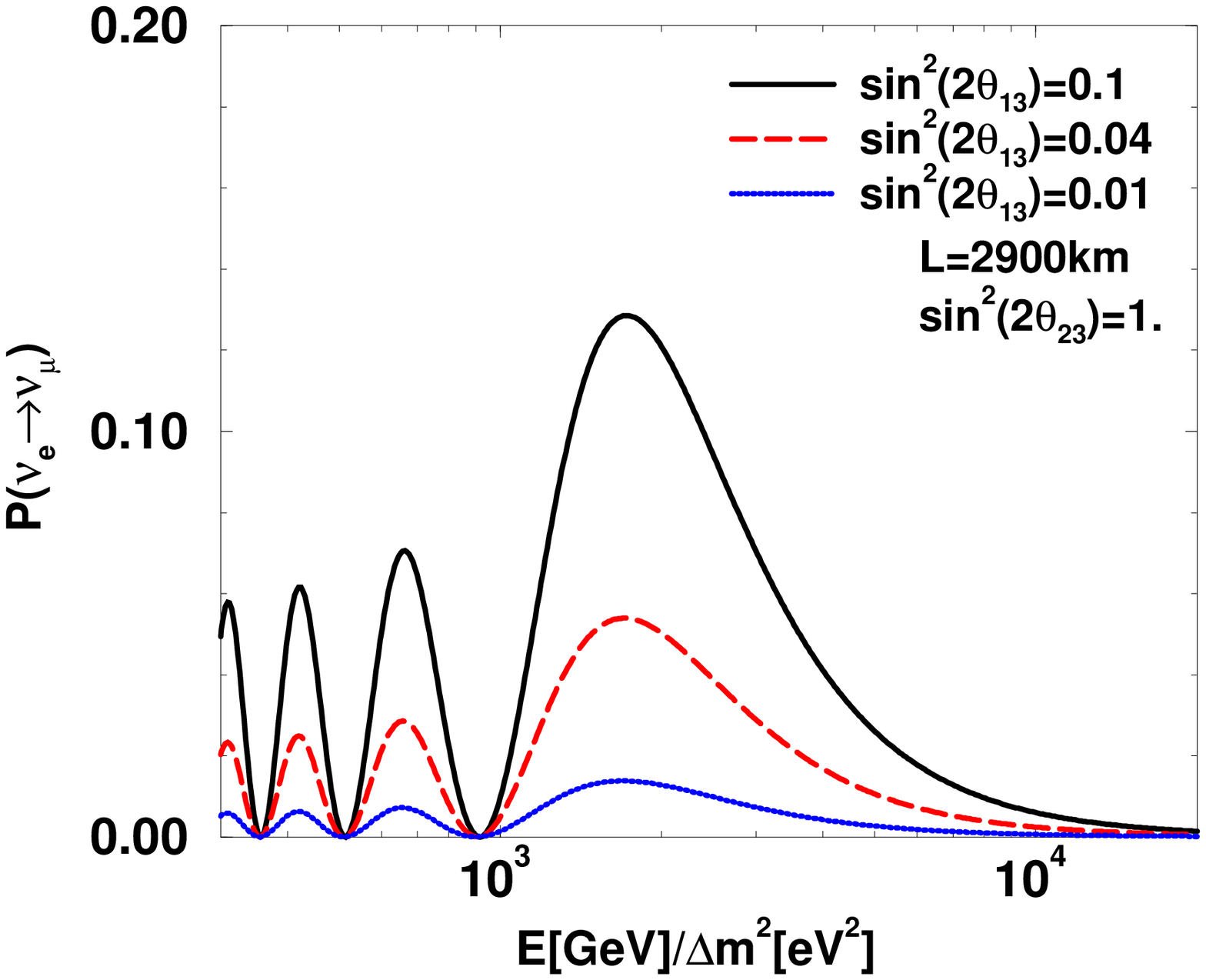}}

\end{center}
\caption{Dependence of $\nu_e\to\nu_\mu$ oscillation probability on neutrino 
energy for some representative oscillation parameters. Plots are from 
Ref.~\ref{lb} and show the effects 
of varying $\delta$ (top plots), matter effects (middle plots), and 
$\sin^22\theta_{13}$ dependence (bottom plots).}
\label{fig:shrock}
\end{figure}

It is instructive to look at the dependence of the oscillation probabilities 
on the neutrino energy as a function of the oscillation parameters and 
the baseline. Some examples from Ref.~\cite{lb} are shown in 
Fig.~\ref{fig:shrock} for $\nu_e\to\nu_\mu$ oscillations. 
Note that for parameters corresponding to the 
large mixing angle MSW solar solution, maximal  CP violation results in 
a small but visible effect. 
Matter effects, which have been computed using the density profile 
from the Earth Model, can have substantial effects, and 
are very sensitive to $\sin^22\theta_{13}$.

\subsubsection{Three Active Flavor Oscillation Scenarios}

We now define some representative three--flavor neutrino 
oscillation parameter sets that can be used 
to establish how well experiments at a neutrino factory 
could determine the oscillation parameters. We begin by 
considering constraints from existing experiments.

If we assume CPT invariance then
the oscillation probability for $\bar{\nu}_e \rightarrow
\bar{\nu}_e$ is equal to that for $\nu_e \rightarrow \nu_e$. 
The CHOOZ results~\cite{chooz} imply:
\begin{equation}
	\sin^2 2\theta_{reac} \equiv 4|U_{e3}|^2 (1 - |U_{e3}|^2) 
        	= \sin^2 2\theta_{13} \le 0.1
\end{equation}

\noindent 
for the range $\delta M^2 \aprge 10^{-3}\ {\rm eV^2}$. 
On the other hand, for the solar neutrino experiments, with $|U_{e3}|^2 \ll 1$,
one finds 

\begin{equation}
 	\sin^2 2\theta_{solar} \equiv 4|U_{e1}|^2|U_{e2}|^2 = 
        	\sin^2 2\theta_{12}\cos^4 \theta_{13} \sim \sin^2 2\theta_{12}
\end{equation}

\noindent
with $\sin^2 2\theta_{12} \sim 0.006$ in the case of the small angle MSW
solution with $\delta m^2_{21} \sim 6 \times 10^{-6}\ {\rm eV^2}$ or 
$\sin^2 2\theta_{12} \sim 1.0$ in the case of the large angle MSW solution 
with $\delta m^2_{21} 
\sim 5 \times 10^{-5}\ {\rm eV^2}$, the LOW solution with $\delta m^2_{21} 
\sim 10^{-7}\ {\rm eV^2}$, or the vacuum solutions with $\delta m^2_{21}
\sim 4 \times 10^{-10}\ {\rm eV^2}$ or $\delta m^2_{21} \sim 8 \times 
10^{-11}\ {\rm eV^2}$.

The atmospheric neutrino oscillation experiments favor $\nu_\mu \rightarrow 
\nu_\tau$ \cite{learned}, and in the one-mass-scale-dominant 
approximation the 
best fit from the SuperK experiment \cite{sk} yields 
\begin{equation}
	\sin^2 2\theta_{atm} \equiv 4|U_{\mu 3}|^2 |U_{\tau 3}|^2 
		= \sin^2 2\theta_{23} \cos^4 \theta_{13} = 1.0
\end{equation}

\noindent
with $\delta m^2_{atm} = 3.5 \times 10^{-3}\ {\rm eV^2}$.  Unpublished
analyses of a substantially enlarged data set by the SuperK experiment have yielded
the same central value for $\sin^2 2\theta_{atm}$ and essentially the same
value of $\delta m^2_{atm}$, $2.5 \times 10^{-3}$ \cite{sk}; we shall use 
the published fits in the following.

Based on these considerations we define the representative three--flavor 
parameter sets shown in Table~\ref{table:3flav}. 
The first three scenarios do not attempt to fit the LSND anomaly. 
These scenarios have
the Atmospheric anomaly explained by $\nu_{\mu} \rightarrow \nu_{\tau}$
oscillation with maximal mixing and the Solar Anomaly explained by
one of the MSW Solar solutions:
\begin{description}

\item{Scenario IA1}  - Large Angle MSW

\item{Scenario IA2} - Small Angle MSW

\item{Scenario IA3} - LOW MSW.

\end{description}

Alternatively we can keep the LSND anomaly, and either drop the 
solar neutrino deficit, or attempt to find a ``fit" (necessarily 
with a poor $\chi^2$) that explains all three neutrino 
anomalies~\cite{allfit}:

\begin{description}

\item{Scenario IB1} - Atmospheric and LSND

\item{Scenario IC1} - Atmospheric, Solar and LSND

\end{description}

For scenario IC1 the Atmospheric anomaly is a mixture of 
$\nu_{\mu} \rightarrow \nu_{\tau}$ and $\nu_{\mu} \rightarrow \nu_{e}$
and the solar electron neutrino flux is reduced by a factor two
independent of energy. There are large contributions to the $\chi^2$ for
this scenario coming from the Atmospheric Neutrino Anomaly as well as the
Homestake (Chlorine) Solar neutrino experiment.

\begin{table}
\caption{Parameters for the three-flavor oscillation scenarios 
defined for the study.}
\vspace{0.1cm}
\begin{tabular}{c|ccccc}
\hline
parameter & IA1 & IA2 & IA3 & 1B1 & 1C1 \\
\hline
$\delta m^2_{32}$ (eV$^2$)&$3.5\times10^{-3}$&$3.5\times10^{-3}$&
$3.5\times10^{-3}$&$3.5\times10^{-3}$&0.3 \\
$\delta m^2_{21}$ (eV$^2$)&$5\times10^{-5}$&$6\times10^{-6}$& 
$1\times10^{-7}$&0.3&$7\times10^{-4}$\\
$\sin^22\theta_{23}$ &1.0&1.0&1.0&1.0&0.53 \\
$\sin^22\theta_{13}$ &0.04&0.04&0.04&0.015&0.036 \\
$\sin^22\theta_{12}$ &0.8&0.006&0.9&0.015&0.89 \\
$\delta$ &0,$\pm\pi/2$&0,$\pm\pi/2$&0,$\pm\pi/2$&
0,$\pm\pi/2$&0,$\pm\pi/2$ \\
\hline
$\sin^22\theta_{atm}$ &0.98&0.98&0.98&0.99& - \\
$\sin^22\theta_{reac}$ &0.04&0.04&0.04&0.03& - \\
$\sin^22\theta_{solar}$ &0.78&0.006&0.88& - & - \\
$\sin^22\theta_{LSND}$ & - & - & - &0.03&0.036\\
$J$ &0.02&0.002&0.02&0.002&0.015\\
\hline
\end{tabular}
\label{table:3flav}
\end{table}

Note that the Jarlskog J-factor
is small for all scenarios. 
It is clear 
that CP violation will be very difficult to observe.

\subsubsection{Three Active and One Sterile Neutrinos}

In order to incorporate the observed 
$\nu_\mu \rightarrow \nu_e$ and $\bar{\nu}_\mu \rightarrow\bar{\nu}_e$
LSND appearance results~\cite{s1LSND}
and achieve an acceptable $\chi^2$ in the fit, it is necessary to
introduce at least one light sterile neutrino.  As discussed earlier, the
theoretical case for sterile neutrinos is unclear, and various neutrino mass 
schemes predict anything from $n_s = 0$ to many.
To admit just one must be
regarded as a rather unnatural choice.
We consider 
this case because it allows us to explain the Atmospheric, 
Solar and LSND anomalies with the fewest number of new parameters.

Scenarios with three nearly degenerate neutrinos (for example 
$m_1 \leq m_2 \leq
m_3 \ll m_4$ or $m_1 \ll m_2 \leq m_3 \leq m_4$) are essentially ruled out
by a Schwarz inequality on the leptonic mixing elements \cite{bilenky}: 
$|U_{\mu 4}U^*_{e4}|^2 \leq |U_{\mu 4}|^2|U_{e4}|^2 \leq 0.008$ which fails
to be satisfied in the allowed LSND region.  Of the two scenarios with
$m_1 < m_2 \ll m_3 < m_4$, the one with $\delta m^2_{21} \sim \delta 
m^2_{solar},\ \delta m^2_{43} \sim \delta m^2_{atm}$ is preferred over 
the other arrangement which is on the verge of being ruled out by the 
Heidelberg-Moscow $\beta\beta_{0\nu}$ decay experiment \cite{h-m} giving 
$\langle m \rangle \leq 0.2$ eV. 

With the three relevant mass scales given by\\
    $$\delta m^2_{sol} = \delta m^2_{21} \ll \delta m^2_{atm} 
	= \delta m^2_{43} \ll \delta m^2_{LSND} = \delta m^2_{32}$$
and the flavors ordered according to $\{s,\ e,\ \mu,\ \tau\}$, the
$4 \times 4$ neutrino mixing matrix depends on six angles and three phases 
and is conveniently chosen to be \cite{donini}

\begin{equation}
\begin{array}{rl}
        U =& \left(\matrix{U_{s1} & U_{s2} & U_{s3}, & U_{s4}\cr
                U_{e1} & U_{e2} & U_{e3}, & U_{e4}\cr
                U_{\mu 1} & U_{\mu 2} & U_{\mu 3}, & U_{\mu 4}\cr
                U_{\tau 1} & U_{\tau 2} & U_{\tau 3}, & U_{\tau 4}\cr}
                \right)\\[0.4in]
          =& R_{14}(\theta_{14},0)R_{13}(\theta_{13},0)
             R_{24}(\theta_{24},0)R_{23}(\theta_{23},\delta_3)
             R_{34}(\theta_{34},\delta_2)R_{12}(\theta_{12},\delta_1)\\
  \end{array}
\end{equation}
\noindent
where, for example, 
$$R_{23}(\theta_{23},\delta_3) = \left(\matrix{ 1 & 0 & 0 & 0 \cr
	0 & c_{23} & s_{23}e^{-i\delta_3} & 0 \cr
	0 & -s_{23}e^{i\delta_3} & c_{23} & 0 \cr
	0 & 0 & 0 & 1 \cr}\right).$$


\noindent
In the limit where the $m_1 - m_2$ and $m_3 - m_4$ pairs are considered 
degenerate, $R_{12}(\theta_{12},\delta_1) = R_{34}(\theta_{34},\delta_{34}) 
= I$, and only four angles and one phase appear in the mixing matrix

\begin{equation}
	U = \left(\matrix{c_{14}c_{13} & -c_{14}s_{13}s_{23}e^{i\delta_3}
        -s_{14}s_{24}c_{23} & c_{14}s_{13}c_{23} -s_{14}s_{24}s_{23}
        e^{-i\delta_3} & s_{14}c_{24}\cr
        0 & c_{24}c_{23} & c_{24}s_{23}e^{-i\delta_3} & s_{24}\cr
        -s_{13} & -c_{13}s_{23}e^{i\delta_3} & c_{23}c_{13} & 0\cr
        -s_{14}c_{13} & s_{14}s_{13}s_{23}e^{i\delta_3} -c_{14}s_{24}c_{23}
        & -s_{14}s_{13}c_{23} -c_{14}s_{24}s_{23}e^{-i\delta_3} & c_{14}c_{24}
        \cr}\right)
\end{equation}

\noindent
with the same angle and phase rotation convention adopted as before.

In this one-mass-scale-dominant approximation with the large mass gap labeled
$\delta M^2 = \delta m^2_{LSND}$, the oscillations are again CP-conserving,
and a short baseline experiment is needed to determine the extra
relevant mixing angles and phase.  The oscillation probabilities of interest
are:

\begin{equation}
\begin{array}{rl}
  P(\nu_e \rightarrow \nu_e) =& 1 - 4c^2_{24}c^2_{23}(s^2_{24} + 
        s^2_{23}c^2_{24})\sin^2\left({{\delta M^2 L}\over{4E}}\right),
        \\[0.1in]
  P(\nu_e \rightarrow \nu_\mu) =& P(\nu_\mu \rightarrow \nu_e) = 
        4c^2_{13}c^2_{24}c^2_{23}s^2_{23}
        \sin^2\left({{\delta M^2 L}\over{4E}}\right),\\[0.1in]
  P(\nu_e \rightarrow \nu_\tau) =& 4c^2_{23}c^2_{24}
        \left[(s^2_{13}s^2_{14}s^2_{23} + c^2_{14}c^2_{23}s^2_{24})\right.
        \\
        & \left. -2c_{14}s_{14}c_{23}s_{23}s_{13}s_{24}\cos \delta_3\right]
        \sin^2\left({{\delta M^2 L}\over{4E}}\right),\\[0.1in]
  P(\nu_\mu \rightarrow \nu_\mu) =& 1 - 4c^2_{13}c^2_{23}(s^2_{23} + 
        s^2_{13}c^2_{23})\sin^2\left({{\delta M^2 L}\over{4E}}\right),
        \\[0.1in]
  P(\nu_\mu \rightarrow \nu_\tau) =& 4c^2_{13}c^2_{23}
        \left[(s^2_{13}s^2_{14}c^2_{23} + c^2_{14}s^2_{23}s^2_{24})\right.
        \\
        & \left. +2c_{14}s_{14}c_{23}s_{23}s_{13}s_{24}\cos \delta_3\right]
        \sin^2\left({{\delta M^2 L}\over{4E}}\right) \; . \\
\end{array}
\end{equation}

If the neutrinos propagate through matter, these expressions 
must be modified. 
Matter effects for the three active and one sterile neutrino scenario 
are similar in nature to those for the three active neutrino case, 
Eq.~(\ref{eq:prop}).
However in Eq.~(\ref{eq:prop}) a flavor diagonal term 
that only contributes to an overall phase has been discarded. 
This term comes from the coherent forward scattering amplitude for the 
active flavors scattering from the electrons, protons and neutrons 
in matter via the exchange of a virtual Z-boson. 
Since the sterile neutrino does not interact with the Z-boson this 
term must be added to the diagonal terms for the active 
neutrinos (or equivalently
subtracted from the diagonal part for the sterile neutrino).
That is in Eq.~(\ref{eq:prop})
\begin{equation}
{A \over 2E_\nu} \delta_{\alpha e} \delta_{\beta e}
\rightarrow
{A \over 2E_\nu} \delta_{\alpha e} \delta_{\beta e}
-{A^\prime \over 2E_\nu} \delta_{\alpha s} \delta_{\beta s}
\end{equation}
\noindent
where $A^\prime$ is given by Eq.~(\ref{eq:defnA}) with
$Y_e$ replaced by $-\frac{1}{2}(1-Y_e)$
for electrically neutral matter.

In order to search for CP violation, at least two mass scales must be 
relevant. 
For simplicity consider 

\begin{equation}
\begin{array}{rl}
        \delta m^2_{21} =& 0, \qquad \delta m^2_{43} = \delta m^2,\\[0.1in]
        \delta m^2_{32} =& \delta m^2_{31} = \delta M^2,\\[0.1in]
        \delta m^2_{42} =& \delta m^2_{41} = \delta M^2 + \delta m^2\\
\end{array}
\end{equation}

\noindent
with five angles and two phases present, since $U_{12}(\theta_{12},\delta_1) 
= I$.  The CP-odd parts of the relevant probabilities are:
\begin{equation}
\begin{array}{rl}
  P_{\rm CP-odd}(\nu_e \rightarrow \nu_\mu) =& 8c^2_{13}c^2_{23}c_{24}c_{34}
        s_{24}s_{34} \sin (\delta_2 + \delta_3)\left({{\delta m^2 L}
        \over{4E}}\right) \sin^2 \left({{\delta M^2 L}\over{4E}}\right)
        \\[0.1in]
  P_{\rm CP-odd}(\nu_e \rightarrow \nu_\tau) =& 4c_{23}c_{24}\left\{2c_{14}
        s_{14}c_{23}s_{23}s_{13}s_{24}(s^2_{13}s^2_{14}-c^2_{14})
        \sin(\delta_2 + \delta_3)\right.\\[0.05in]
        & +c_{14}c_{34}s_{13}s_{14}s_{34}\left[(s^2_{23} - s^2_{24})
        \sin \delta_2 + s^2_{23}s^2_{24}\sin (\delta_2 + 2\delta_3)\right]
        \\[0.05in]
        & + \left. c_{14}c_{24}s_{13}s_{14}s_{23}s_{24}(c^2_{34} - s^2_{34})
        \sin \delta_3 \right\}\\[0.05in]
        & \times \left({{\delta m^2 L}\over{4E}}\right) 
        \sin^2 \left({{\delta M^2 L}\over{4E}}\right)\\[0.1in]
  P_{\rm CP-odd}(\nu_\mu \rightarrow \nu_\tau) =& 8c^2_{13}c^2_{23}c_{24}
        c_{34}s_{34}\left[c_{14}c_{23}s_{13}s_{14} \sin \delta_2 + 
        c^2_{14}s_{23}s_{24} \sin (\delta_2 + \delta_3)\right]\\[0.05in]
        & \times \left({{\delta m^2 L}\over{4E}}\right) 
        \sin^2 \left({{\delta M^2 L}\over{4E}}\right)\\
\end{array}
\end{equation}

\noindent
where only the leading order term in $\delta m^2$ has been kept. The 
CP-even expressions also have such additional small corrections.

The present atmospheric neutrino data favors the $\nu_\mu \rightarrow 
\nu_\tau$ oscillation over the $\nu_\mu \rightarrow \nu_{s}$ oscillation.
On the other hand, if a solar neutrino oscillates significantly into a
sterile neutrino, only the small angle MSW solution is viable since  
the large angle solutions fail to provide enough $\nu + e^- \rightarrow 
\nu + e^-$ elastic scattering to be consistent with SuperK measurements~\cite{sk}.  
Hence 
if it turns out that one of the large angle mixing solutions is the correct
solution to the solar anomaly then
something other than a single light sterile neutrino will be needed
to explaining 
the solar, atmospheric and LSND results.

\subsubsection{Scenarios with Three Active plus One Sterile Neutrino}

We now consider some representative four--flavor neutrino oscillation 
parameter sets that can be used to establish how well experiments 
at a neutrino factory could determine the oscillation parameters.
As was noted earlier, the only viable solutions with one sterile
and three active neutrinos require that there be two sets of
almost degenerate neutrinos separated by the largest $\delta m^2$.
We begin by considering the constraints from CHOOZ and LSND. 
Note that the effective two-component atmospheric 
and solar mixing angles are:
\begin{equation}
\begin{array}{rl}
	\sin^2 2\theta_{atm} =&4|U_{\mu 3}|^2 |U_{\mu 4}|^2
		= c^4_{23}c^4_{13}\sin^2 2\theta_{34}\\[0.1in]
	\sin^2 2\theta_{sol} =&4|U_{e1}|^2 |U_{e2}|^2
		= c^4_{24}c^4_{23}\sin^2 2\theta_{12}\\
\end{array}
\end{equation}
The CHOOZ constraint~\cite{chooz} from $P(\bar{\nu}_e \rightarrow 
\bar{\nu}_e)$ is:

\begin{equation}
        c^2_{23}\sin^2 2\theta_{24} + c^4_{24}\sin^2 2\theta_{23} \le 0.2
\end{equation}

\noindent
while the LSND constraint~\cite{s1LSND} from $P(\nu_\mu \rightarrow \nu_e)$ is:

\begin{equation}
        10^{-3} \leq c^2_{13}c^2_{24}\sin^2 2\theta_{23} \le 10^{-2} \; .
\end{equation}
With this in mind, 
the parameter sets we have defined are summarized in Table~\ref{table:4flav}. 
They are: 
\begin{description}
\item{Scenario IIA1} - Low Mass LSND
\item{Scenario IIB1} - High Mass LSND
\end{description}

\begin{table}
\caption{Parameters for the four-flavor oscillation scenarios 
defined for the study. Note that for these parameter sets 
$\delta m^2_{41}\sim\delta m^2_{31}\sim\delta m^2_{42}
\sim\delta m^2_{32}\equiv\delta M^2$, 
and 
$\sin^22\theta_{14}=\sin^22\theta_{13}=\sin^22\theta_{24}=\sin^22\theta_{23}$}
\bigskip
\begin{center}
\begin{tabular}{c|cc}
\hline
parameter & IIA1 & IIB1 \\
\hline
$\delta m^2_{43}$ (eV$^2$)&$3.5\times10^{-3}$&$3.5\times10^{-3}$ \\
$\delta m^2_{21}$ (eV$^2$)&$6\times10^{-6}$&$6\times10^{-6}$ \\
$\delta M^2$ (eV$^2$)     &0.3&1.0 \\

$\sin^22\theta_{34}$ &1.0&1.0\\
$\sin^22\theta_{12}$ &0.006&0.006 \\
$\sin^22\theta_{14}$ &0.03&0.003 \\
$\delta_1$ &0&0\\
$\delta_2$ &0,$\pm\pi/2$&0,$\pm\pi/2$ \\
$\delta_3$ &0&0\\
\hline
\end{tabular}
\end{center}
\label{table:4flav}
\end{table}

\subsection{Where will we be in 5-10 years ?}
\label{future}

\begin{figure}
  \epsfxsize=0.8\textwidth
  \centerline{\epsfbox{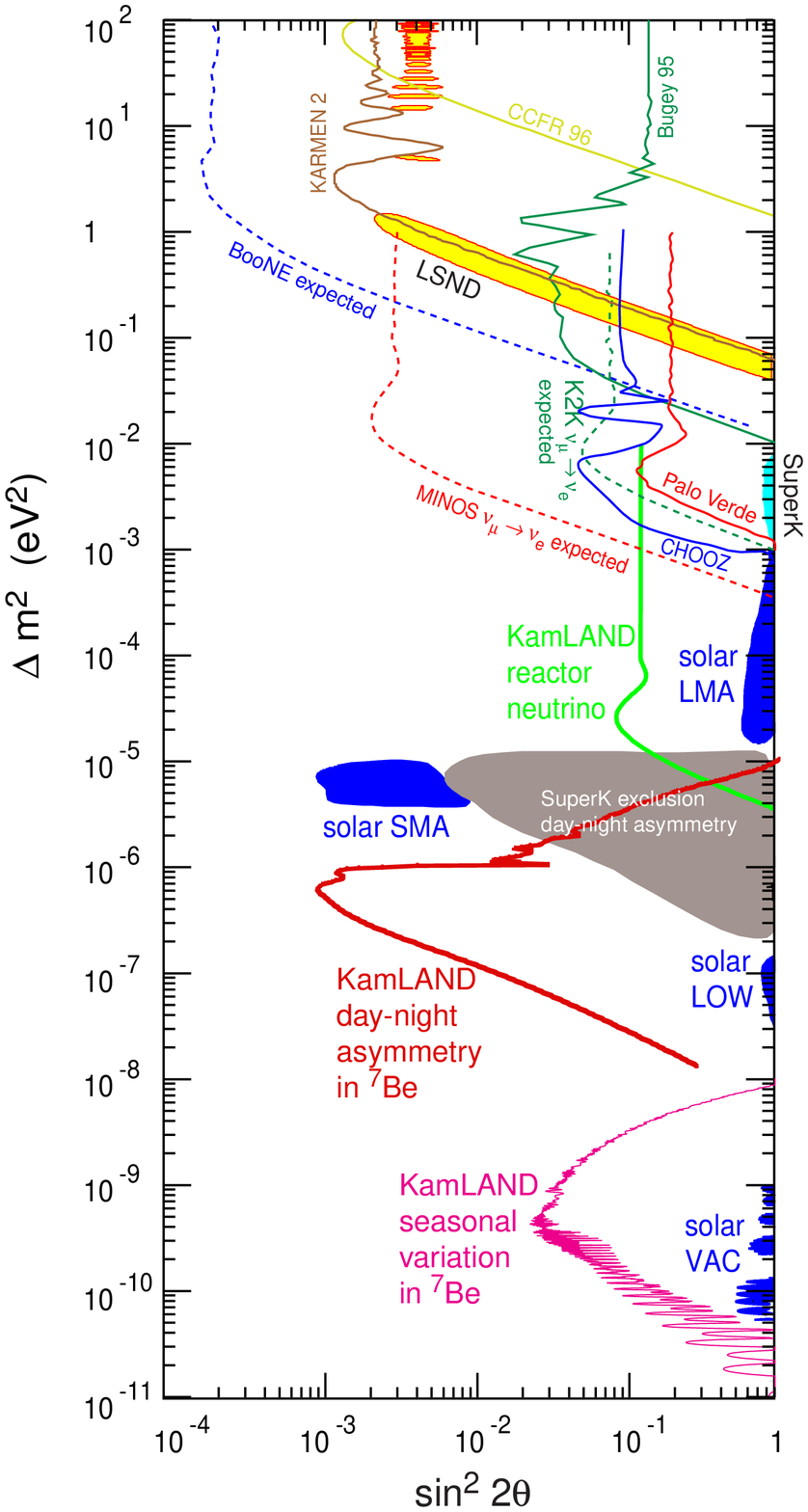}}
  \caption{The current and expected limits at some of the future
    neutrino oscillation experiments.  Note that different 
    oscillation modes are shown together.}
  \label{fig:everything}
\end{figure}

In this section, we briefly discuss the prospects for currently 
operating, planned, 
or proposed experiments exploring neutrino oscillations.  
The discussion will be broken down according to the various oscillation modes.
The current limits and the expected reach of some of the future
experiments are summarized in Fig.~\ref{fig:everything}, and 
Tables~\ref{expt_table} and \ref{3nu_table}.

\subsubsection{$\nu_\mu \rightarrow \nu_\tau$, $\nu_s$}

The evidence for $\nu_\mu$ disappearance in atmospheric neutrinos at
SuperK is convincing \cite{evidence}. 
The preferred region of parameter space is given~\cite{Kajita} by 
$10^{-3} < \delta m^2 < 10^{-2}$~eV$^2$ 
at near maximal mixing ($\sin^2 2\theta \sim 1$). 
The $\nu_\mu \leftrightarrow \nu_e$ interpretation of the atmospheric 
neutrino deficit 
is disfavored by the SuperK data and is ruled out by the CHOOZ~\cite{CHOOZ} 
and PaloVerde \cite{PaloVerde} experiments.  
The two immediate issues are 
(1) the precise determination of $\delta m^2$ and $\sin^2 2\theta$, and 
(2) discrimination between $\nu_\mu\to\nu_\tau$ and $\nu_\mu\to\nu_s$.  

Future SuperK data will probably not shrink the currently preferred
region of parameter space by very much. The precision with which 
$\delta m^2$ can be extracted from the observed event distributions 
depends on the precisions with which the event--by--event values 
of $L$ and $E$ are determined. The greatest sensitivity to $\delta m^2$ 
comes from the sample of events with values 
of $L/E$ corresponding to the region of the first oscillation maximum, 
which in practice are those events 
from neutrinos coming from approximately the horizontal direction. 
However, for these 
events the $L$ precision is limited by the angular resolution of the detector. 

SuperK can discriminate between $\nu_\mu \to \nu_s$ 
and $\nu_\mu \to \nu_\tau$ by looking for matter effects and 
by measuring the number of NC interactions.  
The lack of evidence for matter effects in
up-going muons (in both partially-contained events and in
the NC-enriched multi-ring event sample) already disfavors 
$\nu_\mu\to\nu_s$ at the 99\% confidence level. 
This result is expected to become firmer in the future. 
In addition, if there is 
a significant fraction of incident sterile neutrinos, there will
be fewer NC events detected.  The cleanest sample of NC  
events is the sample of events with a single detected $\pi^0$.  
By comparing the ratio of $\pi^0$ (``two--electron")  
events to $\nu_e$ CC (single--electron) events, SuperK already 
has a statistically significant handle on the NC/CC ratio ($\pm6\%$ for 
848.3 days livetime).  However, the measurement is currently limited by 
large uncertainties on 
the NC single $\pi^0$ cross section 
($23-25\%$ total systematic).  K2K will measure this cross section 
in its near detector, and over the next few years this new 
information may produce  
the most dramatic improvement in $\nu_\tau/\nu_s$ 
discrimination~\cite{learned}.  

The next generation of long-baseline experiments 
have been designed to be sensitive to oscillations with 
parameters that correspond to the SuperK favored region.
The currently running K2K experiment~\cite{K2K} 
will cover $\delta m^2 > 2 \times 10^{-3}~{\rm
  eV}^2$ after 3--5 years of running, and MINOS~\cite{numi} will
cover $\delta m^2 > 0.6 \times 10^{-3}~{\rm eV}^2$ (both at 90\%
CL and for maximal mixing).  These experiments are expected to confirm the
neutrino oscillation interpretation of the atmospheric neutrino data 
by about 2005.  In addition to searching for $\nu_\mu$ disappearance, 
K2K can also look for a distortion in
the neutrino energy spectrum using quasi-elastic events, and MINOS can
study NC/CC event energy distributions.  
In 2005 two experiments, OPERA \cite{OPERA} and ICANOE \cite{ICANOE}, 
are expected to begin taking data at the CNGS beam.
Both OPERA and ICANOE aim primarily
at $\tau$-appearance and will cover $\delta m^2 > 2 \times
10^{-3}~{\rm eV}^2$ after about 5~years of running.  

Unless $\delta m^2 < 2 \times 10^{-3}~{\rm eV}^2$
(allowed at 99\% CL at SuperK), we expect to have a
complete accelerator based experimental 
confirmation of atmospheric neutrino oscillations by 2010, 
and measurements of $\delta m^2$ and $\sin^2 2\theta$ at the 
O(10\%) level (see Table~\ref{3nu_table}). 
If $\delta m^2 < 2 \times 10^{-3}~{\rm eV}^2$  some 
additional experiments will be necessary, such as MONOLITH
(30~kt calorimeter)~\cite{MONOLITH} or AQUARICH (novel 1~Mt Water
Cerenkov)~\cite{AQUARICH}.  These experiments may study atmospheric neutrinos 
and exploit good angular resolution to 
search for dips in the zenith angle (or $L/E$) distribution.

\subsubsection{$\nu_\mu \leftrightarrow \nu_e$}

For large $\delta m^2 \sim 1 ~{\rm eV}^2$ suggested by the 
LSND experiment~\cite{s1LSND}, 
Mini-BooNE~\cite{Mini-BooNE} is expected to cover the entire preferred region of 
LSND parameter space with a wide safety margin.  
In the event of a positive signal,
they plan to build another detector (BooNE) that will be able to
measure the parameters with a precision O(10\%) (see Table~\ref{3nu_table}).
Independent confirmation from ICANOE would also be expected.
Should all of the experimental indications for oscillations 
(LSND, atmospheric, and solar) be confirmed we may be seeing evidence for 
the existence of sterile neutrinos. This would be a very exciting 
discovery, would raise many questions, and would require a 
new round of experiments.

For $\delta m^2 \sim 10^{-3}\mbox{--}10^{-2}~{\rm eV}^2$ we expect
some $\nu_\mu \leftrightarrow \nu_e$ mixing if the heavier of 
the two mass eigenstates
involved in the atmospheric neutrino oscillation contains any 
admixture of $\nu_e$ ({\it i.e.}\/, if $U_{e3} \neq 0$).  Current
limits from CHOOZ~\cite{CHOOZ} and Palo Verde \cite{PaloVerde} require
$|U_{e3}| < 0.1$.  SuperK by itself is unlikely to
improve on this sensitivity.  K2K can look for $\nu_e$ appearance and
improve the sensitivity to a finite $|U_{e3}|$ 
 in some $\delta m^2$ range.  MINOS and ICANOE are expected to
be sensitive to $\sin^2 2\theta_{13} > {\cal{O}}(10^{-2})$  
in the $\delta m^2$ region of interest 
by searching for $\nu_e$ appearance in their predominantly $\nu_\mu$ beams. 
At this time it is not clear what is the interesting range 
for $\sin^2 2\theta_{13}$.  
If this mixing angle is not too small then K2K/MINOS/ICANOE
can make a first measurement. The baselines for these 
experiments are too short, and statistics will be too limited, 
to observe matter effects. 
For very small mixing angles, comparable
with the Small Mixing Angle MSW solution for the solar 
neutrino deficit (see 
\cite{Valle}), an order of magnitude improvement in sensitivity 
beyond these experiments is required to make a first observation 
of $\nu_\mu\to\nu_e$ oscillations.

\subsubsection{$\nu_e \rightarrow \nu_\mu$, $\nu_\tau$, $\nu_s$}

Reactor and solar neutrino experiments can only look for these 
oscillations in the $\overline{\nu}_e$ disappearance mode.  

The SNO~\cite{SNO} detector 
should discriminate between $\nu_e \rightarrow \nu_{\mu,\tau}$
and $\nu_e \rightarrow \nu_s$ solutions to the solar neutrino
deficit by studying the distortion in the $\nu_e$ energy spectrum
and by measuring the NC/CC ratio.
The spectral distortion should occur for the SMA solution 
and for some regions of the VAC solution.
Borexino \cite{Borexino}
(or possibly KamLAND) will study $^7$Be solar neutrinos, and
should see day/night effects for the LOW scenario and seasonal effects for the VAC
solution.  The absence of the $^7$Be electron neutrino 
flux would strongly suggest the SMA
solution.  There are additional experiments proposed to study lower energy
neutrinos (esp. $pp$): HELLAZ, HERON, LENS, etc (see \cite{Lanou} for 
a recent overview).  
KamLAND \cite{KamLAND} will look for the disappearance of 
$\bar{\nu}_e$ from reactors with sensitivity down to $\delta m^2
> 10^{-5}~{\rm eV}^2$ for large mixing angles.
With all of this data in the next 5-10 years we should have convincingly 
tested whether or not any of the current neutrino oscillation
solutions to the solar neutrino problem are correct.

None of the solar neutrino experiments, however, discriminate between 
$\nu_e\rightarrow \nu_\mu$ and $\nu_e\to\nu_\tau$.  MINOS, OPERA, and 
ICANOE can look for $\tau$ appearance but cannot separate 
$\nu_e \to \nu_\tau$ from $\nu_\mu \to \nu_\tau$. 

\subsubsection{Summary}

To summarize, in 5--10~years:
\begin{description}
\item{(i)} $\nu_\mu\to\nu_\tau,\nu_s$. 
If the $\delta m^2$ associated with the atmospheric $\nu_\mu$ deficit 
exceeds $\sim2\times10^{-3}$~eV$^2$ accelerator experiments will 
measure $\delta m^2$ and $\sin^2 2\theta$ with precisions O(10\%). 
If $\delta m^2$ is less than $\sim2\times10^{-3}$~eV$^2$ new 
experiments will be required to accomplish this in the 2010 era.
\item{(ii)} $\nu_\mu \leftrightarrow \nu_e$. 
If the LSND oscillations are confirmed BooNE would measure the 
associated $\delta m^2$ and $\sin^2 2\theta$ with precisions O(10\%). 
However the oscillation framework (sterile neutrinos ?) might be 
complicated. If LSND is not confirmed 
and if $\sin^22\theta_{13} > 10^{-2}$, the first evidence for a finite 
value of $\sin^22\theta_{13}$ would be expected at long baseline 
accelerator experiments. If $\sin^22\theta_{13} < 10^{-2}$ then 
$\nu_\mu \to \nu_e$ will not be observed in the accelerator experiments 
and new experiments with at least an order of magnitude improved 
sensitivity will be needed.
\item{(iii)} $\nu_e \rightarrow \nu_\mu$, $\nu_\tau$, $\nu_s$. 
Either one or none of the current solar neutrino deficit solutions 
will be remaining. If one survives, we will know whether the 
solar neutrino deficit is due to $\nu_e\to\nu_s$. If the 
$\nu_e\to\nu_\mu,\nu_\tau$ mode is favored we 
will not be able to distinguish between $\nu_e\to\nu_\mu$ or 
$\nu_e\to\nu_\tau$. In addition, $\nu_e\to\nu_\tau$ will not have 
been observed at long baseline accelerator experiments.
\item{(iv)} Sterile neutrinos. 
If the LSND, atmospheric, and solar neutrino oscillation results are all 
confirmed we may be seeing evidence for 
the existence of sterile neutrinos. This would be a very exciting 
discovery !
Many new questions will arise requiring new experimental input.
\end{description}

\begin{figure}
\epsfxsize6.0in
\centerline{\epsffile{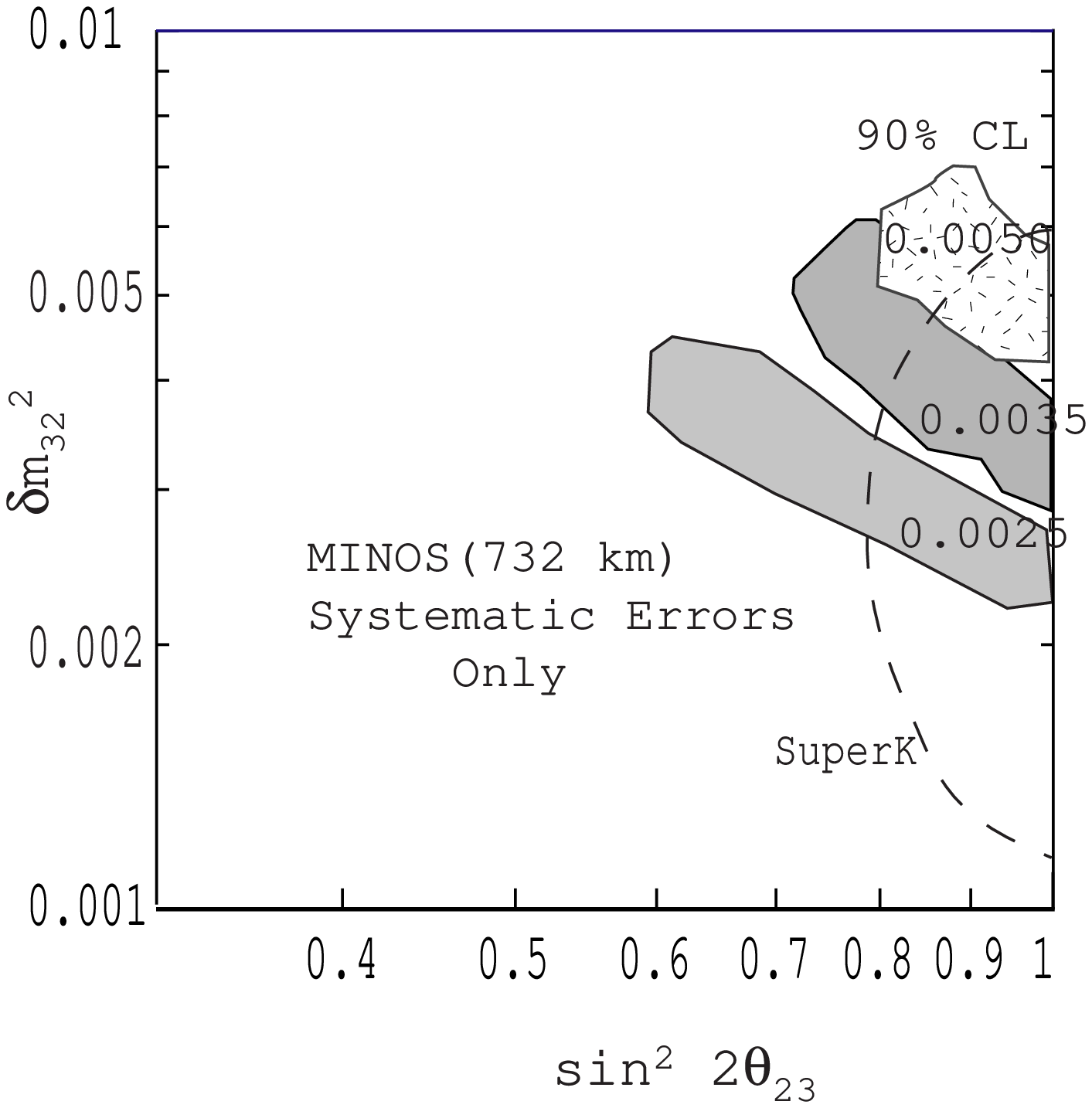}}
 \caption{The expected precision that could be achieved by a 
MINOS--like experiment (low energy, baseline 732 km) in the limit of infinite statistics 
but with conservative estimates of systematic errors.   The calculated 
sensitivities are based only on disappearance measurements. 
The oscillation parameters correspond to scenario IA1, and 
the regions of sensitivity shown are at 90\% CL.}
  \label{fig:bestminos}
\end{figure}

Finally, it is worthwhile considering the possibility that a 
conventional neutrino beam and the corresponding detectors 
undergo significant upgrades within the coming decade. For example, 
a Fermilab proton driver upgrade might enable the acceleration of 
up to about a factor of four more beam in the Main Injector, resulting 
in a corresponding increase of the NUMI beam intensity. With an 
additional factor of 2 - 3 increase in detector mass, the event 
samples might be increased by an order of magnitude. 
However, systematic uncertainties must also be considered. 
For example, there will be limiting systematic uncertainties on the 
measurements of $\delta m^2_{32}$ and $\sin^22\theta_{23}$ with a
MINOS--type experiment that arise from the 
uncertainties on the near/far detector CC reconstruction efficiencies, 
backgrounds to CC events from NC interactions, 
and an assumed 2\% flux uncertainty from the near/far detector 
extrapolation. 
These uncertainties would prevent the precise determination of 
the oscillation parameters, even in the limit of infinite statistics.
The ultimate (infinite statistics) precision that could be achieved 
with a MINOS--type experiment is shown in Fig.~\ref{fig:bestminos}. 
A very--long--baseline neutrino factory experiment would be able 
to make very significant improvements to the precision with which 
$\delta m^2_{32}$ and $\sin^22\theta_{23}$ are determined.

\begin{table}
\caption{Experimental neutrino oscillation observations 
expected in the next 5--10~years at accelerator based experiments.}
\bigskip
\begin{tabular}{cc|ccc|ccc}
\hline
Scenario&Experiment&$\nu_\mu$ Disap. & $\nu_\mu\to\nu_e$&$\nu_\mu\to\nu_\tau$&
$\nu_e$ Disap. & $\nu_e\to\nu_\mu$&$\nu_e\to\nu_\tau$ \\
\hline
IA1 &   K2K&{\bf Y}&n &n &n &n &n \\
    & MINOS&{\bf Y}&n &{\bf Y}&n &n &n \\
    &ICANOE& {\bf Y} &{\bf Y}
&{\bf Y}&n &n &n \\
    & OPERA&n &n &{\bf Y}&n &n & n\\
    & BooNE&n &n &n &n &n &n \\
\hline
IA2 & K2K  &{\bf Y}&n &n &n &n &n \\
    & MINOS&{\bf Y}&n & {\bf Y}&n &n & n\\
    &ICANOE&{\bf Y}& {\bf Y}& {\bf Y}&n &n &n \\
    & OPERA&n &n &{\bf Y}&n &n &n \\
    & BooNE&n &n &n &n &n &n \\
\hline
IA3 & K2K  &{\bf Y}&n &n &n &n &n \\
    & MINOS&{\bf Y}& {\bf Y}& {\bf Y}&n &n &n \\
    &ICANOE&{\bf Y}& {\bf Y}& {\bf Y}&n &n &n \\
    & OPERA&n &n &{\bf Y}&n &n &n \\
    & BooNE&n &n &n &n &n &n \\
\hline
IB1 & K2K& {\bf Y}&n &n &n &n &n \\
    & MINOS& {\bf Y}& {\bf Y}& {\bf Y}&n &n &n \\
    &ICANOE&{\bf Y} &{\bf Y} & {\bf Y}&n &n &n \\
    & OPERA&n &n &{\bf Y}&n &n &n \\
    & BooNE&n &{\bf Y}&n &n &n &n \\
\hline
IC1 & K2K& {\bf Y}&n &n &n &n &n \\
    & MINOS& {\bf Y}& {\bf Y}& {\bf Y}&n &n &n \\
    &ICANOE&{\bf Y} & {\bf Y}& {\bf Y}&n &n &n \\
    & OPERA&n &n &{\bf Y}&n &n &n \\
    & BooNE&{\bf Y}&{\bf Y}&n &n &n &n \\
\hline
IIA1 & K2K& {\bf Y}&n &n &n &n &n \\
    & MINOS& {\bf Y}& {\bf Y}& {\bf Y}&n &n &n \\
    &ICANOE&{\bf Y} &{\bf Y} &{\bf Y} &n &n &n \\
    & OPERA&n &{\bf Y}&{\bf Y}&n &n &n \\
    & BooNE&n &{\bf Y}&n &n &n &n \\
\hline
IIB1 &  K2K& {\bf Y}&n &n &n &n &n \\
    & MINOS& {\bf Y}&n & {\bf Y}&n &n &n \\
    &ICANOE&{\bf Y} &{\bf Y} &{\bf Y} &n &n &n \\
    & OPERA&n &n &{\bf Y}&n &n &n \\
    & BooNE&n &{\bf Y}&n &n &n &n \\
\hline
\end{tabular}
\label{expt_table}
\end{table}

\begin{table}
\caption{Neutrino oscillation mixing angle and leading $\delta m^2$ 
measurements 
expected in the next 5--10~years at accelerator based experiments.
}
\vspace{0.1cm}
\begin{tabular}{cc|ccccc}
\hline
 & & \multicolumn{4}{c}{Parameter} \\
Scenario&Experiment&$\sin^2 2\theta_{12}$&$\sin^2 2\theta_{23}$&
$\sin^2 2\theta_{13}$& $\delta$ & $\delta m^2\ \rm (eV^2) $ \\
\hline
IA1 &   K2K&   & 30\%  &   &   & 50\%  \\
    & MINOS&   & 10\%$^\dagger$  &   &   & 10\%$^\dagger$  \\
    &ICANOE&   &13\% &60\% &   &11\% \\
    & OPERA&   &20\% &   &   &14\% \\
    & BooNE&   &   &   &   &   \\
\hline
IA2 &   K2K&   & 30\%  &   &   & 50\%  \\
    & MINOS&   & 10\%$^\dagger$  &   &   & 10\%$^\dagger$  \\
    &ICANOE&   &13\% &60\% &   &11\% \\
    & OPERA&   &20\% &   &   &14\% \\
    & BooNE&   &   &   &   &   \\
\hline
IA3 &   K2K&   & 30\%  &   &   & 50\%  \\
    & MINOS&   & 10\%$^\dagger$  &   &   & 10\%$^\dagger$  \\
    &ICANOE&   &13\% &60\% &   &11\% \\
    & OPERA&   &20\% &   &   &14\% \\
    & BooNE&   &   &   &   &   \\
\hline
IB1 &   K2K&   & 30\%  &   &   & 50\%  \\
    & MINOS&   & 10\%  &   &   & 15\%  \\
    &ICANOE&   & 13\%  &   &   & 11\%  \\
    & OPERA&   &20\% &   &   &14\% \\
    & BooNE&10\% &   &   &   &10\% \\
\hline
IC1 &   K2K&   & 100\%  &   &   & 100\%  \\
    & MINOS&   & 10\%  &   &   & 15\%  \\
    &ICANOE& 25\%  & 5\%  &   &   & 7\%  \\
    & OPERA&   &5\%  &   &   &7\%  \\
    & BooNE&   &10\% &15\% &   &10\% \\
\hline
\hline
 & & & & & &  \\
 & &$\sin^2 2\theta_{23}$&$\sin^2 2\theta_{34}$&
$\delta m^2_{23}\ \rm (eV^2)$ & $\delta m^2_{34}\ \rm (eV^2) $ & \\
\hline
IIA1&   K2K&   & 30\%  &   & 50\%  &   \\
    & MINOS&   & 10\%  &   & 6\%  &   \\
    &ICANOE& 10\%  & 13\%  & 7\%  & 11\%  &   \\
    & OPERA& 30\%  &20\% & 30\%  & 14\%  & \\
    & BooNE& 10\%  & & 10\%  &   & \\
\hline
IIB1&   K2K&   & 30\%  &   & 50\%  &   \\
    & MINOS&   & 10\%  &   & 6\%  &   \\
    &ICANOE& 50\%  & 13\%  & 50\%  & 11\%  &   \\
    & OPERA&   & 20\%  &   & 14\%   & \\
    & BooNE& 10\%  & & 10\%  &   & \\
\hline
\end{tabular}
$^\dagger$ With $\sin^22\theta_{23}$ constraint from SuperK.
\label{3nu_table}
\end{table}

\clearpage

\subsection{The neutrino factory oscillation physics program}
\label{program}

We now consider the program of neutrino oscillation measurements 
at a neutrino factory in the era beyond the next generation of 
long baseline experiments. The main goals in this era are likely 
to be to precisely establish the oscillation framework, 
determine the pattern of neutrino masses, 
measure matter effects to confirm the MSW phenomenon, 
make precise measurements or place stringent limits 
on all of the mixing--matrix elements (and hence mixing--angles), 
and observe or place stringent limits on CP violation 
in the lepton sector. 
A neutrino factory can address each of these goals:
\begin{description}

\item{(i)} Establishing the oscillation framework. 
This requires measuring as a function of $L/E$, 
or putting stringent limits on,  
all of the oscillation probabilities $P(\nu_e\to\nu_x)$ 
and $P(\nu_\mu\to\nu_x)$. The oscillation framework 
can be established by summing the probabilities 
(a) $P(\nu_e\to\nu_e) + P(\nu_e\to\nu_\mu) + P(\nu_e\to\nu_\tau)$, and 
(b) $P(\nu_\mu\to\nu_e) + P(\nu_\mu\to\nu_\mu) + P(\nu_\mu\to\nu_\tau)$.
In a three--flavor mixing framework, both sums should be unity for all 
$L/E$. If there are sterile neutrinos participating in the oscillations 
one or both of the sums will be less than unity. Part (b) of the test 
will almost certainly be made with conventional neutrino beams, 
although with a precision that will be limited by the 
$\nu_\mu\to\nu_\tau$ statistics and 
by the uncertainty 
on the $P(\nu_\mu\to\nu_e)$ measurement arising from the O(1\%) $\nu_e$ 
contamination in the beam. Part (a) of the test, which includes 
the first observation of (or stringent limits on) $\nu_e\to\nu_\tau$ 
oscillations, can only be made 
with an energetic ($E_\nu > 10$~GeV) 
$\nu_e$ (or $\overline{\nu}_e$) beam, and will therefore 
be a unique part of the neutrino factory physics program.

\item{(ii)} Determining the pattern of neutrino masses.
The present experimental data suggests that, within a three--flavor 
mixing framework, there are two neutrino mass eigenstates separated 
by a small mass difference, and a third state separated from the pair 
by a ``large" mass difference $\delta M^2$. What is unknown is whether there  
is one low state plus two high states, or two low states 
plus one high state. This can be determined by measuring the sign of 
$\delta M^2$. The only way we know of making this 
measurement is to exploit matter effects which, in a very long baseline 
experiment, alter the probabilities for oscillations 
that involve electron neutrinos; the modification being dependent 
on the sign of $\delta M^2$. In principle the measurement could 
be made using a conventional neutrino beam and measuring 
$\nu_\mu\to\nu_e$ and $\overline{\nu}_\mu\to\overline{\nu}_e$ 
transitions over a baseline of several thousand km. However, 
the O(1\%) $\nu_e$ ($\overline{\nu}_e$) 
contamination in the beam will introduce an 
irreducible background that is comparable to, or larger than, 
the $\nu_e$ signal. 
In contrast, at a neutrino factory it appears that 
the measurement can be done with great precision. 
Hence, determining the 
sign of $\delta M^2$ and the pattern of neutrino masses 
would be a key measurement at a neutrino factory.

\item{(iii)} Measuring matter effects to confirm the MSW phenomenon.
The same technique used to determine the sign of $\delta m^2_{32}$ 
can, with sufficient statistics, provide a precise quantitative 
confirmation of the MSW effect for neutrinos passing through the 
Earth. The modification to $P(\nu_e\to\nu_\mu)$, for example, 
depends upon the matter parameter $A$ (Eq.~(\ref{eq:defnA})). 
Global fits to 
appearance and disappearance spectra that are used to 
determine the oscillation parameters can include $A$ as a 
free parameter. The quantitative MSW test would be to recover the 
expected value for $A$. This measurement exploits the clean $\nu_e\to\nu_\mu$ 
signal at a neutrino factory, and would be a unique part of the 
neutrino factory physics program.

\item{(iv)} Making precise measurements or placing stringent limits
on all of the mixing--matrix elements. 
In practice the measured oscillation probability amplitudes 
are used to determine the mixing angles. If any of the angles 
are unmeasured or poorly constrained the relevant entries in 
the mixing matrix will also be poorly determined. At present 
there is only an upper limit on $\theta_{13}$, the angle that 
essentially determines the $\nu_e\to\nu_\mu$ oscillation amplitude. 
A neutrino factory would provide a precise measurement of, or 
stringent limit on, this difficult angle. In fact, because 
all of the $\nu_\mu\to\nu_x$ and $\nu_e\to\nu_x$ oscillation 
amplitudes can be measured at a neutrino factory, global fits 
can be made to the measured spectra to provide a very precise 
determination of the mixing angles. This exploits 
the $\nu_e$ component in the beam.  Finally, it should be 
noted that it is important to test the overall consistency 
of the oscillation framework by determining the mixing 
angles in more than one way, 
i.e. by using more than one independent set of measurements. 
Clearly the $\nu_e$ beam is an asset for this check.

\item{(v)} Placing stringent limits on, or observing, CP violation
in the lepton sector. 
Most of the oscillation scenarios defined for the study 
predict very small CP violating amplitudes. An important test of 
these scenarios would be to place stringent experimental limits 
on CP violation in the lepton sector. The LMA scenario IA1 
might result in sufficiently large CP violating effects to 
be observable at a neutrino factory. The CP test involves 
comparing $\nu_e\to\nu_\mu$ with $\overline{\nu}_e\to\overline{\nu}_\mu$ 
oscillation rates, possible at a neutrino factory because backgrounds 
are very small. A search for CP violation in the lepton sector with 
the required precision cannot be done with a conventional neutrino 
beam, and is therefore a unique part of the neutrino factory 
physics program.

\end{description}

Note that it is the $\nu_e$ ($\overline{\nu}_e$) 
component in the neutrino factory beam 
that drives the oscillation physics program.
A $\nu_e$ beam would 
(a) enable a basic test of the oscillation 
framework that cannot be made with a $\nu_\mu$ beam, 
(b) enable the first observation of (or stringent limits on) 
$\nu_e\to\nu_\tau$ oscillations, 
(c) make a convincing determination of the pattern of neutrino 
masses that would be difficult or impossible with a conventional 
neutrino beam, 
(d) make a quantitative check of the MSW effect 
only possible with a neutrino factory beam, 
(e) enable measurements or stringent limits on all of the 
(three--flavor) 
mixing angles with a precision that requires both $\nu_e$ 
and $\nu_\mu$ beams, and 
(f) measure or put meaningful limits on CP violation 
in the lepton sector, which requires a signal purity only available 
at a neutrino factory. 

A neutrino factory operating in the 
next decade, after the next generation of long baseline experiments, 
would appear to be the right tool at the right time. 
However, before we can quantitatively assess how well a 
neutrino factory might realize the physics program we 
have listed, we must first understand the capabilities 
of neutrino detectors in the neutrino factory era.

\subsection{Detector considerations}
\label{detectors}

We would like to measure the oscillation probabilities 
$P(\nu_\alpha \rightarrow \nu_\beta)$ as a function of the 
baseline $L$ and neutrino energy $E$ (and hence $L/E$) 
for all possible initial and final flavors $\alpha$ and $\beta$.
This requires a beam with a well known initial flavor content, 
and a detector that can identify the flavor of the interacting 
neutrino. The neutrinos interact in the detector via charged 
current (CC) and neutral current (NC) interactions to produce 
a lepton accompanied by a hadronic shower arising 
from the remnants of the struck nucleon. 
In CC interactions the final state lepton 
tags the flavor ($\beta$) of the interacting neutrino. 

\begin{figure}
\epsfxsize3.4in
\centerline{\epsffile{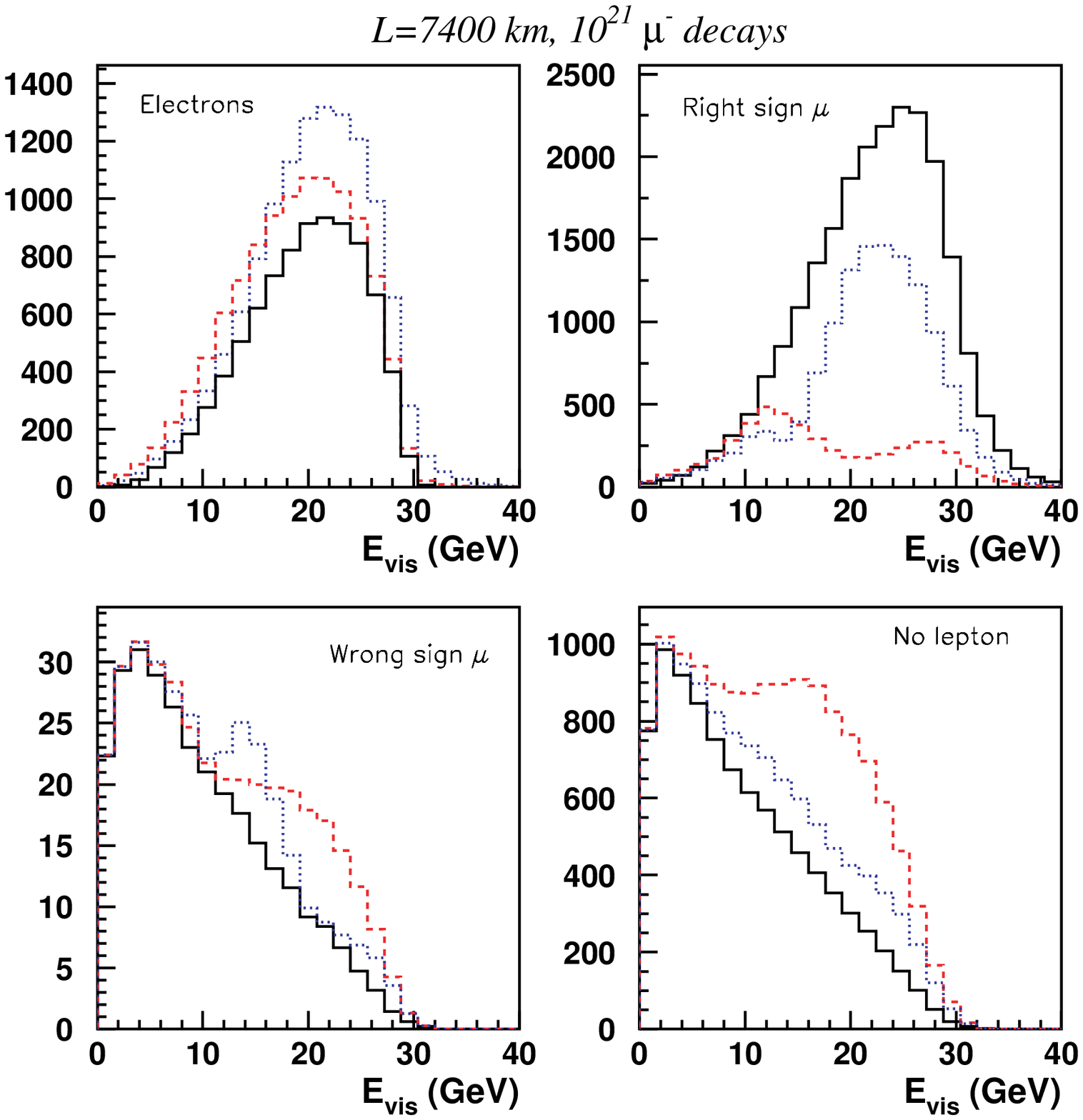}}
\caption{Visible energy spectra for four event classes when 
$10^{21} \mu^-$ 
decay in a 30~GeV neutrino factory at $L = 7400$~km.
Black histogram: no oscillations. 
Blue dotted histogram: $\delta m^2_{32}=3.5\times 10^{-3}$~eV$^2$/c$^4$, 
$\sin^2\theta_{23}=1$. 
Red dashed histogram: $\delta m^2_{32}=7\times 10^{-3}$~eV$^2$/c$^4$, 
$\sin^2\theta_{23}=1$. 
The distributions in this figure and the following figure 
are for an ICANOE-type detector, and are 
from Ref.~\ref{camp00}.}
\label{fig:m1}
\end{figure}

\begin{figure}
\epsfxsize3.4in
\centerline{\epsffile{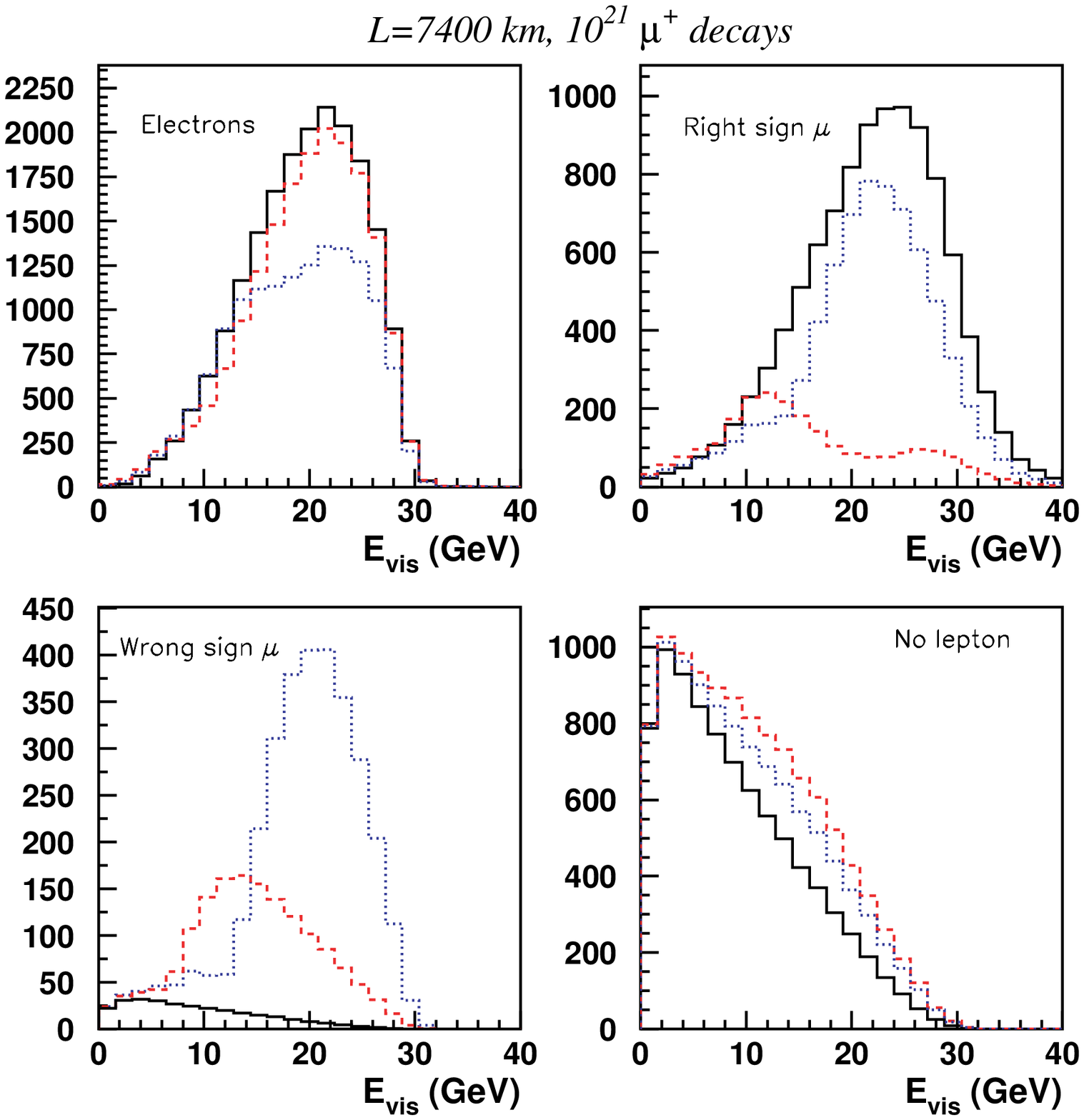}}
\caption{Same as previous figure, but with positive muons circulating in the
storage ring. The difference between the two figures is due to the different
cross section for neutrinos and antineutrinos, and to matter effects.}
\label{fig:m2}
\end{figure}

At a neutrino factory in which, for example, positive 
muons are stored, the initial beam consists of 50\% $\nu_e$ and 
50\% $\overline{\nu}_\mu$. 
In the absence of oscillations, the $\nu_e$ CC interactions 
produce electrons and the $\overline{\nu}_\mu$ CC interactions 
produce positive muons. 
Note that the charge of the final state lepton tags the flavor 
($\alpha$) of the initial neutrino or antineutrino. 
In the presence of 
$\nu_e \rightarrow \nu_\mu$ oscillations the $\nu_\mu$ CC interactions 
produce negative muons (i.e. wrong--sign muons). Similarly, 
$\overline{\nu}_\mu \rightarrow \overline{\nu}_e$ oscillations 
produce wrong--sign electrons, 
$\overline{\nu}_\mu \rightarrow \overline{\nu}_\tau$ oscillations 
produce events tagged by a $\tau^+$ and 
$\nu_e \rightarrow \nu_\tau$ oscillations 
produce events tagged by a $\tau^-$. 
Hence, there is a variety of information that can be used 
to measure or constrain neutrino oscillations at a neutrino factory, 
namely the rates and energy distributions of events tagged by 
(a) right--sign muons, (b) wrong--sign muons, (c) electrons 
or positrons (their charge is difficult to determine in a massive 
detector), 
(d) positive $\tau$--leptons, (e) negative $\tau$--leptons, 
and (f) no charged lepton. If these 
measurements are made when there are alternately positive and negative 
muons decaying in the storage ring, there are a total of 12 spectra 
that can be used to extract information about the oscillations. 
Some examples of the predicted measured spectra are shown as a function of the 
oscillation parameters in Figs.~\ref{fig:m1} and 
\ref{fig:m2} for a 10~kt detector sited 7400~km 
downstream of a 30~GeV neutrino factory. 
Clearly, the high intensity $\nu_e$, $\overline{\nu}_e$, $\nu_\mu$, and 
$\overline{\nu}_\mu$ beams at a neutrino factory would provide a wealth of 
precision oscillation data.


The detectors required at a neutrino factory will have many similarities 
to the detectors that have been designed 
for the next generation of experiments at conventional neutrino beams. 
However, there are some important differences. 
First, we can anticipate more massive detectors. The sensitivity of 
a neutrino factory oscillation experiment is proportional to the 
product of the detector mass and beam intensity. It is likely that 
the cost of increasing the MINOS detector fiducial 
mass (for example) by a factor of a few 
is smaller than the cost of increasing the neutrino factory beam intensity 
by a factor of a few. Therefore, we believe that it is reasonable to assume 
that detectors at a neutrino factory would be a factor of a few to 
a factor of 10 more massive than the generation of neutrino 
detectors presently under construction. 
Second, the presence of both neutrinos and antineutrinos in the same 
beam at a neutrino factory places a premium on measuring the sign 
of the charge of the lepton produced in CC interactions. Charge--sign 
determination may 
not be practical for electrons, but is mandatory for muons and 
highly desirable for $\tau$--leptons. 
Finally, a relatively low energy threshold for the detection and 
measurement of wrong--sign muons is very desirable. This is because 
high muon detection thresholds require high energy interacting 
neutrinos, and hence a high energy neutrino factory. Since the muon 
acceleration system at a neutrino factory is likely to be expensive, 
low energies are preferable.

In the following sections 
we begin by considering general detector issues for the 
measurement of final state muons and $\tau$--leptons, and then consider 
some specific candidate detectors for a neutrino factory.  
Some of these detector types are quite new and 
are just beginning to be studied; for the more mature detectors 
the ``neutrino'' energy resolution, 
the signal efficiency, background rejection, 
and fiducial mass are discussed.

\subsubsection{Muon identification and measurement} 
\label{bkgds} 

The detection and measurement of muons (especially those of 
opposite sign to the muons in the storage ring) is 
crucial for many of the key oscillation physics measurements 
at a neutrino factory. Before considering some specific 
neutrino factory detectors it is useful to consider more generally 
muon backgrounds and related issues. 
Background muons can be produced in NC and CC interactions by:
\begin{description}
\item{(i)} Pions or kaons from the hadronic shower that decay to produce a 
muon.
\item{(ii)} Non-interacting pions which fake a muon signature (punch-through).
\item{(iii)} Charm meson production and muonic decay.
\end{description}

A background muon event can be produced when 
a background ``muon" of the appropriate sign is recorded in (a) 
a NC event or (b) a CC event in which the primary lepton has been lost.
If the background muon has the same charge sign as that in the storage ring 
the resulting event will be a background
for disappearance measurements, but more importantly, if it has 
the opposite sign then the event will be a background for 
wrong--sign muon appearance measurements.  

\begin{figure}[h]
\begin{center}
\epsfxsize6.0in
\centerline{\epsffile{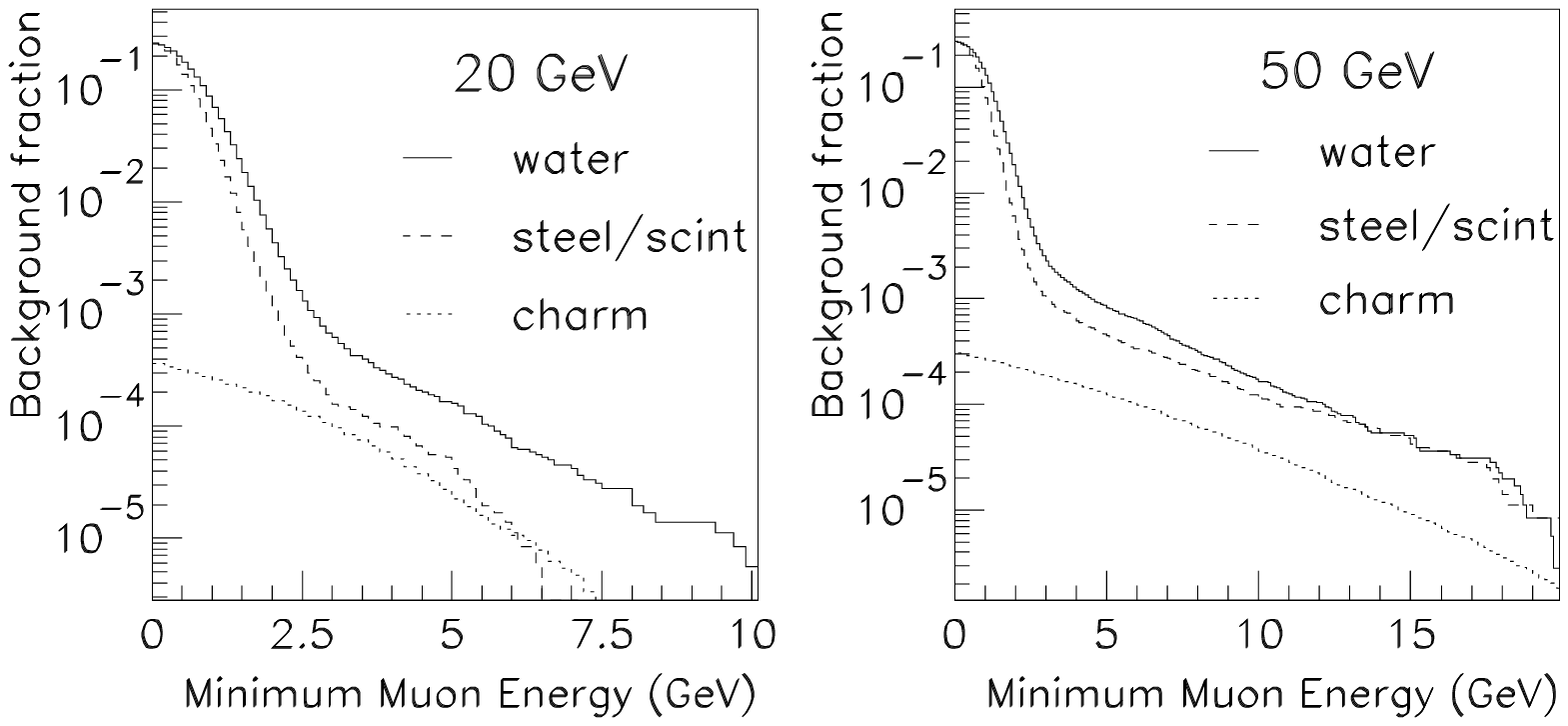}}
\end{center}
\vspace{-0.5cm}
\caption{Background levels from punch through, pion/kaon decay, and 
charm backgrounds for 20~GeV (left) and 50~GeV (right) neutrino 
factories. The fraction of neutrino interactions that produce a 
wrong--sign muon background event is shown as a function of 
the minimum muon energy accepted.}
\label{punch-charm}
\end{figure}

The integrated wrong-sign 
background fraction from the hadronic shower 
is shown in Fig.~\ref{punch-charm} 
as a function of the minimum muon energy accepted 
for Steel/Scintillator and water detectors downstream of 20~GeV 
and 50~GeV neutrino factories.  The charm background comes from 
$\nu_\mu$ CC events where the primary muon was less than 2~GeV.  
The peak at low muon energies is from 
the hadron shower itself and from punch through, while the long tail is from 
shower particles decaying to muons.  

In general there are two different standards for background levels which 
are relevant:  that of a disappearance experiment and that of an 
appearance experiment.  
Background estimates are not trivial, but if the backgrounds for a 
disappearance measurement 
are at the one per cent level, then the uncertainties on those 
backgrounds can be expected to be small compared to the flux uncertainty.
On the other hand, wrong-sign muon appearance measurement uncertainties 
are expected to be dominated by the statistics.  An extremely aggressive 
background 
level requirement would be to have less than of the order of one 
background event.  
If there are several thousand CC events expected, then this 
would require a minimum background rejection factor of $10^{4}$.

\begin{figure}
\epsfxsize3.0in
\centerline{\epsffile{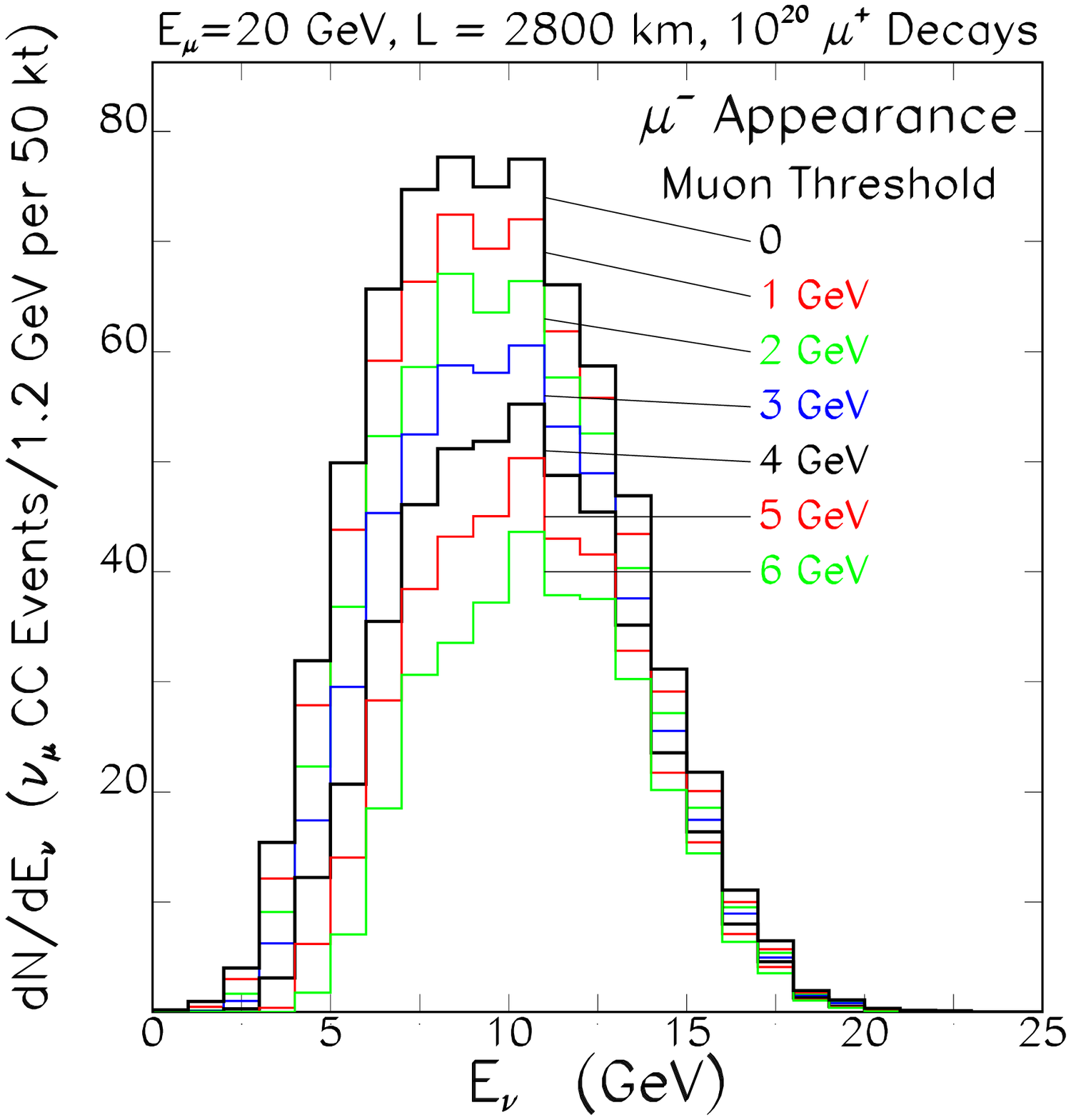}}
\bigskip
\caption{Reconstructed neutrino energy distribution for 
several different minimum muon energy cuts for a 20~GeV ring. 
Result is from Ref.~\ref{bgrw00}.
} 
\label{enucut}
\end{figure}

Backgrounds can be suppressed by imposing a minimum energy requirement 
on the measured muon.  
Figure ~\ref{enucut} shows the effect of several different minimum muon 
energy cuts on a simulated oscillation signal observed in 
a steel-scintillator type detector at a 20~GeV muon storage 
ring, at a baseline length of 2800km \cite{bgrw00}.
A muon threshold energy of 4~GeV for example depletes the 
low energy part of observed measured ``neutrino energy" distribution, 
degrading but not completely removing the information about the 
neutrino oscillation parameters that is encoded in the shape of 
the distribution. A 4~GeV threshold at a 20~GeV neutrino factory 
is probably tolerable. If higher thresholds are needed to reject 
backgrounds, then a higher energy neutrino factory is desirable. 
If a lower energy neutrino factory is to be viable, then lower 
muon thresholds are desirable. 

\begin{figure}[tb]
\centering
\epsfig{file=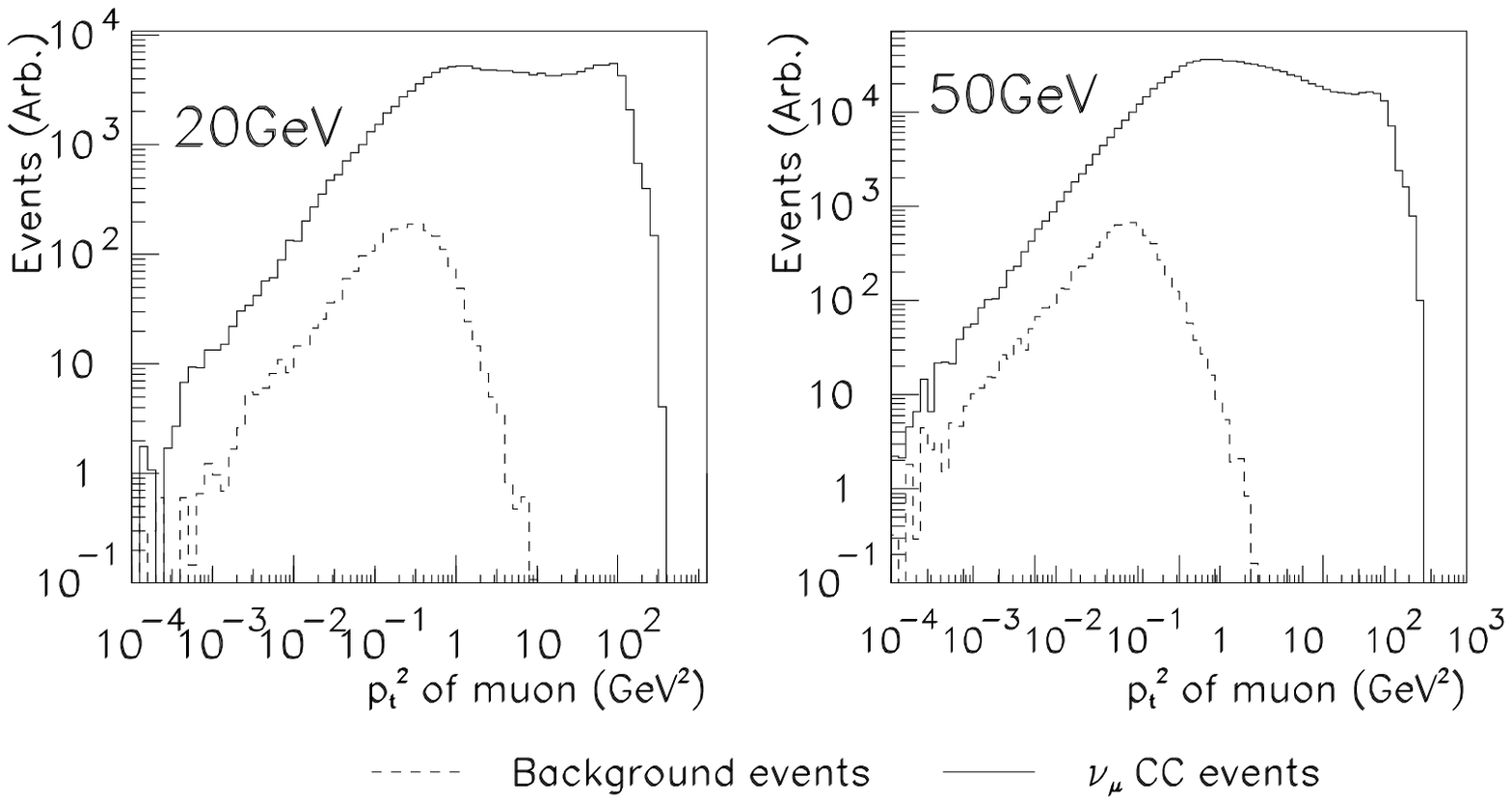,width=\textwidth}
\caption{Distributions of the square of the muon momentum component 
transverse to the hadronic 
shower direction for $\nu_\mu$ charged 
current events compared to background muons for a 20 
and 50~GeV muon storage ring. }
\label{fig:pt2gen}
\end{figure}

As is shown in Fig.~\ref{punch-charm}, to get to a background level 
of $10^{-4}$ one would need a 5 (6.5) GeV muon momentum cut in 
Steel/scintillator (Water) for a 20 GeV muon storage ring, and a 10 (12) GeV 
muon momentum cut in Steel/Scintillator (Water) for a 50 GeV muon storage 
ring.  Clearly more background rejection is desirable. 
Fortunately muons from hadron decay in the hadronic shower 
are likely to be more aligned with the shower direction than 
muons from the leptonic vertex of the CC interaction. 
This provides another handle on the background. 
A useful variable to cut on is the momentum of the muon in the direction 
transverse to the hadronic shower ($p_t$).  Figure~\ref{fig:pt2gen} 
shows the generated $p_t^2$ distribution for background and signal 
events, with no cut on the final state muon momentum.  Note that 
requiring $p_t^2 > 1$ the background is extremely low, while the signal 
efficiency is high.  The resolution with which $p_t^2$ is determined 
is detector dependent, and for detectors with reasonable 
transverse and
longitudinal segmentation is dominated by the hadronic energy resolution. 

\subsubsection{$\tau$--lepton identification and measurement} 

The detection and measurement of $\tau$--leptons is crucial 
for $\nu_\mu \rightarrow \nu_\tau$ and $\nu_e \rightarrow \nu_\tau$ 
measurements at a neutrino factory. Note that $\nu_e \rightarrow \nu_\tau$ 
oscillations will be of special interest since they will not have been 
previously 
observed. The $\nu_e \rightarrow \nu_\tau$ signal can be separated from 
$\nu_\mu \rightarrow \nu_\tau$ ``background" if 
the sign of the $\tau$--lepton charge is measured. 
The majority of $\tau$--lepton decays produce either one charged 
track (electron, muon, of hadron) or three charged tracks (hadrons). 
There are two general techniques that can be used to identify 
$\tau$--leptons. The first technique exploits the one-prong and 
three-prong topologies, and uses kinematic cuts to 
suppress backgrounds. The second technique uses a detector with a high 
spatial resolution to look for the displaced vertex or kink resulting 
from $\tau$--lepton decay.

The advantage of the displaced vertex or kink detection $\tau$--lepton 
technique is that the detailed $\tau$--lepton decay is measured and 
background suppression is therefore large. The disadvantage is that 
detectors that have sufficient spatial resolution are necessarily 
less massive than coarse--grained detectors.  

The advantage of the kinematic technique is that a very massive 
detector can be used. If the $\tau$--leptons decay muonically 
(BR = 17\%) a measurement of the muon charge--sign determines 
the sign of the $\tau$ charge. 
However, there are substantial backgrounds 
that must be reduced.  In the case of muonic $\tau$ decays, 
the backgrounds are from 
(a) $\nu_\mu$  (or $\bar\nu_\mu$) CC interactions
which typically produce muons at high 
momentum and high $p_t^2$, and 
(b) meson decays (discussed earlier) 
which are at low momentum and low $p_t^2$.  
For $\tau \to e$ decays, the main background comes from 
$\nu_e$ and $\bar\nu_e$ CC interactions. 
Fortunately the undetected neutrinos from 
$\tau$ decays result in a larger missing transverse momentum 
than expected for background events. 
Exploiting these kinematic characteristics the backgrounds 
can be reduced by a large factor. 
For example, for an ICANOE--type detector a background rejection 
factor of 200 has been estimated, with a corresponding signal efficiency 
of 30\%. In the electron channel background can also come from
NC interactions which produce photon conversions 
or Dalitz $\pi^0$ decays. These backgrounds can be suppressed 
in detectors with good pattern recognition allowing conversions, 
for example, to be identified and rejected. 
The analysis of hadronic $\tau$ decays requires the identification of the
$\tau$ decay product inside a jet. 
This can only be done with a detector having good pattern recognition.
It has been demonstrated that with an ICANOE--type detector a 
background rejection factor of 200 can be expected for 
$\tau\to$ 1 prong, $\tau\to\rho$, and $\tau\to 3\pi$ decays, with 
a signal efficiency of 8\%.

\begin{figure}[tb]
\centering
\epsfxsize=\textwidth
\epsfbox[60 150 800 490]{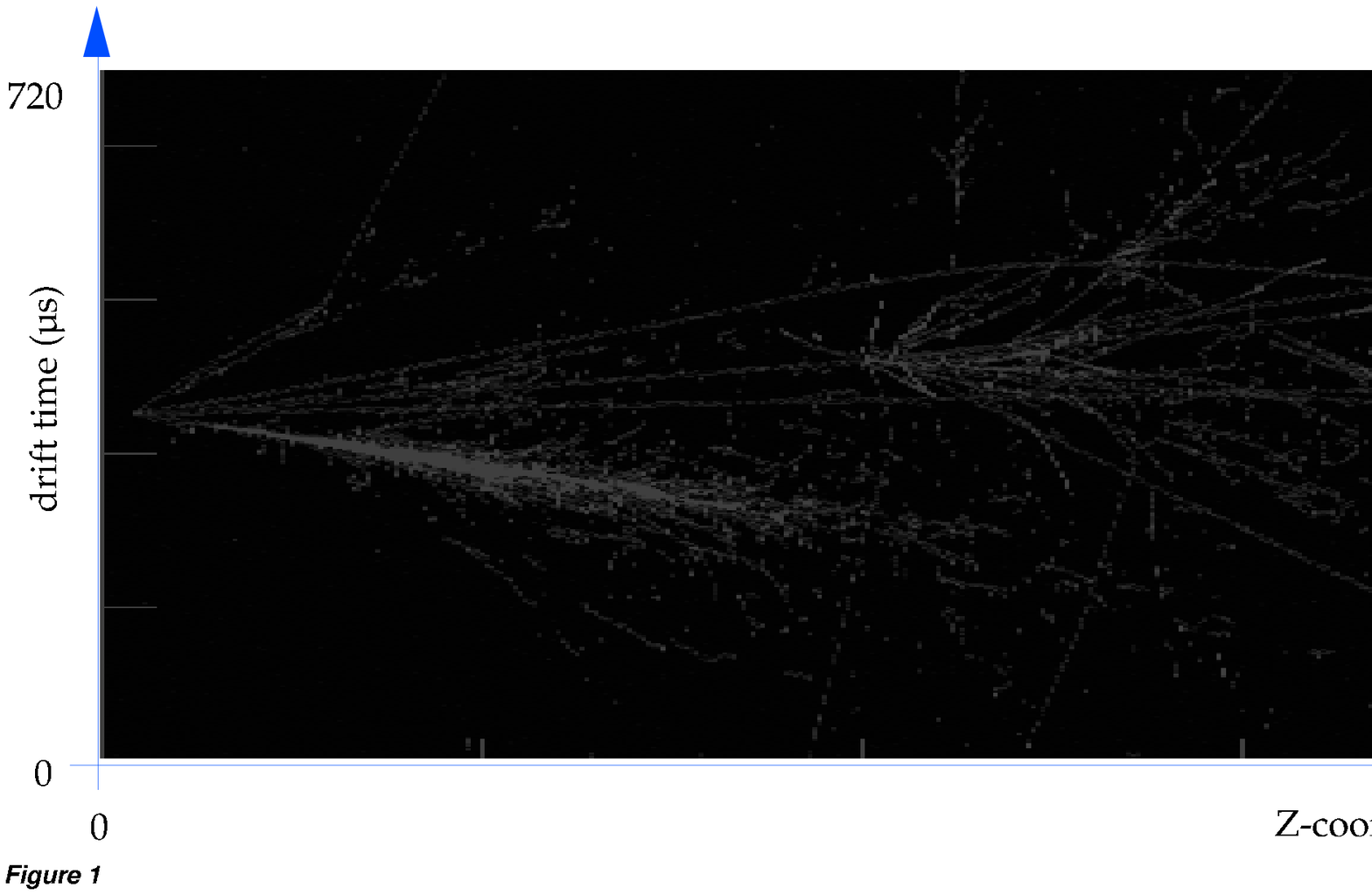} 
\label{fig:icanoe}
\caption{Example of a $\nu_e$ Charged current event from the 
full simulation of the ICANOE detector.} 
\end{figure}

\subsubsection{A Liquid Argon neutrino detector} 

We have studied the performance of a large Liquid Argon 
neutrino detector at a neutrino factory 
using the ICANOE monte carlo program.
One ICANOE detector unit
consists of a liquid argon TPC followed by a magnetic spectrometer. 

The Liquid Argon TPC has extremely fine granularity, producing 
bubble chamber like event images.  
Figure~21 
shows an example of an electron neutrino charged current event--note
the separation between the electromagnetic shower and the 
hadronic shower of the nucleon remnant.  
The TPC is instrumented with 3~mm pitch wires which allow tracking, 
$dE/dx$ measurements, electromagnetic, and hadronic calorimetry.  
Electrons and photons can be identified and their 
energies are measured with a resolution given by 
$\sigma_E/E = 0.03/\sqrt{E} \oplus 0.01$.   
The hadron energy resolution is given by 
$\sigma_E/E = 0.2/\sqrt{E} \oplus 0.05$.  
The magnetic spectrometer is primarily needed to measure muon energy 
and charge, but it is assumed that it will 
also be instrumented as a calorimeter 
to allow the hadron energy of showers 
which leak into the spectrometer to be correctly measured 
(albeit with worse resolution). 
The muon momentum resolution is expected to be $20\%$.  

In the design we have simulated, the 
liquid Argon module is 18~m deep with a cross section of 
$11.3m\times11.3m$.  The active (total) mass of one Liquid Argon module 
is 1.4~kt (1.9~kt).  The magnetized calorimeter module is 2.6~m deep 
with a cross section of $9m\times 9m$, and has a mass of 0.8~kt.  
It consists of 2~m of steel, corresponding 
to $7.4\ \lambda_{int}$ and $59\ X_0$, interleaved with 
tracking chambers.  Four Super-Modules are assumed, yielding 
a total detector length of $82.5$~m and a total active mass 
of 9.3~kt that is fully instrumented. 

ICANOE can reconstruct 
neutrino (and antineutrino) events of all active flavors, and with an
energy ranging from tens of MeV to tens of GeV, for the relevant
physics analyses. The unique imaging capabilities of the liquid 
argon TPC allow one to cleanly determine whether a given  
event is a $\nu_\mu$ CC event, a $\nu_e$ CC event, or a NC 
event. 

For our studies 
the ICANOE fast simulation was used. Neutrino interaction
events are generated, with a proper treatment of quasi-elastic interactions,
resonance and deep-inelastic processes. The 4-vectors for all the
particles generated are smeared, according to the resolutions derived from the 
full simulation. Muonic decays of pions and kaons are also considered, for a 
proper wrong- and right-sign muon background treatment.  Once a 
2-GeV cut is placed on the outgoing muon momentum, the background levels
tend to be about $10^{-5}$ times the actual charged current event rate, 
and are dominated by meson decay in the hadronic shower.  

Examples of simulated oscillation signals in an ICANOE--type detector 
at a neutrino factory are shown in Figs.~\ref{fig:m1} and \ref{fig:m2}. 
More detailed results from a study of the sensitivity that might be 
achieved using an ICANOE--type detector are discussed in the 
oscillation measurements section of this report.

\subsubsection{A magnetized Steel/Scintillator neutrino detector} 

Steel/Scintillator calorimeters have been used extensively in past neutrino 
experiments. Their performance is well understood 
and well simulated. Typically a magnetized Steel/Scintillator (MINOS--like) 
neutrino detector consists of iron plates 
interspersed with scintillator planes. To obtain transverse position 
information the scintillator can be 
segmented transversely, or a separate detector system (e.g. 
drift chambers) used. 
Penetrating charged particles (muon candidates) can then be 
reconstructed. With a reasonable transverse segmentation, 
the transverse position resolution is dominated by multiple 
coulomb scattering. 
The detector performance depends primarily on its 
longitudinal segmentation.  The 
segmentation needs to be fine enough to determine whether a 
charged track has penetrated beyond 
the region of the accompanying hadronic shower. If it has, 
then the penetrating track is a muon candidate. 
The muon momentum resolution is 
determined by the magnetic field and the thickness
of the steel plates.  

Neutrino CC and NC interactions have well defined signatures. 
In a MINOS--like detector NC interactions produce a hadronic shower 
reconstructed as a large energy deposition in a small 
number of scintillator units. A 
$\nu_\mu$ or $\overline{\nu}_\mu$ CC interaction will produce 
a muon in the final state, characterized by a long 
track in addition to the hadronic shower. These events can be 
identified provided the muon penetrates well beyond the hadronic 
shower. This imposes a minimum track-length, and hence minimum 
energy, requirement on muons that can be identified. If the muon 
is not identified the CC interaction will look like a NC event.
A $\nu_e$ or $\bar{\nu_e}$ CC interaction, will produce an 
electron in the final state which cannot be resolved, so these events look 
similar to NC interactions.  
A $\nu_\tau$ or $\bar{\nu_\tau}$ CC interaction will also look like a NC 
interaction unless the $\tau$--lepton decays muonically. 
%

To study the performance of a magnetized Steel/Scintillator detector at a 
neutrino factory we have considered a detector geometry similar to the
CCFR/NuTeV calorimeter~\cite{nutevdet}, but with the addition of a toroidal 
magnetic field of 1T.  
The detector is constructed from $3\times 3 \times 0.3$~m$^3$ modules 
(see Fig.~22).  
The 0.7~kt CCFR detector consists 
of 42 modules. A neutrino factory detector with a mass of 50~kt 
(10 $\times$ the MINOS detector) would require 3000 of these 
modules. 
The ultimate transverse size (and hence module mass) that is 
practical is probably determined by the largest size over which 
a large magnetic field can be generated.

\begin{figure} [h]
\begin{center}
\epsfxsize3.0in
\centerline{\epsffile{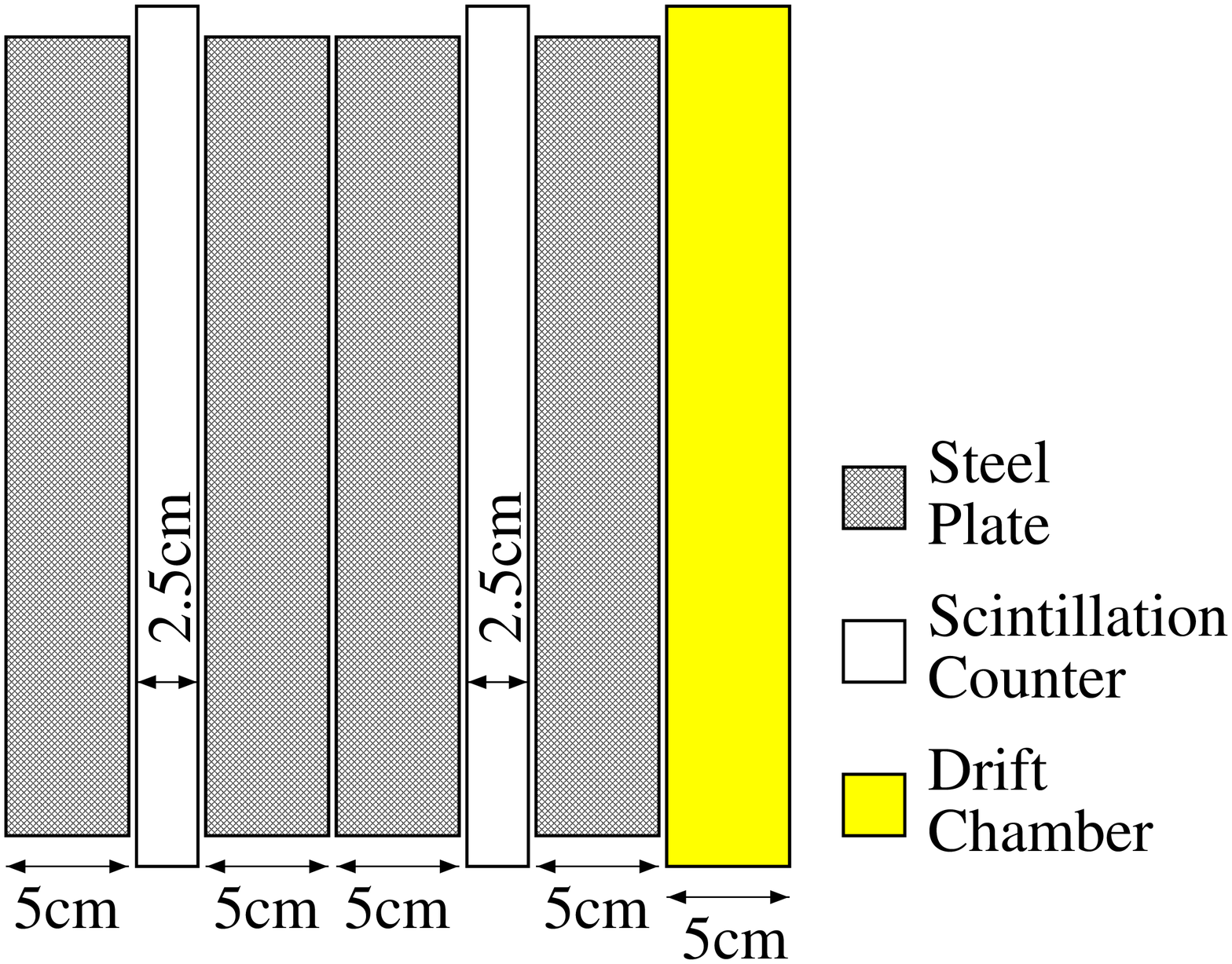}}
\caption{Schematic of a CCFR/NuTeV calorimeter module.}
\end{center}
\label{fig:ccfr}
\end{figure}
In the following we consider how well a magnetized Steel/Scintillator detector 
can identify and measure wrong--sign muon events at a neutrino factory. 
For our simulations, we used the parameterized Monte Carlo developed by the 
NuTeV 
collaboration,
 modified to include particle tracking in the
magnetic field. 
The hadron energy resolution of this detector is described in detail in
~\cite{nutevdet}, and is approximately given by 
$\sigma_E / E = 0.85 / \sqrt{E}$.  
The muon momentum resolution 
depends on the track length in the steel, and whether the muon is contained 
within the detector.  For muons which range out in the detector the
effective 
momentum resolution is $\sigma_P/P = 0.05$, while for tracks which leave the
fiducial volume of the detector the resolution is described by 
$\sigma_P/P \sim \theta_{MCS}\theta_{BdL}$, where the angles 
$\theta_{MCS}$ and $\theta_{BdL}$ describe respectively 
the change in direction 
due to multiple scattering and curvature in the magnetic field.

The simulation includes a detailed parameterization of the hadron-shower 
development, with the inclusion of charm production and $\pi$, $K$ 
decays (the data set on which the decay probability parameterization 
was tuned contained only muons with momentum higher than 4~GeV/c). 
Note that $\pi$ punchthrough was not included in the parameterization, 
but is expected to make only a small contribution to background 
muons above 4~GeV.

\begin{figure} [h]
  \epsfxsize=7.cm
  \epsfbox[40 10 520 520]{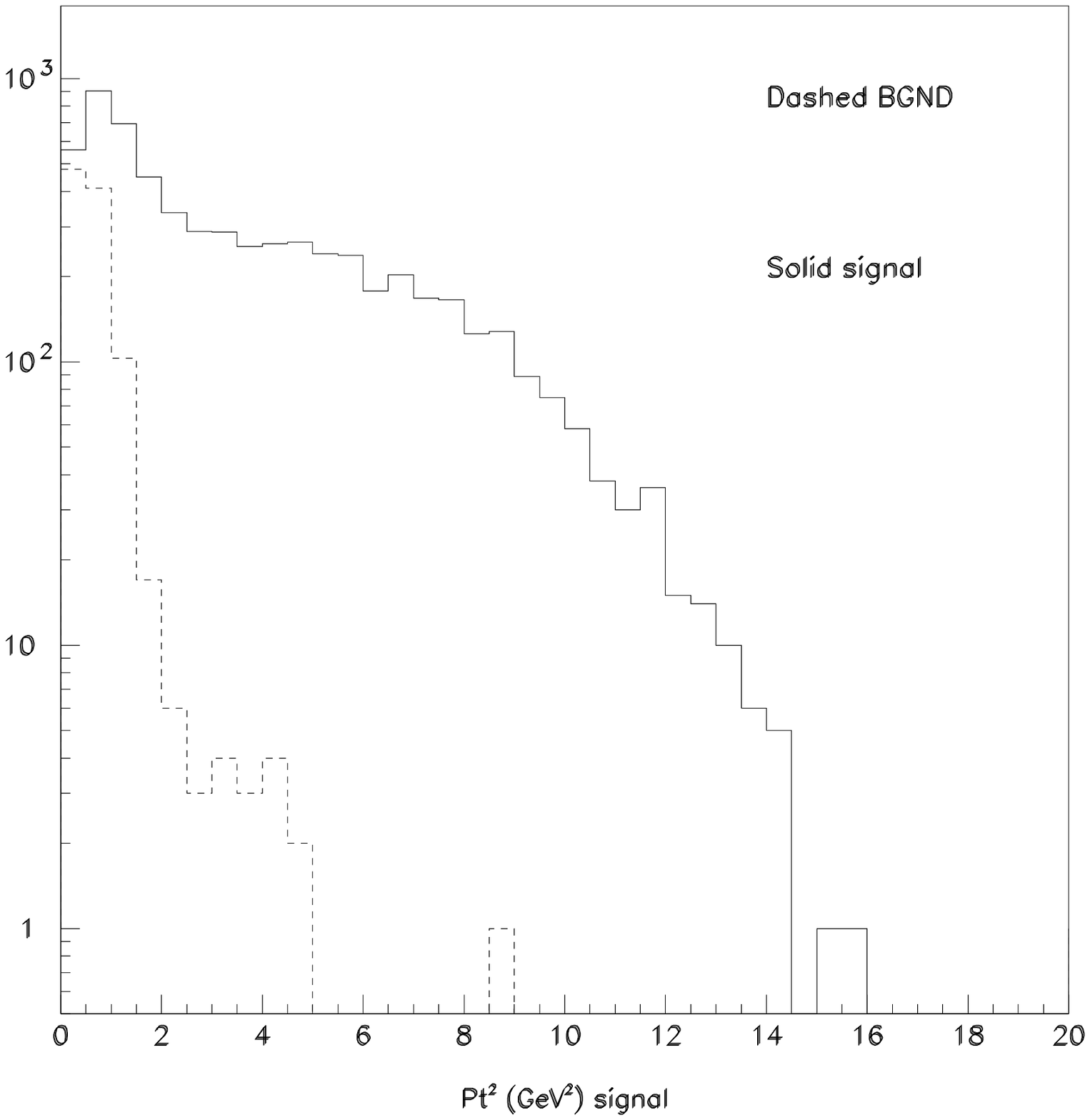}
  \epsfxsize=7.cm
  \epsfbox[40 10 520 520]{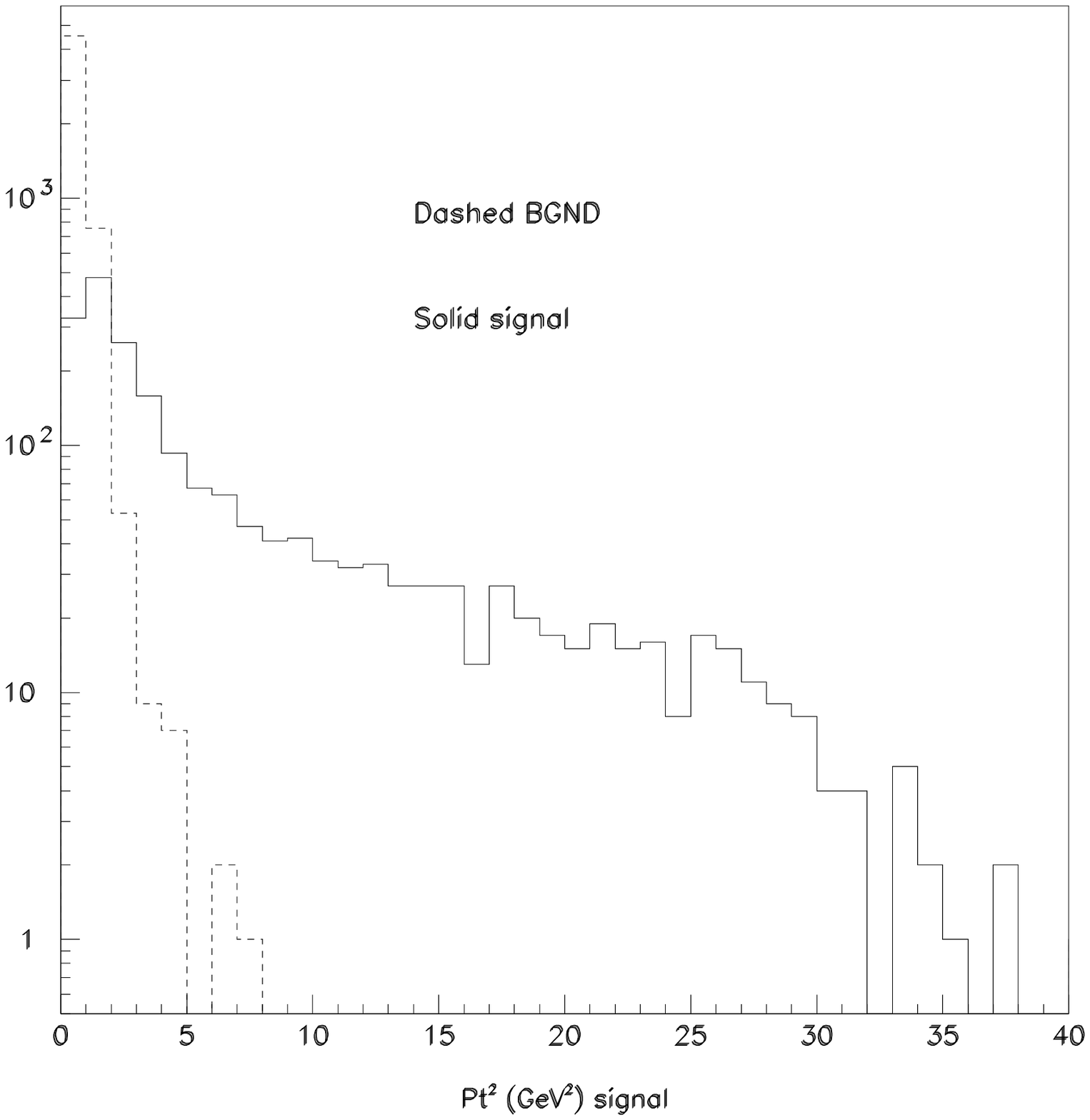}
\caption{Reconstructed $\mu^-$ $P_t^2$ with respect to the shower direction 
for 20~GeV and 50~GeV $\mu^+$ decaying in a neutrino factory. 
The muons are required to have energies exceeding 4~GeV.}
 \label{pt}
\end{figure}

\begin{figure} [h]
  \begin{center}
  \epsfxsize=0.75\textwidth
  \epsfbox{{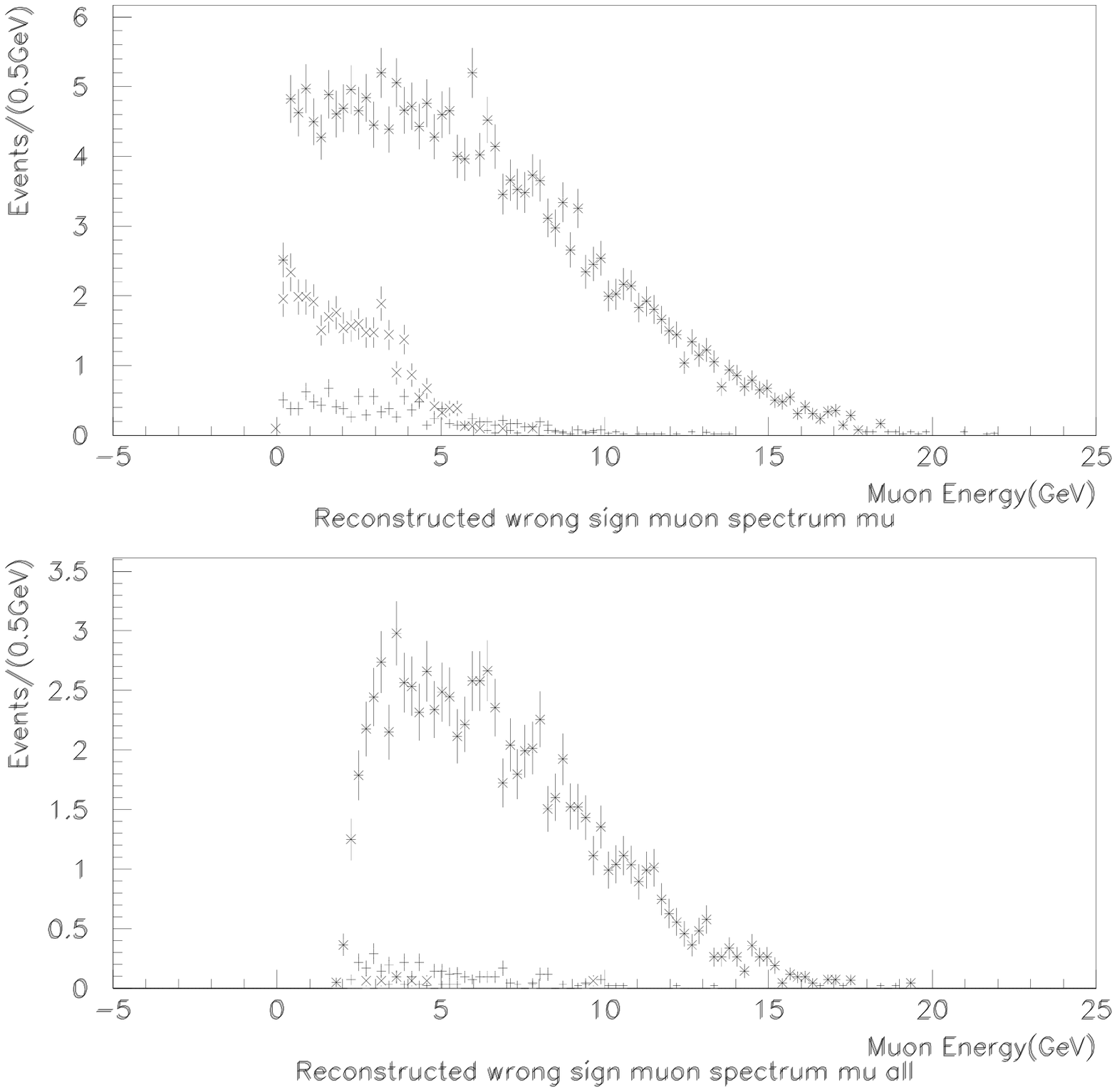}}
\caption{Reconstructed wrong-sign muons as a function of the muon energy for
a $\mu+$ 20 GeV ring. Top plot accepted events for the signal
($\nu_e \rightarrow \nu_\mu$--stars-- and 
$\nu_e \rightarrow \nu_\tau \rightarrow \mu + X$--crosses) and the potential
backgrounds (x).  The bottom plot shows the signal and the
background after cuts.}
\label{data}
\end{center}
\end{figure}

To be conservative, and reduce the dependence of our study on low energy 
processes that may not be adequately described by the Monte Carlo program, 
in our analysis all muons with generated energy below
4~GeV are considered lost. Muons with 
track length in steel less than 50~cm past the hadronic shower 
are also considered lost. All other muons are assumed 
to be identified 
with 100\% efficiency, and measured sufficiently well to determine 
their charge sign. 
For the background events we considered (i) all the $\pi$, $K$ decay events 
producing ``wrong--sign'' muons in NC interactions, 
and (ii) all the charm production
and $\pi$, $K$ decay events producing ``wrong--sign" muons 
in CC events where the primary muon was considered lost. 
To reduce the backgrounds, we cut on $P_t^2$. 
The reconstructed $P_t^2$ distribution is shown in Fig.~\ref{pt} 
for signal and background muons in a 10~kt detector 
2800~km downstream of 20~GeV and 50~GeV neutrino factories which 
provide $10^{20}$ $\mu^+$ decays. The oscillation parameters
corresponding to the LMA scenario IA1. 
As expected, background wrong--sign muons, 
tend to have smaller $P_t^2$ than genuine wrong--sign muons from 
the leptonic vertex. 
The reconstructed wrong-sign muon spectrum is shown in Fig.~\ref{data} 
for a 20~GeV storage ring before (top plot) and after (bottom plot) 
muon energy, track length and  $P_t^2 > 2$~GeV$^2$ cut were applied. 
Signal and background rates are summarized in Table~\ref{thetable}. 
After the cuts the signal/background ratio is above 10 to 1 in 
scenario IA1 for a detector 2800~km away, while $40-50\%$ of 
the $\nu_e \rightarrow \nu_\mu$ signal events are retained.

\begin{table}
\caption{Wrong-sign muon rates after all cuts for a 10~kt 
steel-scintillator detector downstream of a neutrino factory providing 
$10^{20}$ muon decays. The oscillation parameters correspond to  
scenario IA1. The loss of signal acceptance and the background 
rejection are due solely to the 
kinematic and reconstruction cuts.}
\bigskip 
\begin{tabular}{cc|ccccc}
\hline
\multicolumn{2}{c}{$\mu$ Ring}&$\nu_e \rightarrow \nu_\mu$  & 
$\nu_e \rightarrow \nu_\tau$  &  &  &  \\
Energy & Charge  & events &$\rightarrow \mu + X$ & 
background & signal &  background/CC \\
GeV &  &accepted& events &  events & acceptance & rate   \\
 \hline

 50 & + & 268.5 & 15.4  & 21.6 & 0.50$\pm$0.02 &$4.5\times10^{-4}$ \\
 50 & $-$ & 55.2 & 4.7 & 3.5 &0.48$\pm$0.02&$4.0\times10^{-5}$ \\
 20 & + & 85.7 & 3.5 & 0.7 & 0.41$\pm$0.02 & $2.0\times10^{-4}$ \\
\hline
\end{tabular}
\label{thetable}
\end{table}

\subsubsection{A Water Cerenkov detector} 

Preliminary studies have explored the possibility of using a large water
Cerenkov detector as a distant target for a neutrino factory beam.
Traditionally this type of detector has 
been used for measuring much lower energy neutrinos than expected 
at a muon storage ring, but to date water Cerenkov 
neutrino detectors are the only existing neutrino detectors with 
masses already in the 50~kt range.  Water is of course the 
lightest target material under consideration 
in this report, but this type of detector has several advantages 
when extrapolating to large masses, namely 
(i) low cost target material, (ii) only the surface of a very large volume 
needs to be instrumented, and (iii) good calorimetry.  
A large volume guarantees containment of hadronic and
electromagnetic showers (as well as muons up to a certain energy).  
The low density of the target and good 
angular resolution from the Cerenkov cone might yield an 
overall hadron angle resolution that is as good as or better than 
the corresponding resolution obtained with 
steel-scintillator calorimeters.

Water Cerenkov devices as large as 50~kt (SuperK) are already in
operation and is expected to continue data-taking for ten years or more.  
Therefore, the
response of the existing SuperK detector at a baseline 
distance of 9100~km has been
studied as a test case.  Next generation detectors, up to 1~Mton in mass, are
technically feasible and are currently under consideration for proton decay
and neutrino measurements, sited perhaps at the Kamioka mine or elsewhere.

\begin{figure} [h]
  \begin{center}
  \epsfxsize=0.75\textwidth
  \epsfbox[0 310 550 530]{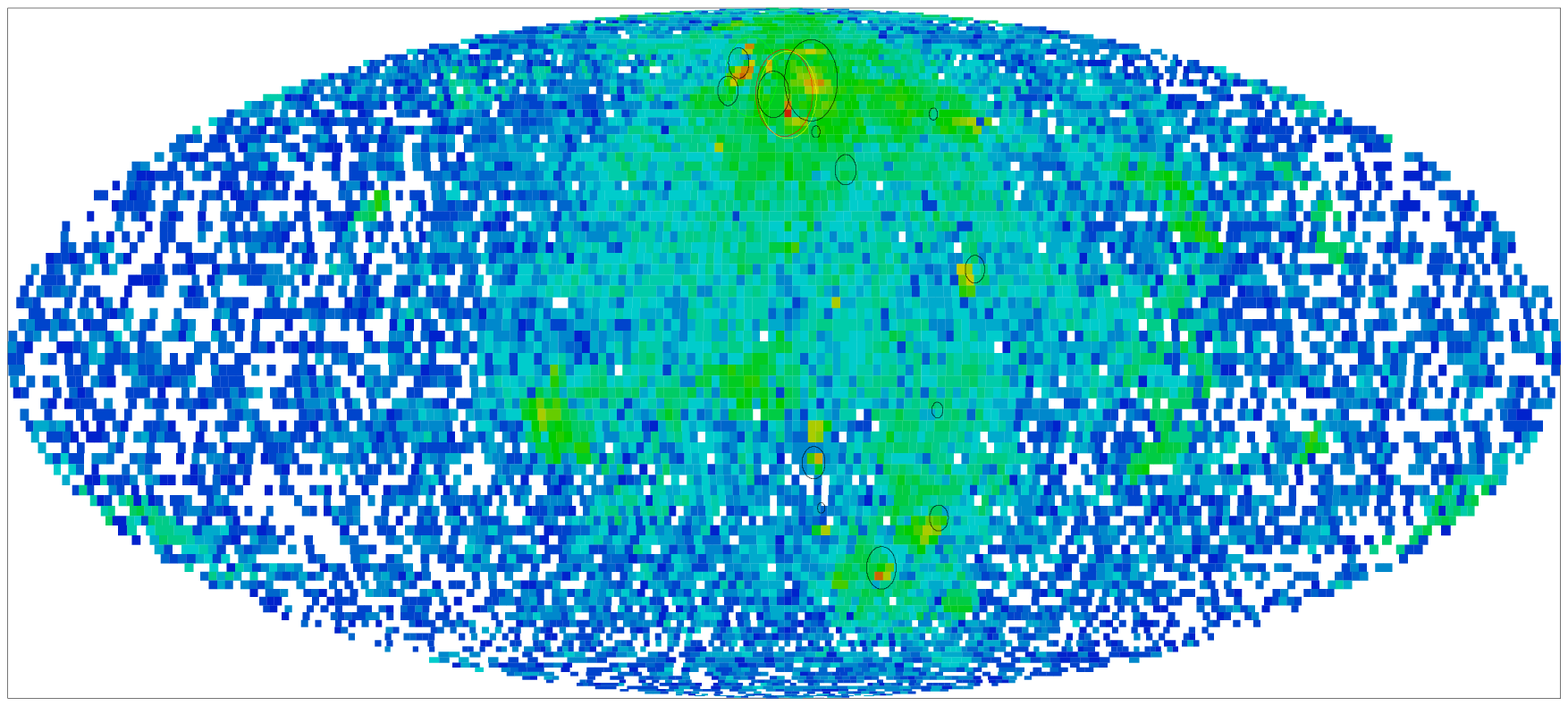}
\caption{Simulated neutrino event from a 50~GeV muon storage ring
in the SuperKamiokande detector.  The rings indicate where the 
reconstruction software found charged particles in the hadronic shower, 
as well as the exiting muon.}
\label{fig:f2kevt}
\end{center}
\end{figure}

For this initial study, the primary question is the suitability of a water 
Cerenkov detector for the higher energy neutrino beam produced by a 10-50
GeV muon storage ring. At these energies, the multiplicity of hadrons is
greater than for typical atmospheric neutrino interactions, and event
topologies are correspondingly more complex.  Figure~\ref{fig:f2kevt} 
shows the Cerenkov light produced in a 
typical neutrino event from a 50 GeV muon storage ring at the SuperK 
detector: the circles in the display are estimates of the outgoing 
angles of different charged particles produced in the hadronic 
shower.  Some particle identification is possible from the pulse-height 
information.  Reconstruction software from the
SuperK experiment must be further optimized to study the detector response to
neutrinos from 10~GeV and 50~GeV muon storage rings.  It is worth noting
that neutrinos produced by a 50~GeV muon beam induce a large number of events
in the material (rock) surrounding any detector (producing an entering muon),
and for a SuperK sized device these events outnumber the
those produced in the detector's water volume.  Both contained and 
entering events have therefore been studied.

The response of a detector the size of SuperK changes drastically as a
function of beam energy. At a 10~GeV neutrino factory, 57\% of the muon CC 
events are fully contained in the inner water volume, whereas only
11\% are fully contained at a 50~GeV neutrino factory. This large difference only
exists for events containing penetrating muons; at 10~GeV (50~GeV) both
${\bar \nu_e}$ and NC events are contained greater than 98\%
(90\%) of the time. The existing $\mu$-like particle identification 
algorithm works to produce a reasonably pure (89\%) $\nu_\mu$-CC sample for
fully contained events in the 10~GeV beam, but $e$-like events are a mixture
of of ${\bar \nu_e}$-CC, NC and $\nu_\mu$-CC contamination. Exiting and 
entering events are pure samples of $\nu_\mu$-CC simply because of 
their penetrating
nature.  The muon angular resolution (3${}^{\circ}$) is much less than the
muon-neutrino angular correlation (15${}^{\circ}$). With $2 \times 10^{20}$
decays at a 50~GeV neutrino factory and a baseline of 9100~km, 
approximately $200,000$
$\nu_\mu$ CC events would be observed entering or exiting the
current SuperK detector.  Combined with muon charge identification 
this sample should be able to provide good oscillation measurements.  

Implementing charge identification in a water Cerenkov 
detector is not trivial.  Two possibilities 
have been proposed: (i) several large water targets, 
each one followed by a thin external muon spectrometer, 
and (ii) a magnetic field 
introduced into the water volume itself.  Although the first 
design would have lower geometrical acceptance and a higher muon 
energy threshold, it would pose much less of a problem for the 
phototubes since the magnetic field would presumably be well-contained
in the spectrometer.  The second proposal could in principle have 
good low energy muon momentum acceptance, but the resolution on the 
muon and hadron shower angles might be compromised.  

For a magnetic field internal to the target, 
0.5-1~kG is sufficient to visually determine
the charge of a several meter-long ($>1$~GeV) 
muon, but no automated algorithms have 
yet been developed.  A number of conceptual magnet designs have been 
studied: solenoidal, toroidal, and concentric current loops in the center or
at the ends of the detector.
A detailed study of one particular design has shown that one can 
immerse the central volume of a SuperK sized
detector in a 0.5~kG magnetic field while leaving only a 0.5~G fringe field
(which may be acceptable with shielding and/or local compensation) in the
region of the PMTs.  Many of the difficulties inherent in placing a field
inside a water detector would be avoided if an alternative light collector
(insensitive to the field) were used.  Work on magnet design and
alternative light collection is ongoing, but the internal magnetic field
option must be considered speculative at this point.

The results we 
will describe in the remainder of this section 
are for a water Cerenkov detector with an external 
magnetic field, because 
neutrino event reconstruction is more straightforward to simulate and the 
spectrometer technology is well-understood.  
Although the studies of this detector are very preliminary, 
they look promising and warrant further investigation.  
We have used a 
LUND/GEANT Monte Carlo program which uses as its geometry a 
$40\times40\times100$~m$^3$  
box of water, followed by a 1~m long muon spectrometer.  This 
simulation can be used to study  
acceptance issues and background contamination for a range of 
geometries and storage ring energies.  

\begin{figure} [h]
  \begin{center}
  \epsfxsize=0.75\textwidth
  \epsfbox[0 0 520 530]{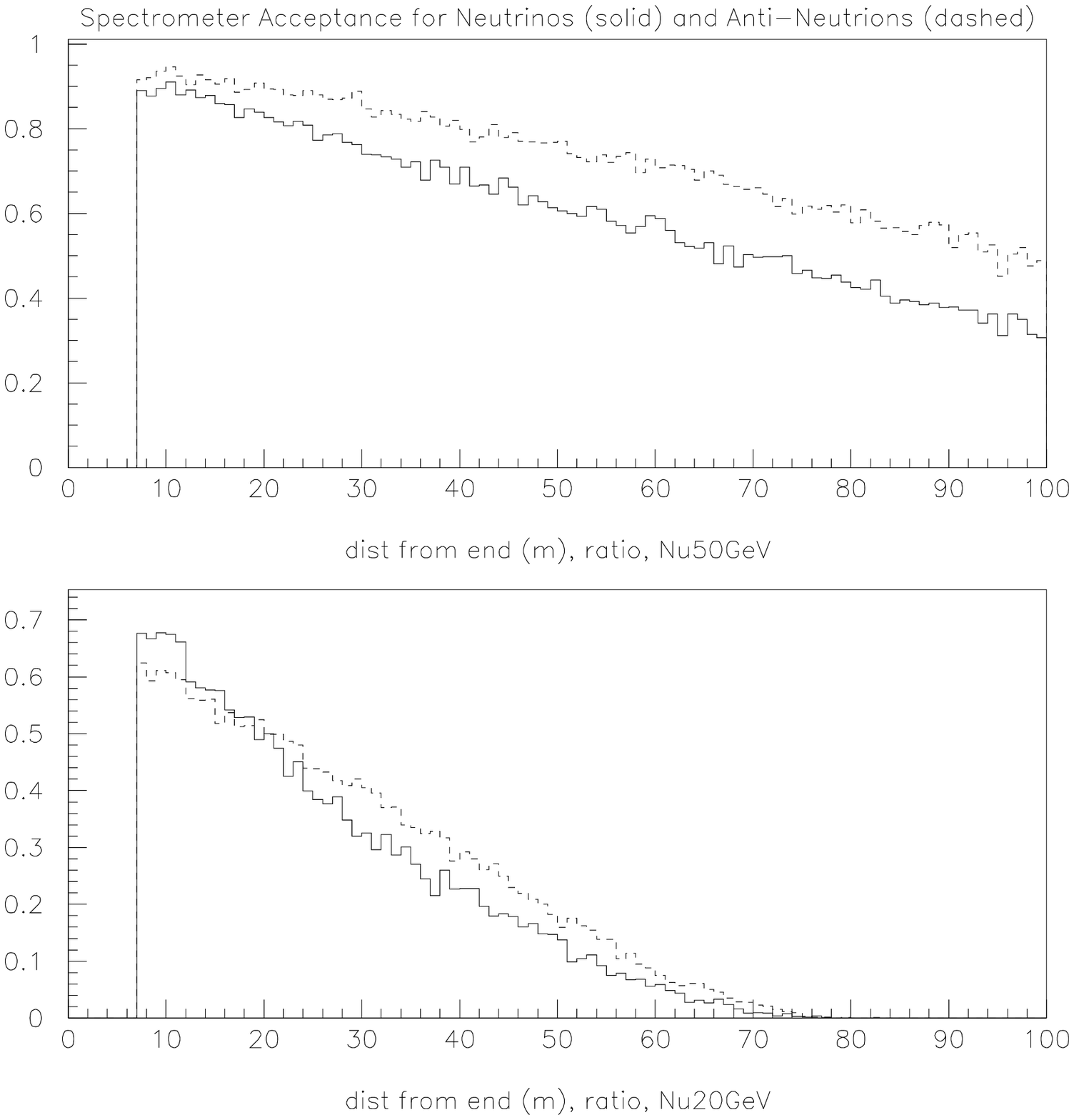} 
\caption{Acceptance in a water target for 
charged current $\nu_\mu$ (solid) and $\overline\nu_\mu$ (dashed) 
events in a 20 and 50 GeV storage ring, as a function of distance of 
the neutrino interaction vertex from the muon spectrometer.}
  \end{center}
\label{fig:wateracc}
\end{figure}

Figure~\ref{fig:wateracc} shows the geometrical acceptance for the box-like 
water Cerenkov detector as a function of distance of the neutrino 
interaction vertex from the spectrometer, for CC 
$\nu_\mu$ events from 20 and 50~GeV storage rings.  The 
loss in acceptance close to the spectrometer is due to rejection of 
events where there is more than one muon which traverses the 
spectrometer (where the extra muon comes from background processes).
It is clear that for a 20~GeV muon storage ring one would want 
a muon spectrometer much more frequently than once every 100~m.  Of 
course, noting that steel has a density of 8 times that of water, 
the smaller the ratio of water thickness to steel thickness 
the more it approximates a magnetized steel/scintillator target 
interspersed with water volumes with fine granularity.

Clearly more work is needed to optimize the design for this 
kind of detector, but 
it might be an inexpensive 
compromise between a coarse-grained sampling calorimeter 
and a very fine-grained liquid argon TPC.  

\subsubsection{Specialized $\tau$--lepton detectors}

The measurement of $\tau$--lepton appearance in large mass 
neutrino detectors is challenging. There are several ideas 
that might lead to viable new $\tau$--appearance detectors 
within the next 5--10~years, and that might be suitable for use at a 
neutrino factory. We briefly describe three examples in the 
following: (i) a perfluorohexane Cerenkov detector, 
(ii) a hybrid emulsion detector, and (iii) a very fine--grained 
micro--strip gas chamber target.

\begin{figure}[h]
  \epsfxsize=\textwidth
  \epsfbox[50 400 580 690]{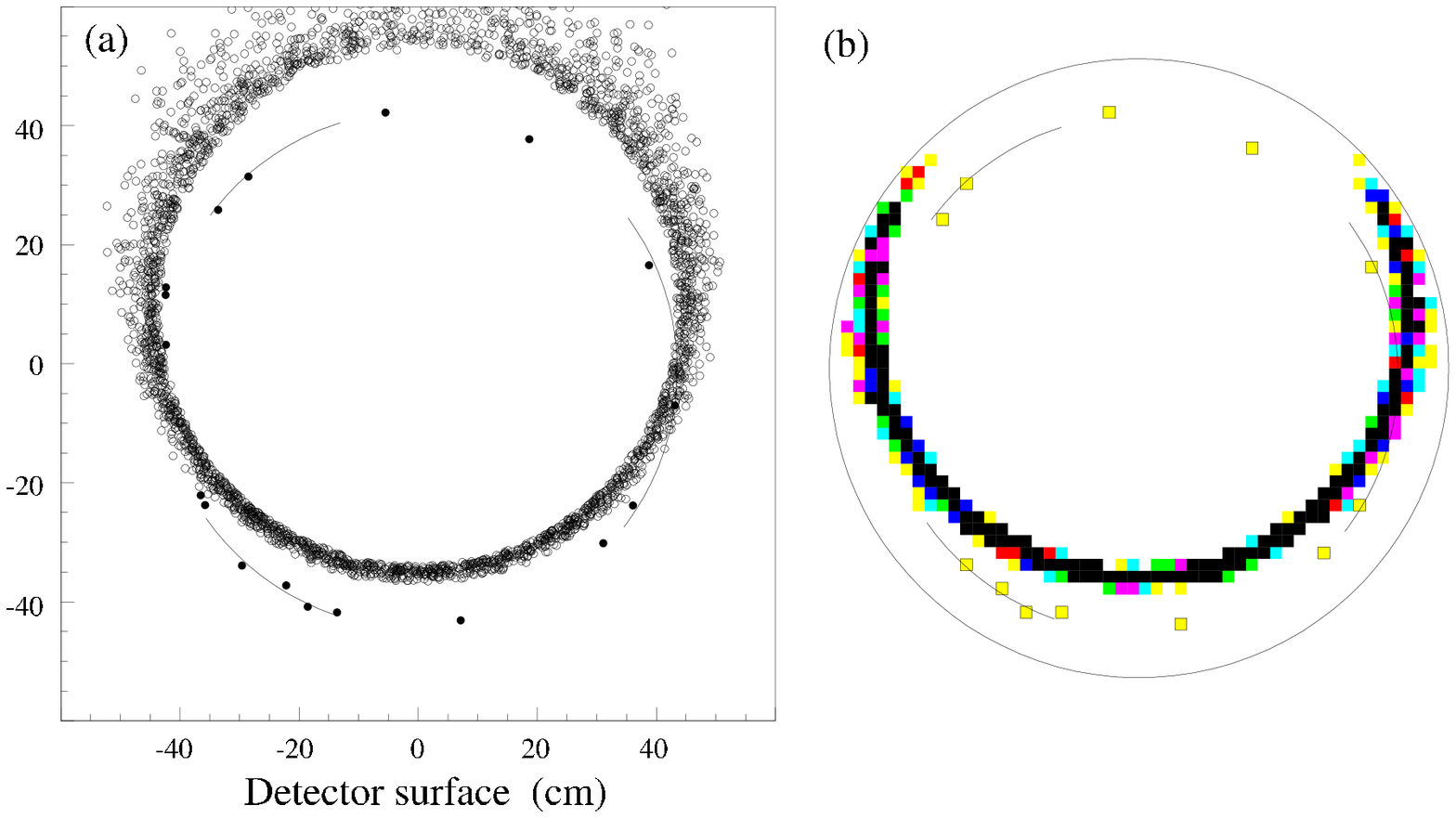}
\caption{Quasi-elastic $\nu_\tau$ event in a perfluoroHexane Cerenkov 
detector:  the ring described by about eight hits 
on the left is from the tau before it decays.}  
\label{fig:forty}
\end{figure}

Consider first a Cerenkov detector filled with perfluorohexane ($C_6F_{14}$), 
which has a density 1.7 times that of water. 
This has been proposed~\cite{forty} for use 
in the CERN to Gran Sasso beamline.  
The detector geometry consists of several target volumes followed by short
muon spectrometer modules.  
A 1~Ton perfluorohexane detector 
(with a very different geometry) exists at DELPHI. 
The $\tau$--lepton signature in this type of detector consists of 
a sparsely populated Cerenkov ring from the $\tau$ before it decays, 
together with a more densely populated ring from the daughter muon. 
The two rings would have offset centers. Figure~\ref{fig:forty} shows a
simulated quasi-elastic $\nu_\tau$ event (no hadron energy)
from this kind of detector. 
This technique would probably 
not work for events with high energy hadron showers 
because of the large 
number of charged particles that would result in overlapping rings near 
the initial $\tau$--lepton ring.

Next consider a hybrid emulsion detector consisting of, for example, 
thin ($\sim 100\mu$m) sheets of emulsion
combined with low-density ($\sim 300\mu$m) spacers. 
The signature for a $\tau$--lepton 1--prong decay would be 
a change in direction of the track measured before and after the 
spacer~\cite{strolin}.  
For charge identification the  
detector could be within a large magnetic field volume. 
With an emulsion track angular resolution of 2~mrad, a 
5$\sigma$ charge--sign determination of a 10~GeV/c  
charged particle could be achieved with a 
2~T field and a 1.2~mm thick spacer \cite{para}.  
An $\sim 20$~kt hybrid emulsion detector of this type might 
consist of 20~kt of steel segmented into 1~mm thick sheets,
and an equal volume of thin emulsion layers plus low density spacers. 
The resulting detector would fit into the ATLAS barrel toroid 
magnet, which has a magnetic field ranging from 2 to 5 Tesla~\cite{atlas}. 
A hybrid emulsion detector with an external downstream muon spectrometer 
will be used by 
the OPERA experiment, which is to be put in the CERN to Gran Sasso beam. 
The muon spectrometer will determine the charge sign for
$\tau\to\mu$ decays provided the muon 
reaches the spectrometer. 
According to the OPERA studies~\cite{opera}, 
with an average neutrino energy of 20~GeV 
the total efficiency for 
seeing the $\tau$ decays is 29\% (including the branching
ratios). 
The efficiency is largely geometric 
and should not be compromised by the addition of a magnetic field, 
provided the bend in the spacer due to the magnetic field is much less
than the "apparent bend" due to the $\tau$--lepton decay.  

Finally, consider a target consisting of a tracking 
chamber constructed from micro--strip gas chambers (MSGCs) and a low $Z$ 
material (for example, nylon) in a large magnetic field volume.
This would be a NOMAD~\cite{nomad}--like detector with a much larger 
O(1~kt) fiducial mass and an improved spatial resolution.  Because of the 
low $Z$ of the material electrons can travel a long distance in the 
detector before showering, and with a high enough field their charge 
can therefore be measured.  Although a kink is not seen, the 
tau decay could be distinguished kinematically. 
For example, nylon has a radiation length of 37 cm.  With a $B$ field of 
1~Tesla and 
MSGC's every 10~cm one would have an $8\sigma$ measurement of a 50~GeV 
electron's charge.  This idea is worthy of further consideration, 
particularly if the LSND signal is confirmed and lower-mass tau 
detectors are warranted.  

\subsubsection{Detector summary} 

In our initial studies we have simulated the performance of 
steel/scintillator and liquid Argon detectors at 
a neutrino factory. Results are encouraging. These technologies could 
provide detectors with masses of order 10~kt (liquid Argon) to 
a few $\times 10$~kt (steel/scintillator) that 
yield good wrong--sign muon identification and adequate 
background rejection. Our simulations of the capabilities 
of water Cerenkov detectors at a neutrino factory are less 
advanced, but initial results are encouraging, and this 
detector technology might permit very large detector masses 
to be realized. Some relevant characteristics of 
steel/scintillator,  liquid Argon, and water Cerenkov detectors 
are listed in Table~\ref{dettab}. It is premature to choose 
between detector types at this early stage. However, some general 
points are worth noting:
\begin{description}
\item{(i)} We believe that a cost optimization of detector mass (cost) 
versus neutrino factory beam intensity (cost) will probably favor 
detectors that are at least a factor of a few to a factor of 10 
more massive than, for example, ICANOE or MINOS. A detector mass 
in the range 10~kt to 50~kt does not seem unreasonable.
\item{(ii)} The minimum energy a muon must have for good identification 
and measurement may well determine the minimum viable 
muon storage ring energy. This threshold is a few GeV, and is 
detector technology dependent. With a steel/scintillator detector 
and a threshold of 4~GeV, for example, the minimum acceptable neutrino 
factory energy appears to be in the neighborhood of 20~GeV.
\end{description}
In this initial study we have not comprehensively considered 
to what extent massive detectors at a neutrino factory need to 
be deep underground. It seems very likely that detectors with 
low detection thresholds (water Cerenkov and liquid argon) 
will need to be well protected from cosmic ray backgrounds, 
regardless of the neutrino factory energy.  For the 
steel/scintillator detector, the cosmic ray backgrounds for 
charged current events with muons in them are likely to be small 
for a detector at the surface of the earth, but 
there will be substantial background to neutral current 
or $\nu_e$ charged current interactions.  
Finally, we note that the development of a new generation of 
very massive detectors 
capable of identifying and measuring the charge--sign of muons and 
$\tau$--leptons, would be of great benefit to a neutrino factory. 
There is a possible area of mutual interest with the nucleon decay 
community in developing the technology for a really massive 
1~Mton scale water Cerenkov detector. This possibility deserves 
further investigation.

\begin{table}[h]
\caption{Comparison of detector parameters for candidate 
detectors at a neutrino factory.}
\bigskip
\begin{center}
\begin{tabular}{|l|l|l|l|} 
\hline
Characteristic  &  \multicolumn{3}{|c|}{Detector Technology } \\ 
        &        Steel/Scint  &   Liquid Argon TPC & Water Cerenkov \\\hline
Resolutions of: & & & \\ 
Electron Energy& $50\%/\sqrt{E}$ &  $3\%/\sqrt{E}\oplus 1\%$ & $0.6\oplus 2.6\%/\sqrt{E}$ \\
Hadron Energy  & $85\%/\sqrt{E}$ &  $20\%/\sqrt{E}\oplus 5\%$ & 20-30\% \\
Muon Energy    &$5\%$ &  $20\%$ & $20\%^\dagger$ \\
Hadron Shower Angle &    &$.13/\sqrt{{\rm p}}$ rad & \\ 
  & &  (each hadron)& \\
Muon Angle &       5\% for  & & $3^\circ$ \\
           &          50cm track & $.02\oplus .21/\sqrt{{\rm p}}$ 
  &                    \\
Maximum mass  &50 kton & 30 kton & 1Mton?\\
What limits size? & & safety, tunnel & tunnel\\
Required Overburden$^{**}$ & 0 m  &50 m & 50-100m\\
Analysis Cuts & $P_{\mu}>4$ GeV &$P_{\mu}>2$ GeV& \\
  & $P_t^2 > 2$ GeV$^2$ & & \\ 
Background level & $10^{-4}$  &$2\times 10^{-5}$ & \\
\hline\hline 
\end{tabular}
\end{center}
$^{**}$ The overburden required for all technologies
depends on the neutrino factory duty factor.  
The overburden required for a steel-scintillator
calorimeter also depends on the energy of the muon storage ring; but
in the past this type of detector has been used 
at ground level with minimal 
contamination in the $\nu_\mu$ charged current sample 
above a neutrino energy of 5~GeV.$^{\dagger}$ The 
muon momentum resolution would be comparable to that of an 
ICANOE detector if the muon spectrometer were separated from the 
water tank volume.
\label{dettab}
\end{table}

\clearpage
\subsection{Oscillation measurements}

Using the oscillation scenarios described in section~\ref{theory}
as examples, we can now assess how well the neutrino oscillation
physics program outlined in section~\ref{program} can be pursued 
at a neutrino factory with the detectors described in section~\ref{detectors}. 
In the following sub-sections the oscillation measurements 
that can be made at a neutrino factory are discussed as a function of 
baseline, 
muon beam energy, and muon beam intensity. 
In particular we consider: 
\begin{description}
\item{(i)} The first observation of $\nu_e \rightarrow \nu_\mu$
oscillations, the measurement of the sign of $\delta m^2$ and hence the 
pattern of neutrino masses (section~3.5.1).

\item{(ii)} The first observation of $\nu_e \rightarrow \nu_\tau$ oscillations 
(section~3.5.2).

\item{(iii)} The measurement of $\nu_\mu \rightarrow \nu_\tau$ oscillations 
(section~3.5.3).

\item{(iv)} Precision measurements of the oscillation parameters 
(section~3.5.4).

\item{(v)} The search for CP violation in the lepton sector 
(section~3.5.5). 
\end{description}
The results are based on the calculations described in more detail in 
Refs.~\cite{cerv00, bgrw00, bern00, camp00}. 
The calculations from Ref.~\cite{camp00} are for an ICANOE type detector, 
and include realistic resolution functions, analysis cuts, and 
background modeling, but use a constant average matter density to 
compute matter effects. 
The calculations from 
Refs.~\cite{bern00, bgrw00, cerv00} all use resolution functions typical 
of steel/scintillator detectors and, unless explicitly stated, 
reasonable thresholds on the detected muon energies. The calculation 
from Ref.~\cite{bgrw00} does not include backgrounds but covers a broad 
range of scenarios, and uses the explicit trans--Earth density profile 
to compute matter effects. In contrast, the calculations from 
Refs.~\cite{bern00,cerv00} have been used to look at only a few 
oscillation scenarios, but 
include backgrounds  
and use respectively the average Earth density and 
the explicit density profile in computing matter effects.
It should be noted that although there are significant differences in the 
details implemented in the calculations, in general all the four groups 
arrive at similar assessments for the measurement sensitivity at a 
neutrino factory as a function of energy, intensity, and baseline.

\subsubsection{Observation of $\nu_e \rightarrow \nu_\mu$ oscillations and 
the pattern of neutrino masses}

At a neutrino factory $\nu_e \rightarrow \nu_\mu$ oscillations would be 
signaled by the appearance of CC interactions tagged by a 
wrong--sign muon~\cite{geer98}. 
Within the framework of three--flavor 
oscillations the $\nu_e \rightarrow \nu_\mu$ oscillation amplitude is 
approximately proportional to $\sin^2 2\theta_{13}$. 
At the present time only an upper limit exists on $\sin^2 2\theta_{13}$. 
The next generation long-baseline oscillation experiments are expected 
to be able to improve the sensitivity to 
$\sin^2\theta_{13}\approx 10^{-2}$, i.e. about one 
order of magnitude below the present bound. If $\sin^2 2\theta_{13}$ 
is in this range we would expect to observe 
$\nu_e \rightarrow \nu_\mu$ oscillations at a relatively low intensity 
neutrino factory, measure matter effects, 
and determine the pattern of neutrino masses~\cite{bgrw99}. 
This is discussed further in the remainder of this sub--section. 

\begin{figure}[h]
\epsfxsize3.0in
\centerline{\epsffile{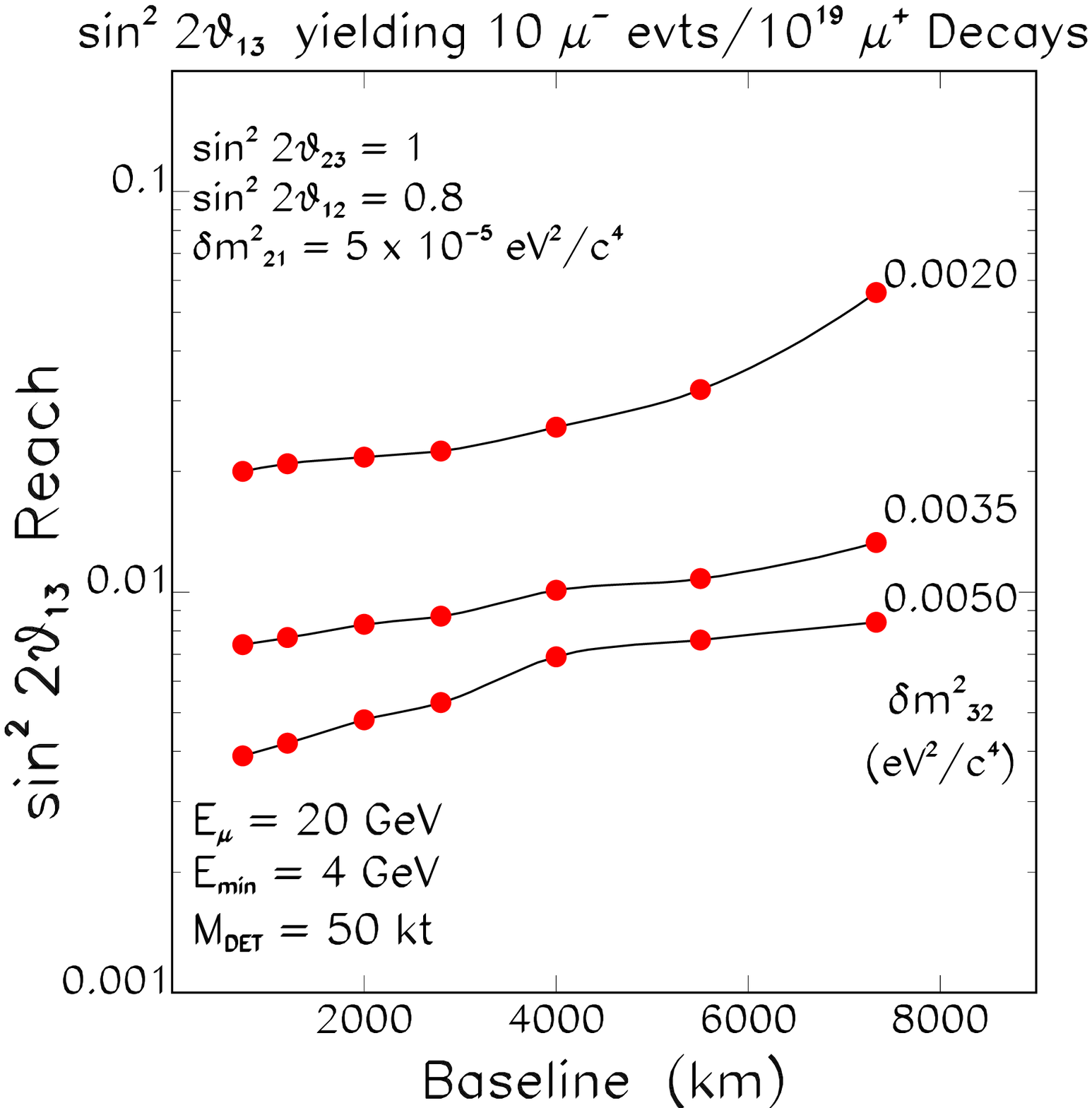}}
\vspace{0.3cm}
\caption{Reach in $\sin^22\theta_{13}$ for the observation of
10 $\mu^-$ events from $\nu_e \rightarrow \nu_\mu$ oscillations,
shown versus baseline for three $\delta m^2_{32}$
spanning the favored SuperK range. The other oscillation parameters
correspond to the LMA scenario IA1.
The curves correspond to
$10^{19} \mu^+$ decays in a 20~GeV neutrino factory with
a 50~kt detector, and a minimum muon detection threshold of 4~GeV. 
Results are from Ref.~\ref{bgrw00}.}
\label{fig:v1}
\end{figure}

\begin{figure}
\epsfxsize5.0in
\centerline{\epsffile{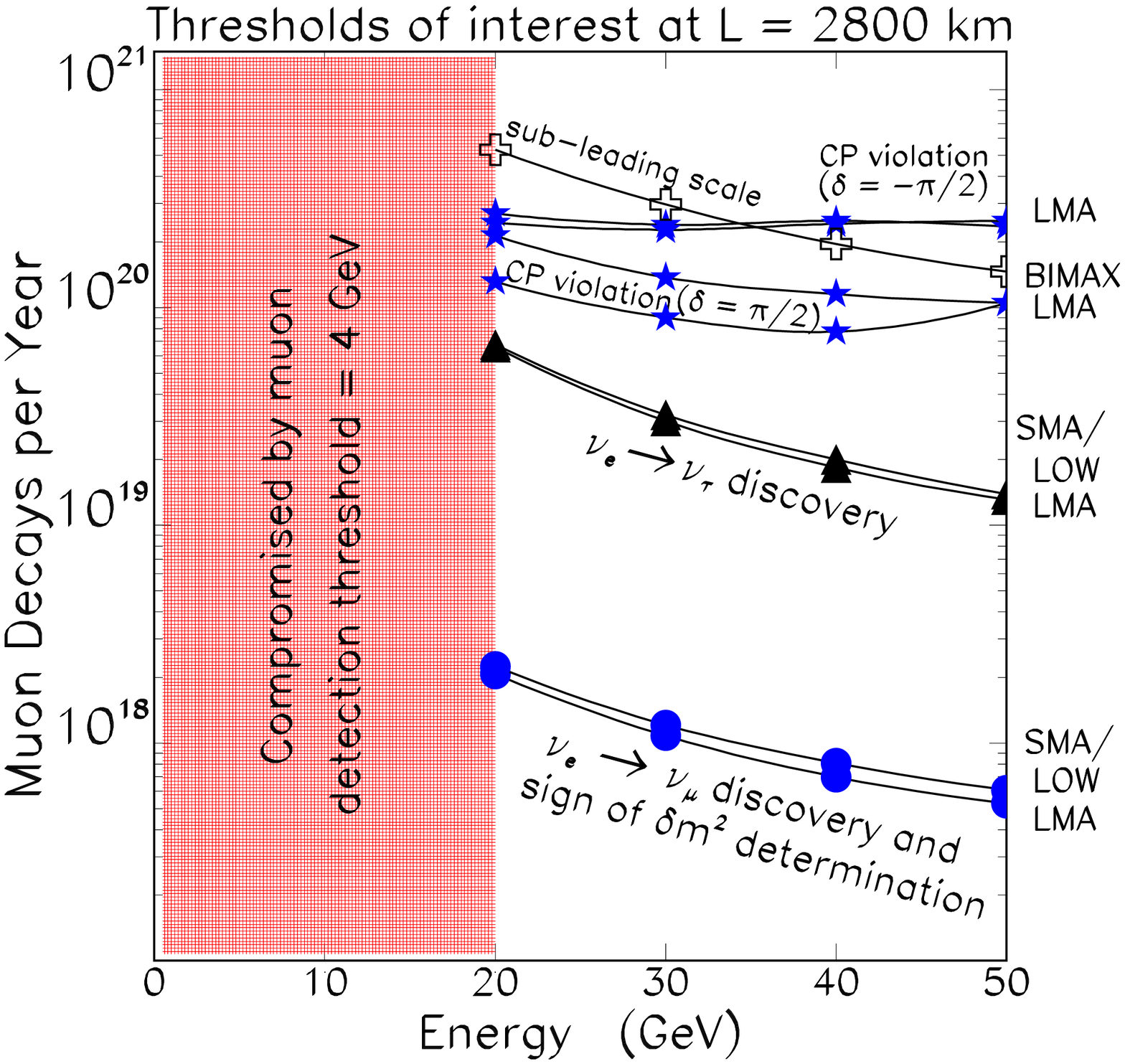}}
\vspace{0.3cm}
\caption{The required number of muon decays needed in the beam--forming
straight section of a neutrino factory to achieve the physics goals described
in the text, shown as a function of storage ring energy for the 
LMA scenario IA1, SMA scenario IA2, LOW scenario IA3, and a bimaximal 
mixing scenario BIMAX.
The baseline is taken to be 2800~km, and
the detector is assumed to be a 50~kt wrong--sign muon
appearance device with a muon detection threshold of 4~GeV or, for
$\nu_e \rightarrow \nu_\tau$ appearance, a 5~kt detector. 
Results are from Ref.~\ref{bgrw00}.}
\label{fig:v2}
\end{figure}

It is useful to define~\cite{bgrw00} the $\sin^22\theta_{13}$ 
``reach" for an
experiment as the value of $\sin^22\theta_{13}$ for which a given physics
goal would be met. We take as our initial goal the observation of 10 
$\nu_e \rightarrow \nu_\mu$ events tagged by a wrong--sign muon. 
Consider first the 
$\sin^22\theta_{13}$ reach for a 50~kt detector sited a distance $L$ 
from a 20~GeV neutrino factory in which there are $10^{19} \mu^+$ decays 
in the beam--forming straight section. The baseline--dependent 
$\sin^22\theta_{13}$ reach is shown in Fig.~\ref{fig:v1} for a three-flavor 
oscillation scenario in which $\delta m^2_{21}, \sin^22\theta_{12}$, 
and $\sin^22\theta_{23}$ correspond to the LMA scenario IA1, 
and the value of $\delta m^2_{32}$ is varied over the favored SuperK range. 
Backgrounds are expected to be less than one event for $L \ge 2800$~km 
(Table~\ref{dm2table}), and are not included in the calculation shown 
in the figure. 
If $\delta m^2_{32}$ is in the center of the SuperK range, 
the $\sin^22\theta_{13}$ reach is about an order of magnitude below 
the currently excluded region, improving slowly with decreasing $L$. 
However, at short baselines ($L < 2800$~km) backgrounds may degrade the 
$\sin^22\theta_{13}$ reach. 
The reach improves (degrades) by a about a factor of 2 (3) if 
$\delta m^2_{32}$ is at the upper (lower) end of the current SuperK 
range. If the oscillation parameters correspond to the LMA scenario IA1 
($\sin^22\theta_{13} = 0.04$), 
then only $2 \times 10^{18}$ muon decays are required at a 20~GeV 
neutrino factory to observe 10~signal events in a 50~kt detector at 
$L = 2800$~km. The calculation~\cite{bgrw00} 
assumes that CC events producing 
muons with energy less than (greater than) 4~GeV are detected with 
an efficiency of 0 (1). 
The number of muon decays needed to observe 10 $\nu_e \rightarrow \nu_\mu$ 
events 
is shown in Fig.~\ref{fig:v2} as a function of $E_\mu$ for the LMA 
scenario IA1, the SMA scenario IA2, and the LOW scenario IA3. 
The required muon beam intensities decrease with increasing $E_\mu$, 
and are approximately proportional to $E_\mu^{-1.5}$. Compared with 
the SMA and LOW scenarios, slightly less 
intensity is needed for the LMA scenario, showing the small but 
finite contribution to the signal rate from the sub--leading $\delta m^2$ 
scale. In all three scenarios (LMA, SMA, LOW) a 20~GeV neutrino factory 
providing $10^{19}$ decays in the beam--forming straight section 
would enable the first observation of $\nu_e \rightarrow \nu_\mu$ 
oscillations in a 50~kt detector provided $\sin^22\theta_{13} > 0.01$.
It should be noted that although  $\sin^22\theta_{13}$ could be very 
small, there are models~\cite{albright} 
that predict $\sin^2 2\theta_{13} \simeq 0.01$. 
\begin{figure}
\epsfxsize3.5in
\centerline{\epsffile{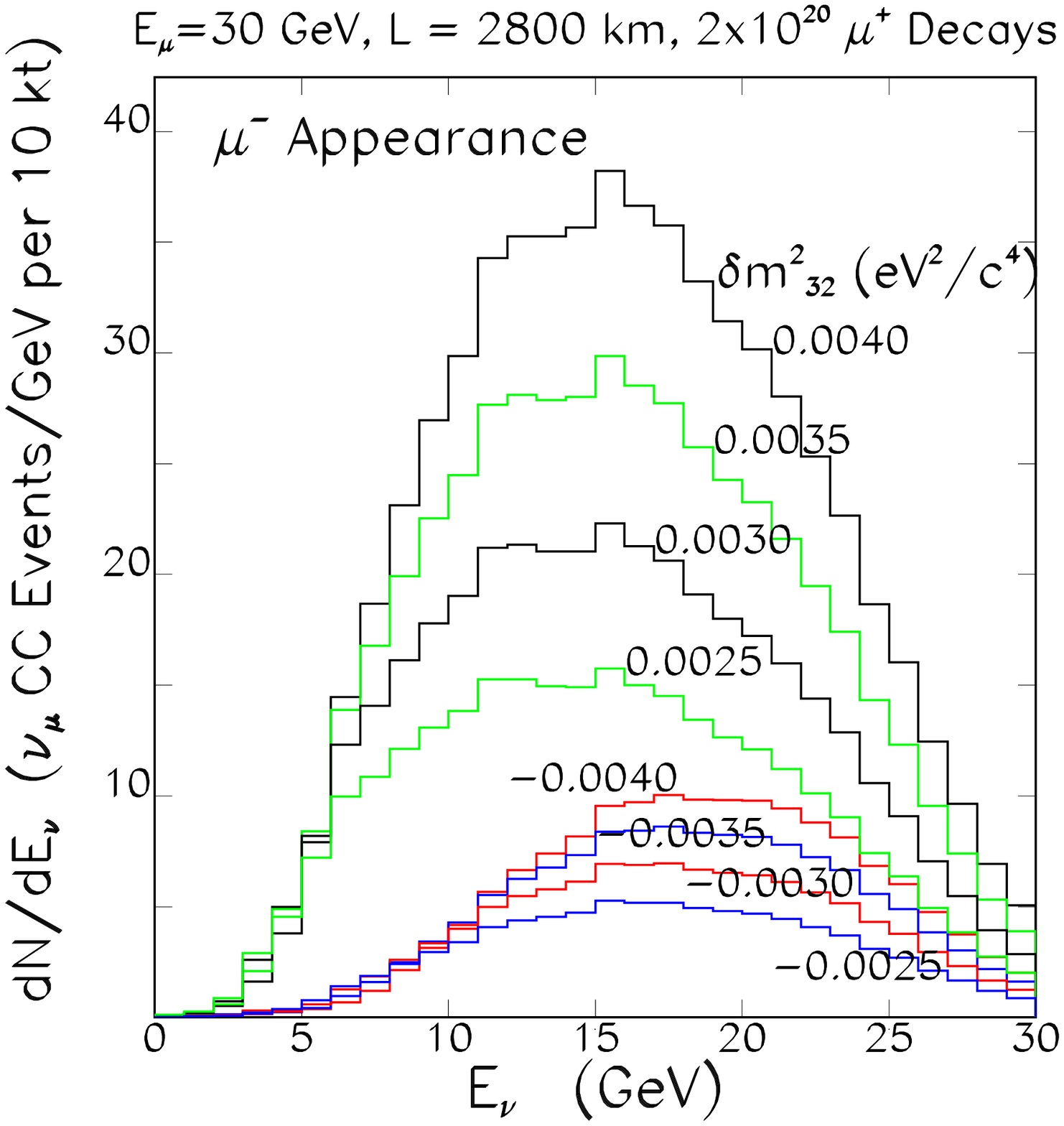}}
\caption{Predicted measured energy distributions for CC events tagged by a
wrong-sign (negative) muon from $\nu_e \rightarrow\nu_\mu$ oscillations 
(no cuts or backgrounds), shown 
for various $\delta m^2_{32}$, as labeled. The predictions correspond to $2
\times 10^{20}$ decays, $E_\mu = 30$~GeV, $L = 2800$~km, with the values for
$\delta m^2_{12}$, $\sin^22\theta_{13}$, $\sin^22\theta_{23}$, 
$\sin^22\theta_{12}$, and $\delta$ corresponding to 
the LMA scenario IA1. Results are from Ref.~\ref{bgrw99}.}
\label{fig:v3}
\end{figure}
\begin{figure}
\epsfxsize3.5in
\centerline{\epsffile{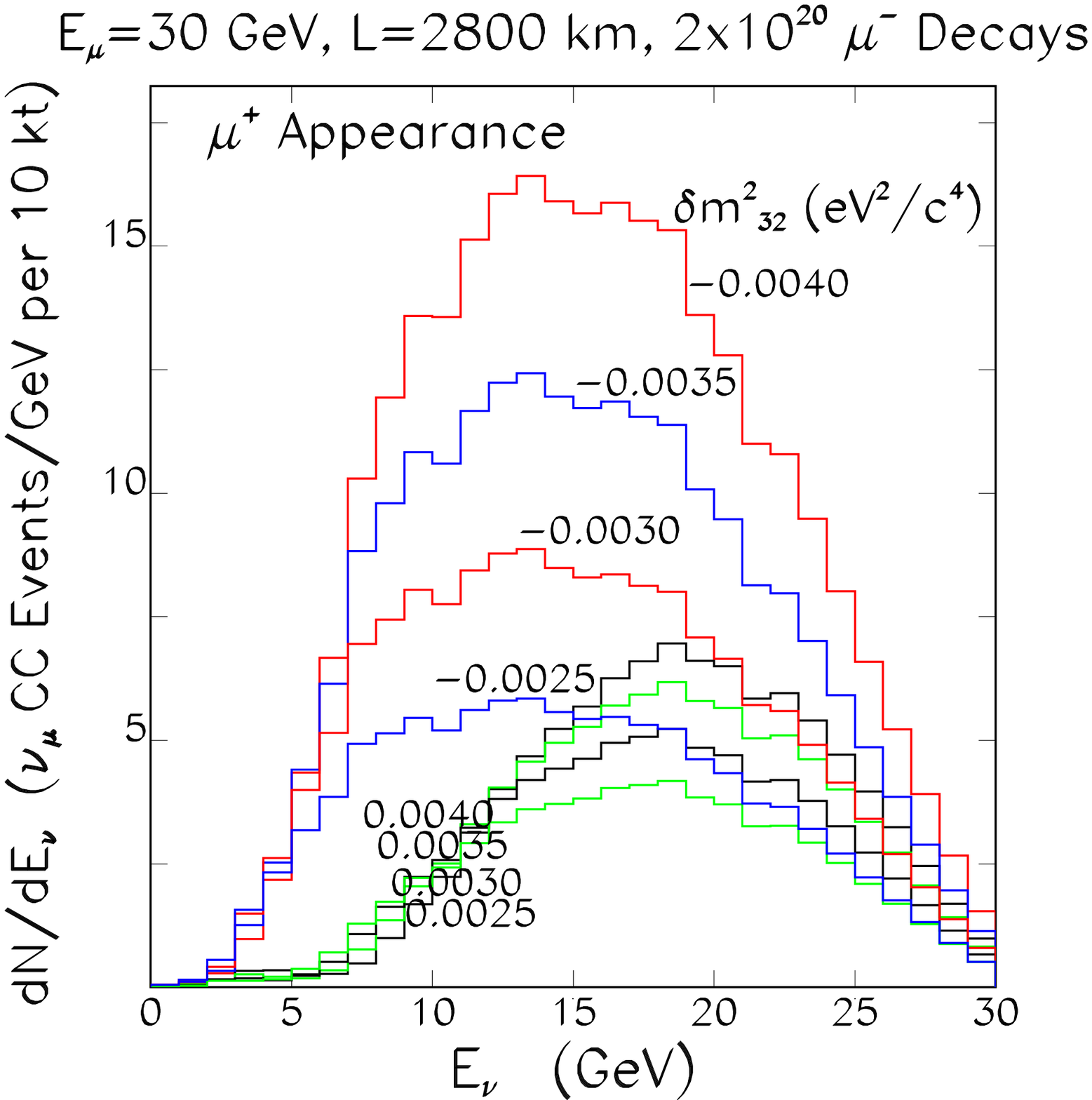}}
\caption{Same as previous figure, for CC events
tagged by a wrong-sign (positive) muon from $\bar{\nu}_e \rightarrow
\bar{\nu}_\mu$ oscillations.
}
\label{fig:v4}
\end{figure}

Having established $\nu_e \rightarrow \nu_\mu$ oscillations, 
further data taking would facilitate the 
measurement of matter effects and the determination of the sign of 
$\delta m^2$, and hence the pattern of neutrino masses. 
To illustrate the effect of matter on the 
$\nu_e \rightarrow \nu_\mu$ oscillation probability,  
the predicted measured energy distributions 2800~km downstream of 
a 30~GeV neutrino 
factory are shown in Figs.~\ref{fig:v3} and \ref{fig:v4} for respectively 
$\nu_e \rightarrow \nu_\mu$ and 
$\overline{\nu}_e \rightarrow \overline{\nu}_\mu$ wrong--sign muon 
events. 
The distributions are shown for 
a range of positive and negative values of $\delta m^2_{32}$. 
Note that for a given $|\delta m^2_{32}|$, if $\delta m^2_{32} < 0$ 
we would expect to observe a lower wrong--sign muon event rate and 
a harder associated spectrum when positive muons are stored in 
the neutrino factory than when negative muons are stored. 
On the other hand, if $\delta m^2_{32} > 0$
we would expect to observe a higher wrong--sign muon event rate and
a softer associated spectrum when positive muons are stored in
the neutrino factory than when negative muons are stored. Hence, 
measuring the differential spectra when positive and negative muons are 
alternately stored in the neutrino factory can enable the sign of 
$\delta m^2_{32}$ to be unambiguously determined~\cite{bgrw99}. 

\begin{table}[t]
\caption{Wrong-sign muon rates for a 50~kt detector 
(with a muon threshold of 4~GeV) a distance $L$ downstream 
of a neutrino factory (energy $E_\mu$) providing $10^{19}$ muon 
decays. Rates are shown for LMA scenario IA1 with both signs of 
$\delta m^2_{32}$ considered separately. The background rates 
listed correspond to an assumed background level of 
$10^{-4}$ times the total CC rates (see section~\ref{detectors}) 
with no energy dependence. Energy dependent cuts might suppress 
backgrounds further. 
Results are from Ref.~\ref{bgrw99}.
}
\bigskip
\begin{tabular}{cc|ccc|ccc} 
\hline
$E_\mu$&$L$&\multicolumn{3}{c}{$\mu^+$ stored}&\multicolumn{3}{c}{$\mu^-$ 
stored}\\
GeV & km & $\delta m^2_{32} > 0$ & $\delta m^2_{32} < 0$ & Backg &
$\delta m^2_{32} > 0$ &$\delta m^2_{32} < 0$ & Backg \\
\hline
20 & 732 & 52. & 36. & 7.3 & 32. & 26. & 6.5 \\
   &2800 & 46. & 9.2 &0.43 & 7.1 & 26. & 0.36\\
   &7332 & 33. & 0.97&0.063&0.55 & 19. & 0.05\\
\hline
30 & 732 &100. & 72. & 25. & 58. & 45. &  24.\\
   &2800 & 90. & 26. & 1.6 & 19. & 43. &  1.4\\
   &7332 & 43. & 3.3 & 0.19& 2.1 & 33. & 0.17\\
\hline
40 & 732 &150. &110. & 60. & 83. & 65. &  58.\\
   &2800 &140. & 48. & 4.0 & 36. & 64. &  3.8\\
   &7332 & 54. & 5.6 & 0.49& 3.1 & 28. & 0.43\\
\hline
50 & 732 &200. & 140.& 120.& 110.& 84. & 120.\\
   &2800 &180. & 71. &  7.9&  53.& 82. &  7.7\\
   &7332 & 56. & 8.0 &  1.1& 5.0 & 34. &  1.0\\
\hline
\end{tabular}
\label{dm2table}
\end{table}

The expected number of wrong--sign muon events are listed in 
Table~\ref{dm2table} for the LMA scenario IA1, and a 50~kt detector 
downstream of a neutrino factory providing $10^{19} \mu^+$ decays 
and the same number of $\mu^-$ decays. The event rates are shown 
for both signs of $\delta m^2_{32}$, and for various storage ring 
energies and baselines. Even at a 20~GeV neutrino factory the 
signal rates at $L = 7332$ and 2800~km are large enough to permit the 
sign of $\delta m^2_{32}$ 
to be determined with a few years of data taking. 

We conclude that for the LMA, SMA, and LOW 
three--flavor mixing scenarios we have considered,
a 20~GeV neutrino factory providing $10^{19}$ decays in the beam--forming 
straight section would be a viable entry--level facility. In particular, 
with a 50~kt detector and a few years of data taking either 
$\nu_e \rightarrow \nu_\mu$ oscillations would be observed and the sign of 
$\delta m^2_{32}$ determined or a very stringent upper limit on 
$\sin^2 2\theta_{13}$ will have been obtained (discussed later).
Long baselines ($>2000$~km) are preferred. The longest baseline we 
have considered (7332~km) has the advantage of lower total event rates 
and hence lower background rates.

\subsubsection{Observation of $\nu_e \rightarrow \nu_\tau$ oscillations}

\begin{figure}
\epsfxsize3.5in
\centerline{\epsffile{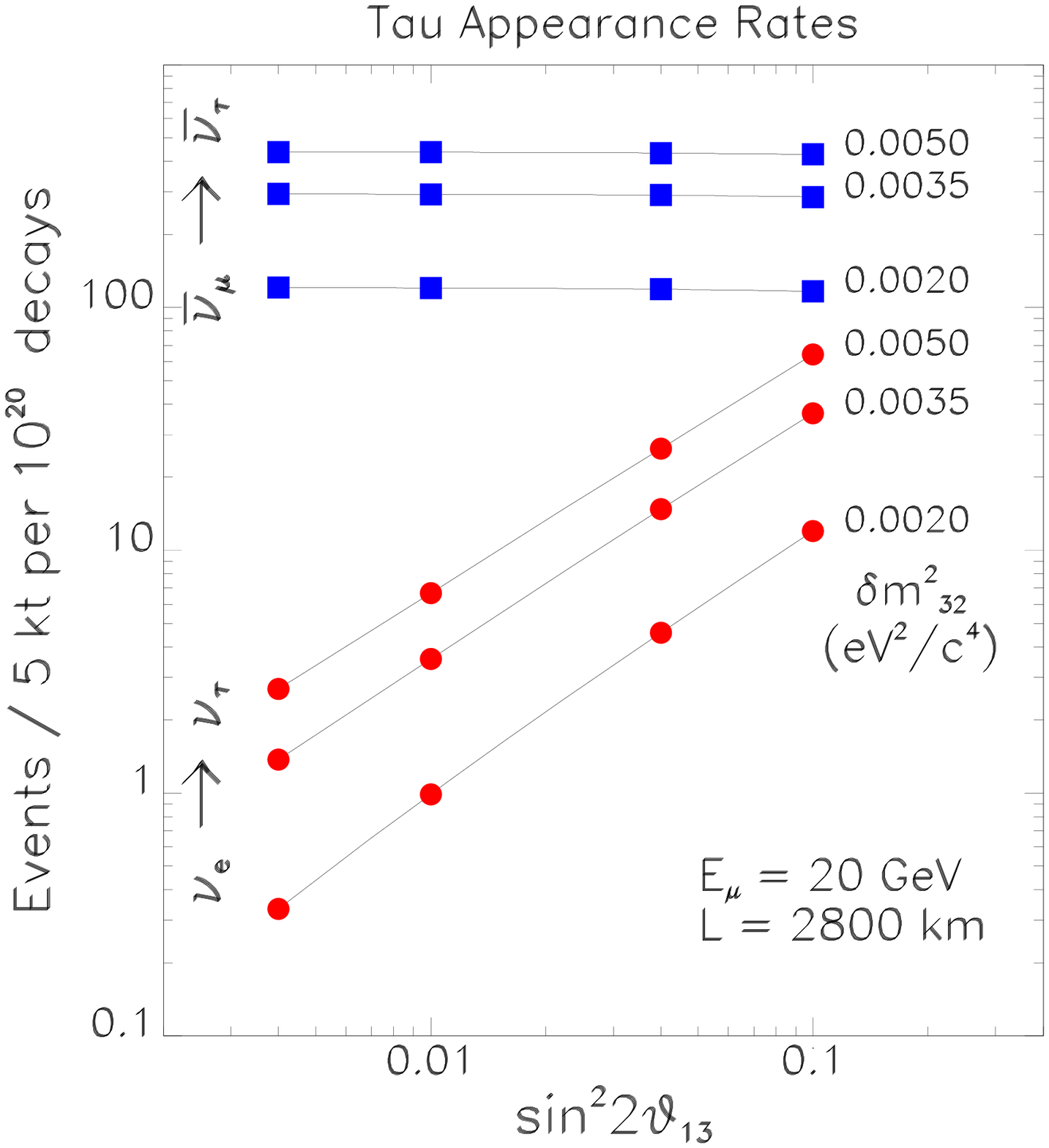}}
\caption{$\nu_\tau$ CC appearance rates in a 5~kt detector 2800~km
downstream of a 20~GeV neutrino factory in which there are
$10^{20} \mu^+$ decays in the beam--forming straight section.
The rates are shown as a function of
$\sin^22\theta_{13}$ and $\delta m_{32}^2$ with the other oscillation
parameters corresponding to the LMA scenario IA1. 
The top 3 curves are the predictions for
$\overline{\nu}_\mu \rightarrow \overline{\nu}_\tau$
events and the lower curves are for $\nu_e \rightarrow \nu_\tau$ events.
Results are from Ref.~\ref{bgrw00}.
}
\label{fig:t1}
\end{figure}

We begin by considering the LMA scenario IA1, and ask: What 
beam intensity is needed to make the first
observation of $\nu_e \rightarrow \nu_\tau$ oscillations in a detector 
that is 2800~km downstream of a 20~GeV neutrino factory ? 
The $\nu_e \rightarrow \nu_\tau$
and the accompanying $\overline{\nu}_\mu \rightarrow \overline{\nu}_\tau$ 
event 
rates are shown in Fig.~\ref{fig:t1} as a function of the 
oscillation parameters 
$\sin^2 2\theta_{13}$ and $\delta m^2_{32}$. The $\nu_e \rightarrow \nu_\tau$ 
signal rate is sensitive to both of these parameters, and hence provides an 
important consistency check for three-flavor mixing: the observation or 
non--observation of a $\nu_e \rightarrow \nu_\tau$ signal must be consistent 
with the oscillation parameters measured from, for example, 
$\nu_e \rightarrow \nu_\mu$, $\nu_\mu \rightarrow \nu_\tau$, 
and $\nu_\mu$ disappearance measurements.
For the LMA scenario IA1 the observation of 
10 signal events in a 5~kt detector (with 30\% $\tau$--lepton efficiency)  
would require 3 years with 
$7 \times 10^{19} \mu^+$ decays per year in the beam forming straight
section. Very similar beam intensities are required for the SMA and LOW 
scenarios (IA2 and IA3). 
Note that, over the $\sin^22\theta_{13}$ range shown in Fig.~\ref{fig:t1}, 
the $\overline{\nu}_\mu \rightarrow \overline{\nu}_\tau$ rates are one to two 
orders of magnitude higher than the $\nu_e \rightarrow \nu_\tau$ rates. 
Hence, we will need a detector that can determine the sign of 
the tau--lepton charge at the $2\sigma-3\sigma$ level, or better.

\begin{figure}
\epsfxsize3.3in
\centerline{\epsffile{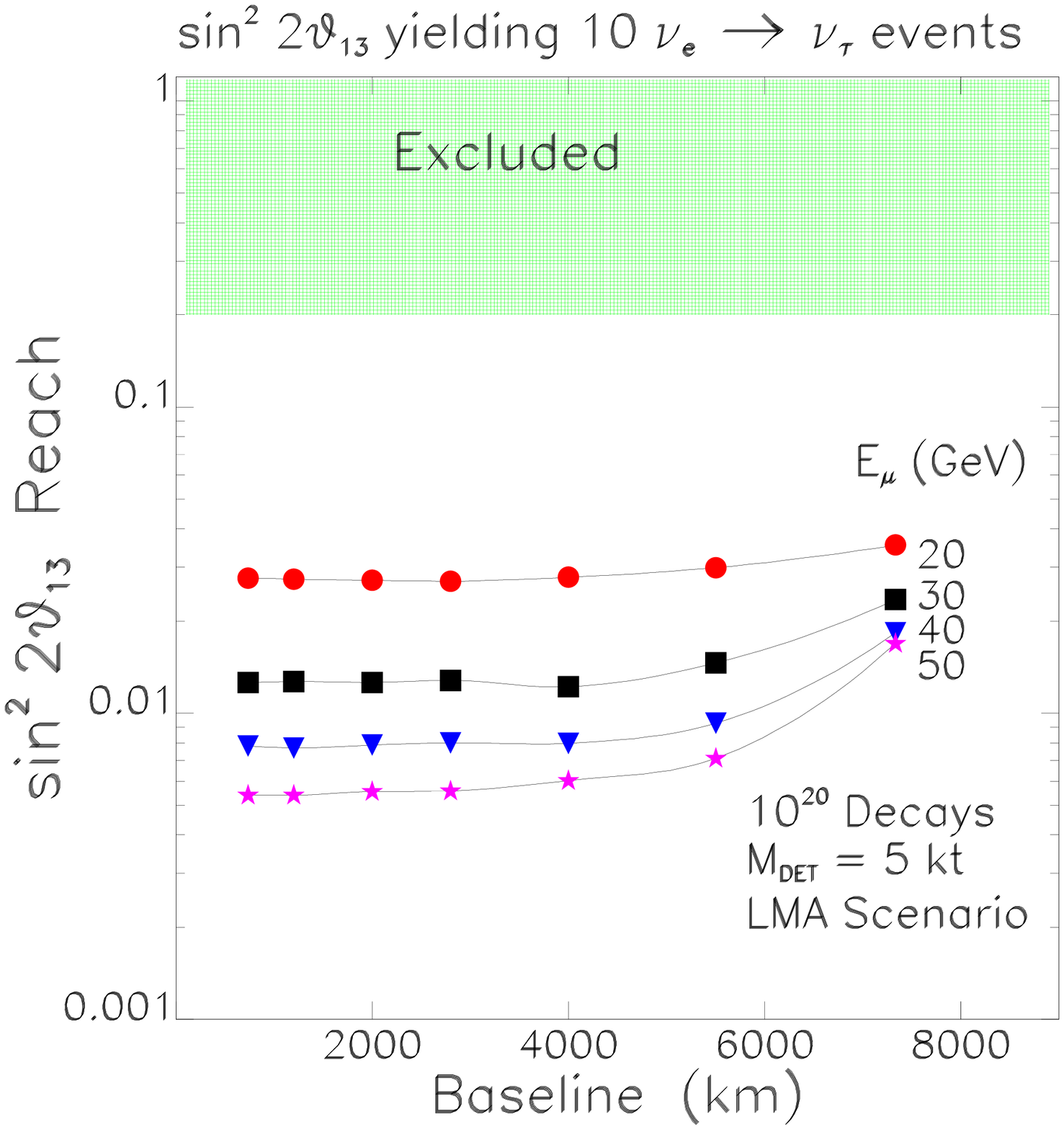}}
\caption{Reach in $\sin^22\theta_{13}$ for the observation of
10 $\nu_e \rightarrow \nu_\tau$ oscillation events,
shown as a function of baseline for four storage ring energies.
The oscillation parameters correspond to the LMA scenario IA1. 
The curves correspond to
$10^{20}$ $\mu^+$ decays in a 20~GeV neutrino factory with
a 5~kt detector. 
Results are from Ref.~\ref{bgrw00}.
}
\label{fig:t2}
\end{figure}

Let us define the $\sin^22\theta_{13}$ ``reach" for an experiment as 
the value of
$\sin^22\theta_{13}$ for which we would observe 10 
$\nu_e \rightarrow \nu_\tau$ events when there are $10^{20}$ muon decays 
in the beam--forming straight section of a neutrino factory.
The $\sin^22\theta_{13}$ reach is shown as a function of the baseline and 
storage ring energy in Fig.~\ref{fig:t2} for a 5~kt detector and an 
oscillation scenario in which all of the
parameters except $\sin^22\theta_{13}$ correspond to scenario IA1. 
The reach improves with energy (approximately $\sim E^{-1.5}$~\cite{bgrw00}) 
and is almost independent of baseline 
except for the highest energies and baselines considered. 
However, backgrounds to a $\nu_e \rightarrow \nu_\tau$ oscillation 
search have not been studied in detail, and are not included in the 
calculation. Background considerations will favor longer baselines. 
The number of muon decays needed to observe 
10 $\nu_e \rightarrow \nu_\tau$ events 
is shown as a function of muon beam energy in Fig.~\ref{fig:v2} for the 
LMA scenario IA1, the SMA scenario IA2 , and the LOW scenario IA3. 
We conclude that within these three--flavor mixing scenarios, 
a 20~GeV storage ring in which there are O($10^{20}$) muon decays per year 
would begin to permit an observation of, or meaningful limits on, 
$\nu_e \rightarrow \nu_\tau$ oscillations provided a multi-kt detector 
with good tau--lepton identification and charge discrimination is 
practical.

\begin{figure}
\epsfxsize3.3in
\centerline{\epsffile{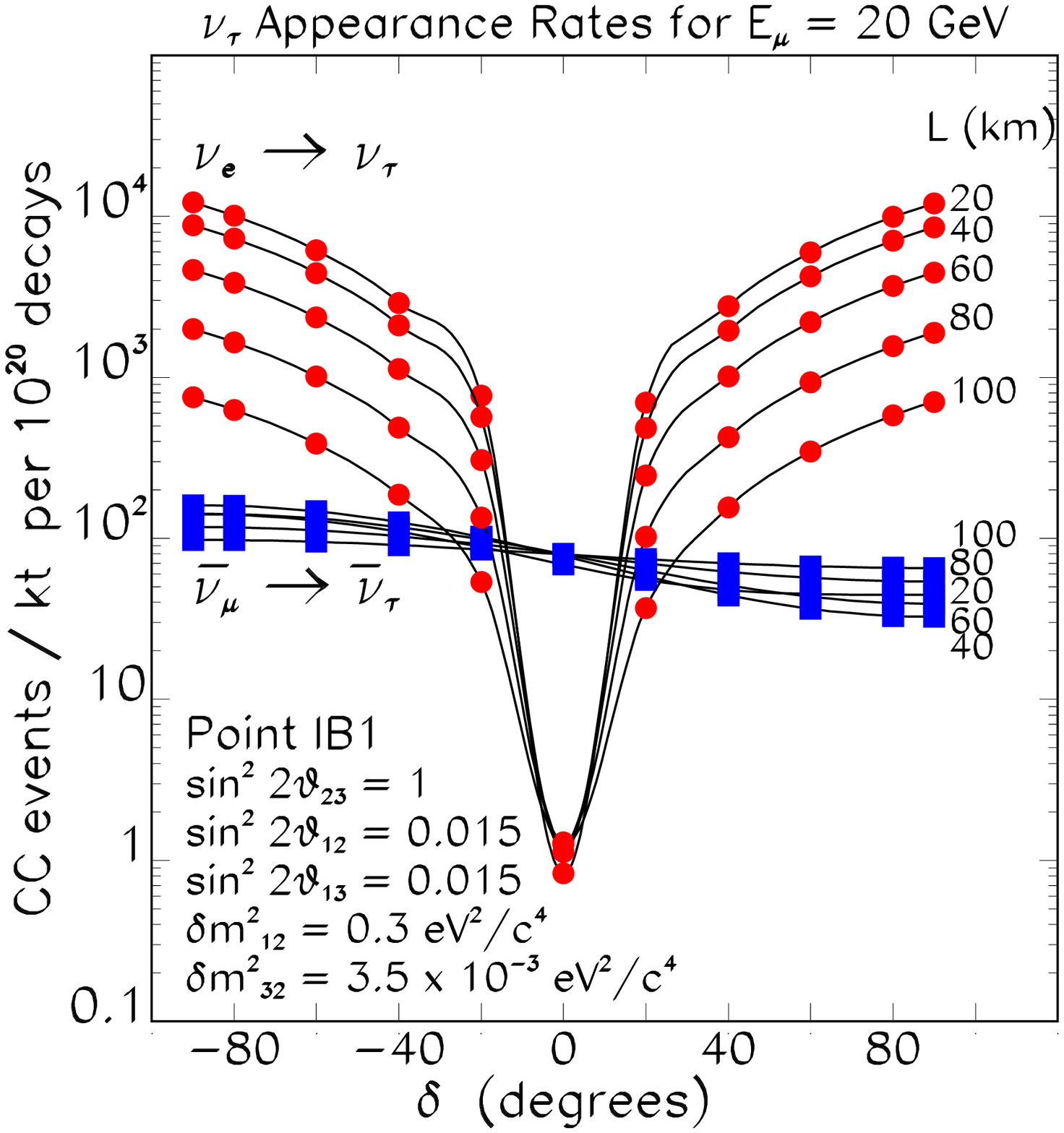}}
\caption{$\nu_\tau$ CC appearance rates in a 1~kt detector 
downstream of a 20~GeV neutrino factory in which there are
$10^{20} \mu^+$ decays. 
Rates are shown as a function of the baseline $L$ and 
phase $\delta$, with the other oscillation 
parameters corresponding to the LSND + Atmospheric scenario IB1.
Predictions for $\nu_e \rightarrow \nu_\tau$ and 
$\overline{\nu}_\mu \rightarrow \overline{\nu}_\tau$ 
are shown separately, as labeled. 
Results are from Ref.~\ref{bgrw_prep}.
}
\label{fig:t3}
\end{figure}

Next, consider the oscillation scenarios IB1 (atmospheric + LSND scales) and 
IC1 (three--flavor with atmospheric, solar, and LSND data stretched). 
In these cases the leading $\delta m^2$ is large (0.3~eV$^2$/c$^4$) 
and medium baseline experiments ($L =$ 10-100~km) become interesting. 
As an example, consider a medium baseline experiment a few $\times 10$~km 
downstream of a 20~GeV neutrino factory in which there are $10^{20} \mu^+$ 
decays. The $\nu_e \rightarrow \nu_\tau$ 
and accompanying $\overline{\nu}_\mu \rightarrow \overline{\nu}_\tau$ event
rates are shown in Fig.~\ref{fig:t3} as a function of the
baseline and the phase $\delta$ with the other oscillation parameters 
corresponding to scenario IB1. In contrast to the 
$\overline{\nu}_\mu \rightarrow \overline{\nu}_\tau$ rates, 
the $\nu_e \rightarrow \nu_\tau$ rates are very sensitive to $\delta$, 
and for $|\delta| > 20^\circ$ can be very large, yielding thousands of 
events per year in a 1~kt detector at $L = 60$~km, for example. 
Note that the corresponding 
$\overline{\nu}_\mu \rightarrow \overline{\nu}_\tau$ rate is 
of order 100~events. For small $|\delta|$ the 
$\overline{\nu}_\mu \rightarrow \overline{\nu}_\tau$ rate 
will dominate the $\tau$ appearance event sample. For 
larger $|\delta|$ the $\nu_e \rightarrow \nu_\tau$ rate dominates. 
Good $\tau$ charge determination will therefore be important 
to measure both $\nu_e \rightarrow \nu_\tau$ and 
$\overline{\nu}_\mu \rightarrow \overline{\nu}_\tau$ oscillations. 

Now consider the $\tau$ appearance rates in scenario IC1. 
In this case the rates are not sensitive to $\delta$ and, 
for a 1~kt detector at $L = 60$~km, there are about 8000 
$\nu_e \rightarrow \nu_\tau$ events and 
93000 $\overline{\nu}_\mu \rightarrow \overline{\nu}_\tau$ 
events~\cite{bgrw_prep}. 
A detector with $3\sigma$ (or better) $\tau$-lepton charge discrimination 
would enable these two rates to be separately measured.

We conclude that measurements of the $\nu_e \rightarrow \nu_\tau$ oscillation 
rate at a neutrino factory would provide an important test of the 
oscillation scenario. In LMA, SMA, and LOW three-flavor oscillation 
scenarios, a 20~GeV neutrino factory providing O($10^{20}$) muon decays 
could permit an observation of, or meaningful limits on,
$\nu_e \rightarrow \nu_\tau$ oscillations. In LSND-type scenarios where the 
leading $\delta m^2$ scale is large, a 20~GeV neutrino factory providing 
O($10^{19}$) muon decays might already permit hundreds of 
$\nu_e \rightarrow \nu_\tau$ events to be measured. 
It should be noted that the feasibility of a multi-kt detector with 
good $\tau$ identification efficiency (for example 30\%) 
and good charge sign determination has not been 
explored in detail at this stage, and further work is required to 
identify the best detector technology for this, and determine the 
expected resolutions and efficiencies.

\subsubsection{Measurement of $\nu_\mu \rightarrow \nu_\tau$ oscillations}

The present SuperK data suggests that the atmospheric neutrino deficit 
is due to $\nu_\mu \rightarrow \nu_\tau$ oscillations. If this is 
correct the next generation of accelerator based long baseline experiments 
are expected to measure these oscillations. 
Nevertheless, 
for a fixed neutrino factory energy and baseline, it is important  
to measure or put stringent constraints on all of the appearance channels 
so that the sum of the appearance modes can be compared with 
the disappearance measurements. Hence, we 
briefly consider $\nu_\mu \rightarrow \nu_\tau$ rates 
at a neutrino factory. Note that at a 20~GeV neutrino factory the 
average interacting neutrino energy is of order 15~GeV, and for 
$\delta m^2$ within the favored SuperK range, the first oscillation 
maximum occurs at baselines of $7000 \pm 3000$~km. At shorter baselines 
the oscillation probabilities are lower and hence the signal/background 
ratio is lower, although the signal rate can be higher.

Consider first a 5~kt detector 2800~km downstream of a 20~GeV 
neutrino factory in which there are $10^{20}$ muon decays. 
The $\overline{\nu}_\mu \rightarrow \overline{\nu}_\tau$ event
rates are shown in Fig.~\ref{fig:t1} as a function of 
$\sin^2 2\theta_{13}$ and $\delta m^2_{32}$, with the other oscillation 
parameters corresponding to the LMA scenario IA1. If negative muons 
are stored in the neutrino factory, the resulting 
$\nu_\mu \rightarrow \nu_\tau$ event rates would be about a factor of two 
higher than the $\overline{\nu}_\mu \rightarrow \overline{\nu}_\tau$ rates 
shown in the figure. A neutrino factory providing 
O($10^{20}$) muon decays would enable $\nu_\mu \rightarrow \nu_\tau$ 
appearance 
data samples of a few hundred to a few thousand events to be obtained. 
Similar rates 
are expected in SMA and LOW three-flavor mixing scenarios. 

Next consider a longer baseline example in which a 10~kt ICANOE--type 
detector is 7400~km downstream of a 30~GeV neutrino factory which 
provides $10^{20}$ muon decays in the beam forming straight 
section~\cite{camp00}. The main advantage of 
a longer baseline is that the total interaction rate, and hence the 
$\tau$--lepton background, is reduced. 
The energy distribution for 
events in which there is no charged lepton can directly reflect 
the presence of a $\nu_\tau$ signal (see Fig.~\ref{fig:m1}). 
The non--$\tau$ events in this event sample can be suppressed 
using topology--dependent kinematic cuts. It is desirable that 
the $\tau$ charge--sign also be determined 
which, with an external muon spectrometer, will be 
possible for the $\tau \rightarrow \mu$ subsample. 
We conclude that the measurement of 
$\nu_\mu \rightarrow \nu_\tau$ oscillations with high statistical 
precision will be possible at a neutrino factory in long and 
very long baseline experiments. A more complete study is warranted. 

\subsubsection{Determination of $\sin^2 2\theta_{13}$, $\sin^2 2\theta_{23}$, 
and $\delta m^2_{32}$}

Consider first the determination of $\sin^2 2\theta_{13}$. 
The most sensitive way to measure $\sin^2 2\theta_{13}$ at a 
neutrino factory is to measure the 
$\nu_e \rightarrow \nu_\mu$ oscillation amplitude, which is 
approximately proportional to $\sin^2 2\theta_{13}$. 
More explicitly, the value of $\sin^2 2\theta_{13}$ is extracted 
from a fit to the spectrum of CC interactions tagged by a wrong--sign muon. 
Background contributions from, for example, muonic decays of charged mesons 
must be kept small, which favors small total event samples and hence 
long baselines.

\begin{figure}[th]
\epsfxsize3.3in
\centerline{\epsffile{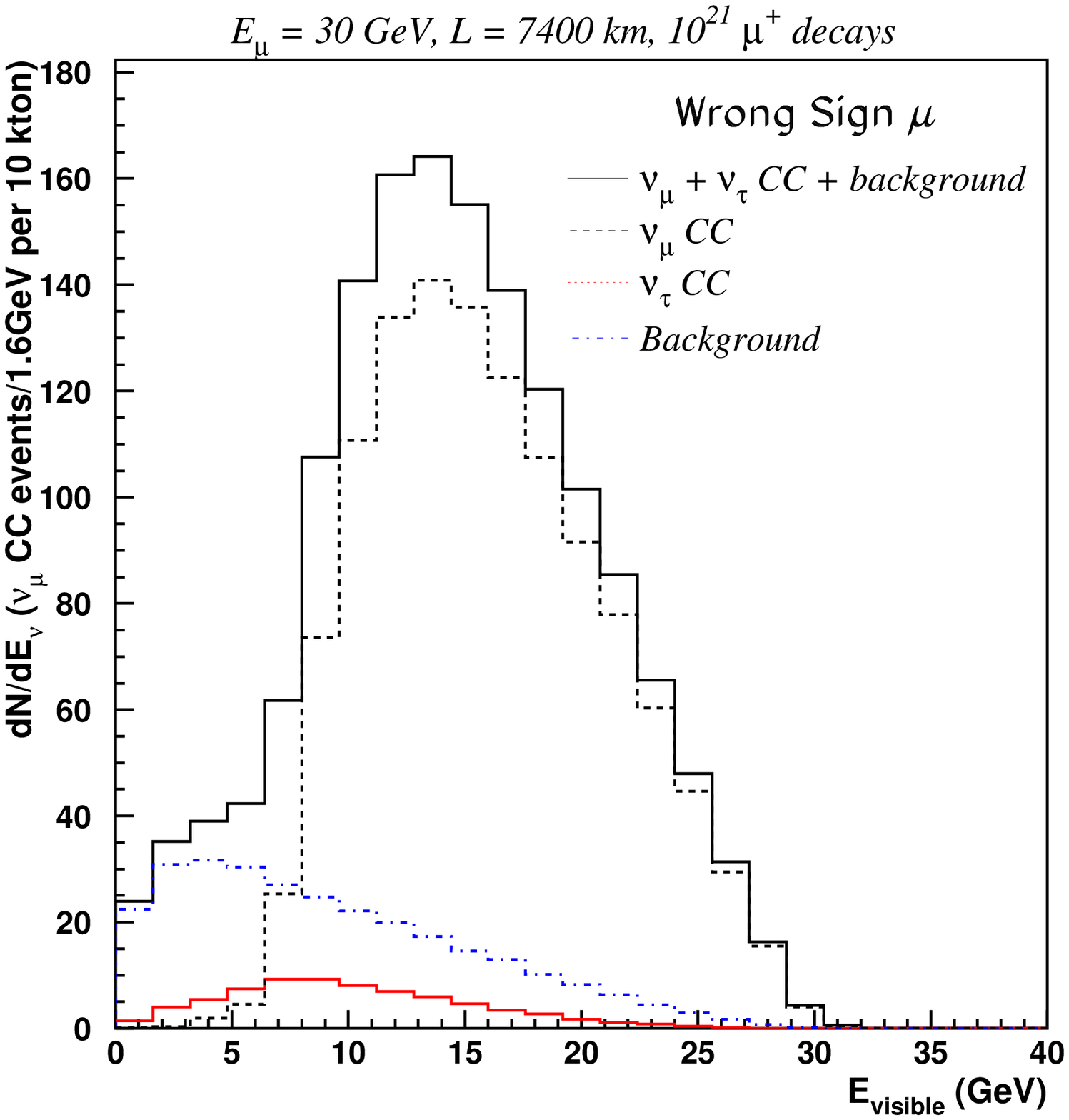}}
\caption{Visible energy spectrum for events tagged by wrong-sign muons 
in an ICANOE--type detector (full histogram). The oscillation parameters 
are $\delta m^2_{32}=3.5\times 10^{-3}$~eV$^2$/c$^4$, 
$\sin^2\theta_{23}=1$, and  
$\sin^2 2\theta_{13}=0.05$. Also shown are the contributions from 
$\nu_e\to\nu_\mu$ oscillations (black dashed curve), 
$\nu_e\to\nu_\tau$, with a subsequent muonic decay of the $\tau$ lepton
(red curve), and background from 
muonic decays of pions or kaons in neutral current or charged current 
events (blue dot-dashed curve). Results are from Ref.~\ref{camp00}.
}
\label{fig:m3}
\end{figure}

\begin{figure}[t]
\epsfxsize3.3in
\centerline{\epsffile{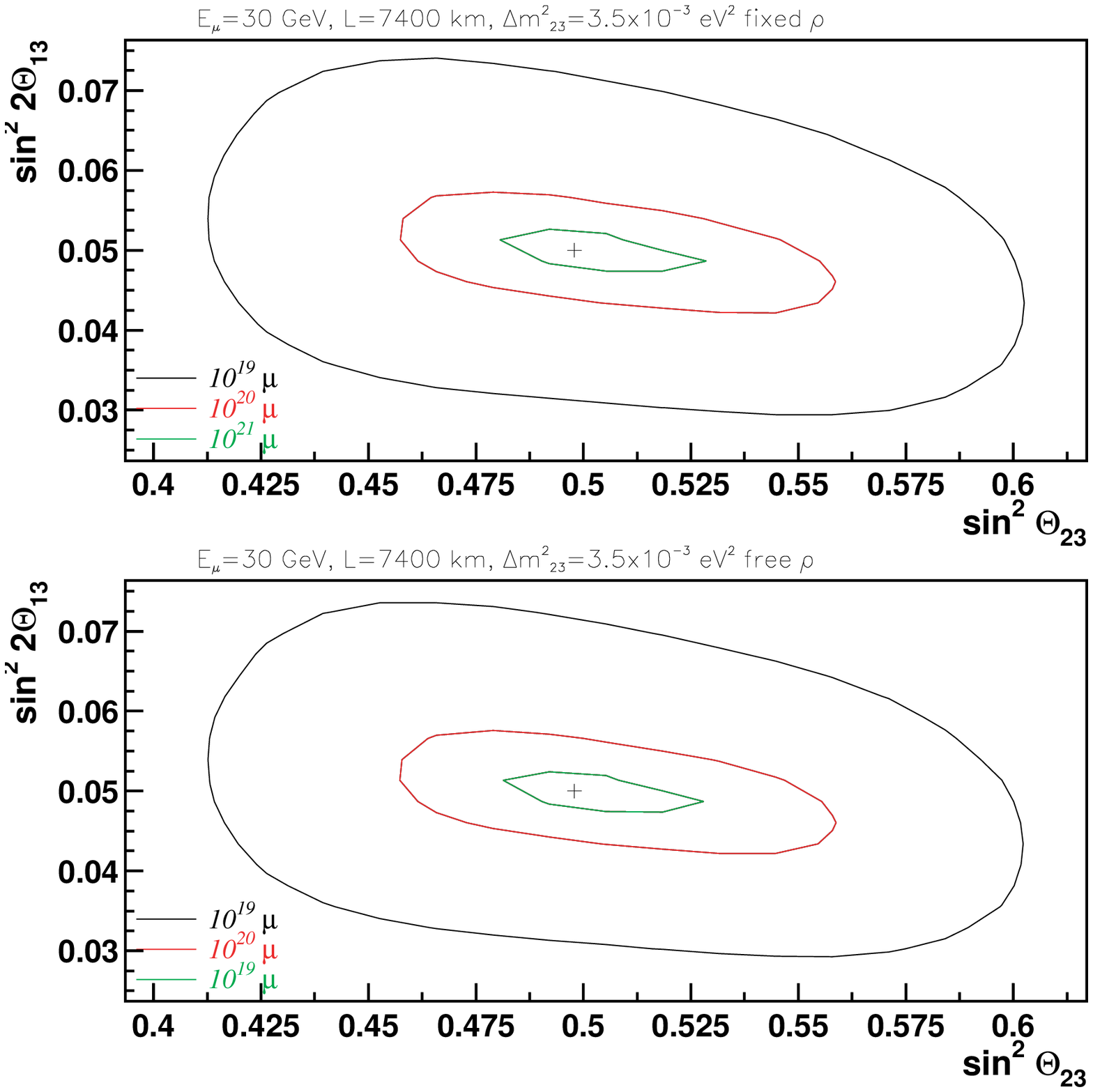}}
\caption{Results from a global fit to the visible energy distributions for 
various event classes recorded in a 10~kt ICANOE--type detector 
7400~km downstream of a 30~GeV neutrino factory. 
The 68\% CL contours correspond to
experiments in which there are $10^{19}$, $10^{20}$, and 
$10^{21} \mu^+$ decays in the neutrino factory (as labeled) 
followed by the same number of $\mu^-$ decays. 
Upper plot: density fixed to its true value. Lower plot: density is a free
parameter of the fit. Results are from Ref.~\ref{camp00}.
}
\label{fig:m7}
\end{figure}

\begin{figure}[t]
\epsfxsize3.3in
\centerline{\epsffile{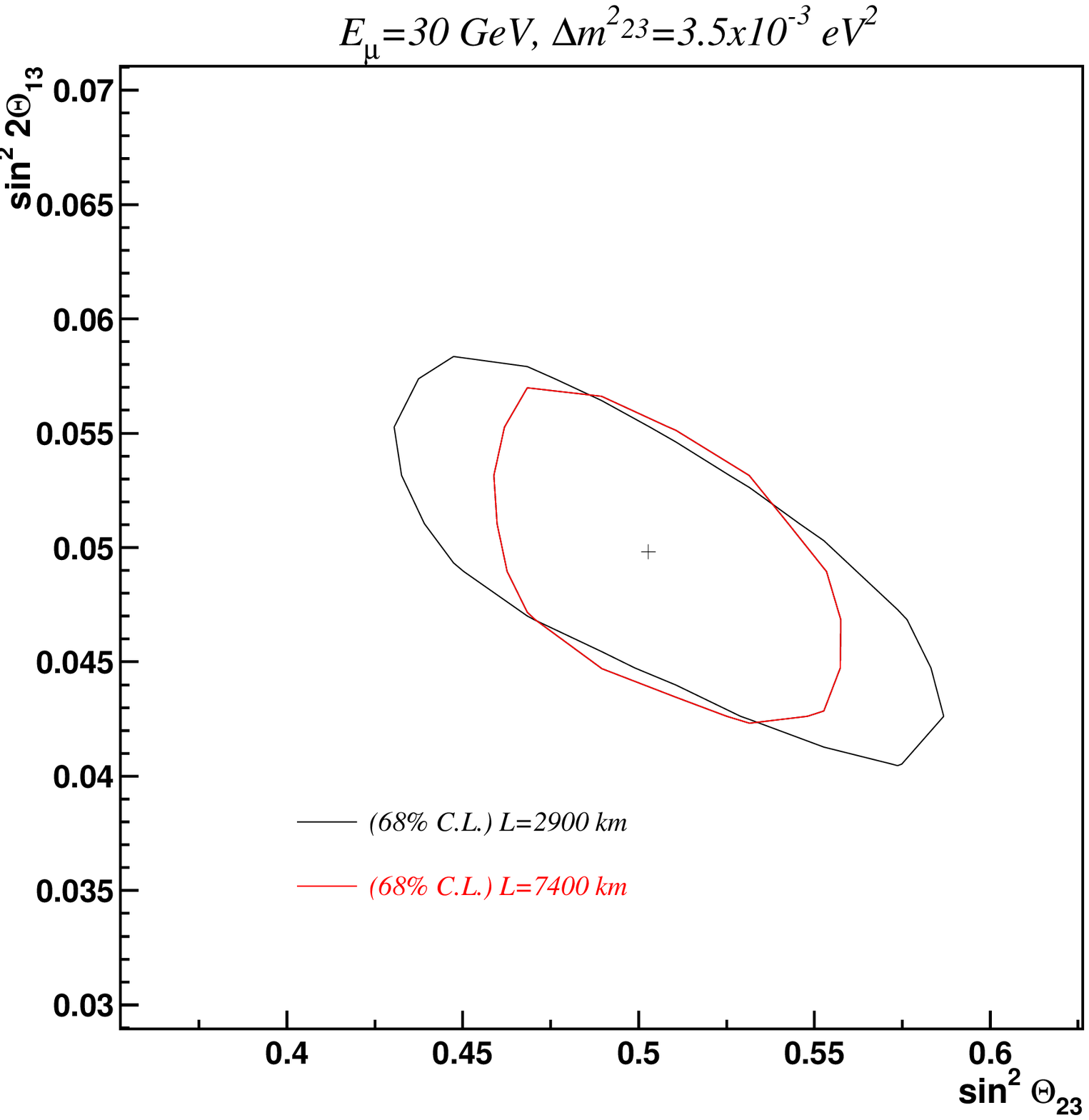}}
\caption{Results from a global fit to the visible energy distributions for 
various event classes recorded in a 10~kt ICANOE--type detector 
downstream of a 30~GeV neutrino factory in which there are 
$10^{20} \mu^+$ decays in the neutrino factory 
followed by the same number of $\mu^-$ decays.
The 68\% CL contours correspond to baselines of 7400~km and 2900~km, 
as labeled. Results are from Ref.~\ref{camp00}.
}
\label{fig:m7a}
\end{figure}

As an example, consider a 10~kt ICANOE--like detector that is 7400~km 
downstream of a 30~GeV neutrino factory~\cite{camp00}.
The simulated energy spectrum of wrong-sign muon events
is shown in Fig.~\ref{fig:m3} for three--flavor oscillations with 
the parameters 
$\delta m^2_{32}=3.5\times 10^{-3}$~eV$^2$/c$^4$, 
$\sin^2 2\theta_{23}=1$, and 
$\sin^2 2\theta_{13}=0.05$. 
Note that the backgrounds predominantly contribute to the low energy 
part of the spectrum. 
To fit the observed spectrum and extract $\sin^2 2\theta_{13}$ 
matter effects must be taken into account. The modification of the 
oscillation probability due to matter effects is a function of the 
profile of the matter density $\rho$ between the neutrino source and 
the detector. The density profile is known from geophysical measurements, 
and this knowledge can either be used in the fit, or alternatively $\rho$ 
can be left as a free parameter. It has been shown that both methods 
give consistent results~\cite{camp00}, and that the uncertainties 
on the fitted values of $\rho$ and $\sin^2 2\theta_{13}$ are not 
strongly correlated. However, the fitted value for 
$\sin^2 2\theta_{13}$ does depend on the assumed values for 
$\sin^2 \theta_{23}$ and $\delta m^2_{32}$. 
The measured right--sign muon ($\nu_\mu$ disappearance) distribution, 
together with the distributions of events tagged by electrons, 
$\tau$--leptons, or the absence of a lepton, 
can be used to constrain these additional oscillation parameters. 
Hence, the best way to extract $\sin^2 2\theta_{13}$ is from a global 
fit to all of the observed event distributions, with the 
oscillation parameters (and optionally $\rho$) left as free parameters. 
If the density profile is left as a free 
parameter, the fit determines its value with an uncertainty of about 
10\%~\cite{camp00,cerv00}. This provides a quantitative test of the 
MSW effect ! Examples of fit results in the 
($\sin^2 2\theta_{13}$,~$\sin^2 \theta_{23}$)--plane are shown in 
Fig.~\ref{fig:m7} for $10^{19}$, $10^{20}$, and $10^{21}$ muon decays 
in the neutrino factory. As the beam intensity increases the measurements 
become more precise. With $10^{19} \mu^+$ and $\mu^-$ decays 
$\sin^2 2\theta_{13}$ and $\sin^2 2\theta_{23}$ are determined with 
precisions of 40\% and 20\% respectively. With $10^{21}$ decays 
these precisions have improved to $\sim5$\%. 
If the baseline is decreased from 7400~km to 2900~km the oscillation 
parameters are determined with comparable (although slightly worse) 
precisions (Fig.~\ref{fig:m7a}). 
We conclude that within the framework of three--flavor mixing, 
provided $\sin^2 2\theta_{13}$ is not too small, 
a global fit to the observed oscillation distributions would enable 
$\sin^2 2\theta_{13}$, $\sin^2 \theta_{23}$, and $\delta m^2_{32}$ 
to be simultaneously determined, and the MSW effect to be measured.

\begin{figure}
\epsfxsize3.3in
\centerline{\epsffile{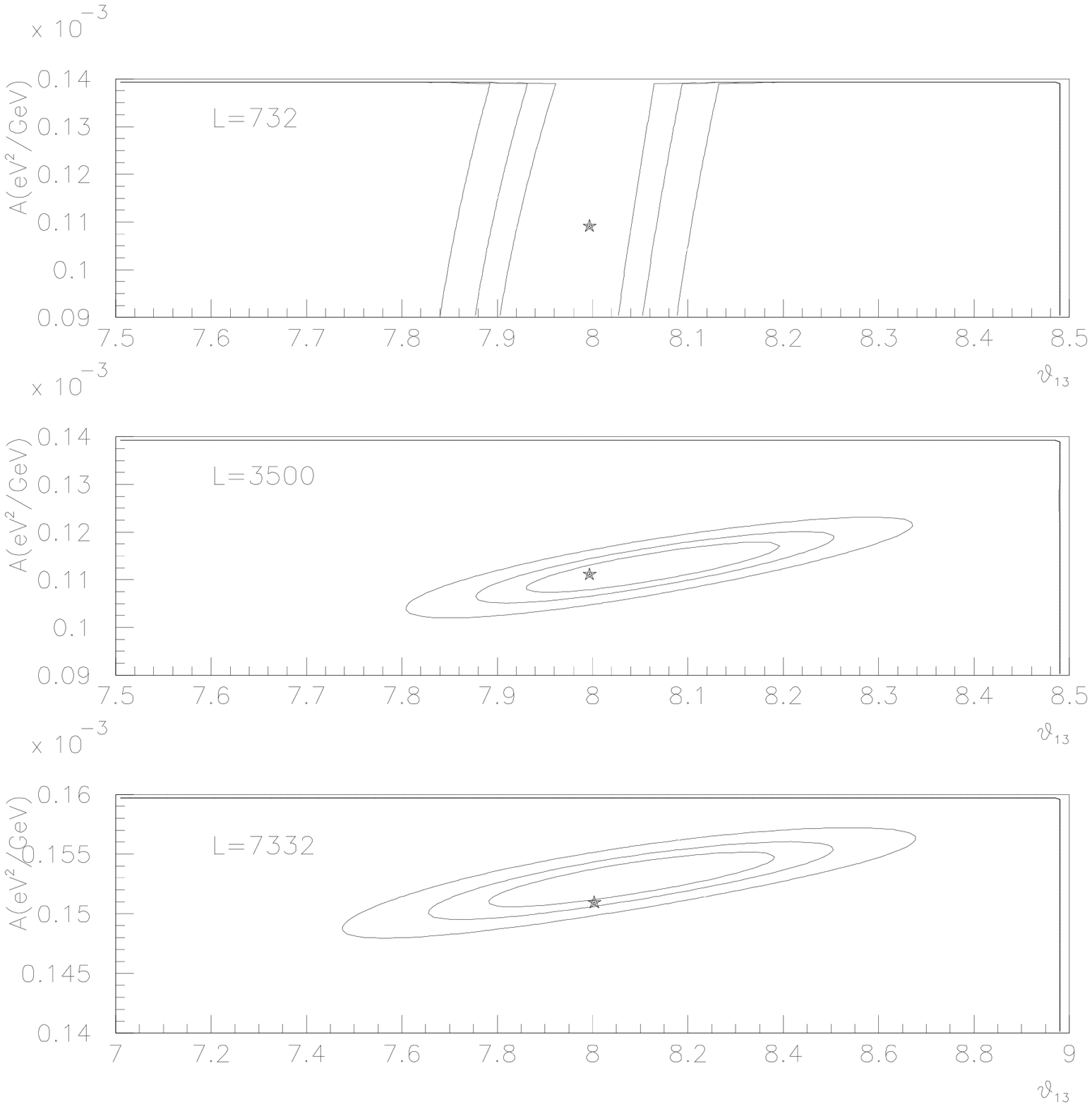}}
\caption{Fit results in the ($A$, $\theta_{13}$)--plane 
for a simulated experiment in which a 40~kt 
Fe-scintillator detector is a distance $L$~km downstream 
of a 50~GeV neutrino factory in which there are 
$10^{21} \mu$ decays. 
The density parameter $A$ is defined in Eq.~(\ref{eq:defnA}).  
The curves are 68.5, 90, and 99\% CL contours. 
Results are from Ref.~\ref{cerv00}.}
\label{fig:s15}
\end{figure}

\begin{figure}
\epsfxsize3.3in
\centerline{\epsffile{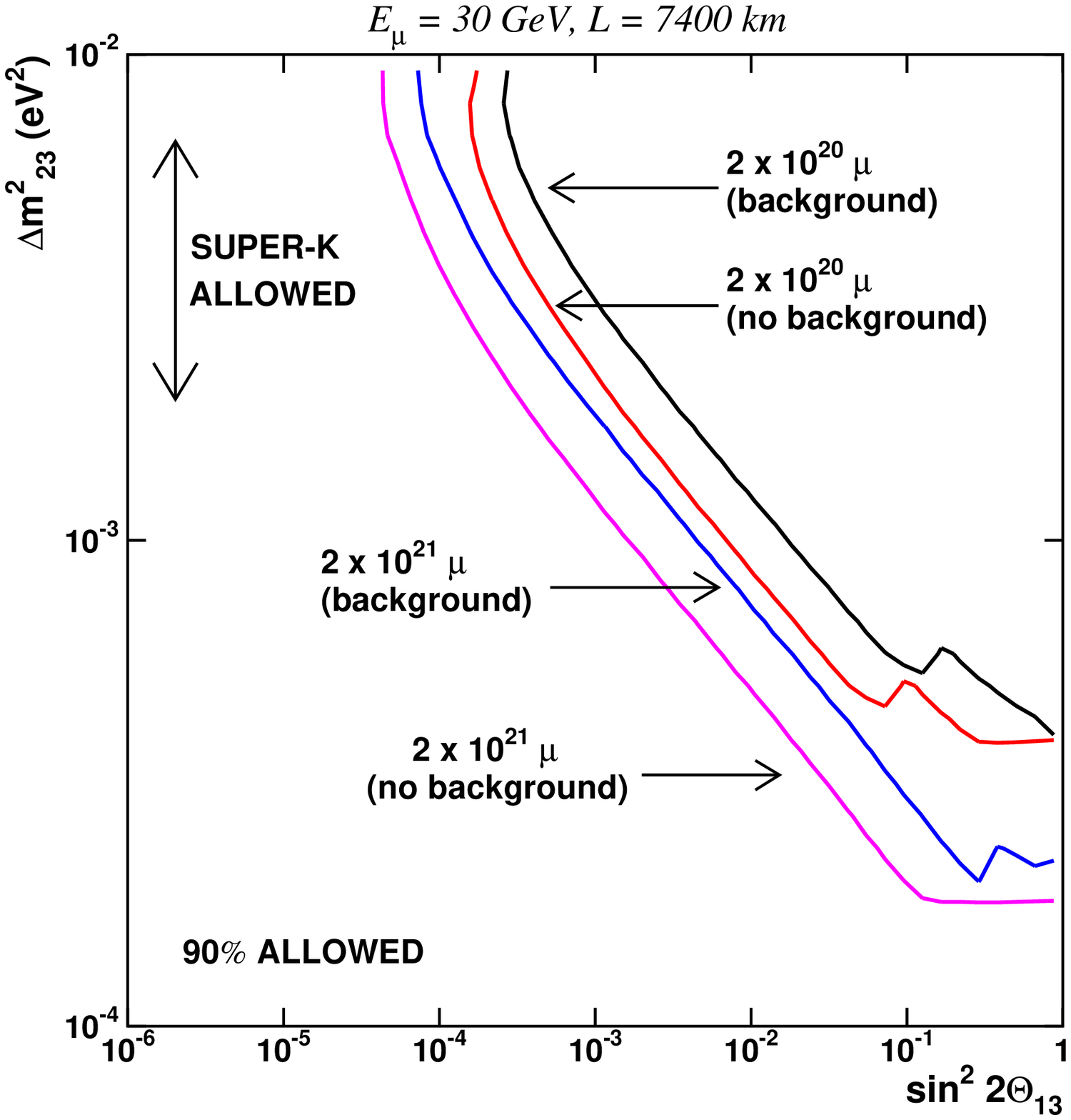}}
\caption{Allowed regions in oscillation parameter space 
calculated for a simulated experiment in which there are $N$~$\mu^+$ decays 
followed by $N$~$\mu^-$ decays in a 30~GeV neutrino factory that is 7400~km 
from a 10~kt ICANOE--type detector. The contours correspond to $N = 10^{20}$ 
and $10^{21}$ with and without backgrounds included in the calculation. 
Results are from Ref.~\ref{camp00}.}
\label{fig:m4}
\end{figure}

\begin{figure}
\epsfxsize3.3in
\centerline{\epsffile{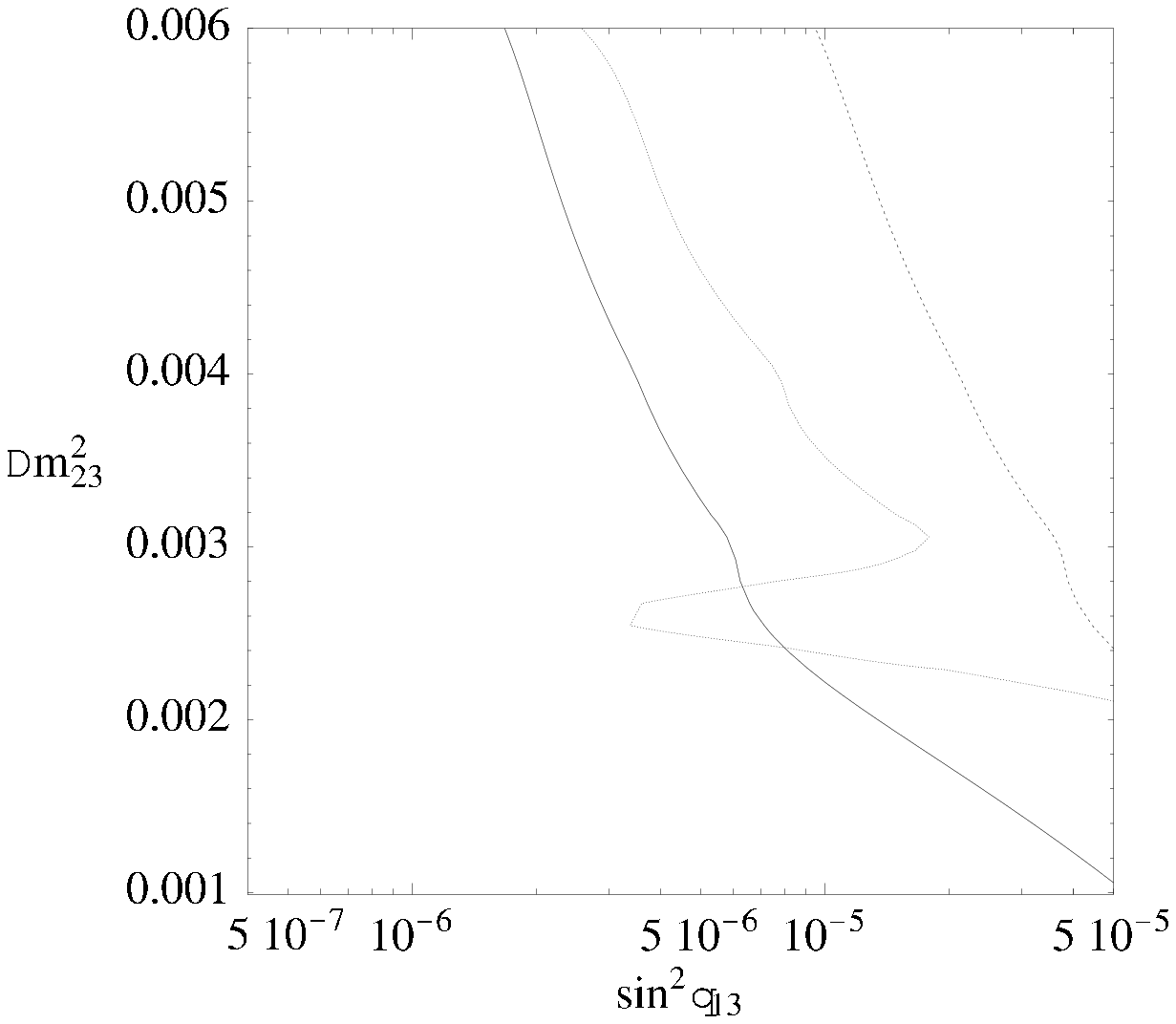}}
\caption{Allowed regions in oscillation parameter space 
calculated for a simulated experiment in which a 40~kt 
Fe-scintillator detector is a distance $L$~km downstream 
of a 50~GeV neutrino factory in which there are 
$10^{21} \mu$ decays. The curves are 90\% CL contours for 
$L = 732$~km (dashed), 3500~km (solid), and 7332~km (dotted).
Results are from Ref.~\ref{cerv00}.}
\label{fig:s13}
\end{figure}

\begin{figure}
\epsfxsize3.0in
\centerline{\epsffile{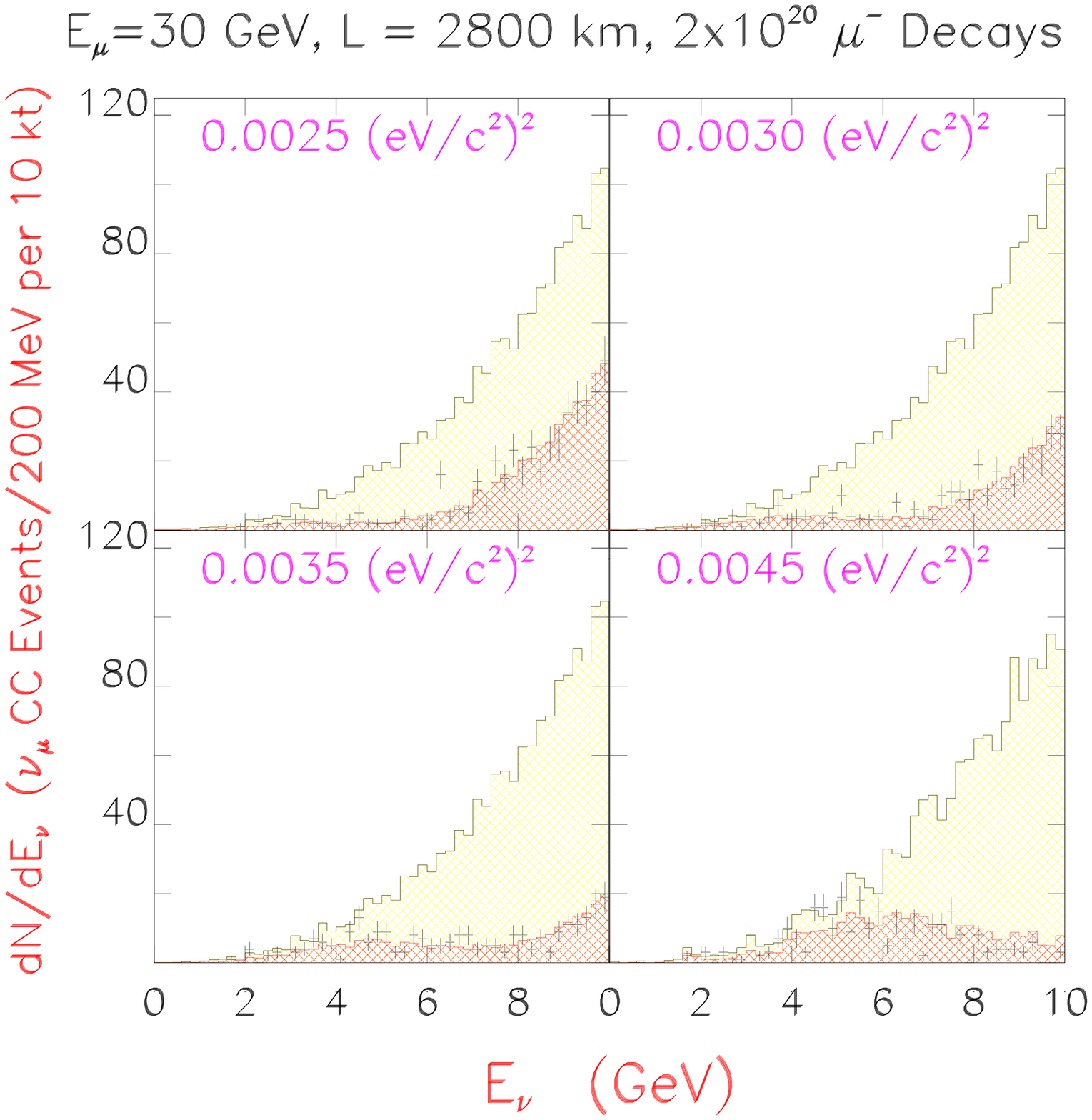}}
\caption{Visible energy distributions for events tagged by a right--sign muon 
in a MINOS--type detector 2800~km downstream of a 20~GeV neutrino factory 
in which there are $2 \times 10^{20} \mu^-$ decays. Predicted distributions 
are shown for four values of $\delta m^2_{32}$, with the other parameters 
corresponding to the LMA scenario IA1. For each panel, the points with 
statistical error bars show an example of a simulated experiment. The 
light shaded histograms show the predicted distributions in the 
absence of oscillations. Results are from Ref.~\ref{bgrw00}.
}
\label{fig:v5}
\end{figure}

\begin{figure}
\epsfxsize2.8in
\centerline{\epsffile{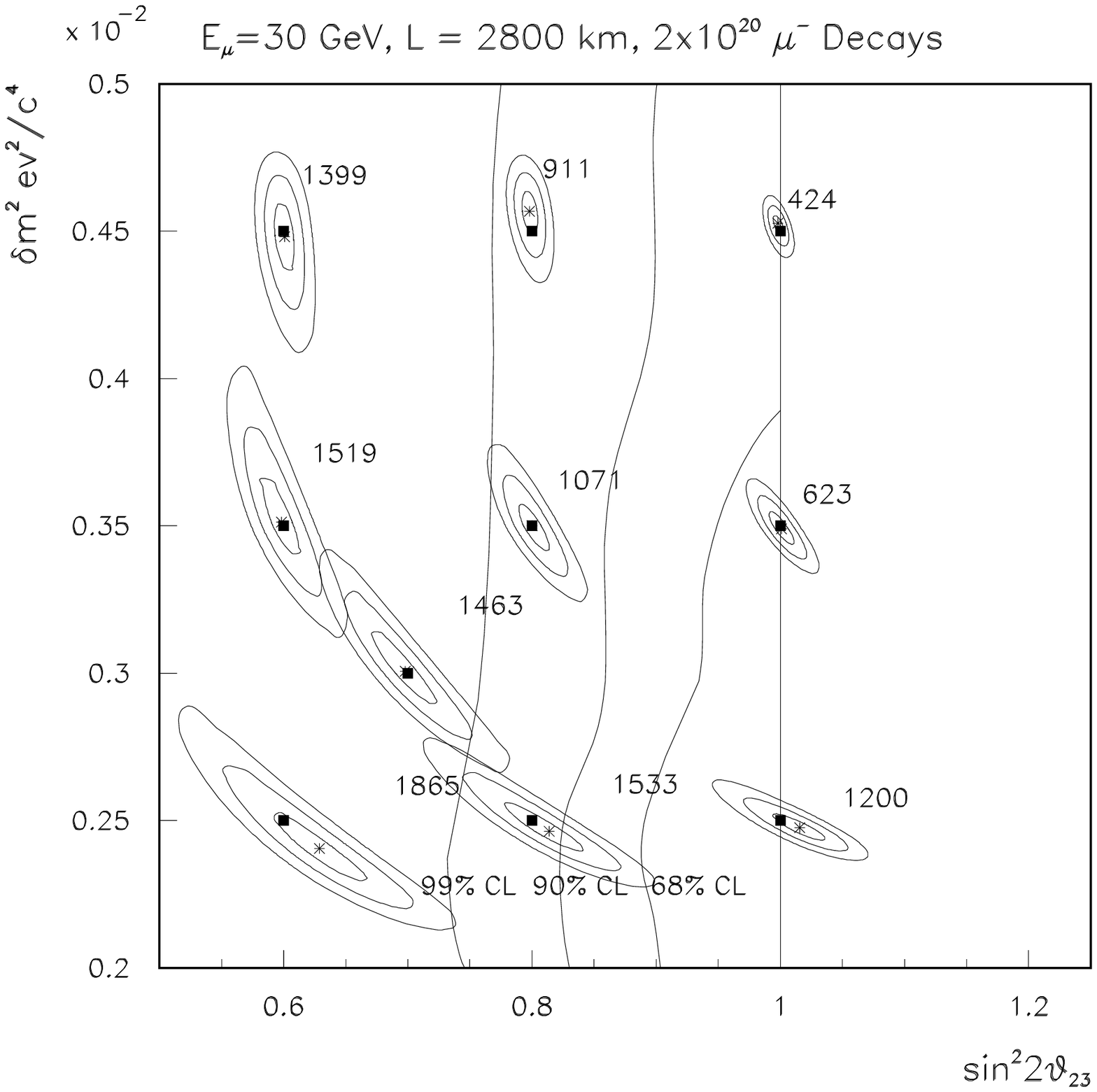}}
\caption{Fit results for simulated $\nu_\mu$ disappearance 
measurements with a 10~kt MINOS-type 
detector 2800~km downstream of a 30~GeV neutrino factory in which 
there are $2 \times 10^{20} \mu^-$ decays. 
For each trial point the $1\sigma$, $2\sigma$, 
and $3\sigma$ contours are shown for a perfect detector 
(no backgrounds) and no systematic uncertainty on the beam flux. 
The 68\%, 90\% and 95\% SuperK regions are indicated. 
Results are from Ref.~\ref{bgrw00}.
}
\label{fig:v6}
\end{figure}


\begin{figure}
\epsfxsize4.8in
\centerline{\epsffile{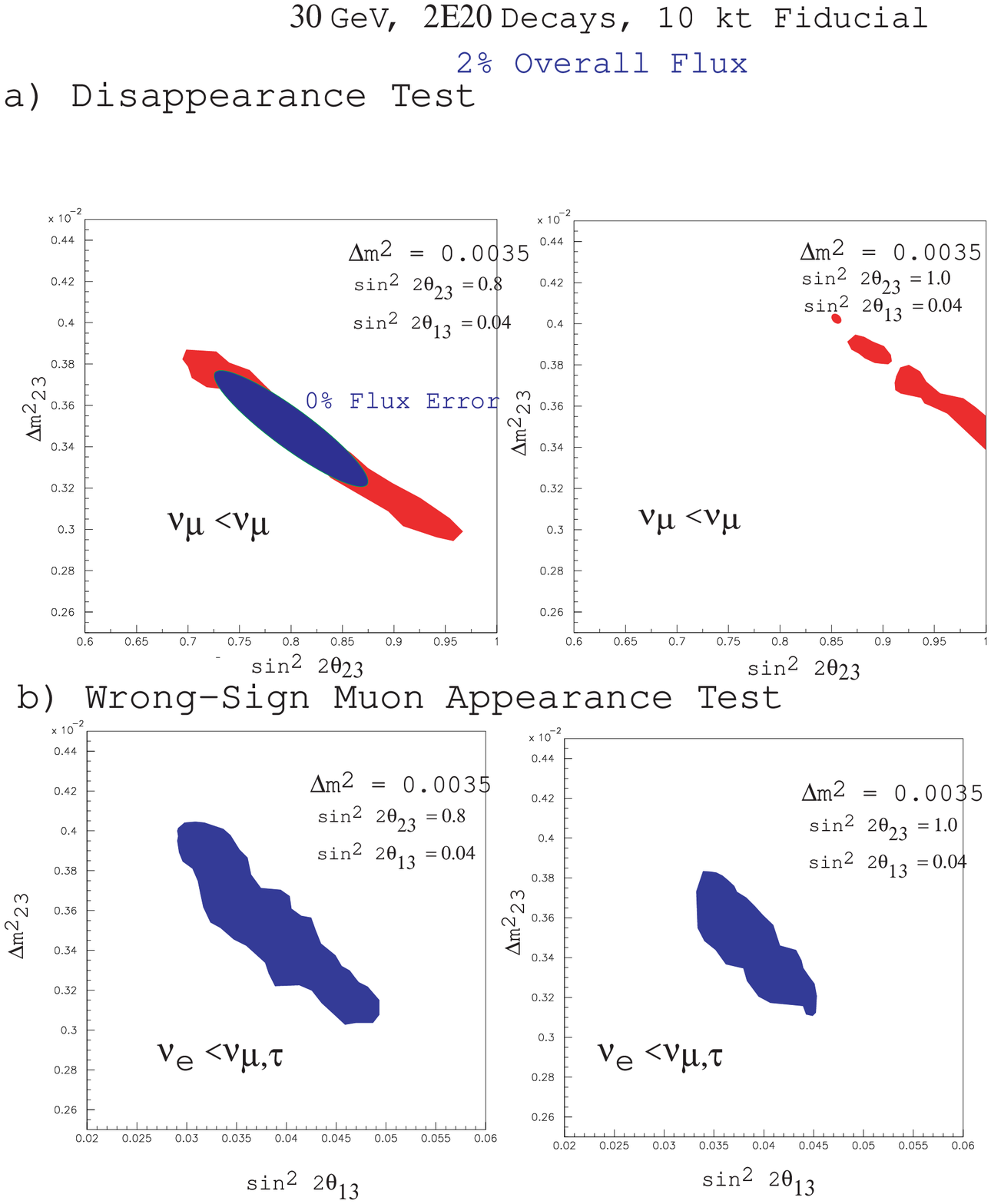}}
\caption{Fit results (1~$\sigma$ contours) 
for (a) simulated $\nu_\mu$ disappearance 
measurements with a 10~kt MINOS-type 
detector 2800~km downstream of a 30~GeV neutrino factory in which 
there are $2 \times 10^{20} \mu^-$ decays, with and without an 
ad hoc 2\% systematic uncertainty on the neutrino flux, and 
(b) wrong--sign muon appearance measurements including an ad hoc 
2\% systematic uncertainty on the flux.  
The acceptance for a muon is 
zero for $p_{\mu} < 4$ GeV and unity for $p_{\mu} \geq 4$ GeV.  Backgrounds 
are included but no $p_{\perp}^2$ cut has been used.
Results are from Ref.~\ref{bern00}.
}
\label{fig:b1}
\end{figure}

Consider as a second example a 20~kt MINOS--type detector 2800~km 
downstream of a 30~GeV neutrino factory providing $10^{20}$ muon decays. 
Some preliminary fit results~\cite{bern00} for a LMA type scenario are 
shown in Fig.~\ref{fig:b1}b. For this example the analysis required 
$p_{\mu} < 4$~GeV/c, but did not use the full set of cuts described 
in section~\ref{detectors}, and therefore tolerated a background level 
a factor of a few greater than shown in Fig.~\ref{data} and 
Table~\ref{thetable}. Nevertheless, the fits to the measured distribution 
of energies for events tagged by wrong--sign muons were able to 
determine $\sin^2 2\theta_{13}$ and $\delta m^2_{32}$ with precisions 
of 14\% and 10\% respectively for scenario IA1.

To illustrate the ultimate sensitivity to the oscillation parameters 
that might be achievable at a high intensity neutrino factory, 
consider next a 40~kt Fe-scintillator
detector downstream of a 50~GeV 
neutrino factory in which there are $10^{21} \mu^+$ decays followed 
by $10^{21} \mu^-$ decays~\cite{cerv00}. Fit results in the 
(matter density, $\sin^2 2\theta_{13}$)--plane
are shown in Fig.~\ref{fig:s15} for three baselines. 
The precision on the $\sin^2 2\theta_{13}$ determination is a few percent. 
Note that the analysis described in Ref.~\cite{cerv00} suggests that 
backgrounds can be suppressed to less than $10^{-5}$ of the total CC rate 
in the detector. This impressive level of background rejection deserves 
further study. 
At the shortest baselines (732~km) matter effects are too 
small to obtain a good determination of the matter density parameter. 

Consider next the precision with which the oscillation parameters 
can be determined if $\sin^2 2\theta_{13}$ is very small, and hence 
no $\nu_e \rightarrow \nu_\mu$ oscillation signal is observed. 
The resulting limits on $\sin^2 2\theta_{13}$ are shown as a 
function of $\delta m^2_{32}$ in Fig.~\ref{fig:m4} 
for a 10~kt ICANOE type detector 7400~km downstream of a 30~GeV 
neutrino factory in which there are $10^{20} \mu^+$ decays followed 
by $10^{20} \mu^-$ decays~\cite{camp00}. 
The resulting upper limit on $\sin^2 2\theta_{13}$ 
would be O($10^{-3}-10^{-4}$), 
about three orders 
of magnitude below the present experimental bound, and 
one to two orders of magnitude below the bound that would be expected at 
the next generation of long--baseline experiments.

The limit would become even more stringent at a higher intensity 
neutrino factory. As an example of the ultimate sensitivity that 
might be achievable, in Fig.~\ref{fig:s13} 
the limits on $\sin^2 2\theta_{13}$ are shown as a
function of $\delta m^2_{32}$ and baseline for a 40~kt Fe-scintillator 
detector downstream of a 50~GeV 
neutrino factory in which there are $10^{21} \mu$ decays~\cite{cerv00}. 
The non--observation of $\nu_e \rightarrow \nu_\mu$ oscillations could 
result in an upper limit on $\sin^2 2\theta_{13}$ below $10^{-5}$ !
With this level of sensitivity $\nu_e \rightarrow \nu_\mu$ oscillations 
driven by the sub--leading $\delta m^2$ scale might be observed~\cite{bgrw00}. 
For example, the number of muon decays required to produce 10 
$\nu_e \rightarrow \nu_\mu$ events in a 50~kt detector 2800~km 
downstream of a neutrino factory is shown for a bimaximal mixing 
scenario ($\sin^22\theta_{13} = 0$) 
in Fig.~\ref{fig:v2} as a function of the stored muon energy. 
Approaching $10^{21}$ muon decays might be sufficient to observe 
oscillations driven by the sub--leading scale, but would require 
background levels of the order of $10^{-5}$ of the total CC rate, 
or better. 

With a vanishing or very small $\sin^2 2\theta_{13}$ 
only the $\nu_\mu\to\nu_\tau$ 
oscillations will have a significant rate, and the oscillation 
parameters $\sin^2 2\theta_{23}$ and $\delta m^2_{32}$ can be 
determined by fitting the right--sign muon ($\nu_\mu$ disappearance) 
spectrum. Good sensitivity can be obtained provided the baseline 
is chosen such that the first oscillation maximum occurs in the 
middle of the visible energy spectrum. 
As a first example, 
spectra of events tagged by right--sign muons are shown in 
Fig.~\ref{fig:v5} as a function of $\delta m^2_{32}$ for a 10~kt MINOS--type 
detector 2800~km downstream of a 30~GeV neutrino factory in which 
there are $2 \times 10^{20} \mu^-$ decays in the beam--forming straight 
section~\cite{bgrw00}. 
The position of the oscillation maximum (resulting in a dip in 
the observed distributions) is clearly sensitive to $\delta m^2_{32}$. 
The depth of the observed dip is sensitive to the oscillation amplitude, 
and hence to $\sin^2 2\theta_{23}$. 
The visible energy spectrum of right--sign muon events can be fit 
to obtain $\sin^2 2\theta_{23}$ and $\delta m^2_{32}$. 
We begin by considering the statistical precision that could be 
obtained with a perfect detector having MINOS--type resolution functions, 
no backgrounds, no selection requirements, 
and no systematic uncertainty on the neutrino flux. 
Fit results are shown in Fig.~\ref{fig:v6}. 
For $\delta m^2_{32}=3.5\times 10^{-3}$~eV$^2$/c$^4$ 
the fit yields statistical 
precisions of a few percent on the the values of the oscillation parameters. 
If $L$ is increased to 7332~km, the statistical precision improves to 
about 1\%. 
With this level of precision it is likely that 
systematic uncertainties will be significant~\cite{bern00}. 
To illustrate this in Fig.~\ref{fig:b1}a the $1\sigma$ contours are shown 
in the ($\delta m^2_{32}$, $\sin^2 2\theta_{23}$) from fits which 
include backgrounds together with 0\% and 2\% systematic uncertainties 
on the beam flux. With a 2\% flux uncertainty the precision on 
$\delta m^2_{32}$ and $\sin^2 2\theta_{23}$ are respectively 
11\% and 14\%. 

\begin{figure}[t]
\epsfxsize3.5in
\centerline{\epsffile{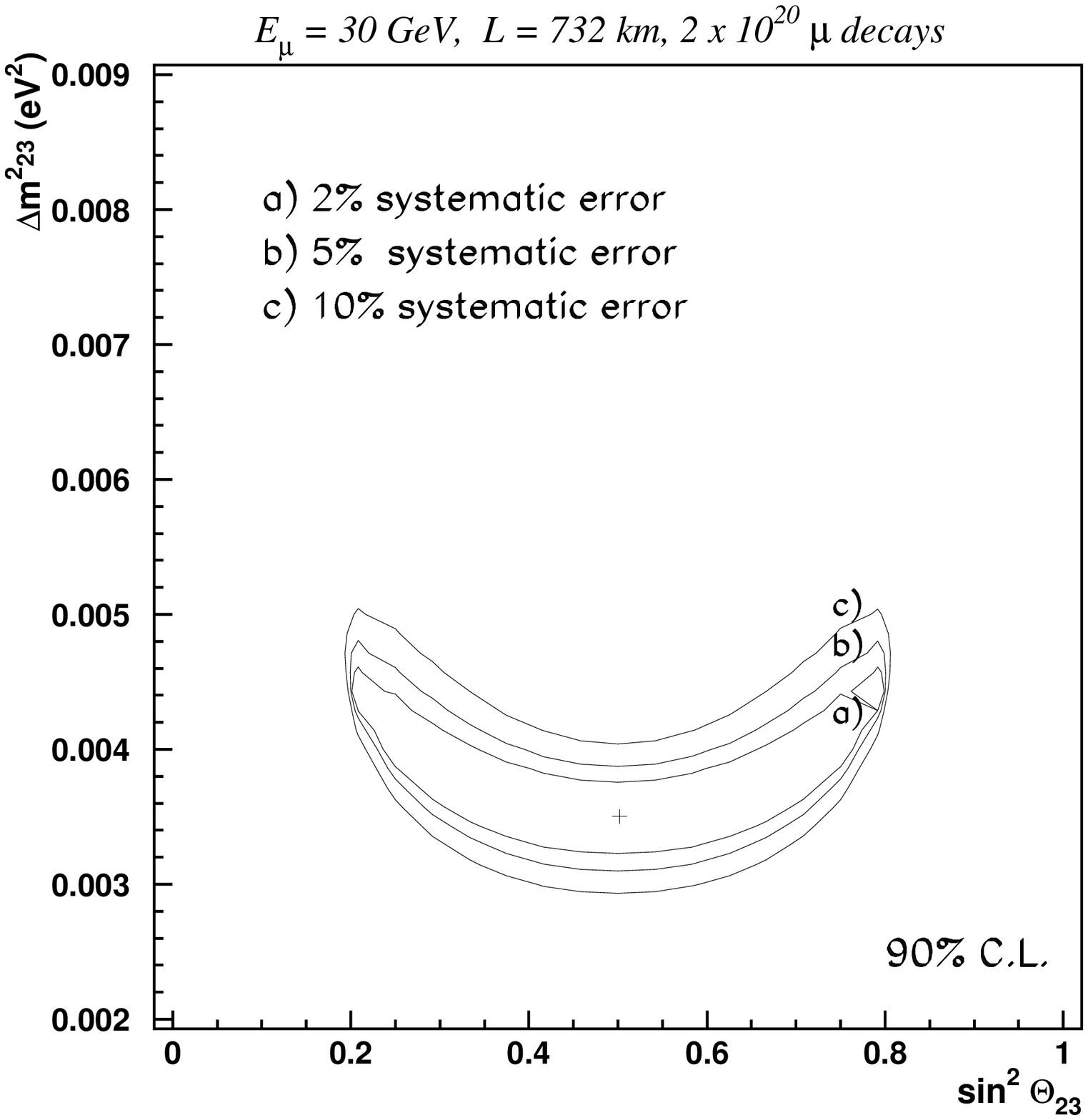}}
\caption{Fit results for simulated $\nu_\mu$ disappearance 
measurements with a 10~kt ICANOE type 
detector 732~km downstream of a 30~GeV neutrino factory in which 
there are $10^{20} \mu$ decays. The effect 
of a systematic uncertainty on the neutrino flux is shown.
Results are from Ref.~\ref{camp00}.
}
\label{fig:m8}
\end{figure}

\begin{figure}
\epsfxsize3.5in
\centerline{\epsffile{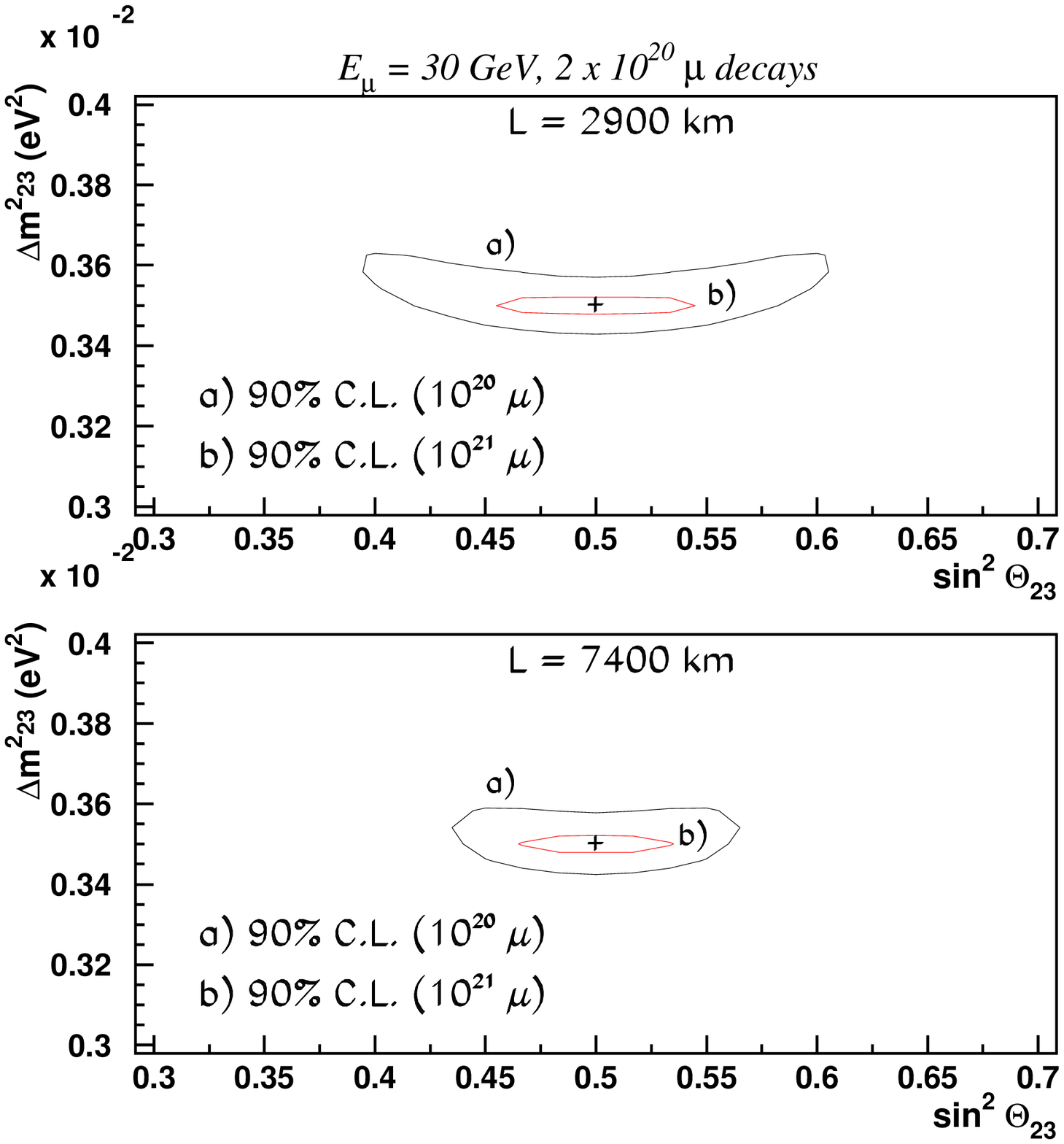}}
\caption{Fit results for simulated $\nu_\mu$ disappearance 
measurements with a 10~kt ICANOE type 
detector 2900~km (top plot) and 7400~km (bottom plot) 
downstream of a 30~GeV neutrino factory in which 
there are (a) $10^{20} \mu$ decays and (b) $10^{21} \mu$ decays. 
Results are from Ref.~\ref{camp00}.
}
\label{fig:m9}
\end{figure}

\begin{figure}[t]
\epsfxsize3.5in
\centerline{\epsffile{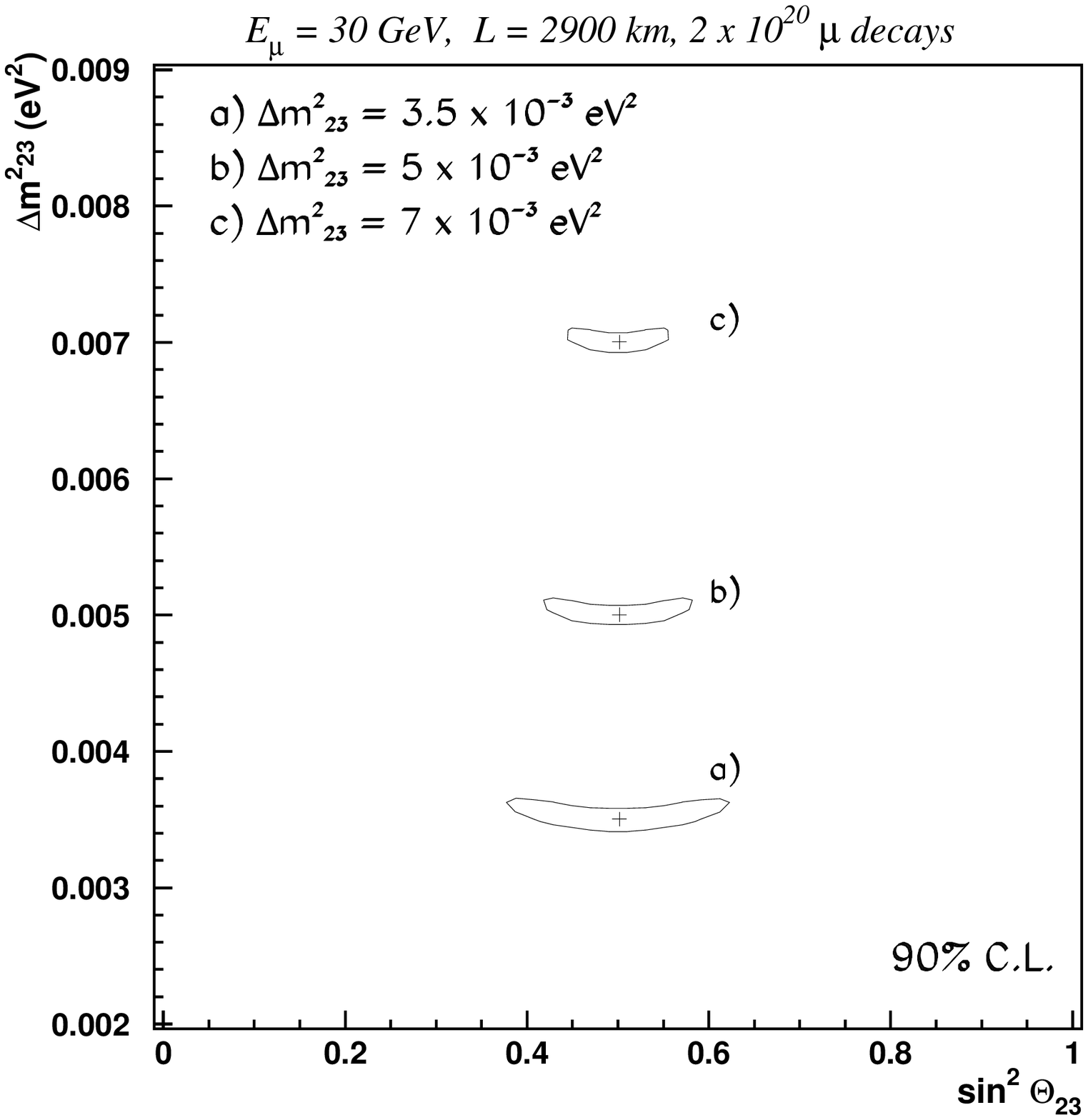}}
\caption{Fit results for simulated $\nu_\mu$ disappearance 
measurements with a 10~kt ICANOE type 
detector 2900~km 
downstream of a 30~GeV neutrino factory in which 
there are $10^{20} \mu^-$ decays followed by $10^{20} \mu^+$ decays. 
Results are shown for 3 values of $\delta m^2_{32}$, and are from 
Ref.~\ref{camp00}.
}
\label{fig:m10}
\end{figure}

\begin{figure}
\epsfxsize3.5in
\centerline{\epsffile{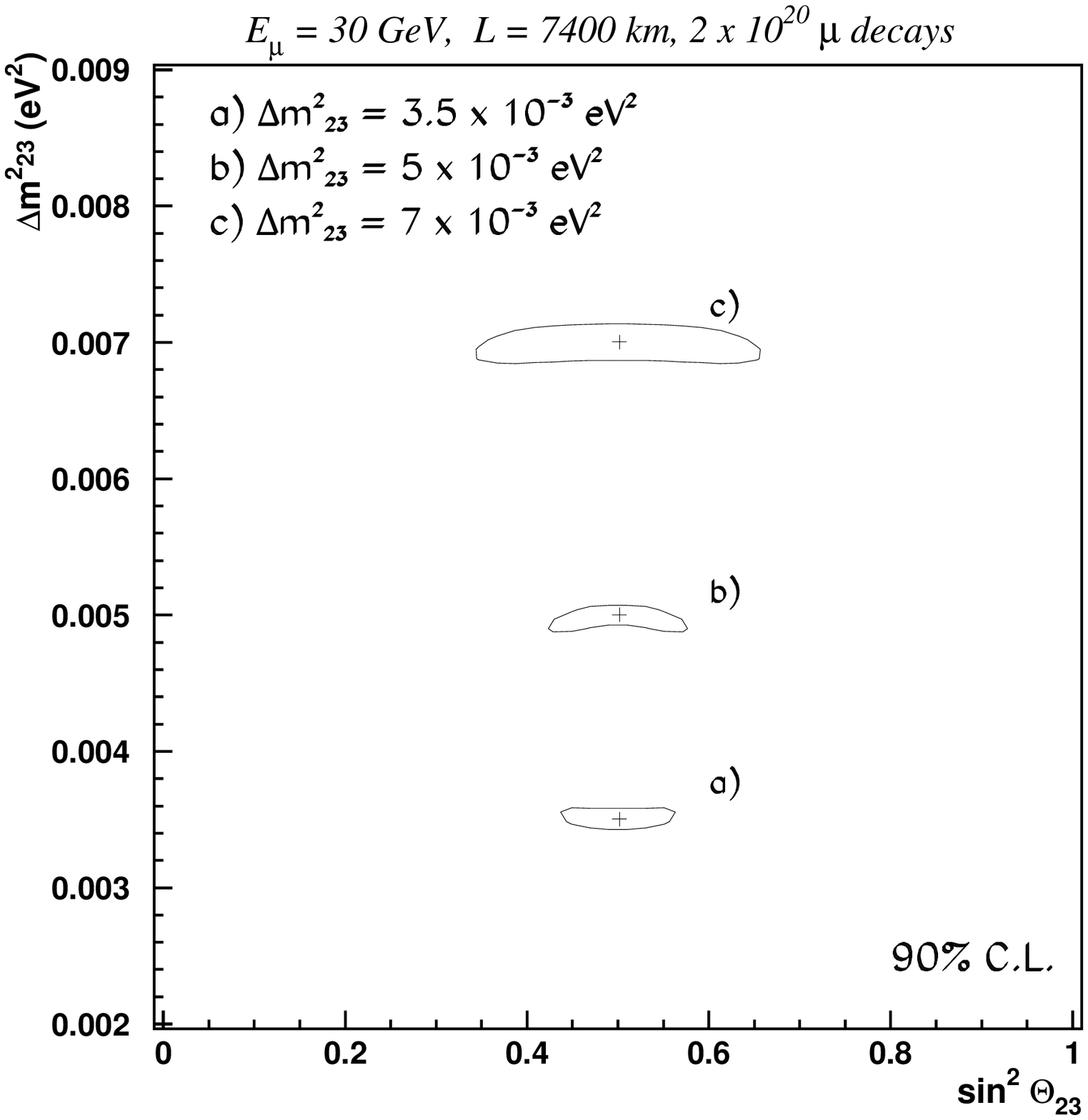}}
\caption{Fit results for simulated $\nu_\mu$ disappearance 
measurements with a 10~kt ICANOE type 
detector 7400~km 
downstream of a 30~GeV neutrino factory in which 
there are $10^{20} \mu^+$ decays followed by $10^{20} \mu^+$ decays. 
Results are shown for 3 values of $\delta m^2_{32}$, and are from 
Ref.~\ref{camp00}.
}
\label{fig:m11}
\end{figure}

As a second example, consider a 10~kt ICANOE--type detector 
that is downstream of a 30~GeV neutrino factory in which there are 
$10^{20} \mu^+$ decays in the beam--forming straight section 
followed by $10^{20} \mu^-$ decays~\cite{camp00}. The sensitivity to the 
oscillation parameters has been studied by fitting simulated 
visible energy distributions for events tagged by a right--sign muon. 
The analysis includes a 2\% bin-to-bin uncorrelated systematic error on 
the number of neutrino interactions which takes into account 
the uncorrelated uncertainties on neutrino flux, the cross section, 
and the selection efficiency. To reduce background from charged meson decays, 
the events entering the fit are those with muons having momenta $>2$~GeV/c. 
Figures~\ref{fig:m8}-\ref{fig:m11} show fit results in the 
($\sin^2 2\theta_{23}$,~$\delta m^2_{32}$)--plane as a function 
of the oscillation parameters and baseline. 
Note that for the ``short'' baseline ($L =732$~km) 
the first oscillation maximum for the reference value of 
$\delta m^2_{32}=3.5\times 10^{-3}$~eV$^2$/c$^4$ 
occurs at a neutrino energy of about 2~GeV. 
This is too low to produce a clear dip in the visible energy spectrum,
and as a result $\sin^2 2\theta_{23}$ and $\delta m^2_{32}$ can only 
be determined with relatively low precision (Fig.~\ref{fig:m8}), and 
the fit results are sensitive to systematic uncertainties on the 
neutrino flux.
At the longer baselines ($L = 2900$~km and 7400~km) 
the oscillation dip is visible, and the oscillation 
parameters can be measured with a precision that is mostly determined 
by the statistical uncertainty (Fig.~\ref{fig:m9}). 
For a 30~GeV neutrino factory and 
$\delta m^2_{32} = 3.5 \times 10^{-3}$~eV$^2$/c$^4$ 
the longer baseline (7400~km) yields the most precise result.
Specifically, for $10^{20} \mu$ decays 
the statistical precisions on $\sin^2 2\theta_{23}$ and $\delta m^2_{32}$ 
are respectively about 10\% and 1\%. 
With $10^{21} \mu$ decays the  $\sin^2 2\theta_{23}$ 
precision improves by about a factor of 2. 
It should be noted that the best 
baseline choice depends on $\delta m^2_{32}$ 
(Figs.~\ref{fig:m10}-\ref{fig:m11}), 
or more specifically $\delta m^2_{32}/E$. 

We conclude that, within the framework of three--flavor mixing, 
the oscillation parameters $\sin^2 2\theta_{13}$, $\sin^2 2\theta_{23}$, 
and $\delta m^2_{32}$ can be determined at a neutrino factory 
by fitting the observed visible energy distributions for various 
event types. A comprehensive study of the expected precisions of 
the measurements as a function of the oscillation parameters, baseline, 
and neutrino factory parameters has not yet been undertaken. 
However, detailed studies have been made for some examples in 
which there are  $10^{20} \mu^+$ decays 
followed by $10^{20} \mu^-$ decays in a 30~GeV neutrino factory.
For these examples 
we find that (i) if $\sin^2 2\theta_{13} >$~O($10^{-2}$) global fits 
can be used to determine its value, (ii) if $\sin^2 2\theta_{13}$ 
is too small to observe $\nu_e \rightarrow \nu_\mu$ oscillations 
then we would expect to place the very stringent upper limit on its 
value of $10^{-3}$ or better, and (iii) the values of 
$\sin^2 2\theta_{23}$, and $\delta m^2_{32}$ could be determined 
with precisions of respectively better than or of order 10\% and 
of order 1\%, provided the baseline is chosen so that the dip corresponding 
to the first oscillation maximum is in the middle of 
the visible energy distribution. 
At a high--intensity neutrino factory (for example with $10^{21}$ decays 
of 50~GeV muons) the mixing angles could be measured with a precision 
of a few percent, and 
if $\sin^2 2\theta_{13}$ is vanishingly small, the 
resulting upper limit could be at the O($10^{-5}$)--level.

\subsubsection{Search for CP violation}

\begin{figure}
\epsfxsize2.8in
\centerline{\epsffile{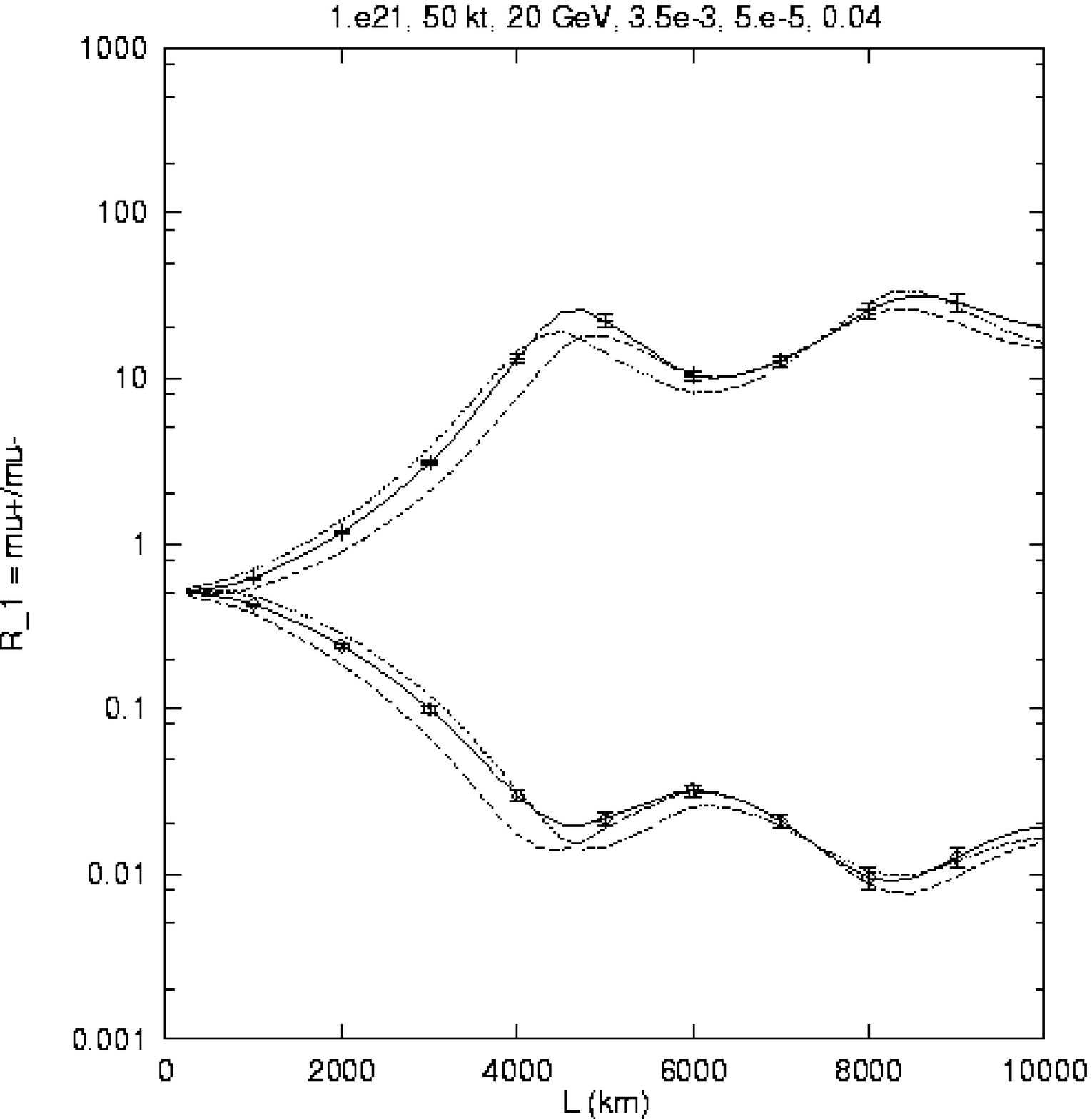}}
\caption{The ratio $R$ of 
$\bar\nu_e \to \bar\nu_\mu$ to $\nu_e \to \nu_\mu$  event 
rates at a 20~GeV neutrino factory 
for $\delta = 0$ and $\pm\pi/2$. The upper group of curves
is for $\delta m^2_{32} < 0$, the lower group is for
$\delta m^2_{32} > 0$. The statistical errors correspond to
$10^{21}$ muon decays of each sign and a 50~kt detector. 
The oscillation parameters correspond to the LMA scenario IA1. 
With no matter or CP effects $R\sim0.5$ for all baselines.
Results are from Ref.~\ref{bgrw00}.
}
\label{fig:cp1}
\end{figure}

\begin{figure}
\epsfxsize3.2in
\centerline{\epsffile{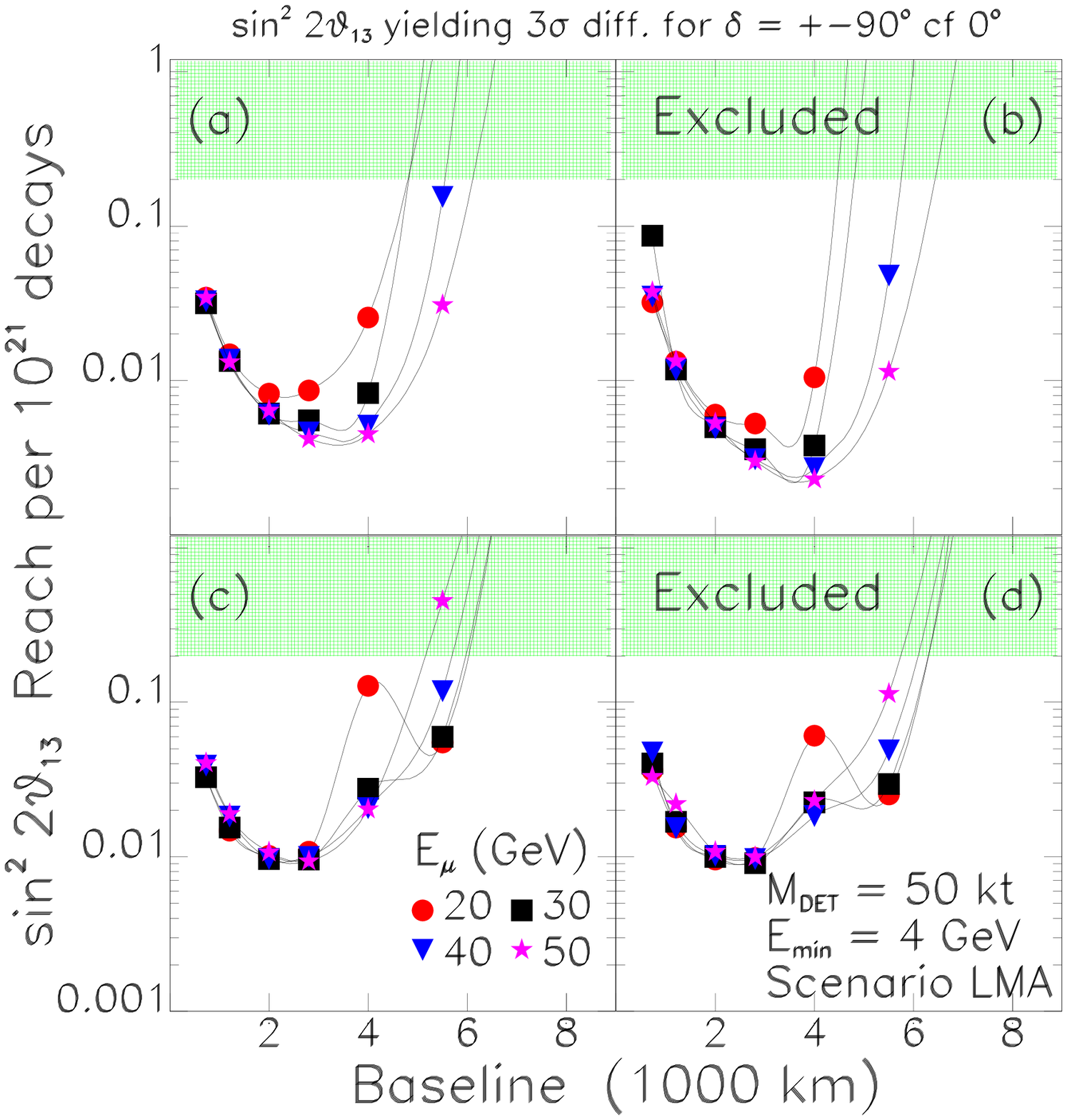}}
\caption{Reach in $\sin^22\theta_{13}$ that yields a $3\sigma$
discrimination between 
(a) $\delta = 0$ and $\pi/2$ with $\delta m^2_{32} > 0$, 
(b) $\delta = 0$ and $\pi/2$ with $\delta m^2_{32} < 0$,
(c) $\delta = 0$ and $-\pi/2$ with $\delta m^2_{32} > 0$, and 
(d) $\delta = 0$ and $-\pi/2$ with $\delta m^2_{32} < 0$. 
The discrimination is based on a comparison of wrong--sign 
muon CC event rates in a 50~kt detector when $10^{21}$ positive and 
negative muons alternately decay in the neutrino factory.
The reach is shown versus baseline for four storage ring energies. 
The oscillation parameters correspond to the LMA scenario IA1.
Results are from Ref.~\ref{bgrw00}.
}
\label{fig:cp2}
\end{figure}

\begin{figure}
\epsfxsize3.2in
\centerline{\epsffile{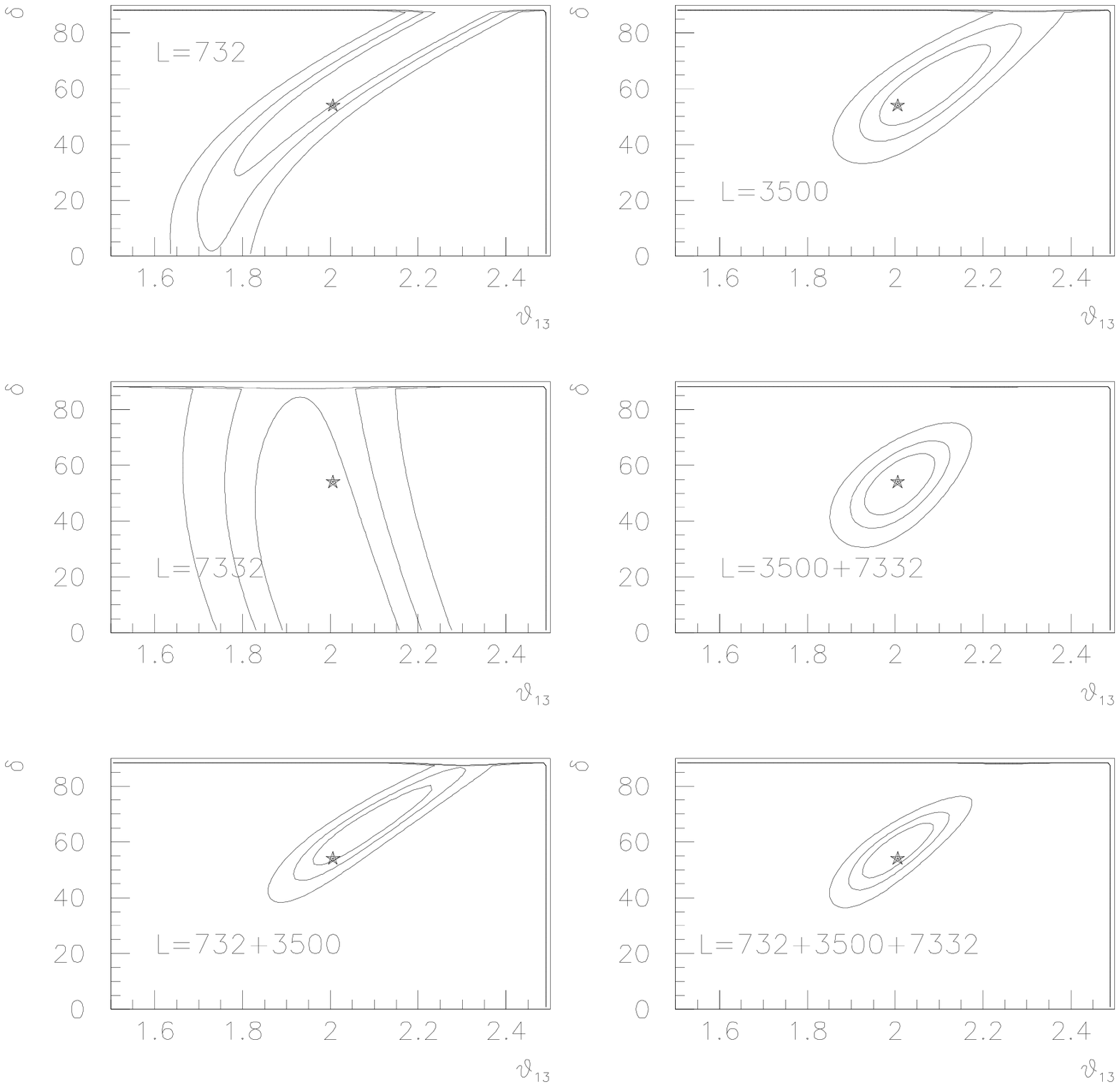}}
\caption{Fit results in the CP phase $\delta$ versus $\theta_{13}$ plane 
for a LMA scenario with $\delta m^2_{21} = 1 \times 10^{-4}$~eV$^2$/c$^4$. 
The 68.5, 90, and 99\% CL contours are shown for a 40~kt detector a distance 
$L$~km  
downstream of a 50~GeV neutrino factory in which there are $10^{21} \mu^+$
and $10^{21} \mu^-$ decays.
Results are from Ref.~\ref{cerv00}.
}
\label{fig:cp3}
\end{figure}

\begin{figure}
\epsfxsize3.2in
\centerline{\epsffile{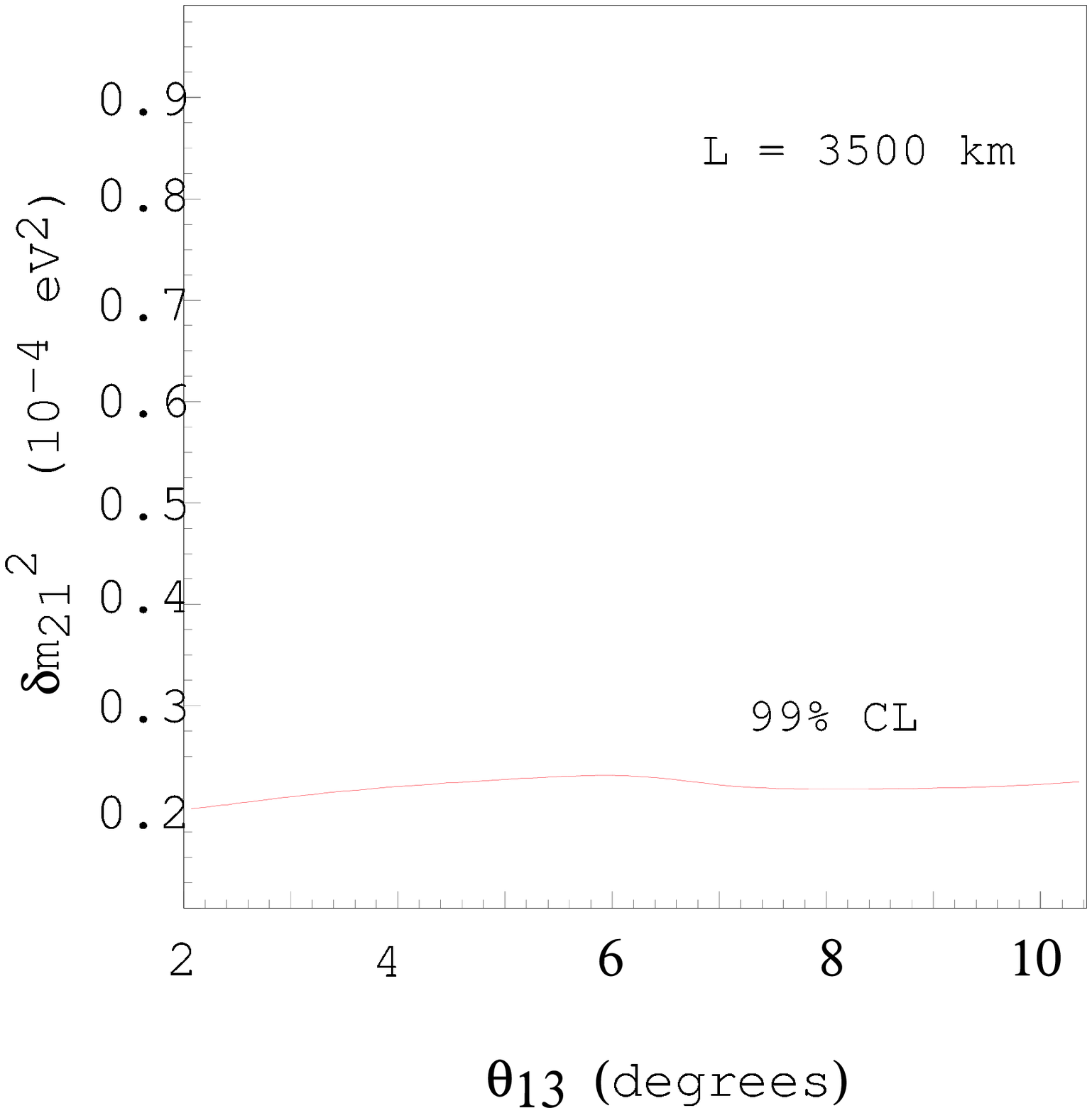}}
\caption{The lowest value of $\delta m^2_{21}$, shown as a function of
$\theta_{13}$, for
which the maximal CP phase $\delta = \pi/2$ can be distinguished
from a vanishing phase in a LMA oscillation scenario. 
The curve corresponds to a 40~kt detector 3500~km 
downstream of a 50~GeV neutrino factory in which there are $10^{21} \mu^+$ 
and $10^{21} \mu^-$ decays. 
Results are from Ref.~\ref{cerv00}.
}
\label{fig:cp4}
\end{figure}

In the majority of the three--flavor oscillation scenarios described in 
section~\ref{theory} the CP violating amplitude is too small to 
produce an observable effect. Nevertheless, in these cases stringent limits 
on CP violation would provide an important check of the overall interpretation 
of the oscillation data. If however the LMA scenario provides the 
correct description of neutrino oscillations, CP violating effects 
might be sufficiently large to be observable at a high--intensity 
neutrino factory~\cite{cerv00,bgrw00}. 
This is illustrated in Fig.~\ref{fig:cp1} which shows, 
as a function of baseline at a 20~GeV neutrino factory, the ratio $R$ 
for $\delta = 0$ and $\pm\pi/2$, where $R$ is defined as the 
$\overline{\nu}_e \rightarrow \overline{\nu}_\mu$ event rate divided by 
the $\nu_e \rightarrow \nu_\mu$ event rate. 
The upper group of curves 
is for $\delta m^2_{32} < 0$, the lower group is for 
$\delta m^2_{32} > 0$, and the statistical errors correspond to 
$10^{21}$ muon decays of each sign with a 50~kt detector. 
If $L$ is a few thousand~km a non--zero $\delta$ can produce a 
modification to $R$ that is sufficiently large to 
be measured ! 

Since the $\nu_e \rightarrow \nu_\mu$ oscillation rates 
are to a good approximation proportional to $\sin^22\theta_{13}$, it 
is useful to define the $\sin^22\theta_{13}$ reach as that value of
$\sin^22\theta_{13}$ that will produce a 3$\sigma$ 
change in the predicted ratio $R$ when 
$\delta$ is changed from $0$ to $\pm\pi/2$. 
The $\sin^22\theta_{13}$ reach is shown as a function of baseline and 
stored muon energy in Fig.~\ref{fig:cp2} for a 50~kt detector 
at a neutrino factory in which there are $10^{21} \mu^+$ decays followed by 
$10^{21} \mu^-$ decays. With an optimum baseline of about 
3000~km (for $\delta m^2_{32} = 3.5 \times 10^{-3}$~eV$^2$/c$^4$) 
the $\sin^22\theta_{13}$ reach is approximately $10^{-2}$, an order 
of magnitude below the current experimental bound. 
Thus, in a LMA scenario, CP violation in the lepton sector might 
be measurable at a neutrino factory providing O($10^{21}$) muon decays. 

As an example, consider a 40~kt Fe-scintillator detector downstream 
of a 50~GeV neutrino factory providing $10^{21} \mu^+$ decays followed 
by $10^{21} \mu^-$ decays in the beam--forming straight section~\cite{cerv00}. 
The results of fits to the simulated wrong--sign muon event distributions, 
with $\delta$ and $\sin^22\theta_{13}$ left as free parameters, are 
shown in Fig.~\ref{fig:cp3} for various baselines, with the 
sub--leading scale $\delta m^2_{21} = 1 \times 10^{-4}$~eV$^2$/c$^4$. 
The analysis includes 
the detector resolutions, reasonable event selection criteria, and 
backgrounds. As might be expected from Fig.~\ref{fig:cp1} at $L = 7332$~km 
there is little sensitivity to $\delta$, and at the ``short" baseline 
$L = 732$~km the fit has difficulty untangling $\delta$ from 
$\sin^22\theta_{13}$. However, at a baseline of $L = 3500$~km for the 
example shown $\delta$ and $\sin^22\theta_{13}$ can be determined with 
precisions of respectively about $15^\circ$ and a few percent. Note 
that a combination of baselines can yield a modest improvement in 
the precision of the measurement.
The sensitivity to CP violation decreases with decreasing 
$\delta m^2_{21}$. Figure~\ref{fig:cp4} shows as a function of 
$\sin^22\theta_{13}$ the lowest value of $\delta m^2_{21}$ for 
which the maximal CP phase $\delta = \pi/2$ can be distinguished 
from a vanishing phase at $L = 3500$~km. This limiting $\delta m^2_{21}$ 
is below the current central value for the LMA parameter space suggested 
by solar neutrino deficit, and is about $2 \times 10^{-5}$~eV$^2$/c$^4$, 
independent of $\sin^22\theta_{13}$. 

Finally we note that the sensitivity of short and medium baseline 
experiments to CP violation in a three--active 
plus one sterile neutrino scenario has been considered in Ref.~\cite{doninietal}. 
They concluded that a 1~kt detector and a 100~km baseline could provide a 
clean test of CP violation, particularly in the $\tau--lepton$ appearance 
channel.

\subsection{Summary}

The oscillation physics that could be pursued at a neutrino factory 
is compelling. In particular, experiments at a neutrino factory 
would be able to simultaneously measure, or put stringent limits on, 
all of the appearance modes $\nu_e \rightarrow \nu_\tau$, 
$\nu_e \rightarrow \nu_\mu$, and $\nu_\mu \rightarrow \nu_\tau$. 
Comparing the sum of the appearance modes with the disappearance 
measurements would provide a unique basic check of candidate 
oscillation scenarios that cannot be made with a conventional neutrino 
beam. 
In addition, for all of the specific oscillation 
scenarios we have studied, the 
$\nu_e$ component in the beam can be exploited to enable 
crucial physics questions to be addressed. These include 
(i) the pattern of neutrino masses (sign of $\delta m^2$) and 
a quantitative test of the MSW effect, 
(ii) the precise determination of (or stringent limits on) all of the 
leading oscillation parameters, which in a three--flavor mixing 
scenario would be  $\sin^22\theta_{13}$, $\sin^22\theta_{23}$, 
and $\delta m^2_{32}$, and 
(iii) the observation of, or stringent limits on, CP violation in 
the lepton sector.

To be more quantitative in assessing the beam energy, intensity, and 
baseline required to accomplish a given set of physics goals it is 
necessary to consider two very different experimental possibilities: 
(a) the LSND oscillation results are not confirmed by the MiniBooNE 
experiment, or (b) the LSND results are confirmed. 
\begin{description}
\item{(a) LSND not confirmed.}
Fairly extensive neutrino factory studies have been made within 
the framework of three--flavor oscillation scenarios in which there is 
one ``large" 
$\delta m^2$ scale identified with the atmospheric neutrino deficit 
results, and one small $\delta m^2$ identified with the solar neutrino 
deficit results. A summary of the energy dependent beam intensities 
required to cross a variety of ``thresholds of interest" is provided by 
Fig.~\ref{fig:v2}. A 20~GeV neutrino factory providing $10^{19}$ muon 
decays per year is a good candidate ``entry--level" facility which would 
enable either (i) the first observation of $\nu_e \rightarrow \nu_\mu$ 
oscillations, the first direct measurement of matter effects, 
and a determination of the sign of $\delta m^2_{32}$ and 
hence the pattern of neutrino masses, or (ii) a very stringent limit 
on $\sin^22\theta_{13}$ and a first comparison of the sum of all 
appearance modes with the disappearance measurements. 
The optimum baselines for this entry--level 
physics program appears to be of the order of 3000~km or greater, 
for which matter effects are substantial. 
Longer baselines also favor the precise determination of $\sin^22\theta_{13}$. 
A 20~GeV neutrino factory providing $10^{20}$ muon 
decays per year is a good candidate upgraded neutrino factory (or 
alternatively a higher energy facility providing a few $\times 10^{19}$ decays 
per year). This would enable the first observation of, or meaningful 
limits on, $\nu_e \rightarrow \nu_\tau$ oscillations, and precision 
measurements of the leading oscillation parameters. In the more distant 
future, a candidate for a second (third ?) generation neutrino factory might 
be a facility that provides O($10^{21}$) decays per year and enables the 
measurement of, or stringent limits on, CP violation in the lepton sector.

\item{(b) LSND confirmed.}
Less extensive studies have been made for the class of scenarios that 
become of interest if the LSND oscillation results are confirmed. 
However, in the scenarios we have looked at (IB1 and IC1) we find that 
the $\nu_e \rightarrow \nu_\tau$ rate is sensitive to the 
oscillation parameters 
and can be substantial. With a large leading $\delta m^2$ scale medium 
baselines (for example a few $\times 10$~km) are of interest, and 
the neutrino factory intensity required to effectively exploit the 
$\nu_e$ beam component might be quite modest ($< 10^{19}$ decays per 
year).
\end{description}

The neutrino factory oscillation physics study we have pursued 
goes beyond previous studies. In particular we have explored the 
physics capabilities as a function of the muon beam energy and 
intensity, and the baseline. Based on the representative 
oscillation scenarios and parameter sets defined for the study, 
it would appear that a 20~GeV neutrino factory providing 
O($10^{19}$) decays per year would be a viable entry--level 
facility for experiments at baselines of $\sim3000$~km or greater.
There are still some basic open questions that deserve further 
study: 
(1) We have sampled, but not fully explored, the beam 
energy and intensity required to explore the scenarios that become 
relevant if the LSND oscillation results are confirmed. 
(2) Possible technologies for a very massive neutrino factory detector 
have been considered, but these considerations deserve to be 
pursued further. The chosen detector technology will determine 
whether it is necessary to go deep underground. 
(3) We have developed tools that can explore the utility of having 
polarized muon beams. The physics payoff with polarization is a 
detailed issue. It deserves to be studied in the coming months.

Based on our study, we believe that a neutrino factory in 5--10~years 
from now would be the right tool 
at the right time.


\newcommand{\ignore}[1]{}
\section{Non--Oscillation Physics}

Due to the theoretically clean nature of weak interactions, conventional
neutrino scattering experiments have always provided precise
measurements of fundamental parameters.  These include: 
a crucial role in the extraction of parton distribution functions, 
measurements of the Weinberg angle\cite{NuTeV:prelim}, 
and the strong coupling constant \cite{Seligman}
$\alpha_s$, which are competitive with any other methods.  Perhaps because
of this success, we forget how crude existing neutrino experiments are.
The high statistics experiments such as CDHSW\cite{CDHSWsf} and CCFR/NuTeV\cite{Seligman,NuTeV:prelim}, in order
to obtain samples of more than 10$^5$ events, rely on 
coarsely segmented massive iron/scintillator calorimeters weighing
close to 1000 tons.  Measurements on proton targets and detailed studies
of the final state have been confined to very low statistics bubble chamber
experiments.  As a result we have virtually no precise measurements
of neutrino-proton scattering and no measurements on polarized targets
which could offer  new insights into the spin structure of the nucleon.

The advent of a neutrino factory, with neutrino fluxes of
10$^{20}$/year instead of the 10$^{15-16}$ at existing facilities would
open a new era in conventional neutrino physics. We would be able
to use low mass targets and high resolution detection technologies and
still achieve better statistical power than present-day experiments. 
For example a 50 GeV muon storage ring at the above rate 
would produce around 18 M
neutrino charge-current interactions per year in a 10 kg hydrogen target.  
This is 5-10 times the statistics of the CCFR and 
NuTeV experiments with 600 ton detectors.  
Better understanding of neutrino fluxes from the decay of monochromatic
muons  will also reduce many of the dominant systematic errors. 

In this study we have concentrated on measurements that are only possible
with higher fluxes rather than repeating older measurements with thousands
of times the statistics.  As a result, the statistical errors shown are
often not negligible, but without the high flux at a neutrino factory the
measurements themselves would
be impossible.

\subsection*{Outline}

Due to the breadth of the  field we are unable to give a complete
survey  and instead highlight a few
of the areas where the high flux beam at a neutrino factory allows
new measurements:

\begin{itemize}

\item A description of a low mass target/detector and typical rates in such a detector. 

\item Nucleon deep inelastic scattering measurements and a proposed
detector design.

\item Neutrino cross section measurements, a topic
of great interest to the nuclear physics community and also needed to
understand normalization at a far neutrino oscillation detector.

\item Spin structure functions, which have never been measured in neutrino
beams. 

\item The potential of the neutrino factory as a clean source
of single tagged charm mesons and baryons.

\item Electroweak measurements in both the hadronic and purely leptonic sectors.

\item Use of the very clean initial state to search for exotic interactions.

\item Searches for anomalous neutrino interactions with electromagnetic fields.
\end{itemize}

 
\subsection{Possible detector configurations and statistics}

For studies of charged current deep-inelastic scattering on proton
targets, the optimal detector system is probably a target followed by
precision magnetic tracking sytems, an electromagnetic calorimeter
and a muon detection system.  Such detectors have been used in 
muon scattering experiments at CERN and FNAL and in the new generation
neutrino scattering experiments CHORUS \cite{CHORUS} and NOMAD\cite{nomad}.  A low mass target
followed by tracking and electromagnetic calorimetry makes the
electron anti-neutrinos in the beam a source of additional statistics
rather than backround, except in the case of neutral current studies.

The target itself should be thin enough that particles produced within
it have a small probability of interacting before they reach the tracking
systems.
In this study we considered liquid hydrogen and deuterium targets --
both polarized and unpolarized --  and heavier solid nuclear targets.
The hydrogen and deuterium targets are 1m  long while the polarized target
is 50 cm long. All targets
are 20 cm in radius, to fit the central beam spot at 50 GeV.  For
lower beam energies the beam spot grows in size as $\sim 1/E$.
 Nuclear targets are scaled so that the interaction length
in the material is constant at 14\%.  The charged current muon neutrino
interaction rates are summarized in table \ref{rates}.  

The numerical estimates in this study use, unless otherwise noted,
$10^{20}$ 50 GeV muon decays in a 600 m straight section.

\begin{table}
\caption{\label{rates} Charged current muon-neutrino scattering
rates in a small target located near a muon storage ring. Rates
are per $10^{20}$ muon decays.  The detector is located ($1\times E_{\mu}$, GeV)meters away from the ring to assure that primary muons have ranged out
before the detector.}\begin{center}
\begin{tabular}{|c|c|r|r|}
\hline
Machine& Target & Thickness,cm & Events \\

\hline
50 GeV neutrino factory &Liquid H$_2$& 100 & 12.1M\\
&Liquid D$_2$& 100 & 29.0M\\
&solid HD & 50 &9.3M\\
 &C&5.3&20.7M\\
&Si&6.3&25.4M\\
&Fe&2.3&31.6M\\
&Sn&3.1&39.1M\\
&W&1.3&44.3M\\
&Pb&2.4&46.5M\\
\hline
CCFR/NuTeV&Fe& 600& $\sim$ 2M\\
\hline
\end{tabular}\end{center}
\end{table}
\begin{figure} 
\epsfxsize=3.0in
\centerline{
\epsffile{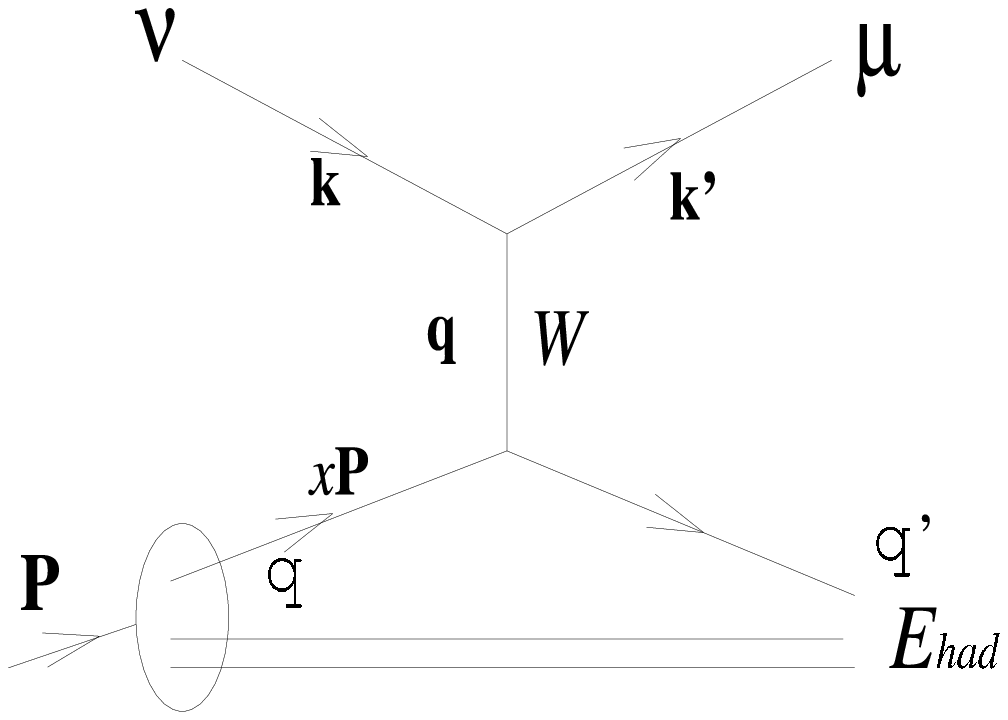}}
\caption{Kinematics of neutrino scattering in the parton model.
The energy-momentum tranfer from the leptons to the proton
is $\fv{q}$ and the fraction of the proton momentum carried
by the struck quark is approximately $x$.
\label{fig:kinematics}}
\end{figure}

These are the total event rates for charged current  muon-neutrino scattering.  The anti-neutrino
rates are half as large. Kinematic cuts reduce the statistics
by less than a factor of two.
We have only considered muon-neutrino charge current scattering for
structure function measurements, although for such thin targets, electron
neutrino scatters should also be reconstructable with high precision.

\subsection{Neutrino Scattering Kinematics}
\newcommand{\Elep}[0]{E_{\lepton}}

The kinematic variables for neutrino deep inelastic scattering are
illustrated in figure \ref{fig:kinematics}:  
\newcommand{\Enu}[0]{E_{\nu}}
\newcommand{\mlep}[0]{m_{\lepton}}
\begin{eqnarray}
 \fourv{q} &=&{\fvk{\nu} - \fvk{\lepton}}, \hskip .4 in Q^2 = -\fourv{q}^2 = 2
\Elep\Enu -\mlep^2 - 2 \Enu p_{\lepton} \cos\theta_{lab},\\
 \nu &=&(\fv{p}\fourv{q})/M  \simeq \Elep - \Elep^{\prime},\\
 x   &=&Q^2/2 \mtarget \nu,\\
 y &=& \mtarget \nu/(\fvk{\nu} \fv{p}) = (1 + cos\theta_{CM})/2 \approx
\nu/\Elep,\\
  W^2 &=&  2 \mtarget  \nu + \mtarget^2   -Q^2,
\end{eqnarray}
where the $\fvk{}$ are the neutrino and final state four vectors,
$\fv{p}$ is the proton four-vector, $M$ is the target nucleon mass, $\Enu$ is the incoming neutrino
energy $\Elep, p_{\lepton}$ are the outgoing lepton energy and momentum
$\theta_{lab}$ is
the lepton angle with respect to the incoming beam. $\fourv{q}$ is the
four-momentum transfer to the target, $\nu$
is the energy transfer, $x$ is the Bjorken $x$ variable, $y$ is the scaled
energy transfer and $W^2$ is the invariant mass of the final state hadronic
system 
squared.

\begin{figure} 
\epsfxsize=5.0in
\centerline{
\epsffile{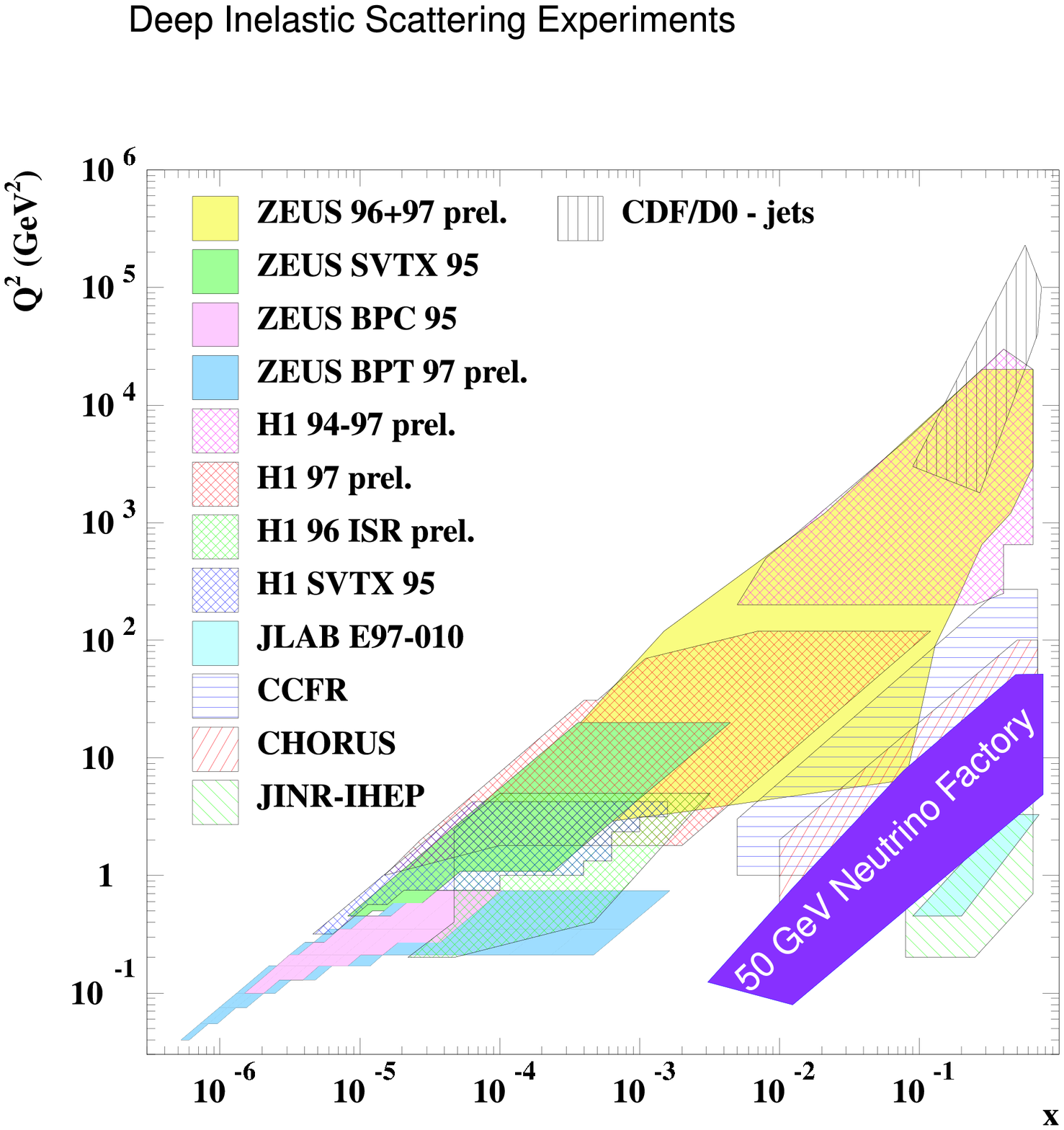}}
\caption{Comparison of kinematic ranges for present DIS experiments
with a 50 GeV Neutrino factory. }
\label{fig:comparejlab}
\end{figure} 
Fig. \ref{fig:comparejlab} shows the kinematic region for a neutrino
factory as compared to other deep-inelastic scattering experiments.



For $Q << E$ and $s << M_W$ the  the unpolarized neutrino 
(anti-neutrino) scattering cross section is:

\begin{eqnarray}
{d\sigma^{\nu(\antinu)}\over dx dy}& = &
{ G_F^2 M E_{\nu} \over 2\pi }\biggr[ [F_2^{\nu(\antinu)}(x,Q^2) \pm xF_3^{\nu(\antinu)}(x,\qsq)]   +\\ \nonumber
& & \ \ \ \ [F_2^{\nu(\antinu)}(x,Q^2)\mp xF_3^{\nu(\antinu)}(x,Q^2)] (1-y)^2 - \\\nonumber
&& \ \ \ \  2 y^2 F_L(x,Q^2) \; ,
\end{eqnarray}

\noindent
where the $F_i$ are 
Structure Functions.  $F_L = F_2 - 2xF_1 $ is a purely longitudinal
structure function.
The $xF_3$ contribution changes sign for anti-neutrino scattering.
There are additional structure functions $F_4$ and $F_5$ which are
suppressed by factors of the lepton mass over the
proton mass  squared.  For $\nutau$ and $\numu$
scattering at very low energies, these terms can become
quite important.

If the target is longitudinally polarized with respect to the 
neutrino polarization, then the cross section difference\cite{Bodo}:
\begin{eqnarray} 
 {d^2(\sigma_{\Rightarrow}^{\leftarrow}
-\sigma_{\Leftarrow}^{\leftarrow})^{\nu(\bar\nu) } \over dxdy}& =& {G_F^2 M E_{\nu} \over \pi  }
\bigl\{\pm
y(1-{y \over 2}-{xyM \over 2E})xg_1
\mp
{x^2yM \over E}g_2
+\\
\nonumber && \ \ \ \ 
y^2x(1+{xM \over E})
g_3+(1-y-{xyM \over 2E})
[(1+{xM \over E})g_4+g_5]
\bigr\},
\label{pol_lon}
\end{eqnarray}

\noindent
is described by  two parity conserving  Polarized Structure
Functions $g_1$ and $g_2$, and  by three parity
violating Polarized Structure Functions $g_3, g_4$ and $g_5$.
However, if the nucleon is transversely polarized, the cross
section difference is:

\begin{eqnarray}
 {d^2(\sigma_{\Uparrow}^{\leftarrow}
-\sigma_{\Downarrow}^{\leftarrow})^{\nu(\bar\nu) } \over dxdy} =
\frac{ G_F^2 M}{ 16 \pi^2  }
 \sqrt{xyM \left[ 2(1-y)E-xyM \right] }
\bigr\{\mp2xy({y\over 2}g_1+g_2)\\\nonumber
+xy^2g_3
+ (1-y-{xyM \over 2E})g_4
-{y\over 2}g_5\bigr\} \/.
\label{pol_tra}
\end{eqnarray}

\noindent The transverse cross section is suppressed by
${M/Q}$ with respect to the longitudinal cross section.

\subsection{Total cross section Measurements} 

A measurement of the total CC neutrino 
scattering cross section is both of intrinsic
interest and essential to precision measurements at a neutrino factory.  
We currently know the cross sections for neutrino scattering at the 2-3\%
level \cite{CCFRsigma} at energies above 30~GeV but 
at energies approaching the resonance region (2~GeV) the uncertainty 
increases considerably.  Because muon decay is so well-understood, 
the flux and hence the total cross section should be measureable across the full energy 
spectrum to the $1\%$ level.

The yield $Y$ of neutrino interactions
observed in any detector can be written as:

$$Y = n_0(E,r) \sigma(E) \epsilon(E,r) N$$

\noindent
where $n_0$ is the flux of incident neutrinos as a funtion of the 
neutrino energy $E$ and distance from the center of the detector, $r$.  
$\sigma$ is the cross section, $\epsilon(E,r)$ is the 
detector efficiency, and N is the number of target particles.  

For a far detector, $n_0(E,r)\sim n_0(E)$ is mainly determined 
by the beam divergence
and the muon decay kinematics and can probably be estimated from
the machine parameters and decay model with  precisions
at the $1\%$ level.  For a near detector, $n_0(E,r)$ depends
mainly on the muon decay kinematics and geometry, with contributions
from  beam size and divergence at the few percent level. 

Without a measurement of the absolute number of neutrinos, 
the best way to determine the flux is  
to normalize to a very well-understood process.  By comparing 
the yield of the normalization process and the total event rate, 
one then has a measurement of the total cross section.  
Inverse muon decay ($\numu + e^- \gt \mu- + \nue$  
and $\nuebar + e^- \gt \mu- + \numubar$) 
provides a clean channel for mapping the beam flux $n_0(E,r)$ 
at a near detector
and for normalizing the total cross section measurement.  
It has the limitation of an energy threshold of $\sim 11$~GeV and no
corresponding channels for the opposite sign beam.  
Quasi-elastic scattering is an additional normalization mode since it 
has a much lower  energy threshold and occurs for beams made with muons 
of either charge.  
Finally, scattering from atomic electrons is suppressed by
a factor of order $m_e/m_p$ relative to the normal neutrino nucleon 
interactions, but still yields an event rate of $\simeq 10^4 $ interactions per
 gr/cm$^2$ for 10$^{20}$ 50 GeV $\mu^-$ decays.

If the ratio of flux shapes at far and near detectors can be understood at
the 1\% level, then measurements of $n_0(E,r)\sigma(E)$ at a near
detector can be used to precisely predict the number of events
expected in the absence of oscillations  at a far 
detector.  Such precise flux measurements are also important for the suite
of measurements described in the remainder of this chapter.

\subsection{Structure function  measurements}
In principle, the structure functions can be extracted by fits to the $y$ dependence of the
cross section.  To date this
has proven very difficult as the data must be binned in $x$, $y$ and $Q^2$ and
no experiment has had sufficient statistics to perform such an analysis with high accuracy\cite{CDHSWsf,UnkiThesis}.

Instead, high statistics experiments\cite{CCFRsigma} such as CHARMII, CCFR and CDHSW have  relied on massive targets (Iron, Calcium) which are
approximately iso-scalar and have combined neutrino and anti-neutrino information in
order to extract average structure functions.
The structure functions averages have naive parton model interpretations:
\begin{eqnarray}
\overline{F}_2^N(x,Q^2) &\simeq &\sum (x\quark(x,Q^2) + x\antiquark(x,Q^2)),\\
\overline{F}_3^N(x,Q^2) &\simeq &\sum(x\quark(x,Q^2) - x\antiquark(x,Q^2)),\\\nonumber
\end{eqnarray}
where $\overline{F}_2(x,Q^2)$ and $\overline{F}_3(x,Q^2) $ are the average of neutrino
and antineutrino structure functions measured on a target which is an average of neutron and proton and $\quark(x,Q^2)$ and $\antiquark(x,Q^2)$ represent the
parton distribution functions or total probability of finding a quark
or antiquark in the proton:  

\begin{eqnarray}
\quark(x,Q^2) &=& \uquark(x,Q^2) +  \dquark(x,Q^2)+\squark(x,Q^2)+\cquark(x,Q^2)...\\
\antiquark(x,Q^2) &=& \antiuquark(x,Q^2) +  \antidquark(x,Q^2)+\antisquark(x,Q^2)+\anticquark(x,Q^2)...\\\nonumber 
\end{eqnarray}


Given the expectation of 12 M (24 M) events/year in a 1 m  hydrogen (deuterium) target at a 50 GeV
muon storage ring we can do a complete analysis of each channel $\nu p, \nu n, \antinu p, 
\antinu n$ without averaging.  Such an analysis allows a unique extraction of individual quark flavor parton
distribution functions.

For example, in the case of $\nu$p scattering, a $W^{+}$ is exchanged and the reaction is only
sensitive to negatively charged quarks. Due to the helicity dependence of the interaction
only left-handed $\dquark$ type and right handed $\antiuquark$ quarks will be involved.

The leading order parton model cross section is simply
\begin{eqnarray}
{d\sigma^{\nu p}\over dx dy} &\simeq& {4 G_F^2 M E_{\nu}\over \pi }x [(\dquark_L (x,Q^2) + \squark_L(x,Q^2)) +(\antiuquark_R(x,Q^2) +\\
&&\hskip 2 in \anticquark_R(x,Q^2))(1-y)^2],\\\nonumber 
\label{partonformula}\end{eqnarray}
and the different contributions can be extracted from the $y$ dependence of this
cross section and the corresponding anti-neutrino cross section. The relative
$\squark$ and $\dquark$ quark contributions can be measured in charm production.  

For an unpolarized target
$\quark_L(x) = \quark_R(x) = \onehalf \quark(x)$.  For
a polarized quark $\quark_L(x) = \onehalf (\quark(x) + \delta\quark(x))$
and $\quark_R(x) = \onehalf(\quark(x) - \delta\quark(x))$ where
$\delta\quark(x)$ is the degree to which the quark spin
is aligned with the proton spin\footnote{ The traditional $\Delta q$ spin distributions from
electron and muon scattering measure the sum $\Delta \quark = \delta \quark + \delta \antiquark$ as
photon probes cannot tell quarks and anti-quarks apart.}.
Thus a $\sigma_{\nu p}$ measurement on an unpolarized target can determine $\dquark+\squark$ and $\antiuquark +
\anticquark$ by averaging over the proton spin, while by measuring the polarization
asymmetry one can measure $\delta \dquark+\delta\squark$ and $\delta\antiuquark +
\delta\anticquark$. 

Scattering on neutrons can be related to scattering on protons by an isospin
transformation which exchanges $\uquark$ and $\dquark$ quarks and anti-quarks.
Differences of neutron and proton cross sections can then be used to cancel the
$\uquark$ and $\dquark$ components leaving observables
sensitive only to  $\squark$ and $\cquark$ 
distributions.

\subsection{Perturbative QCD}

Neutrinos do not couple directly to gluons.  As a result, QCD effects
appear in neutrino scattering as higher order corrections to
the leading order parton model. Measurements
of the $\qsq$ dependence of neutrino cross sections are some of the most
sensitive measurements of the strong coupling constant $\alpha_s$\cite{Seligman} and
 information on the gluon distribution can be obtained from its coupling
to the structure functions via the DGLAP\cite{DGLAP} evolution equations.
The neutrino structure functions can be divided into two types, singlet
and non-singlet, depending on their sensitivity to gluon effects in
their evolution.  

The structure functions $2x F_1$, $F_2$ and $g_1$ are singlet functions
and are directly coupled to the gluon distribution via the evolution equations.
The structure functions $x F_3 + x F_3$, $2x g_3,g_4$ and $g_5$
averaged over neutrino and anti-neutrino are non-singlet
functions and their evolution is independent of the gluon distribution.
The combination  $F_2^p - F_2^n$  also cancels the
gluon contributions and is thus non-singlet in nature.

To date, extractions of $\alpha_s$ from non-singlet distributions have
been statistics limited and strongly affected by flux uncertainties.
The additional factor of 10-100 in statistics
and improved flux understanding
at a neutrino factory should allow vastly improved measurements of strong
interaction parameters in this very clean channel.

Once the quark distributions and strong interaction effects have been thoroughly studied in the non-singlet structure function, that knowledge can be used for
improved constraints on the gluon distributions via the evolution of the
singlet structure functions.

\subsection{Nuclear Effects}

Experiments at a neutrino factory of nuclear effects in the distribution of partons within nuclei relative
to protons and deuterons are
of interest to both the nuclear and high energy communities.
 These nuclear effects have been studied extensively using muon
and electron beams but have only been observed in low-statistics bubble
chamber experiments\cite{BEBCPACS} using neutrinos.  If we consider the behavior of the
structure functions $F_{2}(x,Q^{2})$ measured on a nucleus (A) to
$F_{2}(x,Q^{2})$ measured on a nucleon as a function of $x$ we pass
through four distinct regions in going from $x$  = 0 to $x$  =
1.0:

\subsubsection*{Shadowing Region $x < 0.1$}

   In the shadowing region ($x < 0.1$)  there are several
effects that should yield a different ratio $R_{A}\equiv  F_{2(A)}/F_{2(N)}$ 
when using neutrinos as
the probe.  In the limit $Q^{2} \gt 0$, the vector current is conserved and
goes to 0. The axial-vector part of the weak current is only
partially conserved (PCAC) and $F_{2}(x,Q^{2}) \rightarrow$ a non-zero constant as
$Q^{2}  \gt 0$.  According to the Adler theorem \cite{Adler} the cross
section of $\nu_{\mu} - N$ can be related to the cross section for $\pi - N$
at $Q^{2}$ = 0.   This relation can be studied in both proton and in heavy
nucleii.

\begin{figure} 
\epsfysize=2.5in
\epsfxsize=5.0in
\centerline{
\epsffile{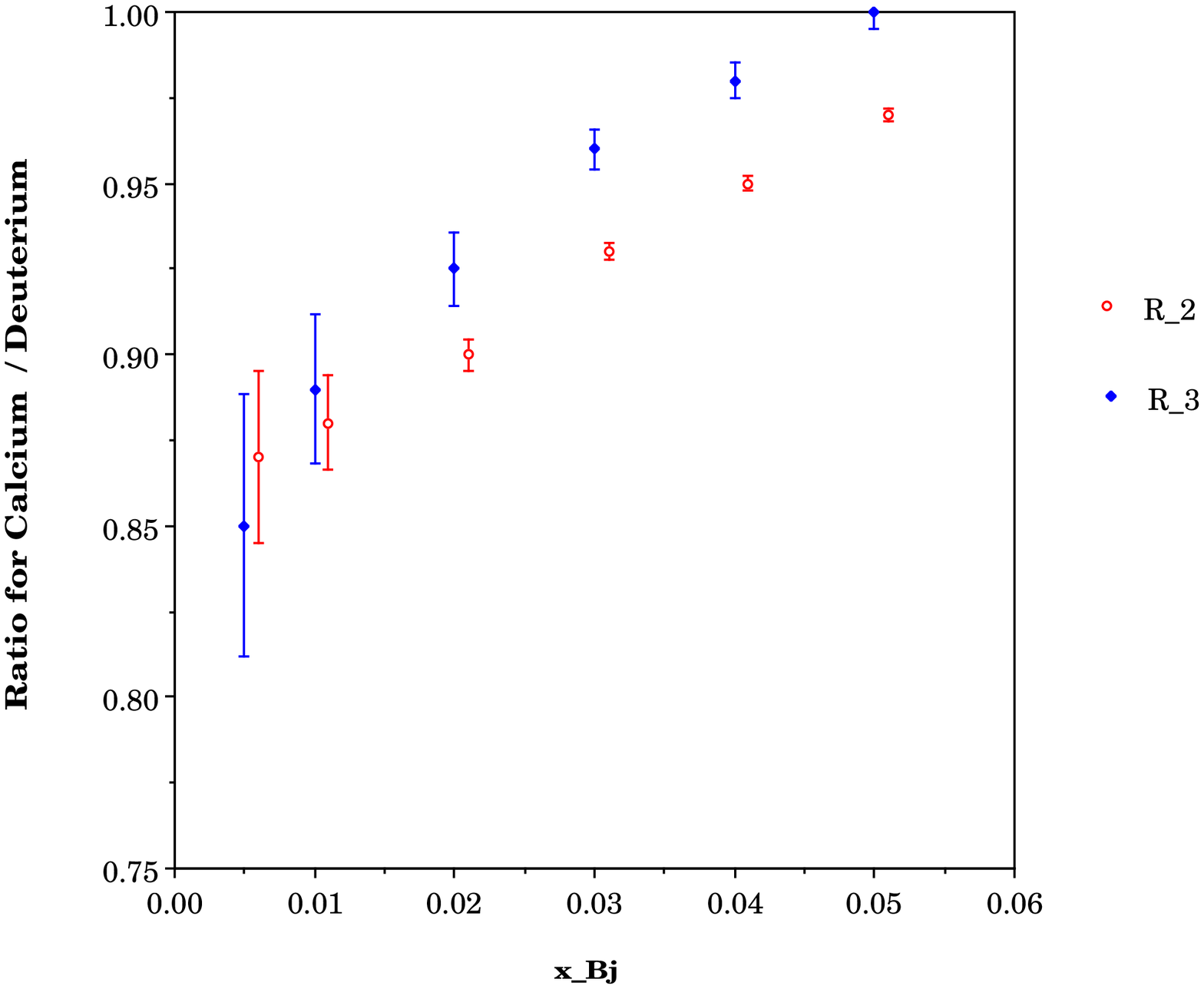}}
\caption{$R_{Ca:D_{2}}$ for both $F_{2}$ and $xF_{3}$ as measured with
14 M events on each target. }
\label{fig:shadow}
\end{figure}

    As we increase $Q^{2}$  from 0 but keep it under 10 GeV$^{2}$  in the
shadowing region we enter the region of vector meson dominance (VMD) in
$\mu/e-A$ scattering.  The physics concept of VMD is the dissociation of
the virtual boson into a quark/antiquark pair,  one of which interacts
strongly with the `surface' nucleons of the target nucleus (thus the
`surface' nucleons `shadow' interior nucleons). In $\nu - A$ scattering
there is an additional contribution from axial-vector mesons that is not 
present in $\mu/e - A$ scattering.  Boros et al. \cite{Boros} predict
that the resulting shadowing effects in $\nu - A$ scattering will be
roughly 1/2 that measured in $\mu/e - A$ scattering.
In a more quantitative analysis, Kulagin \cite{Kulagin} used a
non-perturbative parton model to predict shadowing effects in $\nu - A$
scattering.  At 5 GeV$^{2}$ he predicts equal or slightly more shadowing in
$\nu - A$ scattering than in $\mu$/e - A scattering.  He also attempts to
determine quark flavor dependence of shadowing effects by separately
predicting the shadowing observed in $F_{2}(x,Q^{2})$ (sum of all quarks)
and $xF_{3}(x,Q^{2})$ (valence quarks only).  Fig.\ \ref{fig:shadow} shows
the results of a run with 14M events/target using predictions of Kulagin's model for
$F_{2}$ and $xF_{3}$.   As can be seen,
the predicted difference between the shadowing on sea and valence quarks is
clearly visible down to $x \simeq 0.03$.

\subsubsection*{Anti-shadowing Region ($0.1 < x < 0.2$) }

Drell-Yan experiments have also measured nuclear effects.  Their results 
are quite similar to DIS experiments in the shadowing region.  However, in 
the anti-shadowing region, where $R_{A}$  makes a brief but statistically
significant excursion above 1.0 in DIS, Drell-Yan experiments see no 
effect.  This could be an indication of  difference in nuclear effects 
between valence and sea quarks.  Eskola et al \cite{Eskola} quantified this
difference by using a leading order/leading twist DGLAP model.
\noindent
Taking the work of Kulagin and Eskola together implies that nuclear effects
in $xF_{3}(x,Q^{2})$  should be quite dramatic with more shadowing than
$F_{2}(x,Q^{2})$ at lower $x$ and then $R_{A}$ rising fairly rapidly to yield
significant antishadowing around $x = 0.1$.  With 14 M events on each target
 we should be able to measure antishadowing effects and the
difference between shadowing effects in $F_{2}(x,Q^{2})$ and
$xF_{3}(x,Q^{2})$ to the 6 $\sigma$ statistical level.

\subsubsection*{EMC-effect Region($ 0.2 < x < 0.7$)}

Determination of  individual quark contributions to the EMC-effect will be
challenging since the participation of sea quarks, and thus the difference
between $F_{2}(x,Q^{2})$ and $xF_{3}(x,Q^{2})$, shrinks rapidly with
increasing $x$. However, Eskola's predictions for this region indicate
that the contribution of $\overline{u}$ and $\overline{d}$ to $R^{(2)}_{A}$  in the
$Q^{2}$ range of this experiment remains well below unity so that the quantity
$(R^{(2)}_{A}$ - $R^{(3)}_{A})$  should remain negative well into the EMC-effect
region.

\subsubsection*{Behavior of $F_{2}(x,Q^{2})$ as $x \rightarrow 1$ in a Nuclear 
Environment}

When working in the fermi-motion region it has been shown that we need to 
add more than the Fermi gas model to a simple nucleon to reproduce the 
behavior of $F_{2}(x,Q^{2})$ at high $x$.  Few-nucleon-correlation
models and multi-quark cluster models allow quarks to have a higher
momentum which translates into a high-$x$ tail.  In this region
$F_{2}(x,Q^{2})$ should behave as $e^{-ax}$.  There have been analyses
of this behavior in similar kinematic domains using $\mu  + $C and $\nu +$
Fe interactions.  The BCDMS \cite{BCDMShighx} muon experiment finds $a = 16.5~\pm 0.5$ while the CCFR\cite{Massoud}  neutrino 
experiment finds a = 8.3~$\pm$~0.7~$\pm$~0.7 (systematic).  Is the value of $a$
dependent on the nucleus?  One would expect any few nucleon correlation or
multi-quark effects to have already saturated by Carbon.  Is $a$ dependent
on the probe?

\subsubsection*{Summary}

There is a rich program of studying nuclear effects with a neutrino 
probe in a high statistics  neutrino factory experiment.  The effects could 
be measured to statistically significant accuracy in a 2 year exposure to
the beam in the near-detector experiment described above.  The data
gathered would allow separate measurements of the effects on valence quarks
and sea quarks across much of the $x$  range.

The nuclear community would surely be excited by this valuable tool for 
nuclear research at a neutrino factory.

\newcommand{\be}{\begin{equation}}
\newcommand{\ee}{\end{equation}}
\newcommand{\bea}{\begin{eqnarray}}
\newcommand{\eea}{\end{eqnarray}}
\newcommand{\ba}{\begin{array}}
\newcommand{\ea}{\end{array}}
\newcommand{\bmat}{\left(\ba}
\newcommand{\emat}{\ea\right)}
\def\3{\ss}
\def\p{p\llap{/}}
\def\d{\delta}
\def\ga{\gamma}
\def\Ga{\Gamma}
\def\s{s\llap{/}}
\def\k{k\llap{/}}
\def\g5{\gamma_5}
\def\mn{\mu\nu}
\def\rs{\rho\sigma}
\def\b{\beta}
\def\a{\alpha}
\def\ve{\varepsilon}
\def\r{\rho}
\def\si{\sigma}
\def\as2{\alpha^2_s}
\def\ha{{1\over 2}}
\def\pa{\partial}
\def\du{\delta u}
\def\GeV{{\rm GeV}}
\def\Pon{P^{(0)n}}
\def\hPon{\hat P^{(0)n}}
\def\Q2{(Q^2_0)}
\def\zweib{\frac{2}{\beta_0}}
\def\vph{\varphi}
\def\nspm{NS\pm}
\def\gen{\gamma^{(1)n}}
\def\aspi{\frac{\a_s}{2\pi}}
\def\Pen{P^{(1)n}}
\def\hPen{\hat P^{(1)n}}    
\def\tolimit_#1{\mathrel{\mathop{\longrightarrow}\limits_{#1}}}
\def\tosim_#1{\mathrel{\mathop{\thicksim}\limits_{#1}}}

\subsection{Spin Structure}

An intense neutrino beam at a neutrino factory would create significant
event rates in compact detectors.  This opens the possibility of 
using a polarized target, and hence a completely new class of neutrino
measurements becomes possible.  At present we know
very little about the  spin structure functions $g_1^\nu - g_5^\nu$
introduced in Equations \ref{pol_lon} and \ref{pol_tra}. In particular,
the parity violating functions have only been explored via weak-interference
measurements of proton form factors by the SAMPLE collaboration 
\cite{SAMPLE} with  low statistics. A neutrino factory
would allow direct   high-statistics measurements of all of these
structure functions and should be able to answer many
unresolved questions about the spin structure of the nucleon.

\ignore{
In the naive parton model, 
 
\begin{eqnarray}
g_1^{\nu p}(x,Q^2)&=&\delta \dquark(x,Q^2) +\delta \squark(x,Q^2) + \delta
\antiuquark(x,Q^2) + \delta \anticquark(x,Q^2),\\
g_1^{\bar\nu p}(x,Q^2)&=&\delta \uquark(x,Q^2) +\delta \cquark(x,Q^2) + \delta
\antidquark(x,Q^2) + \delta \antisquark(x,Q^2).
\label{pol_g1}
\end{eqnarray}

\nonindent Note that $g_2$ has a  a twist--2
($g_2^{WW}$) and a twist--3 ($\bar g_2$) contribution and  has no
simple parton model interpretation,
\begin{eqnarray}
g_2&=&g_2^{WW}+\bar g_2\\
 g_2^{WW}(x,Q^2)&=&-g_1(x,Q^2)+ \int_x^1 {dy
\over y}g_1(y,Q^2).
\label{pol_g2}
\end{eqnarray}

For $g_3$ and $g_4+g_5$ the parton model predictions are:

}

\subsubsection*{Formalism}

The nucleon spin ($\frac{1}{2}$)
can decomposed in terms of quark and gluon contributions:
\begin{equation}
\frac{1}{2}= \frac{1}{2}\Delta\Sigma + \Delta g + L_q + L_g,
\end{equation}
where $\Delta\Sigma \equiv \Delta u+ \Delta d+ \Delta s+\Delta c $
is the net quark helicity  and 
$\Delta g$  is the net gluon helicity along the nucleon
spin direction,  while  $L_i$ are their relative orbital angular
momentum.( We use $\Delta \quark$ as a shorthand for the integral $\int \Delta \quark(x) dx$.)

To date, the only experiments which have studied the 
spin structure of the nucleon are low energy charged
lepton polarized deep-inelastic scattering experiments (PDIS) where only
the parity conserving polarized structure functions $g_1^l$ and $g_2^l$ 
can be measured.

$g_1^\lepton$ can be written in the leading order parton model as a sum of a nonsinglet and singlet part\cite{bfr95b}:

\begin{eqnarray}
g_1^\lepton(x,Q^2) &=& g_{1,NS}^\lepton(x,\qsq) + g_{1,S}^\lepton(x,\qsq) \\
&=&\onehalf \sum (e_i^2 - <e^2>)\Delta \quark_i(x,Q^2) +  
\onehalf \sum  <e^2>\Delta \quark_i(x,Q^2)
\end{eqnarray}

The first non-singlet term evolves independently of the gluonic spin contribution while the second is coupled to, and thus depends on,  the gluon spin
contribution $\Delta g$.

The integral structure functions have the following relation
to the parton spin contributions: 
\begin{eqnarray} \nonumber
\Gamma_1^{\lepton}(Q^2)& =& \int dx g_1^{l} (x,Q^2)\\ 
&=& \Gamma^{\lepton }_{1,NS}(\qsq) + \Gamma^{\lepton }_{1,S}(Q^2)\\ 
\Gamma_1^{\lepton p}(Q^2)&=& \frac{C_1^{NS}(\qsq)}{6}\biggr [\onehalf a_3 + \frac{1}{6} a_8\biggl ] +
\frac{C_1^S}{9} a_0\\
\Gamma_1^{\lepton n}(Q^2)&=& \frac{C_1^{NS}(\qsq)}{6}\biggr [-\onehalf a_3 + \frac{1}{6} a_8\biggl ] +
\frac{C_1^S}{9} a_0\end{eqnarray}

\begin{figure} 
\centerline{\epsffile{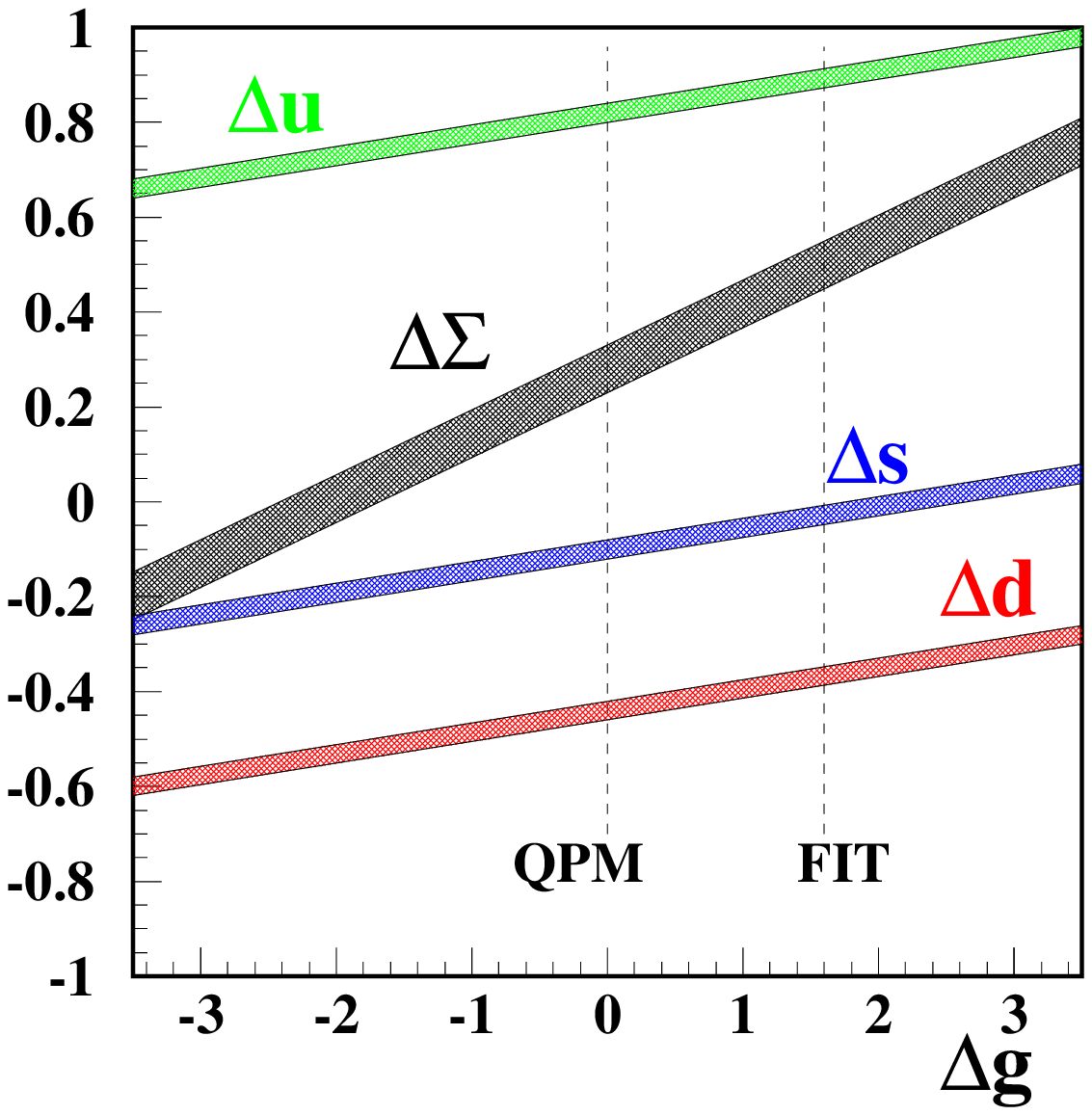}}
\caption{
Model dependent decomposition of singlet term into quarks and gluon based 
on PDIS data, 
$a_0 \rightarrow \Delta q - 3{\alpha_s \over 2 \pi} \Delta g$.
The QPM expectation and the results from a NLO fit of the 
$Q^2$  evolution of most of the available data on $g_1^l$
are also shown.  From the fit it was found that $\Delta g = 1.6 \pm 0.3\pm 1.0$, where 
the error is dominated by theoretical
uncertainties.
\label{gluon}}
\end{figure}

Where  the $C_1$ are coefficient functions and the axial charge matrix elements 

\begin{eqnarray}
a_3 &\equiv& F+D \simeq  \Delta \uquark - \Delta \dquark \\
a_8 &\equiv& 3F-D \simeq \Delta \uquark + \Delta \dquark - 2 \Delta \squark \\
a_0 &\equiv& \Delta \uquark + \Delta \dquark + \Delta \squark  = \Delta \Sigma \\
\end{eqnarray}

\noindent
can be expressed in terms of coupling constants $F$ and $D$ obtained from
neutron and hyperon beta decays \cite{EJ74}.
Because the interaction between $\Delta g $ and $\Delta \Sigma $ 
in the evolution of the singlet ($a_0$) component, interpretation of 
$\Gamma_1^\lepton$ in terms of the quark spin is problematic.
Fig. \ref{gluon} shows NLO QCD predictions for $\Delta \Sigma$ as a function of $\Delta g$.
 The data in the NLO fit are from \cite{SMCp93PRD}.

Neutrino beams introduce both additional parity violating spin 
structure functions $g_3, g_4 $ and $g_5$ and new combinations based
on sums and differences of neutrino and anti-neutrino scattering.

For example\cite{Bodo} the sums

\begin{eqnarray}\Gamma_1^{\nu p } + \Gamma_1^{\antinu p} &=&\int dx [g_1^{\nu p }(x,Q^2) + g_1^{\nubar p}(x,Q^2)] \\
\Gamma_1^{\nu n } + \Gamma_1^{\antinu n} &=&\int dx [g_1^{\nu n }(x,Q^2) + g_1^{\nubar n}(x,Q^2)]
\end{eqnarray} 

for both proton and neutron targets are only sensitive to the singlet $a_0$ term and no input from 
beta decay is necessary.

The parton model interpretation of these new structure functions is:
\begin{eqnarray}
g_{4+5}^{\nu p}(x,Q^2)&=&2xg_3^{\nu p}(x,Q^2)\\\nonumber&=&-x[\delta d(x,Q^2)+\delta
s(x,Q^2)-\delta\antiuquark(x,Q^2)-\delta \anticquark(x,Q^2)],\\
g_{4+5}^{\bar\nu p}(x,Q^2)&=&2xg_3^{\bar\nu p}(x,Q^2)\\\nonumber&=&-x[\delta
u(x,Q^2)+\delta c(x,Q^2)-\delta \antidquark(x,Q^2)-\delta \antisquark(x,Q^2)].
\label{pol_g3}
\end{eqnarray}

On a deuterium target, the $\uquark$ and $\dquark$ contributions
to $g_3$ can be cancelled leading to a direct measurement of
the strange sea contribution to the nucleon spin \cite{Bodo}

\begin{eqnarray}
g_3^{\nu (np)} - g_3^{\nubar (np)} = -2 (\delta \squark + \delta \antisquark)+2 (\delta \cquark + \delta \anticquark), \end{eqnarray}

\noindent
which can also be studied via polarization asymmetries in charm production
from strange quarks\cite{debbie}.

The structure functions $xg_3, g_4$ and  $g_5$, like $F_3$ are non-singlet functions
in which contribution from gluons cancel.  Comparison of the non-singlet
functions with the single functions $g_1$ and $F_2$ is an indirect way
of measuring the contribution of gluons $\Delta g$.

\subsubsection*{Experimental Setup}

A  promising target technology is the `ICE' target \cite{ICE},
a solid hydrogen-deuterium compound in which  the protons or the
deuterons can be polarized independently.
 The expected polarization and dilution are 
$P_H$=80\% and $f_H=1/3$ for H, and $P_D$=50\% and $f_D=2/3$ for deuteron.
A 7 kg ($\rho_t$=1.1\ gr/cm$^2$) polarized target with the qualities mentioned 
above would be 20~cm in radius and 50~cm thick, similar to
the other light targets proposed for structure
 function studies.  Raw event rates in 
such a detector would be around 20M per 10$^{20}$ muon decays.

If such a data sample is analyzed in 10 in $x$ bins, 
 the error 
in each $x$ bin would be
 $\delta {g_1}\simeq (fP_T\sqrt{N})^{-1} \sim 1\%$.


If the neutrino beam intensities and polarized target described above are 
feasible, the physics motivations would be very strong.  We 
will be able  to do high precision 
measurements where we can cleanly separate singlet
(gluon-dependent) from non-singlet (gluon-free) terms.  Furthermore,
due to the nature of the neutrino charged current interactions it will be 
possibility to perform  a measurement of the polarization of the
proton's quarks by flavor, with sea and valence contributions separated.

\newcommand{\stw}{\mbox{$\sin^2\theta_W$}}
\newcommand{\nub}{\overline{\nu}}
\newcommand{\qbar}{\overline{q}}
\newcommand{\nubmu}{\overline{\nu_{\mu}}}
\newcommand{\nube}{\overline{\nu_{e}}}
\newcommand{\muebar}{\numu\nube}
\newcommand{\mubare}{\nubmu\nue}
\newcommand{\ubar}{\antiuquark}        
\newcommand{\dbar}{\antidquark}
\newcommand{\alps}{\mbox{$\alpha_s$}}
\newcommand{\asop}{\mbox{$\frac{\alpha_s}{\pi}$}}
\newcommand{\qnsq}{\mbox{$Q_0^2$}}
\newcommand{\mztwo}{\mbox{$M_Z^2$}}
\newcommand{\mz}{\mbox{$M_Z$}}
\newcommand{\mw}{\mbox{$M_W$}}
\newcommand{\mtop}{\mbox{$M_{\rms top}$}}
\newcommand{\mhiggs}{\mbox{$M_{\rms Higgs}$}}
\newcommand{\lmsb}{\mbox{$\Lambda_{\overline{MS}}$}}
\newcommand{\avgth}{\left< \theta_\nu\right> }

\subsection{Charm Production and $\dzero- \dzerobar$ Mixing}

\begin{figure}[tpb]
\begin{center}
\epsfxsize=5 in 
\epsfbox{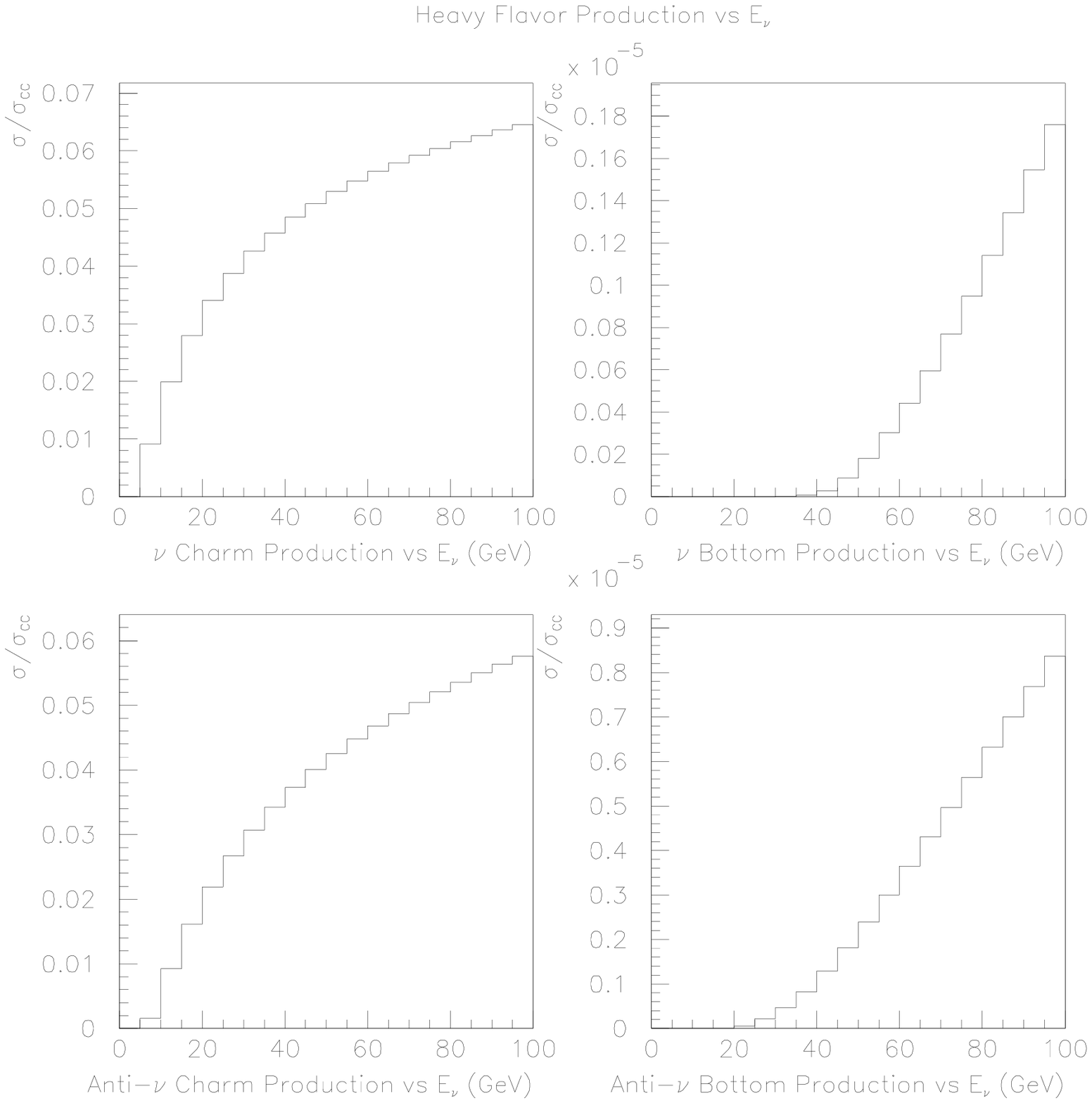}
\end{center}
\caption{Charm and bottom quark production as a fraction of the total
  cross-section as a function of $E_\nu$. }
\label{fig:charmrate}
\end{figure}

\begin{figure}[tpb]
\begin{center}
\epsfxsize=5 in
\epsfbox{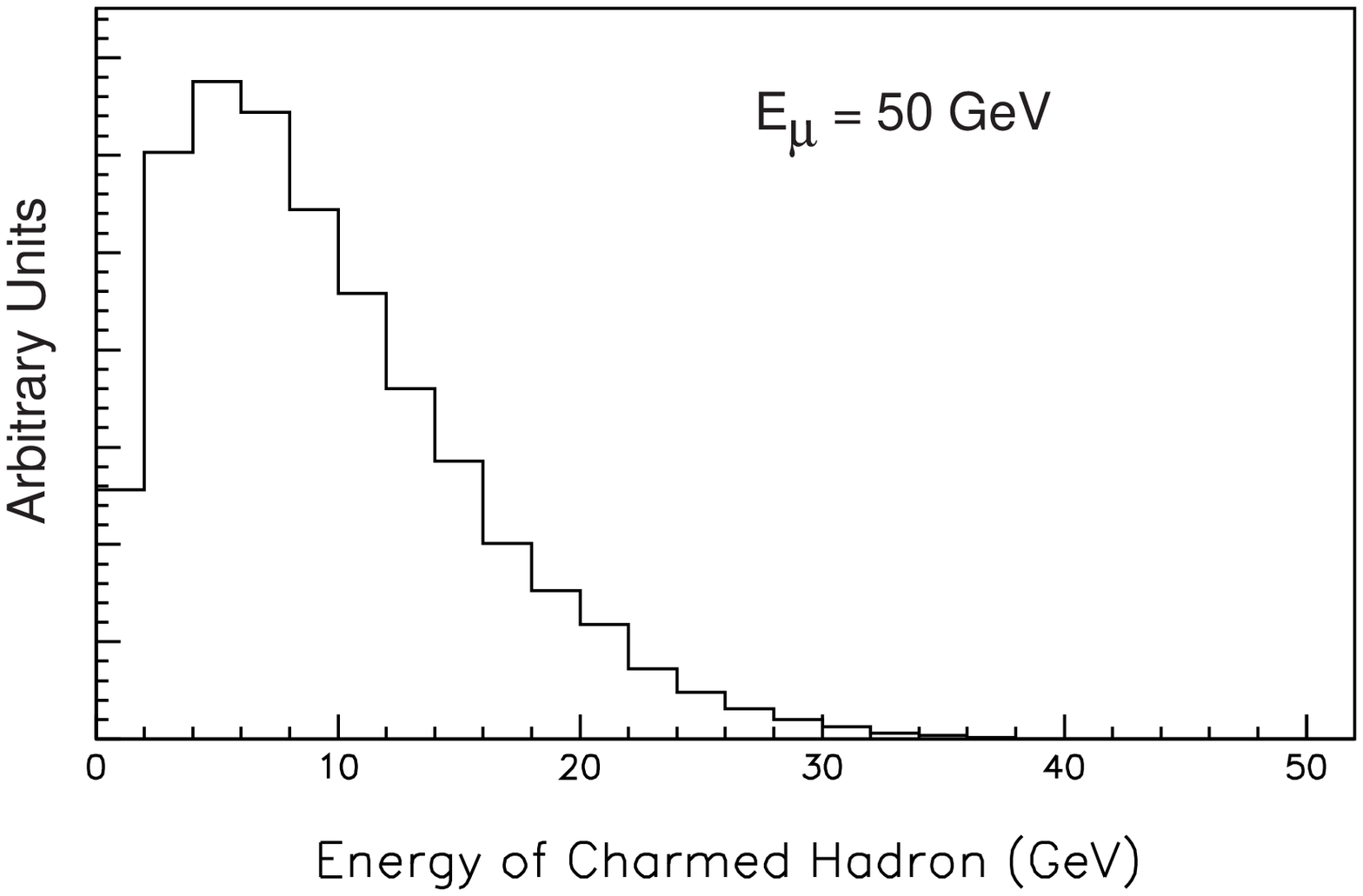}
\end{center}
\caption{Charmed hadron spectra from neutrino interactions in a near detector
  from a $50$~GeV muon storage ring.}
\label{fig:charmspect}
\end{figure}

Neutrino interactions are a very good source of clean, sign-tagged charm 
particles.  Single charm quarks are produced
via the processes

\begin{eqnarray}
\nu \squark &\gt& \lminus \cquark \hbox{\ \ \ Cabbibo favored} \\
\nu \dquark &\gt& \lminus \cquark \hbox{\ \ \ Cabbibo suppressed} \\
\nu \antisquark &\gt& \lplus \anticquark \hbox{\ \ \ Cabbibo favored} \\
\nu \antidquark &\gt& \lplus \anticquark \hbox{\ \ \ Cabbibo suppressed} 
\end{eqnarray}

 The fraction of heavy flavor produced as
a function of $E_\nu$ is shown in Fig.~\ref{fig:charmrate}. 
 An experiment
at a 50 GeV muon storage ring with 10$^{20}$ muon decays and a
 a one ton (fiducial) target made up of silicon strip detectors
interleaved with heavier material would observe $\approx 3\times10^{9}$ muon-neutrino charged-current
interactions and around $1.2\times 10^8$ charm hadrons with energies
above 4 GeV/year.
All of these charmed
hadrons are flavor tagged at the point of production by the charge of the
outgoing primary lepton ($c$ production with $\ell^-$ and $\overline{c}$
production with $\ell^+$).

There are several interesting physics motivations for charm studies at muon
storage rings, including measurements of the strange contribution
to proton structure and spin; however, the primary motivation for producing
charm by this method is the cleanliness of the final state relative to
hadroproduction and the flavor tagging in production.  This experimental fact
compliments the theoretically ``clean laboratory'' of charm in searches for
FCNC, CP asymmetries and ${\rm D^0 \rightarrow \overline{D^0}}$ oscillations,
all of which are  small in the standard model because of the lack of
coupling of charm to the heavy top quark.

Although this study has concentrated on neutrino energies below 50 GeV, we
note that similar arguments hold for bottom production and that for machines
with energies above 100 GeV, single B meson production rates can reach 100 per gr/cm$^2$ of target.   Because nuclei mainly consist of u quarks rather than
c quarks, the $\uquark \gt \bquark$ rate will be enhanced and a clean measurement of $V_{ub}$ without final state effects may be possible.

As an example of the physics reach of a neutrino charm factory, consider the
example of $D^0-\overline{D^0}$ mixing measurements.  The most sensitive
current searches for time-integrated mixing place limits on the process of
$\sim 5\times 10^{-3}$ \cite{E791,CLEO}.  BaBar expects to have sensitivity to
mixing at the $\sim 5\times10^{-4}$ level after several years at design
luminosity \cite{BaBar}.  These measurements are ultimately limited by
tagging mistakes and backgrounds to final state $D^0$ or $\overline{D^0}$
identification from doubly-Cabibbo suppressed decays, such as
$D^0\to K^+\pi^-$ which occur at the few part per thousand level.

At a $50$~GeV muon storage ring, with a high mass detector, 
one could probe $D^0-\overline{D^0}$ mixing
{\em via}
\begin{eqnarray*}
  \nu N \to & c \ell^- X\hspace*{8ex} \\
            & \hookrightarrow \ell^+ X\hspace*{5ex} \\
            & \hookrightarrow \overline{c} \to \ell^- X,
\end{eqnarray*}
and its charge conjugates.  The appearance of like-signed leptons would
indicate mixing, where opposite-signed leptons are expected.  Assuming $50\%$ of the charm produces hadronizes as a $D^0$ or
$\overline{D^0}$, this would result in the observation of $2\times 10^6$
tagged neutral charm meson semi-leptonic decays in either the muon or electron
channel.

\subsection{Precision Electroweak Measurements}
Precision measurements of electroweak parameters from neutrino experiments
have played an important role in testing the Standard Model and in 
searching for new physics. Even with the wealth of on-shell $W$ and 
$Z$ bosons today at colliders, the neutrino data remains important and 
is complimentary to collider studies. 
The intense flux of neutrinos from a neutrino factory opens up a 
new era of precision electroweak measurements  previously
limited by statistics.


Three interesting precision measurements can be contemplated at a neutrino
factory, and each can be cast as a precision measurement of the weak
mixing angle, $\siniiW$.  The experiment that most dramatically highlights new
capabilities at a neutrino factory is the study of
neutrino-electron cross sections.  The second possible experiment is the
extraction of $\siniiW$ from the neutral-current to charged-current deep
inelastic scattering (DIS) cross-sections, a measurement which currently is
the most precise test of weak interactions in neutrinos and which is
currently limited by statistics\cite{nutev-wma}, although difficult
theoretical systematics are not far behind.  A final possibility
is the study of weak boson scattering from photons in the so-called
neutrino ``trident'' process.

Only the first two possibilities are discussed here in detail. Trident 
processes are not considered here because of the difficult theoretical 
systematics that will ultimately limit the measurements\cite{numcbook}.


\subsubsection*{Neutrino-electron scattering}

Neutrino-electron elastic scattering,
\begin{equation}
\nu e^- \rightarrow \nu e^-,
\end{equation}
is perhaps the most promising reaction for the precise probe of electroweak
 un\-ification from neutrino interactions. Because the target particle is point-like, its
structure does not introduce uncertainties in extracting parameters of the
electroweak interaction from observed cross-sections.  These measurements
will therefore be limited only by statistical and experimental uncertainties.  The best previous
measurements of $\nu e$ scattering come from the CHARM II experiment which
observed approximately $5000$ events \cite{CHARMII}.

Two distinct measurements are possible at a muon storage ring
because neutrino-electron
scattering includes different diagrams for the beams from positive
and negative muons.  The processes in the $\muebar$ and $\mubare$ beams
are
\begin{eqnarray}
\bar{\nu}_{\mu} e^-\to\bar{\nu}_{\mu} e^-,~ & \nue e^-\to\nue e^-, \nonumber\\
\numu e^-\to\numu e^-,~ &
\bar{\nu}_e e^-\to\bar{\nu}_e e^-,
 \label{reac:nu-e}
\end{eqnarray}
where the electron-neutrino and electron-antineutrino scattering process
includes a charged-current ($W^\pm$ exchange) in the t-channel and s-channel,
respectively.  The differential cross-section with respect to $y=E_e/E_\nu$
is given by
\begin{equation}
\frac{d\sigma(\nu e^-\to\nu e^-)}{dy}=\frac{2G_F^2m_eE_\nu}{\pi}
                                    \left[ g_L^2+g_R^2(1-y)^2\right] ,
\label{eqn:nue-sigma}
\end{equation}
where terms of ${\cal O}(m_e/E_\nu)$ are neglected and expressions
for the left-handed and right-handed coupling constants,
$g_L$ and $g_R$, for each of the processes shown in 
Eq.~(\ref{reac:nu-e}) are given
in Table~\ref{tab:glgr}. The numerical values for the cross-sections
after integrating over y are:
\begin{equation}
\sigma(\nu e^-\to\nu e^-) = 1.72 \times 10^{-41} \times E_\nu[GeV]
                            \times \left[ g_L^2+\frac{1}{3}g_R^2\right] ,
\end{equation}
\label{eqn:nue-sigmaval}
where the values for the final term are also given in
Table~\ref{tab:glgr}.
Note that neutrino-electron scattering has a much lower cross-section
than DIS, roughly down by the ratio of $m_e$ to $m_p$.
Experimentally, it should be noted that the observed neutrino-electron
scattering rate will be summed over both beams since the observed final
states are identical.

\begin{table}
\begin{center}
\caption{$g_L$ and $g_R$ by $\nu-e$ scattering process}
\label{tab:glgr}
\bigskip
\begin{tabular}{|c|c|c|c|}
\hline
Reaction & $g_L$ & $g_R$ & $g_L^2+\frac{1}{3}g_R^2$ \\ \hline
$\numu e^-\to\numu e^-$ & $-\frac{1}{2}+\siniiW$ & $\siniiW$ & 0.0925 \\ 
$\nue e^-\to\nue e^-$ & $\frac{1}{2}+\siniiW$ & $\siniiW$ & 0.5425 \\ 
$\bar{\nu}_{\mu} e^-\to\bar{\nu}_{\mu} e^-$ & $\siniiW$ & 
$-\frac{1}{2}+\siniiW$ & 0.0758 \\ 
$\bar{\nu}_e e^-\to\bar{\nu}_e e^-$ & $\siniiW$ & $\frac{1}{2}+\siniiW$ & 0.2258 \\ 
\hline
\end{tabular}
\end{center}
\end{table}

The experimental signature for $\nu-e$ scattering is a single negatively
charged electron with a small transverse momentum relative to the incoming
neutrino,
\begin{equation}
p_t^{(e-\nu)}<\sqrt{2m_eE_\nu}.
\end{equation}
The normalization mode is the appearance of a single muon with similarly low
$p_t^{\mu}$.  Of course, the neutrino beam itself has a characteristic divergence
from decay kinematics of $\frac{\pi}{4\gamma_\mu}$, and therefore the observed
lepton $p_t^{(e)}$ relative to the mean beam direction is given by
\begin{equation}
\left< p_t^{(e)}\right> 
\approx\sqrt{\frac{\pi^2}{16\gamma_\mu^2}+\frac{m_e E_\nu}{2}}.
\end{equation}
For a $50$~GeV storage ring, this factor is dominated by the fundamental
$p_t$ of the interaction and is typically $\sim90$~MeV.  For a lower
energy storage ring of about $15$~GeV, these factors become equal.

\begin{figure}
\begin{center}
\mbox{\epsfxsize=5.0in\epsffile{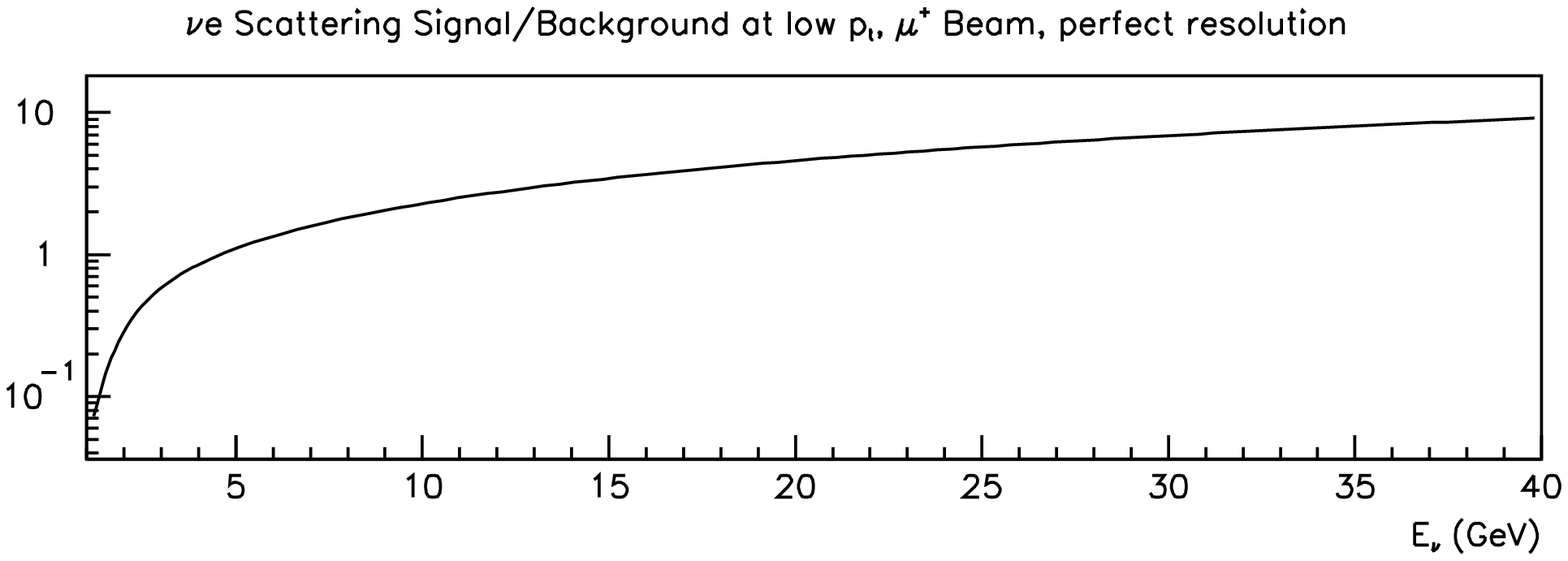}}
\caption{Signal to noise in the low $p_t^{(e)}$ region
  ($p_t^2<\frac{\pi^2}{16\gamma_\mu^2}+\frac{m_eE_\nu}{2}$) as
   a function of $E_\nu$.}\label{fig:nue-sn}
\end{center}
\end{figure}

The primary background
to this measurement is from quasi-elastic $\nue-N$ or 
$\bar{\nu}_e-N$ scattering events which occur at $p_t^{(e)}$ up to $\sqrt{m_N E_\nu}$. 
  Fig.~\ref{fig:nue-sn} shows the estimated signal to background
ratios expected in the low $p_t$ region.

Because of the exceptionally low cross section, the target must be
very massive.
The detector  must therefore be capable of resolving the
 $p_t$ with much better resolution than the background spread.
This favors the use of a fully active, high resolution tracking detector with
sub-radiation length sampling in order to resolve the $p_t$ of the single
electron before it is significantly broadened by shower development.  A
liquid Argon TPC, such as the one proposed for the ICANOE
experiment\cite{ICANOE} might be ideal for such a measurement.  Another
possibility would be a scintillating fiber/tungsten calorimeter.


\begin{figure}
\begin{center}
\mbox{\epsfxsize=5.0in\epsffile{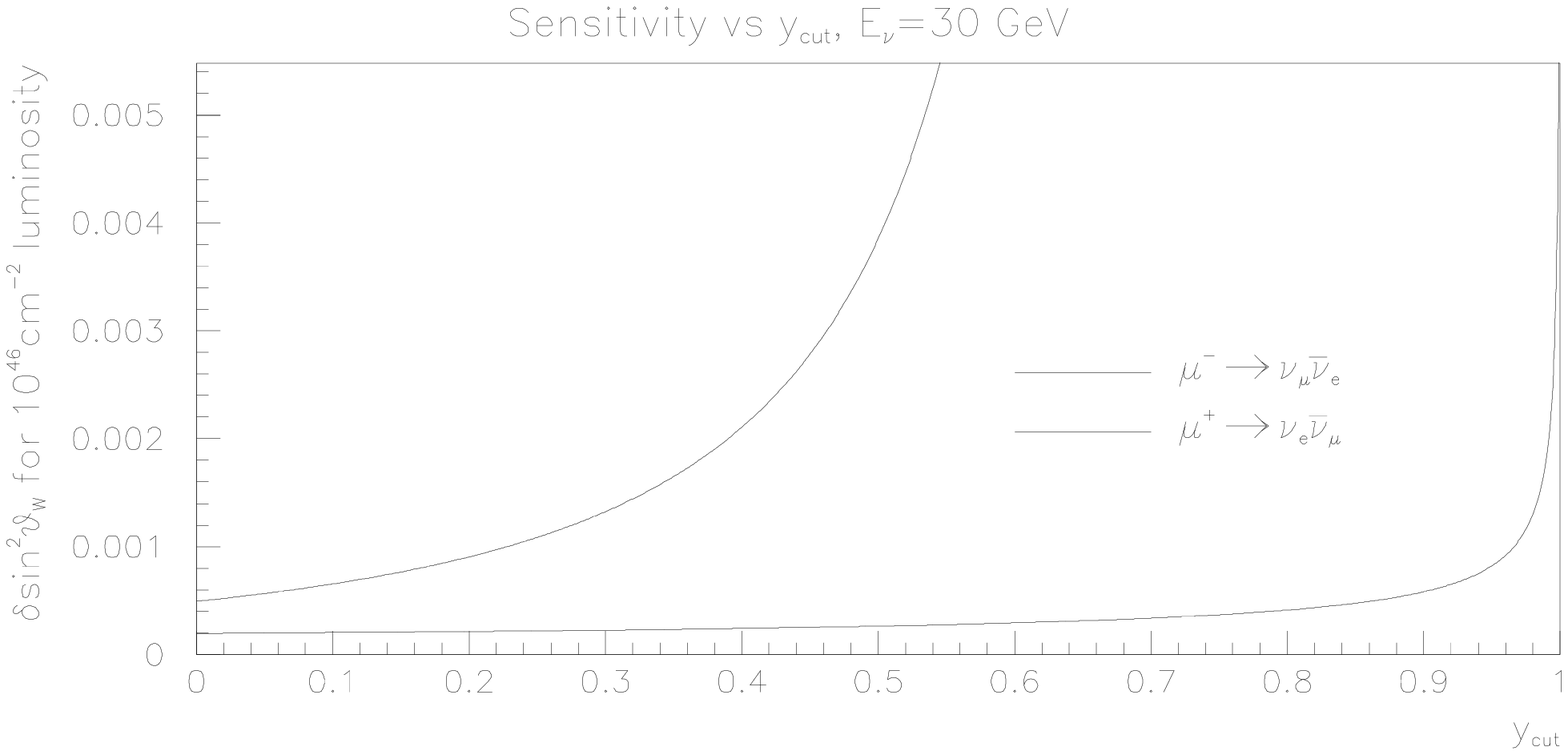}}
\caption{Statistical uncertainty in $\siniiW$ for a luminosity of
  $10^{46}$~cm$^{-2}$ as a function of $y_{\rm cut}$ for a $30$~GeV
  neutrino beam.  Note that the $\mu^-$ produced beam is much less sensitive
  to $\siniiW$ due to nearly exact cancellation in the $\siniiW$ dependence of
  the two neutrino species in the beam.}\label{fig:nue-sens}
\end{center}
\end{figure}

The largest experimental challenge
for such a  measurement is likely to be the normalization of the absolute
neutrino flux. Despite the precise knowledge of muon decays, it
would be extremely difficult to predict the precise neutrino flux
at the $10^{-4}$ level merely from monitoring the parent muon beam.
Instead, the signal processes will probably have to be normalized
to the theoretically predictable processes of inverse muon decay,
$\numu e^-\to\nue \mu^-$, and muon production through annihilation,
$\antinu_e e^-\to \antinu_{\mu}\mu^-$, both of which occur only in the 
$\muebar$ beam.  Normalization of the $\nue$ beam may be possible
through comparison to neutrino-nucleon
scattering, $\nu N \to  \l^\pm N'$, in the $\mubare$ and $\muebar$
beam.

For a 20 ton detector at a 50 GeV muon storage ring, 
with $2\times 10^{20}$ $\muminus$ decays/year there will be approximately 
$1.5\times 10^{10}$ DIS  charged current events and 8.5M $\numu/\antinue$-electron
scatters per year.  This leads to an estimated sensitivity of $\delta\siniiW^{\rm (stat)}\sim0.0002$


\subsubsection*{Neutrino-nucleon scattering}

 There were two dominant systematic uncertainties in present-day meaurements
of the weak mixing angle in neutrino nucleon scattering, 
 $\nu_{e}$ contamination in the $\nu_{\mu} $ beam  and the kinematic
suppression of scattering from strange quarks in the charged current channel.

For an isoscalar target, the neutral current rates can be related to the charged current rates via \cite{neutralcurrent}:
\begin{eqnarray}\label{NCCC}
\nonumber R_{\nu} - \Delta R_s &=& (\onehalf - \siniiW  +{5\over 9} \sin^4\theta_W)[ 1 - \Delta R_c] + \\
&&\hskip 2 in{5\over 9} \sin^4\theta_W [r - r\Delta \overline{R}_c]\\
\nonumber R_{\nubar} - \Delta \overline{R}_s&=& (\onehalf - \siniiW  +{5\over 9} \sin^4\theta_W)[ 1 - \Delta\overline{R}_c]+\\
& & \hskip 2 in{5\over 9}\sin^4\theta_W[ {1\over r}- {1\over r} \Delta R_c]
\end{eqnarray}

\noindent
where $R_{\nu/\nubar}$ is the ratio of neutral to charged current
cross sections,  $r \sim 0.5$ is the ratio of charged current anti-neutrino to neutrino cross sections, and  $\Delta R_s$ and $\Delta R_c$ are small corrections for
the kinematic suppression  of $\squark \gt \cquark$  in charged current scattering where the neutral current process $\squark\gt \squark$ has no suppression.
The charm corrections can  be eliminated by a judicious subtraction
of neutrino and anti-neutrino rates \cite{th:paschos} but with a consequent
reduction in statistical power.

Present-day experiments \cite{CDHSWew, CHARMII, NuTeV:prelim}  have had integrated fluxes of 10$^{15}$-10$^{16}$ neutrinos
and have relied on dense nuclear targets. In such targets neutral current events are distinguished from charged current events by 
the presence or absence of a muon in the final state. In a dense
calorimeter,   electron neutrino
charged current
induced events  look  similar to  neutral current events 
as the electron is lost in the hadronic shower.   They are a significant background for precision measurements with conventional beams
produced by pion and kaon decay and would be even more significant at a neutrino
factory.

The most precise measurement to date is from the NuTeV collaboration \cite{NuTeV:prelim} of
\begin{equation}\label{eq:nutev-stw}
\siniiW=0.2253\pm0.0019{ (stat)}\pm0.0010{ (syst)}.
\end{equation}

At a neutrino factory, the neutrino flux will be several orders of
magnitude higher but the beam will consist of approximately equal numbers
of $\numu$ and $\antinue$. This makes a detector capable of
distinguishing electron charged current events from neutral current
events desirable and implies a low density detector such as those
considered for the deep-inelastic scattering studies.

We have considered several possible observables for a neutrino factory 
measurement and propose:

\begin{eqnarray}\label{eq:Rmuebar}
R_e^{\muminus}=\frac{\sigma(\nu_{\mu},NC)+\sigma(\nubar_{e},NC)}
{\sigma(\nu_{\mu},CC)-\sigma(\nubar_{e},CC)}
&=&{R^{\nu}+grR^{\nubar}\over {1-gr}}
\end{eqnarray}
or 
\begin{eqnarray}\label{eq:Rmuebarhat}
\hat{R}^{\muminus}=\frac{\sigma(\nu_{\mu},NC)+\sigma(\nubar_{e},NC)+\sigma(\nubar_{e},CC)}
{\sigma(\nu_{\mu},CC)}
&=&{R^{\nu}+grR^{\nubar} + gr}
\end{eqnarray}
for the $\muebar$ beam, and

\begin{eqnarray}\label{eq:Rmubare}
R_e^{\muplus}=\frac{\sigma(\nubar_{\mu},NC)
+\sigma(\nu_{e},NC)}
{\sigma(\nu_{e},CC)-\sigma(\nubar_{\mu},CC)}
={R^{\nu}+g^{-1}rR^{\nubar}\over{1-g^{-1}r}}
\end{eqnarray}

or

\begin{eqnarray}\label{eq:Rmubarehat}
\nonumber \hat{R}^{\muplus}&=&{\sigma(\numubar,NC)
+\sigma(\nue,NC)+\sigma(\nue),CC)
\over\sigma(\numubar,CC)}\\
&=&{{g\over r}R^{\nu}+R^{\nubar} + {g\over r}}
\end{eqnarray}

\noindent
        for the $\mubare$ beam, where 
 $R_{\nu/\nubar}$ is the ratio of neutral to charged current cross sections
 from Eq.~(\ref{NCCC}). The observable $R_e^{\mu}$ requires electron identification while
$\hat{R}^{\mu}$ requires only muon identification.

The variable $g$ is the energy-weighted flux ratio between $\nu_{\mu}$ 
and $\overline{\nu}_{e}$ or, equivalently, between $\overline{\nu}_{\mu}$ 
and  $\nu_{e}$:

The flux ratio for neutrinos and anti-neutrinos $g$ is:

\begin{eqnarray}\label{eq:little-g}
g\equiv\frac{\int \Phi(E_{\overline{\nu}_{e}})E_{\overline{\nu}_{e}}dE_{\overline{\nu}_{e}}} 
{\int\Phi(E_{{\nu}_{\mu}})E_{{\nu}_{\mu}}dE_{{\nu}_{\mu}}}
&=&\frac{\int\Phi(E_{{\nu}_{e}})E_{{\nu}_{e}}dE_{{\nu}_{e}}}
{\int \Phi(E_{\overline{\nu}_{\mu}})E_{\overline{\nu}_{\mu}}
dE_{\overline{\nu}_{\mu}}} \simeq {6\over7 }. 
\end{eqnarray}

\noindent and is well determined by the muon decay kinematics. However, the relative detection efficiencies for muons
and electrons must be known at the $2\times 10^{-4}$ level in order to
determine $\siniiW$ to 10$^{-3}$ by the first method.
In addition, the charm contributions are not cancelled in this observable
and must be measured directly in the same experiment.


For the $R_e$ measurement, which requires electron identification,
an active target of 20~cm radius, 10 gr/cm$^2$ thick consisting of either CCD's or silicon strip detectors ($\sim$ 140 300-$\mu$m detectors)
spaced over a meter and 
followed by the
 tracking, electromagnetic and hadron calorimetry and muon identification proposed above
for structure function measurements
would yield 15M muon and 8M electron charged current 
 deep-inelastic scattering events/10$^{20}$ $\muminus$ decays and would yield a statistical precision
of 0.0004 in $\siniiW$.  The charm corrections partially cancel in this
observable and would also be measured directly
 via the 2M charm events/year  produced in such a detector.

The $\hat{R}$ measurement, which relies only on muon identification 
can be done with a much denser target, perhaps an iron/silicon sandwich
calorimeter.  Such a calorimeter 200 gr/cm$^2$ thick would have a
statistical sensitivity of $\Delta \siniiW \sim 0.0001$ per year at a 50 GeV machine. This method is quite similar to the method used in the NuTeV \cite{NuTeV:prelim}
measurement and would be dominated by systematic errors.


\def\pl#1#2#3   {{ Phys. Lett.} {\bf#1}, #2 (#3). }
\def\prev#1#2#3 {{ Phys. Rev. } {\bf#1}, #2 (#3). }

\subsection{Heavy Lepton Mixing}

A muon storage ring offers ample opportunities to search for new 
phenomena in yet unexplored physical regions.  One such opportunity 
is the ability to search for the
existence of neutral heavy leptons. Several models describe heavy isospin
singlets 
that interact and decay by
mixing with their lighter neutrino counterparts \cite{GLR,ShrockMM}. The
high intensity neutrino beam created by the muon storage ring provides an
ideal setting to search for neutral heavy leptons with a mass below
the muon mass, 105.6 MeV$/c^2$.

It is postulated that neutral heavy leptons ($L_0$) could be produced
from muon decay when one of the neutrinos mixes with its heavy,
isospin singlet partner. Neutral heavy leptons can be produced via one of
two channels:

\begin{equation}
\mu^- \rightarrow L_0 + \overline{\nu}_{e} + e^-
\end{equation}

\begin{equation}
\mu^- \rightarrow \nu_{\mu} + L_0 + e^-
\end{equation}

The branching ratio for each of these reactions is given by:

\begin{equation}
BR(\mu\rightarrow L_0 \mu e) = |U_i|^2 (1 - 8x_m^2 + 8x_m^6 - x_m^8 +
12x_m^4\ln{x_m^2})
\end{equation}

\noindent Here $x_m \equiv m_{L_0}/m_{\mu}$ and $|U_i|^2$ is
the mixing constant between the specific type of neutrino
and the neutral heavy lepton: $U_i \equiv \langle L_0 | \nu_i \rangle$.  
Note that $|U_{\mu}|^2$ and $|U_{e}|^2$ need not be identical.

Once produced, a neutral heavy lepton of such low mass can either
decay via $L_0 \rightarrow \nu \nu \nu$, $L_0 \rightarrow \nu e e$, or
$L_0 \rightarrow \gamma \nu$.  The most viable mode for detection is
the two-electron channel.  For this particular decay mode, the
$L_0$ can decay either via charged current or charged and neutral
current interactions. The branching ratio for this decay process has
been previously calculated \cite{Bolton}. Since the decay
width is proportional to $U_j^2$, the number of $L_0$'s detectable is
proportional to $U_i^2\cdot U_j^2$ in the limit where the
distance from the source to the detector is short compared to the
lifetime of the $L_0$.

Using the above model, one can estimate the number of neutral heavy
leptons produced at the muon storage ring which later decay within a
given detector:

\begin{equation}
N_{L_0} = N_{\nu}*
BR(\mu\rightarrow L_0 \nu e)*\epsilon
*e^{-L / \gamma c \tau}*
BR(L_0 \rightarrow detectable)*
(1 - e^{- \delta l / \gamma c \tau})
\end{equation}

\noindent Here $N_{\nu}$ is the number of neutrinos produced from muon
decay, $BR(\mu\rightarrow L_0 \nu e)$ is
the branching ratio of muons decaying into neutral heavy leptons
versus ordinary muon decay, $L$ is the distance from the beamline to
the detector, $\delta l$ is the length of the detector, $\epsilon$ is
the combined detector and geometric efficiency, $\tau$ is the $L_0$
lifetime, and $BR(L_0 \rightarrow \hbox{detectable})$
is the branching ratio for the neutral heavy lepton decaying via a
detectable channel (presumably $L_0 \rightarrow \nu e e$).

In estimating the sensitivity to $L_0$ production at the muon storage
ring, we make a few underlying assumptions.  We assume that the
storage ring utilizes a pure, unpolarized muon beam with straight
sections such that 25 percent of the muons will decay to neutrinos
pointing towards the detector. We assume that the fiducial volume is 3 meters
in diameter and 30 meters in length, (which is probably compatible
with the need for empty space before a conventional detector) 
and that the detector has sufficient 
tracking resolution to detect
the $e^+e^-$ vertex from the $L_0$ decay.  We assume for now that the
background is negligible.  These parameters correspond to the 
fiducial volume of the decay channel used for the $L_0$ search at 
E815 (NuTeV) \cite{NuTeVNHL,NuTeVQ0}.

\begin{figure}
\begin{center}
\mbox{\epsfxsize=6.0in\epsffile{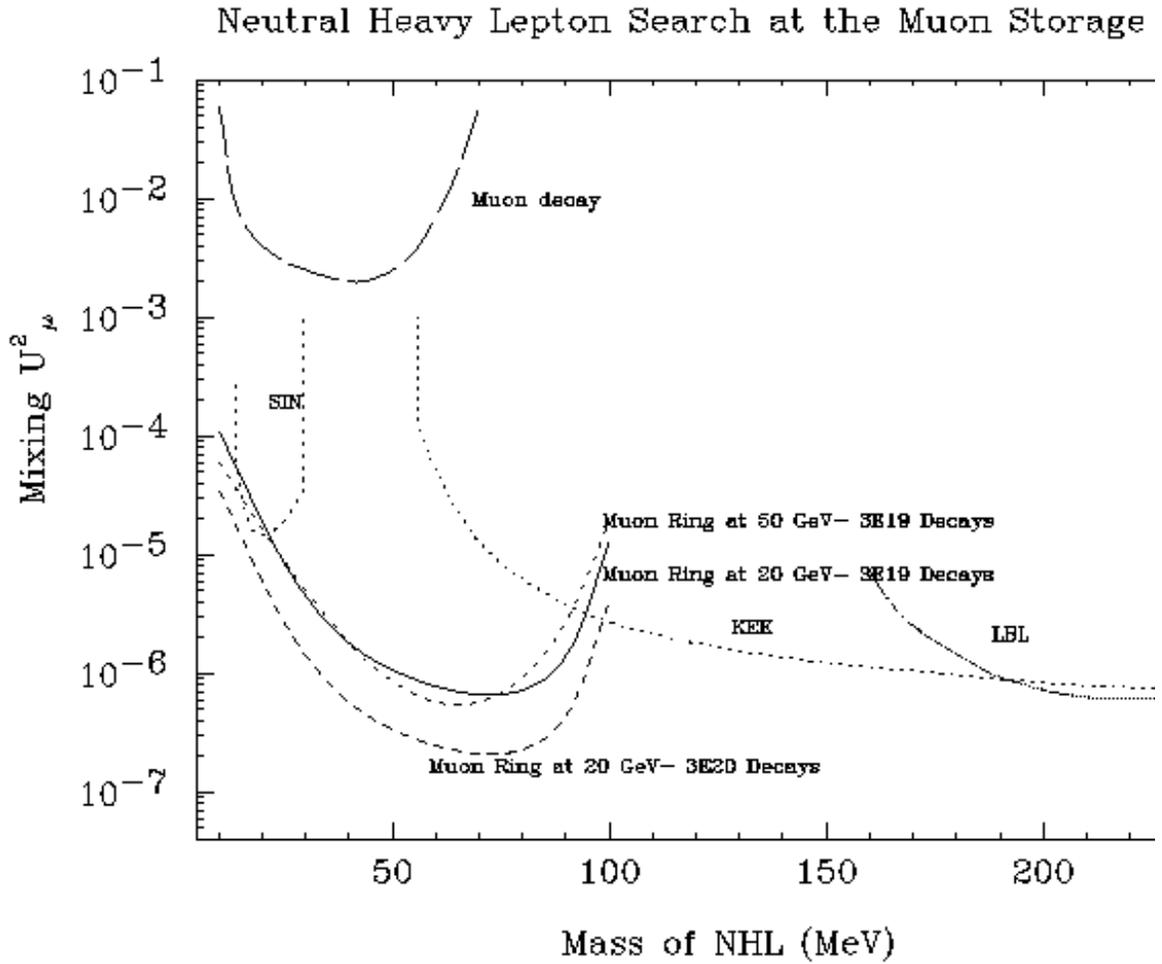}}
\caption{Limits on $|U_{\mu L}|^2$ as a function of $L_0$ mass for one
year of running.  The curves show sensitivities for 20 GeV and 50 GeV
muon energies.  Sensitivities assume no background events in signal
region.}
\label{nhl_mustore}
\end{center}
\end{figure}

The sensitivity of the detector has been calculated for a number of
different muon energies and beam intensities. Fig. \ref{nhl_mustore} shows
limits on the $L_0$-$\nu_\mu$ mixing as a function of $L_0$ mass.  One
achieves the best limits from using relatively low energy/high
intensity muon beams. This is a major improvement over previous
neutral heavy lepton searches, where limits do not reach below
$6.0\times 10^{-6}$ in the low mass region \cite{PDB,ShrockMM}. 

The single event sensitivity quoted here depends  on having
minimal background levels in the signal region.  Part of this can be
achieved by kinematic cuts which discriminate against neutrino interactions
in the detector material.  However, it will probably be necessary to 
reduce the amount of material in the fiducial region  compared
to NuTeV. We estimate that even if the
decay region is composed only of helium gas, the number of
neutrino interactions will approach a few thousand.  The ideal
detector, therefore, would consist of a long vacuum or
quasi-vacuum pipe with appropriate segmentation for tracking.  The
decay pipe could be used in conjunction with larger neutrino detectors
adapted for the muon storage ring.

The muon storage ring would to be an ideal location to continue
the search for neutral heavy leptons.  The high intensity neutrino
beam allows for a neutral heavy lepton search to be sensitive to
the 10 -- 100~MeV/$c^2$ mass range. In addition, such a neutral
heavy lepton program is very compatible with a
neutrino detector which uses the same neutrino beam.  It is also clear,
however, that a neutral heavy lepton search would receive the most
benefit at lower muon energies, and thus would yield best results at
the earlier stages of the muon storage ring program.

\subsection{Neutrino Magnetic Moments}

Although neutrino oscillation searches focus on the mass differences between
neutrino eigenstates, neutrinos can possess other observable properties
such as a magnetic moment. A measurement of the neutrino magnetic
moment (NMM) would not only have great impact in the field of cosmology,
particularly the development of stellar models, but would also help
constrain several Standard Model extensions. An important experimental 
advantage is that a NMM search can run parasitically as the front-end of a
typical long baseline detector.

Despite their lack of charge,  neutrinos can possess a non-zero magnetic
moment that can arise through loop diagrams. In the Standard Model, extended
to include a right--handed neutrino or with left--handed neutrinos which have mass, the expected  magnitude of the \cite{ShrockMM} neutrino magnetic moment
 is given by
\begin{equation}
\label{krane}
\mu _{\nu }\simeq 3\times 10^{-19}\;\mu _{B}\cdot ( \frac{m_{\nu }}{%
1\;\hbox{eV}}) ,
\end{equation}
where $\mu _{B}$ is the Bohr magneton. Although quite minuscule, several
extensions to the Standard Model could dramatically increase $\mu_\nu$ : supersymmetric
models can produce $10^{-14}\mu_{B}$ to $10^{-12}\mu _{B}$
\cite{frank} and calculations that invoke large extra dimensions easily
yield $10^{-11}\mu _{B}$ or larger \cite{ng}.

Relative to the Standard Model expectation, the excluded values of NMM\ are
not at all stringent, being seven to nine orders of magnitude larger. The
current limits on neutrino magnetic moment from laboratory experiments are
$\mu _{\nu }\leq 1.5$ to $1.8\times 10^{-10}\mu _{B}$ for electron
neutrinos \cite{beacom}\cite{mu_e_limit} and $\mu _{\nu }\leq 7.4\times
10^{-10}\mu _{B}$ for muon neutrinos \cite{mu_mu_limit}. Astrophysical
limits are stronger:\ the slow rate of plasmon decay in horizontal branch
stars \cite {star1} implies $\mu _{\nu }\le  10^{-11}\mu _{B}$,
while the neutrino energy loss rate from supernova 1987a \cite{star2}
yields $\mu _{\nu }\le $ $10^{-12}\mu _{B}$. Note, however, that
several assumptions are implicit in the astrophysics limits, including the
core temperature of the stars; if stellar models omit important processes,
these limits might be overestimates. The supernova limit applies only
to Dirac neutrinos and not to the Majorana case.

Existing search schemes possess a weakness that sharply limits their
ultimate sensitivity: the formulae for the hypothesized effect are quadratic
in $\mu _{\nu }$ but linear in terms of the experimenter--controlled
parameters. In contrast, the following scheme is quadratic in terms of the
product of the NMM$\;$and a magnetic field strength, $\mu _{\nu }\cdot $B;
hence a carefully designed and executed experiment could improve the limits
from current experiments and possibly the limits from astrophysics
calculations, or actually detect a NMM.

The energy $E$ of a neutrino with a magnetic moment in a magnetic field B
gains a new term $\mu _{\nu }\cdot $B. Consider a B field along the
$x$-axis, and a neutrino with momentum and helicity along the $z$-axis at
$t=0$. The eigenstates of the neutrino are projections along the $x$-axis,
and the state of the neutrino is expressed as:
\begin{equation}
\left| \uparrow \right\rangle =\frac{e^{-i\left( E+\mu _{\nu }B\right) t}}{%
\sqrt{2}}\left| \leftarrow \right\rangle +\frac{e^{-i\left( E-\mu _{\nu
}B\right) t}}{\sqrt{2}}\left| \rightarrow \right\rangle .  \label{eq_split}
\end{equation}
As the neutrino propagates, the relative phase of the two components
changes, corresponding to a rotation to a sterile state in the case of a
Dirac neutrino or to an antineutrino in the Majorana case. At a far
detector, the signal would be a deficit in the number of neutrinos detected 
or increase in the number of antineutrinos detected with the B field in
place compared to the sample detected with no B field turned on.

In this phase rotation scheme, the energy splitting occurs as the neutrino
passes through a field gradient and experiences a force $F=\nabla (\mu _{\nu
}\cdot $B$)$.  To preserve this energy difference, which drives the phase
difference in the absence of the B-field, the field must be turned off
instead of allowing the neutrino to experience the reverse gradient as it
exits the field region.
(The principle of changing the energy of neutral dipolar molecules
with time--varying electric fields has been demonstrated in the laboratory
\cite{stark_decel}.)

 To be successful, there are  two basic requirements for the
magnetic field:
\begin{itemize}
\item[1) ]  The magnetic field must oscillate such that the neutrino
experiences only one sign of the gradient. This study assumes that the
neutrino bunch length is small compared to the oscillation length. If this
assumption is not true, the effects discussed here will be diluted but the
basic conclusions will still apply.


\item[2) ]  The magnetic field must be as strong as possible.
\end{itemize}

We have explored the possibility \cite{norbert} of using two existing
technologies for the B field: resonant cavities and kicker magnets. In both
cases the maximum magnetic field is too small to yield improved magnetic
moment limits given realistic equipment.  We are currently exploring configurations
involving a series of pulsed current sheets.

The formula for the number of events lost to sterile states may be expressed
very simply as: 
\begin{equation}
N_{lost}=N*\sin ^{2}\left( \mu _{\nu }\hbox{B}t\right)   \label{eq_nlost}
\end{equation}
\begin{figure}
\begin{center}
\mbox{\epsfxsize=4.5in\epsfbox{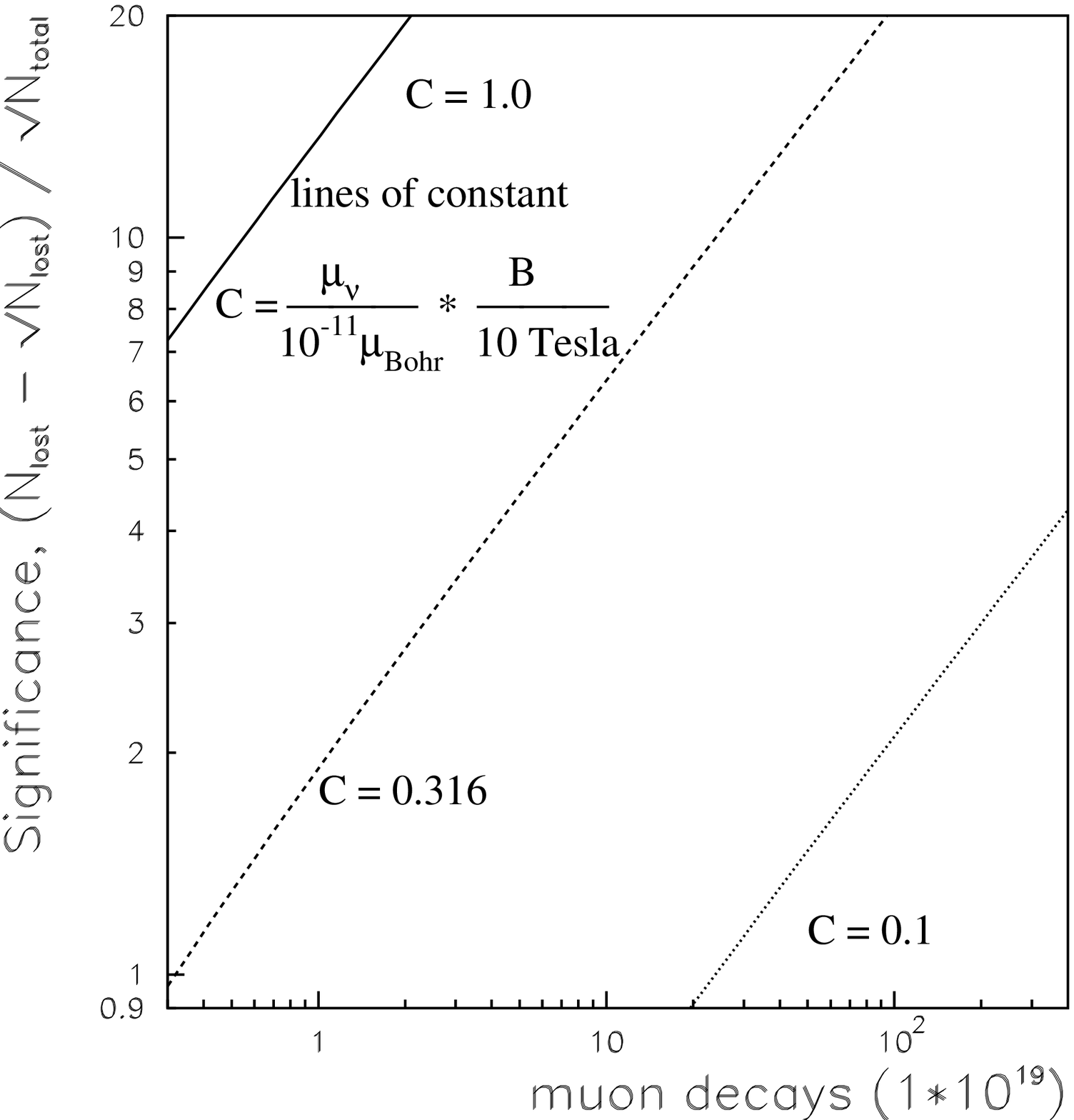}}
\caption{Significance for several scenarios.}
\label{mag_mom_signif}
\end{center}
\end{figure}

\noindent
where $t$ is the neutrino flight time from entering the magnetic field to 
detection.  We note that, in contrast to an oscillation disappearance
signal, this effect is explicitly independent of the neutrino energy. Fig.
\ref{mag_mom_signif} compares the number of events that vanish 
because of phase rotations to the expected statistical fluctuations in the
number of CC events ($ N$) observed in a 50 kton \cite{geer_2day} far
detector. We see that for a cummulative 10 T field gradient and $10^{19}$ muon decays
we expect a $> 10 \sigma$ significance for a NMM of
$10^{-11}\mu _{B}$. With a  3T gradient, the limit drops below
two $\sigma$. The sensitivity can be greatly increased  by increasing the
field strength  and more weakly by
increasing the number of events in the far detector. Because the detector
distance deterimines both $t$ and $N$ in Eqn. \ref {eq_nlost}, the
``significance'' in the figure is linearly dependent on distance.

To conclude, we have discussed a novel neutrino magnetic moment search
technique that uses oscillating magnetic fields at the source of a long
baseline detector's neutrino beam. This is the only technique we know of
that is quadratic in both $\mu _{\nu }$ and a controllable parameter, and 
thus has the potential for improved sensitivity as we improve our ability 
to create oscillating magnetic field gradients.

\subsection{Exotic processes}

Neutrino factories offer the possibility of searching for 
exotic processes resulting in production of $e^-$, $\mu^+$,
or $\tau$--lepton of either charge.
While these searches are interesting in their own right, they are also useful
in ruling out exotic contributions to long-baseline neutrino oscillation
signals.
The neutrino beam from a muon storage ring would consist of a virtually pure
combination of $\bar{\nu}_e$ and $\nu_\mu$ (or charge-conjugate).  
At very short baselines 
the $\bar{\nu}_e$ and $\nu_\mu$ will not have
had time to oscillate into other flavors:  For a 
20~GeV muon storage ring with a
700~m straight section, and neutrino oscillations with 
$\Delta m^2 \ = \ 3.5 \times 10^{-3} \ \rm eV^2$, the oscillation probability
is $\approx 5 \times 10^{-9}$. 

One could distinguish between exotic processes and the beginning of a 
neutrino oscillation
by exploiting their differing dependence on energy and distance.
Specifically,
these exotic processes would probably have a flat or rising dependence on
the neutrino energy $E_\nu$.  In contrast, a neutrino oscillation would
have a $1/E_\nu^2$ dependence.  Also, if the distance $L$ of the experiment
changes, the rate of exotic events would decrease with the flux as $1/L^2$.
In contrast, the neutrino oscillation probability would increase as $L^2$
(for $L$ small compared to  the oscillation period),
and so the rate of oscillated events would be independent of $L$.  

Current understanding of muon interactions allows for exotic
processes in two forms.
Anomalous lepton production could occur if muons decay to neutrino flavors
other than those in the usual decay $\mu \to e \bar{\nu}_e \nu_\mu$, and the
anomalous neutrinos then interact in the target.  Alternatively, they could
be produced if a $\bar{\nu}_e$ or $\nu_\mu$ interacts with the target 
via an exotic process.

The only direct experimental limit on exotic
$\mu \to e \bar{\nu}_x \nu_y$ decays is
$BR(\mu \to e \bar{\nu}_\mu \nu_e) < 1.3\%$\cite{PDG}.  Indirect limits are
also very weak.  The contribution of non- $V-A$ interactions to the muon
decay rate has been limited to 8\%\cite{PDG}.
Also, the total muon decay rate is one of the main measurements used to 
constrain electroweak parameters\cite{PDG}.  To first order,
\begin{equation}
\frac{1}{\tau_\mu} = \frac{G_F m_\mu^5}{192\pi^3} .
\end{equation}
Assuming the standard model, $G_F$ is determined to 1 part in $10^6$
from muon lifetime measurements.  If there are exotic contributions to
the muon lifetime, the measured value of $G_F$ would be shifted from
the true value.
Since
\begin{equation}
m_W \propto G_F^{-1/2} ,
\end{equation}
the 0.1\%
uncertainty on $m_W$ corresponds to a 0.4\%
shift in the muon lifetime.
Finally, the CKM matrix element $V_{ud}$ is determined from the rate of
nuclear $\beta$-decays relative to the muon lifetime.  The assumption
of unitarity on the CKM matrix gives us the following constraint on
the first row:
\begin{equation}
|V_{ud}|^2 + |V_{us}|^2 + |V_{ud}|^2 = 1 .
\end{equation}
The experimental determination is\cite{PDG}:
\begin{equation}
|V_{ud}|^2 + |V_{us}|^2 + |V_{ud}|^2 = 0.991 \pm 0.005 .
\end{equation}
The uncertainty on this constraint corresponds to a 0.5\%
shift in the muon lifetime.  Additional contributions to the muon
decay rate would lead to a downward shift in 
the determined value of $|V_{ud}|^2$ from the true
value.  We conclude that exotic decay modes of the muon with branching
ratios totaling 0.5\%
are possible without contradicting current measurements or tests of the
standard model.

\begin{figure}
\epsfysize=2.0in
\centerline{
\epsffile{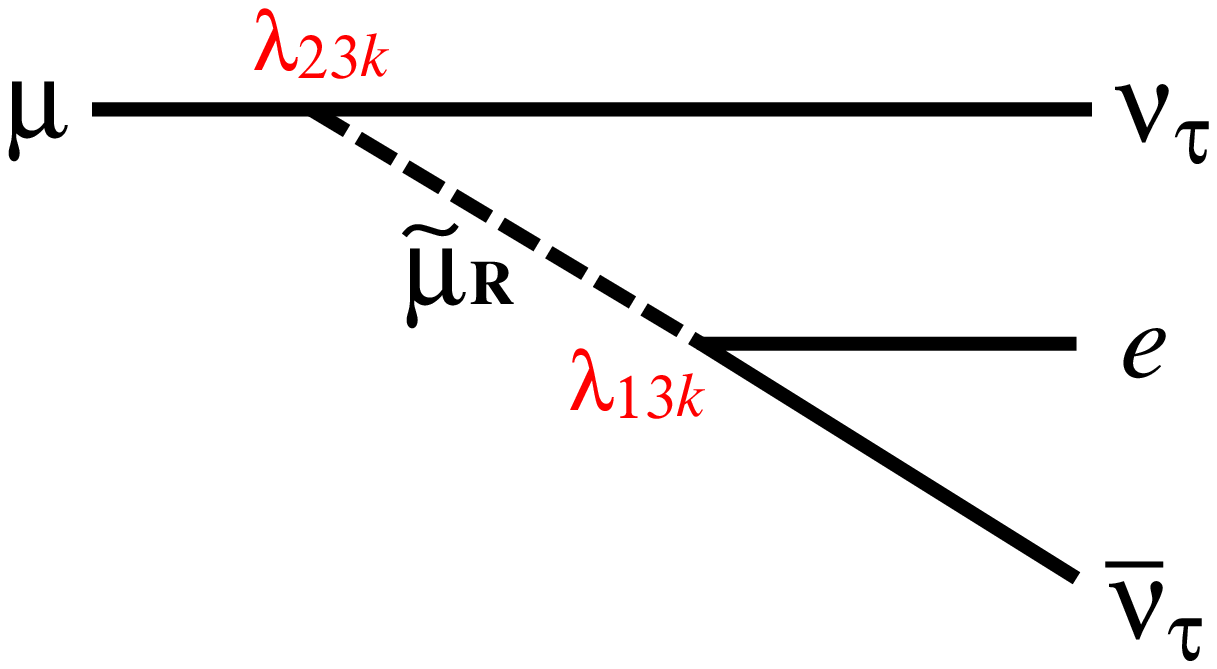}}
\caption{Example of exotic muon decay in R-parity-violating SUSY.}
\label{exotics:decay}
\end{figure}

As a concrete example of such an exotic process we consider 
R-parity-violating supersymmetric models.  These models lead to
lepton-number-violating vertices with couplings $\lambda$, and muon
decay processes such as $\mu \to e \bar{\nu_\tau} \nu_\tau$ as shown
in Fig.~1.  The matrix element for these decays turns out to have
the same form as for the standard W-exchange.  The current constraints
on the couplings $\lambda$ are reviewed in Ref.~\cite{dreiner}.
These constraints allow a branching ratio of 0.4\%
for the process in Fig.~\ref{exotics:decay}.

\begin{figure}
\epsfysize=2.0in
\centerline{
\epsffile{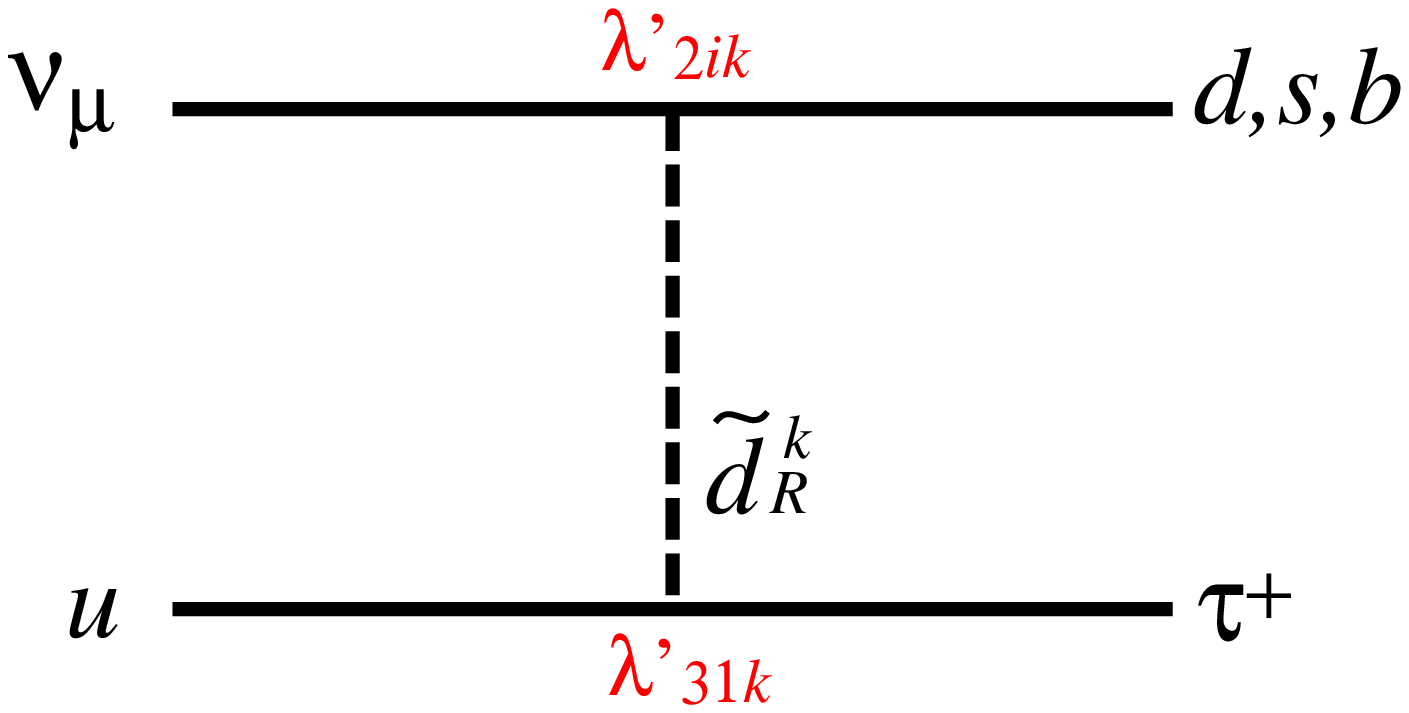}}
\caption{Example of an exotic neutrino interaction in R-parity-violating SUSY.}
\label{exotics:interaction}
\end{figure}

Similar processes are allowed for anomalous
lepton production as shown for example in 
Fig.~\ref{exotics:interaction}.  
Estimates for allowed rates are in progress~\cite{quigg}.
These diagrams
involve the $\lambda '$ couplings.  Currently, the best limit on one of
these couplings,
$\lambda'_{231}$, is from $\nu_\mu$ deep-inelastic scattering, so existing
neutrino data is already providing constraints!
The search for these types of effects at the muon storage ring could
be input into a decision on whether to build a muon-proton collider
where they could be studied in more detail.

\begin{figure}
\epsfysize=2.0in
\centerline{
\epsffile{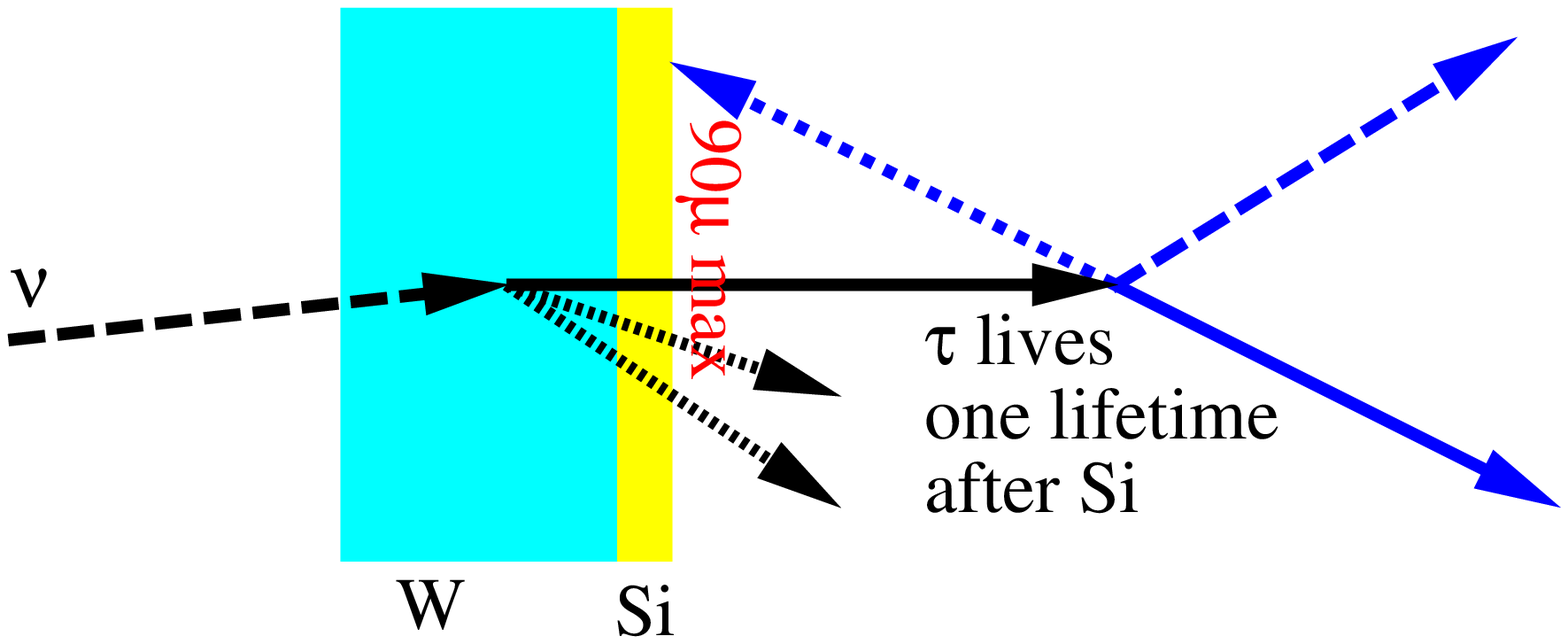}}
\caption{One plane of a detector for $\tau$ production.}
\label{exotics:detector}
\end{figure}

As a start on estimating the capabilities of an experiment at the
neutrino source, we consider the detector concept illustrated in 
Fig.~\ref{exotics:detector}.
This concept consists of a repeating sequence of 1.5 mm-thick Tungsten
sheets with Silicon tracking, separated by
4 mm.  Tungsten, being dense, provides a high
target mass while being thin enough for a 
produced $\tau$ to have a high probability
of hitting the Silicon.  The impact parameter of the $\tau$ decay products
is typically 90 microns with a broad distribution, so we would like a 
hit resolution of 5 microns or better.  Although there is a lot of
multiple scattering in the tungsten, the short extrapolation distance
provides for a good impact parameter resolution on the $\tau$ decay
products.  This configuration has been optimized for a 50 GeV muon beam.
For lower energy beams, the planes should be spaced more closely, and the
Tungsten thickness perhaps reduced.  
Studies of detectors with passive target mass and tracking with
emulsion sheets~\cite{emulsion} suggest that we can expect $\tau$ 
reconstruction efficiencies as high as 30\%.

We would propose placing such a detector in a magnetic field, and
measuring the momentum of muons and hadrons should be straightforward.
However, each Tungsten sheet is 0.4 radiation lengths thick, and while
we should obtain good energy resolution for electromagnetic showers,
it will not be feasible to measure the charge of an electron before it
showers.

In summary, with a
total mass of 6 tons of Tungsten, 200 $\rm m^2$ of Silicon
tracking, located close to a muon storage ring  withy $5\times 10^{20}$ muon decays at 50 GeV, we expect a total
of 35 billion neutrino interactions, 4 orders of magnitude
above present neutrino interaction samples. Thus, there
is much potential for detecting
very rare exotic processes if we can adequately reduce backgrounds.
Detailed simulations and
studies of possible Silicon tracking technologies are needed to quantify
this.  

\subsection{Summary}

We have investigated possible conventional neutrino physics studies done
at a detector located near a muon storage ring.  We emphasized novel methods
rather than extensions of existing experiments with additional statistics.

For a reference machine with 50 GeV stored muons and 10$^{20}$ muon decays
per year we find that it is possible to:

\begin{itemize}
\item Measure individual parton distributions within the proton for all
light quarks and anti-quarks.  
\item Determine the effects of a nuclear environment on individual quark species.
\item Measure the spin contributions of individual quark species, including
strange quarks and do precision studies of the QCD evolution of spin 
effects without a need for data from beta decay measurements.
\item Measure charm production with raw event rates of up to 100 million
charm events/year with $\simeq$ 2M double tagged events.
\item Measure the Weinberg angle in both hadronic and purely leptonic 
modes with a precision of 0.0001 to 0.0002.
\item Search for  the production and decay of neutral heavy leptons
with mixing angle sensitivity two orders of magnitude better than
present limits in the 30-80 MeV region.
\item Search for a neutrino magnetic moment which may be much larger
than the Standard Model prediction in
some supersymmetric theories.
\item Search for  anomalous tau production predicted
by some R-parity violating supersymmetric models.
\end{itemize}

We note that the event rates at a near detector increase
linearly  with neutrino energy. In addition, the acceptance
of small detectors is better for the narrower beam produced
by higher energy machines.  Almost
all of the above measurements, with the exception of the neutral heavy
lepton search, lose sensitivity if the beam energy is less than 50 GeV
and gain if it is greater.

If the storage ring  beam energy is lowered to  20 GeV, the statistical power of almost all of the
measurements considered here would drop a factor of 2.5 or more.  The number
of deep-inelastic scattering events with $\qsq$ high enough for perturbative
QCD to be meaningful  drops even further and the minimum $x$ rises to 0.05.
Measurements involving charm or tau production in the final
state would be have lower statistics due to threshold effects, 
as would the inverse muon
decay normalization for $\nu-e$ scattering, 
which has a threshold of $\sim$ 11 GeV.

However, it should be remembered that a 50 GeV  neutrino factory will produce 
neutrino fluxes of order 10$^4$ higher than existing neutrino beams.  At
20 GeV the improvement for most physics processes  is still greater than a 
factor of a thousand.


\section{Summary and Recommendations}

The main goal of the physics study presented in this report 
has been to answer the question: Is the physics program at a 
neutrino factory compelling ? The answer is a resounding yes, 
provided there are $10^{19}$ or more muon decays per year in the 
beam forming straight section and the
muon beam energy is $\sim20$~GeV or greater. 
Based on our study, we believe that a neutrino factory in the next 
decade would be the right tool at the right time.

\subsection*{The neutrino oscillation physics program}

The recent impressive atmospheric neutrino results from the 
Super-Kamiokande 
experiment have gone a long way towards establishing the existence 
of neutrino oscillations. 
This is arguably the most dramatic recent development in particle physics. 
Up to the present era, neutrino oscillation 
experiments at accelerators were searches for a phenomenon that might 
or might not be within experimental reach. The situation now is quite 
different. The atmospheric neutrino deficit defines the 
$\delta m^2$ and oscillation amplitude to which 
future long baseline oscillation 
experiments must be sensitive to, namely $\delta m^2 = $~O($10^{-3}$)~eV$^2$ 
and $\sin^2 2\theta =$~O(1). 
Experiments that achieve these 
sensitivities are guaranteed an excellent physics program that addresses 
fundamental physics questions. 
Furthermore, should $all$ of the experimental indications for oscillations 
(LSND, atmospheric, and solar) be confirmed, we may be seeing evidence for 
the existence of sterile neutrinos. This would be a very exciting 
discovery which would raise many new questions requiring new experimental 
input.

A neutrino factory would be a $unique$ $facility$ for oscillation 
physics, providing beams containing high energy electron 
neutrinos (antineutrinos) as well as muon antineutrinos (neutrinos). 
These beams could be exploited to 
provide answers to the questions that we are likely to be asking 
after the next generation of accelerator based experiments.

The oscillation physics that could be pursued at a neutrino factory 
is compelling. Experiments at a neutrino factory 
would be able to simultaneously measure, or put stringent limits on, 
all of the appearance modes $\nu_e \rightarrow \nu_\tau$, 
$\nu_e \rightarrow \nu_\mu$, and $\nu_\mu \rightarrow \nu_\tau$. 
Comparing the sum of the appearance modes with the disappearance 
measurements would provide a unique basic check of candidate 
oscillation scenarios that cannot be made with a conventional neutrino 
beam. 
In addition, for all of the specific oscillation 
scenarios we have studied, the 
$\nu_e$ component in the beam can be exploited to enable 
crucial issues to be addressed. These include: 
\begin{description}

\item{(i)} A precise determination of (or stringent limits on) all of the 
leading oscillation parameters, which in a three--flavor mixing 
scenario would be  $\sin^22\theta_{13}$, $\sin^22\theta_{23}$, 
and $\delta m^2_{32}$.

\item{(ii)} A determination of the pattern of neutrino masses. 

\item{(iii)} A quantitative test of the MSW effect. 

\item{(iv)} Stringent limits on, or the observation of, CP violation in 
the lepton sector.
\end{description}

To be more quantitative in assessing the beam energy, intensity, and 
baseline required to accomplish a given set of physics goals it is 
necessary to consider two very different experimental possibilities: 
(a) the LSND oscillation results are not confirmed, 
or (b) the LSND results are confirmed. 
\begin{description}
\item{(a) LSND not confirmed.}
A 20~GeV neutrino factory providing $10^{19}$ muon 
decays per year is a good candidate ``entry--level" facility which would 
enable either (i) the first observation of $\nu_e \rightarrow \nu_\mu$ 
oscillations, the first direct measurement of matter effects, 
and a determination of the sign of $\delta m^2_{32}$ and 
hence the pattern of neutrino masses, or (ii) a very stringent limit 
on $\sin^22\theta_{13}$ and a first comparison of the sum of all 
appearance modes with the disappearance measurements. 
The optimum baselines for this entry--level 
physics program appears to be of the order of 3000~km or greater, 
for which matter effects are substantial. 
Longer baselines also favor the precise determination of $\sin^22\theta_{13}$. 
A 20~GeV neutrino factory providing $10^{20}$ muon 
decays per year is a good candidate ``upgraded" neutrino factory (or 
alternatively a higher energy facility providing a few $\times 10^{19}$ decays 
per year). This would enable the first observation of, or meaningful 
limits on, $\nu_e \rightarrow \nu_\tau$ oscillations, and precision 
measurements of the leading oscillation parameters. In the more distant 
future, a candidate for a second (third ?) generation neutrino factory might 
be a facility that provides O($10^{21}$) decays per year and enables the 
measurement of, or stringent limits on, CP violation in the lepton sector.

\item{(b) LSND confirmed.}
Less extensive studies have been made for the class of scenarios that 
become of interest if the LSND oscillation results are confirmed. 
However, in the scenarios we have looked at we find that 
the $\nu_e \rightarrow \nu_\tau$ rate is sensitive to the 
oscillation parameters 
and can be substantial. With a large leading $\delta m^2$ scale medium 
baselines (for example a few $\times 10$~km) are of interest, and 
the neutrino factory intensity required to effectively exploit the 
$\nu_e$ beam component might be quite modest ($< 10^{19}$ decays per 
year). If sterile neutrinos play a role in neutrino oscillations, 
we will have an exciting window on physics beyond the SM, and 
we anticipate that a neutrino factory would enable crucial measurements 
to be made exploiting the electron neutrino beam component. 

\end{description}

\subsection*{The non-oscillation physics program}

A neutrino factory could provide beams that are a 
factor of $10^4-10^5$ more intense than conventional 
neutrino beams. This would have an enormous impact 
on the detector technology that could be used for 
non--oscillation neutrino experiments. For example, 
the use of silicon pixel targets and hydrogen or deuterium 
polarized targets would become feasible. 
Hence, a neutrino factory would offer experimental 
opportunities that do not exist with lower intensity 
conventional beams.

We have looked at a few explicit examples of interesting 
experiments that might be pursued at a neutrino factory:

\begin{itemize} \item
Precise measurements of the detailed structure 
of the nucleon for each parton flavor, including the 
changes that occur in a nuclear environment.
\item A first measurement of the nucleon spin structure with neutrinos.
\item Charm physics with several million tagged particles. 
Note that charm production becomes 
significant for storage ring energies above 20 GeV.
\item Precise measurements of standard model parameters -
$\alpha_s$, the weak mixing angle, and the CKM matrix elements.
\item Searches for exotic phenomena such as neutrino magnetic
moments, anomolous couplings to the tau and additional neutral
leptons.
\end{itemize}

The non-oscillation measurements benefit from higher beam energies
since event rates and the kinematic reach 
scale with energy, and perturbative calculations become 
more reliable in the kinematic regions accessed by higher energies.


\subsection*{Recommendations}
The physics program we have explored for a neutrino factory is compelling.
  We recommend a sustained effort to study both the physics 
opportunities and the machine realities.
\begin{description}
\item{(i)} We encourage support for the R\&D needed 
to learn whether a neutrino factory can be a real option in the next
decade. 
\item{(ii)} We propose further studies of 
 detector technologies optimized for a neutrino factory, including
both novel low mass detectors
for near experiments and very high mass detectors for long baselines.
For long baseline experiments  detectors should have
 masses of a few times 10~kt or more that are able to detect and 
measure wrong--sign muons, and detectors of a few kt or more able to 
observe tau--lepton appearance with high efficiency.  It is also 
desirable to identify electrons, and if possible measure the 
sign of their charge. 
Both the detector technologies
themselves and the civil engineering issues 
associated with the construction of such massive
detectors need to be addressed.

\item{(iii)} We recommend continued studies to better compare 
the physics potential of upgraded conventional neutrino beams with 
the corresponding potential at a neutrino factory, and also 
studies to better understand the benefits of muon polarization.

\item{(iv)} The present study concentrated on the muon storage ring as a
 neutrino source and did not cover the additional physics  programs
 which would use the proton driver and the high intensity muon beams. 
We recommend a further study directed at these other facets of physics at a muon 
storage ring facility.
\end{description}

\subsection*{Acknowledgments}

We would like to thank Mike Witherell, Mike Shaevitz, and Steve Holmes 
for initiating the two companion 6 month neutrino factory studies, 
and their continued support that enabled these studies to be productive. 
We thank Mike Shaevitz in particular for our charge. 
We would also like to thank the members of the Neutrino Source/Muon Collider 
Collaboration, whose enthusiastic efforts over the last few years 
have enabled us to seriously contemplate a facility that requires 
an intense source of muons. Finally we would like to thank the 
participants of neutrino factory and related physics studies initiated 
in Europe and Japan that have given us encouragement and shown interest 
in the present study.

\clearpage



\begin{thebibliography}{99}
\bibitem{geer_2day}  See Steve Geer's talk at the 2-day meeting. Number of
events (5000) was quoted for a 50 kTon detector at L=2800 km and a 20 GeV
muon storage ring.
\end{thebibliography}
\end{document}